\documentclass[12pt,epsf,amssymb]{article}

\usepackage{graphicx}
\usepackage{amsfonts}
\usepackage{amsmath}
\usepackage[usenames]{color}
\usepackage{amssymb}
\usepackage{amsthm}
\usepackage[mathscr]{eucal} 
\usepackage{url}

  \usepackage[all]{xy}

\def\Z{\mathbb{Z}}
\def\Q{\mathbb{Q}}
\def\R{\mathbb{R}}
\def\C{\mathbb{C}}
\def\P{\mathbb{P}}

\begin{document}

\baselineskip 0.6cm
\newcommand{\nequiv}{\mbox{\ooalign{\hfil/\hfil\crcr$\equiv$}}}
\newcommand{\nsupset}{\mbox{\ooalign{\hfil/\hfil\crcr$\supset$}}}
\newcommand{\nni}{\mbox{\ooalign{\hfil/\hfil\crcr$\ni$}}}
\newcommand{\nin}{\mbox{\ooalign{\hfil/\hfil\crcr$\in$}}}

\newcommand{\vev}[1]{ \left\langle {#1} \right\rangle }
\newcommand{\bra}[1]{ \langle {#1} | }
\newcommand{\ket}[1]{ | {#1} \rangle }
\newcommand{\Dsl}{\mbox{\ooalign{\hfil/\hfil\crcr$D$}}}
\newcommand{\Slash}[1]{{\ooalign{\hfil/\hfil\crcr$#1$}}}
\newcommand{\EV}{ {\rm eV} }
\newcommand{\KEV}{ {\rm keV} }
\newcommand{\MEV}{ {\rm MeV} }
\newcommand{\GEV}{ {\rm GeV} }
\newcommand{\TEV}{ {\rm TeV} }

\def\diag{\mathop{\rm diag}\nolimits}
\def\tr{\mathop{\rm tr}}

\def\Spin{\mathop{\rm Spin}}
\def\SO{\mathop{\rm SO}}
\def\O{\mathop{\rm O}}
\def\SU{\mathop{\rm SU}}
\def\U{\mathop{\rm U}}
\def\Sp{\mathop{\rm Sp}}
\def\SL{\mathop{\rm SL}}

\def\change#1#2{{\color{blue}#1}{\color{red} [#2]}\color{black}\hbox{}}

\theoremstyle{definition}
\newtheorem{thm}{Theorem}[subsection]
\newtheorem{defn}[thm]{Definition}
\newtheorem{exmpl}[thm]{Example}
\newtheorem{props}[thm]{Proposition}
\newtheorem{lemma}[thm]{Lemma}
\newtheorem{rmk}[thm]{Remark}
\newtheorem{notn}[thm]{Notation}
\newtheorem{qstn}[thm]{Question}
\newtheorem{excs}[thm]{Exercise}
\newtheorem{cor}[thm]{Corollary}
\newtheorem{anythng}[thm]{}

\begin{titlepage}
  
 
\vskip 1cm
\begin{center}
  
{\large \bf Supplementary Notes on VOA and 2D SCFT}
  
 \vskip 1.2cm

 Taizan Watari
 
 \vskip 0.4cm
  {\it
%
    Kavli Institute for the Physics and Mathematics of the Universe, 
    University of Tokyo, Kashiwa-no-ha 5-1-5, 277-8583, Japan
   }

\vskip 1.5cm
   
\abstract{Sections 2--3 of this note were prepared primarily for the author's private use, as a means of organizing the materials in (generally non-holomorphic) conformal field theory (CFT) on 2-dimensions in the language of the vertex operator algebra, and also of collecting subtle details.
  The choice of topics is intended to supplement introductory physics textbooks on the subject.
  
Section 4 deals with (generally non-holomorphic) Type 0 CFTs in the language of vertex operator superalgebra (SVOA).
Formulations of crossing symmetry and intertwining superoperators [resp. CFT vertex superoperators] of states in the Ramond (SVOA-twisted) sector [resp. the Ramond--Ramond sector] are presented by using two auxiliary parameters. The formulation allows systematic generalization to the class of simple current extensions characterized by four conditions: the {\it vanishing} third cohomology, {\it skew-symmetric} bimultiplicative form, {\it fixed-point-free} action of simple currents, and one additional condition described in the main text. } 
\end{center}
\end{titlepage}

 \tableofcontents

 \section{Introduction}
\label{sec:Intro}

Sections \ref{sec:review-VOA}--\ref{sec:CFT} of this note were prepared primarily for the author's private use,
as a means of organizing materials from the literature on vertex operator algebras
(VOAs) and conformal field theory (CFT) in 1+1 dimensions,
and collecting subtle details. It is not intended as a textbook or a set of lecture notes. The choice of topics is intended to supplement 
standard physics textbooks on this subject;
popular examples are omitted almost entirely,
and mainstream materials are kept to a minimum.
Instead, we elaborate on results that are often skipped or
only briefly mentioned in introductory physics texts. 

This note adopts the logical construction of the VOA community as its backbone.
The intention is not to restate everything in mathematical language with full rigor of the VOA community;
rather, we aim to highlight structures and subtleties that tend to be obscured when treated using physical intuition. This approach will also assist in translating physics literature into a language natural to VOA experts. 
Although VOA and 2D CFT are treated as nearly synonymous in some literature,
this note introduces them as distinct objects.
2D CFTs are not necessarily holomorphic, while VOAs always are;
intertwining operators of a VOA form a vector space,
while we need to specify a choice of CFT vertex operator as part of the data of a 2D CFT, and there are a couple of consistency conditions on the choice
(see sections \ref{ssec:review-VOA}, \ref{ssec:review-vct-sp-intrtw-op}, 
\ref{ssec:CFT-data}, \ref{ssec:SVOA} and \ref{ssec:fCFT}). 

Here are materials in sections \ref{ssec:review-VOA} and \ref{ssec:review-vct-sp-intrtw-op}--\ref{ssec:algrthm-int.op} that are not typically included in introductory texts for physicists. 
One is the transformation law (\ref{eq:FBZv-Lemma656-globalTrnsf}) under a change of {\it global} time coordinates.
Section \ref{ssec:review-vct-sp-intrtw-op} collects useful formulas (\ref{eq:xing-iso-by-FnOmega-a}--\ref{eq:xing-iso-by-FnOmega-c}) and (\ref{eq:ThetaTheta-after-fusion}) that involve the crossing-symmetry isomorphisms.
Section \ref{ssec:algrthm-int.op} reviews algorithms to compute intertwining operators (rather than chiral three-point functions) that are general enough to handle any unitary VOA, including those that are not necessarily rational.

Section \ref{ssec:g=1} is about what we refer to as ``genus-1 equations'' of a rational VOA in \cite{Moore:1988uz}, \cite{Moore:1988qv}. 
This section offers detailed calculations for both sides of the equations in one specific example (the Ising VOA). Additional explanations are also provided in section \ref{ssec:g=1} on subtle points where the original references were more concise.

Section \ref{ssec:ribbon-MTC} reviews the relation between VOAs and ribbon categories / modular tensor categories. There are two objectives for including this review in this note. One is to digest the mathematics behind this relation and present it in a way natural to physicists without compromising the material. The other is to explore the extent to which this relation holds for non-rational VOAs. 

To experts in the VOA community, a large fraction of expositions in sections \ref{sec:review-VOA}--\ref{sec:CFT} may appear as a reiteration of established results. 
A possible exception in section \ref{sec:review-VOA} is a characterization of the CPT anti-unitary map in section \ref{ssec:mode-op-cyl}, specifically in (\ref{eq:Hconj-VOA-inCylMdOp}); we have adopted a characterization that is faithful to physical intuition (and also to Ref. \cite[\S12]{MR996026}), and yet remains fully general.
Section \ref{sec:CFT} attempts to formulate the physicists' conception of (non-holomorphic) 2D CFTs in a language closer to that of the VOA community, which is by no means the first such attempt, however (cf \cite{Kapustin:2000aa}, \cite{MR1933066}, \cite{MR2130840}).

\vspace{5mm}

The goal of section \ref{sec:Z2} is to write down a formulation of 2D (generally non-holomorphic) Type 0 CFTs in the language of super vertex operator algebra (SVOA).
In the eyes of physicists, that is all about inserting the $\pm$ sign properly, or that at least appears so.
It is subtler than the present author had expected to do so in practice, and to describe the structure and concept in precise language. 
Formulations of intertwining operator [resp. CFT vertex operator] of states in the Ramond (SVOA-twisted) representation [resp. in the Ramond--Ramond representation],
and crossing symmetry isomorphisms are presented in section \ref{ssec:SVOA} [resp. \ref{ssec:fCFT}]. We do so in a way a sysematic generalization is possible to the class of simple current extensions characterized by four conditions: the {\it vanishing} third cohomology $[h]=0 \in H^3({\cal G}; \C^\times)$, a {\it skew-symmetric} bi-multiplicative form $c: {\cal G} \times {\cal G} \rightarrow S^1$ on a finite abelian group ${\cal G}$, {\it fixed-point free} action of ${\cal G}$ on a set of simple representations of the underlying VOA, and one more (see (\ref{eq:reltn-eps-c-h})).
The intertwining operators of an SVOA that involve Ramond representations are formulated  (see \ref{statmnt:int.op-S-wR}) by using formal power series along with\footnote{
Section \ref{sec:Z2} grew out of a practical needs of the present author elsewhere \cite{Nishikawa:2026gep}.
Although he made his best effort as a non-expert on SVOA to look for literature on such formulations, he could not find one that suits his needs.
He would appreciate it very much if he were informed of relevant works. 
} %
auxiliary parameters ($i_\beta, r$), and we study canonical identification between those with different auxiliary parameters (\ref{eq:temp-iso-vvct-sp-int.op-S-diff-StoZ2}, \ref{eq:temp-iso-vvct-sp-int.op-S-diff-brnch}).
This approach works in the application to Type 0 CFTs (\ref{statmnt:data-4-fCFT-2}, \ref{statmnt:out-going-states+can-iso}).
Section \ref{ssec:fCFT} does not cover Heterotic or Type II 2D CFTs,
or CFTs with fully general simple current extensions on both the left-moving and right-moving sectors. 
The presentation may lack the level of rigor required by VOA experts;
nonetheless, we hope that section \ref{sec:Z2} provides a stimulus for
further progress. 

\newpage


\section{Vertex Operator Algebra}
\label{sec:review-VOA}

\subsection{Algebraic data, modules and coordinate transformation}
\label{ssec:review-VOA}

\begin{anythng}
\label{statmnt:def-VOA}
{\bf algebraic data of a VOA and a unitary VOA:}   
The defining data of a {\it vertex operator algebra} (VOA)
consists of $(V, Y, {\bf 1}, \omega)$
%
%
satisfying the conditions (a)--($\ell$) below. $V$ is a vector space over $\C$,
and $Y : V \rightarrow {\rm End}(V)[[z,z^{-1}]]$ a homomorphism
({\it state-operator correspondence}), where
$z$ is a dummy variable to write down a formal power series;
both ${\bf 1}$ and $\omega$ are in $V$. First, (a)
the vector space $V$ has a $\Z$-grading,
\begin{align}
  V \cong \oplus_{h \in \Z_{\geq 0}} V_h, \qquad \dim_\C(V_h) < \infty.
\end{align}
(b) $V_0 \cong \C {\bf 1}$ and $\omega \in V_h|_{h=2}$, and (c)
$Y({\bf 1})$ is the ($z$-independent) identity operator ${\rm id}_V \in
 {\rm End}(V)$. For any $v \in V_h$ for some $h \in \Z_{\geq 0}$
 ({\it homogeneous state}), we call this\footnote{
The VOA community often adopts a convention that $Y(v) = \sum_{n \in \Z}
[v]_n z^{-1-n}$ for $v \in V$, regardless of the conformal weight $h=h(v)$
of $v \in V_h$. We stick to the physics-community convention in this note,
however. 
 } %
\begin{align}
  V_h \ni v \longmapsto Y(v) =: Y(v,z) =: \sum_{n \in \Z} [v]_n z^{-h-n}
  \label{eq:mode-expns-radial}
\end{align}
the mode expansion of $Y(v)$, and $[v]_n \in {\rm End}(V)$ the mode
operators of $v \in V_h$. (d) The mode operators of $\omega \in V_2$,
denoted by $[\omega]_n =: L_n$ for $n\in \Z$, satisfy the relation
\begin{align}
  [L_m,L_n] = (m-n) L_{m+n} + \frac{c}{12}m(m^2-1) {\rm id}_V
\end{align}
for some $c \in \R_{\geq 0}$. The homomorphism $Y$ satisfies (e)
\begin{align}
  Y(L_{-1}\cdot v) = \partial_z Y(v,z), 
\end{align}
(f) $[v]_n$ of a homogeneous state $v \in V_h$ maps $V_m$ to $V_{m-n}$, and
(g)
\begin{align}
  [v]_n \cdot {\bf 1} = 0 \quad {\rm if~}h+n > 0, \qquad
  [v]_n|_{n=-h} \cdot {\bf 1} = v \qquad
  \left( {\rm i.e.,~} \lim_{z\rightarrow 0}Y(v,z) {\bf 1} = v\right). 
\end{align}
(h) The operator $L_0 \cdot v = h v$ for any $v \in V_h$; the eigenvalue
$h$ is called the {\it conformal weight}. 

The property called {\it skew-symmetry} (i) is that
\begin{align}
  Y(u) \cdot v = e^{z L_{-1}} \cdot Y(v,z)|_{z \rightarrow -z} \cdot u
\end{align}
in $V[[z]][z^{-1}]$. There are also properties called (j) {\it commutativity}
({\it locality} in some references), (k) {\it associativity}, and
($\ell$) the Jacobi identity. We do not write down those properties of
a data set
of a VOA precisely here, partly because they are found in many references
on VOA (e.g., \cite[\SS8.5--8.10]{MR996026}, 
\cite[\S3]{MR1142494}, 
\cite[Prop.2.1--Prop.2.9]{MR1363971} 
\cite[\S1.3, \S3]{MR2023933}, 
\cite[\S1.2, \S3.2]{MR2082709}
), and partly because those properties are reviewed later
(\ref{statmnt:def-VOA-mod} and \ref{statmnt:def-VOA-inttw-op}) in a more
general form than for VOAs. Not all of the properties (i--$\ell$) are
independent from one another; it is best to see \cite[\S2]{MR1363971},
\cite[\S3.6]{MR1142494} and \cite[Prop.1.2--Rmk.1.8]{MR1817614}.

A VOA $(V, Y, {\bf 1}, \omega)$ is said to be {\it simple}, if it does not
contain a non-trivial ideal.

\vspace{5mm}

A data set of a {\it unitary} VOA is $(V, Y, {\bf 1}, \omega)$ with
additional data $(\Upsilon, \langle - | - \rangle)$, where 
\begin{align}
  \langle - | - \rangle : V \times V \ni (u,v) \longmapsto
  \langle u | v \rangle \in \C 
\end{align}
is a Hermitian positive definite inner product, $\C$-linear in $v$ and
$\C$-anti-linear (= conjugate linear) in $u$, and the other piece of data  
$\Upsilon: V \rightarrow V$ is a $\C$-anti-linear isomorphism.
Here are a few conditions they should satisfy. 
First, $\Upsilon$ is an involution,
(m) $\Upsilon^2 = {\rm id}_V$. They are normalized
(n) so that $\langle {\bf 1} | {\bf 1} \rangle = 1$,
(o) $\Upsilon({\bf 1}) = {\bf 1}$ and $\Upsilon(\omega) = \omega$.  
The Hermitian inner product and $\Upsilon$ are related (p) by 
\begin{align}
  \langle Y(\Upsilon(a)) \cdot u | v \rangle =
  \langle u | Y\left( e^{x L_1} \cdot
  (-x^{-2})^{L_0} \cdot a, \; z \right)_{z\rightarrow 1/z, x\rightarrow z} \cdot v \rangle
  \label{eq:cond-o-4VOA-onUps}
\end{align}
for any $u,v,a \in V$; here, the dummy variable $z$ in $Y(\Upsilon(a))$
in (\ref{eq:cond-o-4VOA-onUps}) is treated as it is
(without replacing it by $\bar{z}$ or $z^{-1}$ (see Rmk. \ref{rmk:treat-z})),
while the anti-linearity applies to all the rest of
$Y(\Upsilon(a))\cdot u$ in (\ref{eq:cond-o-4VOA-onUps}). 
Finally, it is required to be (q) compatible with the operator product expansion
(OPE):
\begin{align}
  Y(u_1)_{@z} Y(u_2)_{@0} = \sum_I \frac{Y(v_I)_{@0}}{z^{h_1+h_2-h_I}}
  \Rightarrow
  Y(\Upsilon(u_1))_{@z} \; Y(\Upsilon(u_2))_{@0} = \sum_I
  \frac{Y(\Upsilon(v_I))_{@0}}{z^{h_1+h_2-h_I}}
\end{align}
for any $u_1 \in V_{h_1}$, $u_2\in V_{h_2}$ and $v_I \in V_{h_I}$;
equivalently,
\begin{align}
 \Upsilon\left( [u_1]_n \cdot u_2 \right) = [\Upsilon(u_1)]_n \cdot \Upsilon(u_2)
 \qquad {}^\forall n \in \Z 
  \label{eq:Hconj-VOA-wOPE}
\end{align}
for any $u_2 \in V$ and a homogeneous $u_1 \in V$; 
see \cite{MR3119224}, \cite{MR3796433} for more.
%
%
A little more explanation on the notations such as $Y(u)_{@x}$
is provided in \ref{statmnt:prscr-oprtr-trnstnFcn-coordCh}.  

A unitary VOA with $c=0$ is always $\omega = 0 \in V_2$. It thus
follows that $V_n = \{0\}$ for all $n>0$, and $V = \C {\bf 1}$.  
\end{anythng}
\begin{rmk}
\label{rmk:treat-z}
An alert reader (or physics audience) will find potential danger in the
treatment of the dummy variable $z$ in the anti-linear argument of the
Hermitian inner product stated just below (\ref{eq:cond-o-4VOA-onUps}).
The dummy variable $z$ is given interpretations as a holomorphic coordinate
of a Riemann surface (cf \ref{statmnt:prscr-oprtr-trnstnFcn-coordCh}).  
Think of a coordinate change $z' = e^{i\alpha} z$ by a phase
($e^{i\alpha}\in U(1) \subset {\rm SU}(2)$), for example.  
Whether we do the transformation inside $\bra{-}$ or outside
leads us to different expressions. Also in algebraic language, we  have,
on one hand, 
\begin{align*}
   \bra{u} & \; Y(e^{zL_1}(-z^{-2})^{L_0} a, 1/z) \cdot v \rangle =
  \bra{e^{i\alpha L_0} u} e^{i\alpha L_0} Y(e^{zL_1}(-z^{-2})^{L_0} a, 1/z) \cdot v \rangle \\
  & \; = \bra{e^{i\alpha L_0} u} Y\left(e^{e^{-i\alpha} z L_1}
  (-(e^{-i\alpha} z)^{-2})^{L_0} (e^{-i\alpha L_0} a), e^{i\alpha}/z \right)
  \cdot e^{i\alpha L_0} v\rangle  \\
  & \; = \langle \left. Y\left( \Upsilon(e^{-i\alpha L_0} a), e^{-i\alpha} z \right) \cdot e^{i\alpha L_0} u \right| e^{i\alpha L_0} v \rangle 
\end{align*}
using (\ref{eq:cond-o-4VOA-onUps}) at the last equality with
$ze^{-i\alpha}$ as the variable $z$ in (\ref{eq:cond-o-4VOA-onUps}); on the
other hand,
\begin{align*}
  \langle Y(\Upsilon(a),z) & \; \cdot u | v \rangle =
  \langle e^{i\alpha L_0} Y(\Upsilon(a),z) \cdot u | e^{i\alpha L_0} v \rangle \\
  & \; = \langle Y(e^{i\alpha L_0}  \Upsilon(a),e^{i\alpha } z) \cdot
  e^{i\alpha L_0} u | e^{i\alpha L_0} v \rangle
  = \langle Y(  \Upsilon( e^{- i\alpha L_0} a),e^{i\alpha } z) \cdot
  e^{i\alpha L_0} u | e^{i\alpha L_0} v \rangle. 
\end{align*}
There is a discrepancy, depending on whether we apply
(\ref{eq:cond-o-4VOA-onUps}) with the variable $z$ or $z e^{-i\alpha}$. 
Physicists avoid this problem by employing
the treatment $\bra{z \cdots}=z^{-1} \bra{\cdots}$, and $\Upsilon( z \cdots)
= z^{-1} \Upsilon(\cdots)$, so $\bra{ \Upsilon(z \cdots)\cdots} = z \bra{\Upsilon(\cdots) \cdots}$; with the physicists' treatment, the condition
(\ref{eq:cond-o-4VOA-onUps}) would be written as
\begin{align}
  \bra{Y((\zeta/i)^{L_0} \Upsilon(a),\zeta) \cdot u} v \rangle
  & \; = \bra{u} Y(e^{\zeta^{-1} L_1} (\zeta/i)^{L_0} \cdot a, \zeta) \cdot v \rangle  \label{eq:cond-o-4VOA-onUps-phys}  \\
   \bra{Y(\Upsilon(a),\zeta) \cdot u} v \rangle
& \; = \bra{u} Y(e^{\zeta^{-1} L_1} (\zeta/i)^{2L_0} \cdot a, \zeta) \cdot v \rangle, \nonumber 
\end{align}
where the dummy variable $\zeta$ corresponds to $z^{-1}$. 

This treatment in the physics literature is based on quantum field theory.
There is a good reason (known as {\it Wick rotation})
to think that the coordinate $w$
related to $z$ through $z=e^{-i w}$ is \underline{real} than complex when we deal with
the vector space of states and its Hermitian inner product, so
$\bra{e^{-iw} \cdots}
= e^{iw}\bra{\cdots}$ and $\Upsilon(e^{-iw}\cdots) = e^{iw}\Upsilon(\cdots)$.
The present author believes that physicists' treatment is better, but this
article follows the conventional treatment in the VOA community to avoid
confusion. We encounter a potential danger in this article only in the
evaluation of (\ref{eq:temp-unitary-in-cylCoord-LHS}), but we only need to
deal with $\bra{ \Upsilon(z \cdots) \cdots}$ there luckily.
\end{rmk}

\begin{anythng}
({\bf a few consequences} (on unitarity)) 
  It also follows from the condition $\Upsilon(\omega) = \omega$ that
$\Upsilon(v) \in V_h$ if $v \in V_h$.
The $\C$-anti-linear involution $\Upsilon$ is called the
{\it CPT isomorphism}. 

The operator $L_n$ is the Hermitian conjugate of $L_{-n}$,
which follows as an immediate consequence of $\Upsilon(\omega) = \omega$
and the condition (p) with $a=\omega$.
If $a \in V_{h_a}$ is {\it quasi-primary} (i.e., $L_1 \cdot a=0$), then
\cite[(85)]{MR3796433} 
(here, $(-x^{-2})^{L_0}$ in (\ref{eq:cond-o-4VOA-onUps}) is
treated as $(e^{-\pi i}/x^2)^{L_0}$ as in (\ref{eq:cond-o-4VOA-onUps-phys}))
\begin{align}
  \langle [\Upsilon(a)]_n \cdot u | v \rangle = (e^{-\pi i})^{h_a} \langle u |
   [a]_{-n} \cdot v \rangle, \qquad u,v \in V;
  \label{eq:Hconj-VOA-psPrmry}          
\end{align}
see also (\ref{eq:Hconj-VOA-inCylMdOp}) in section \ref{ssec:mode-op-cyl}
later.  
When both $u_0,v_0 \in V_h$ are Virasoro-primary, then
\[
\langle \Upsilon(u_0) | v_0 \rangle = \langle \Upsilon(v_0) | u_0 \rangle
= (-1)^h C_{u0,v0}
\]
where $C_{u0,v0}$ is the leading OPE coefficient
$Y(u_0)_{@z} Y(v_0)_{@0} \sim C_{u0,v0} z^{-2h} {\rm id}_V + {\cal O}(z^{1-2h})$.
\end{anythng}

\begin{anythng}  
\label{statmnt:unitarity-refined-VOA}
({\bf more perspectives on unitarity structure}) (cf \cite[\S1.2]{MR4418715})
Given a VOA (subject to the conditions (a--$\ell$)), one can define
a bilinear form  
\begin{align}
  B_V: V\times V \ni (u,v) \mapsto
  B_V(u,v) := {\rm Res}_z \left(z^{-1}  \langle {\bf 1}^\vee, \;
   Y\left( e^{z L_1} (-z^{-2})^{L_0} \cdot u, 1/z\right) \cdot v \rangle \right) \in \C;
\end{align}
Here, $\langle - , \; - \rangle$ is the natural pairing between
$V^\vee := \oplus_{h\in \Z_{\geq 0}} {\rm Hom}_\C(V_h,\C)$ and $V$;
the linear functional ${\bf 1}^\vee \in {\rm Hom}(V_{h=0}, \C) \subset V^\vee$ 
is the dual basis of ${\bf 1} \in V_{h=0} = \C {\bf 1}$
(cf the condition (b)), so  
$(s)_{L_0=0} =: \langle {\bf 1}^\vee, \; s\rangle
\; {\bf 1}$  for $s \in V$. 
It is possible to see that this bilinear form is symmetric
(($B_V(v,u) = \epsilon B_V(u,v)$ with $\epsilon = +1$) by
using the skew-symmetry condition (i) and
computatations reviewed in \ref{statmnt:prscr-oprtr-trnstnFcn-coordCh}. 
One $\C$-linear isomorphism $\lambda_0: V^\vee \rightarrow V$ is determined
uniquely by the condition that $\langle \lambda_0^{-1}(u), v\rangle = B_V(v,u)$
for $u, v \in V$. The relation $B_V(v,u) = \epsilon B_V(u,v)$ implies that
$\lambda_0^T: V^\vee \rightarrow V$ is equal to $\epsilon \lambda_0 = \lambda_0$. 

In the rest of \ref{statmnt:unitarity-refined-VOA} we use this $\lambda_0$
to reformulate the unitarity structure reviewed in \ref{statmnt:def-VOA}.
This reformulated version is better suited to introducing unitary
structure on modules of a VOA, as we review in \ref{statmnt:cntrG-mod}. 
So, think of adding a Hermitian inner product $\langle - | - \rangle$ on $V$
and an anti-linear $\Upsilon_{00}:V \rightarrow V$ satsifying
the condition (p) but not necessarily (m,n,o); let us also treat
$B_V(u,v) = \epsilon B_V(v,u)$ and $\lambda_0^T = \epsilon \lambda_0$
with $\epsilon \in \{ \pm 1\}$ although we know that $\epsilon = +1$
because of the skew-symmetry condition (i) of the VOA; we make three
independent arguments in the following by using this weaker set of
properties; this serves as a warming up exercise for application to 
\underline{modules} of a VOA later than the VOA itself.  First,
\begin{align}
  \langle \lambda_0^{-1}(b), \; a \rangle = B_V(a,b) = \epsilon B_V(b,a) =
  \frac{\epsilon \bra{\Upsilon(b)} a \rangle}{\langle {\bf 1} | {\bf 1} \rangle}
  = \frac{\epsilon \langle \Upsilon_0 \cdot \Upsilon (b), \; a \rangle }
  {\langle {\bf 1} | {\bf 1} \rangle} ; 
\end{align}
here, $\Upsilon_0: V \rightarrow V^\vee$ is the $\C$-antilinear map given 
by $\Upsilon_0: V \ni u \mapsto \bra{u} -\rangle \in V^\vee$;
the condition (p) is used at the equality in the middle. This means that
$\epsilon (\lambda_0 \cdot \Upsilon_0) \cdot \Upsilon = {\rm id}_V \langle
{\bf 1} \ket{\bf 1}$. Secondly,
$\langle \Upsilon(a) | \Upsilon(b) \rangle = B_V(a,\Upsilon(b)) \; \bra{\bf 1} {\bf 1} \rangle = \epsilon B_V(\Upsilon(b),a) \; \bra{\bf 1} {\bf 1} \rangle
= \epsilon \langle \Upsilon^2(b) | a \rangle$.
Finally, the fact that $\lambda_0({\bf 1}^\vee) = {\bf 1}$ by construction
implies that $\lambda_0$ can be unitary at most in the form of
$\langle u | \lambda_0(m) \rangle_{V} /\bra{\bf 1} {\bf 1}\rangle_V =
\langle \lambda_0^{-1}(u) | m \rangle_{V^\vee} \bra{\bf 1} {\bf 1} \rangle_V$
for $u\in V$ and $m\in V^\vee$ 
when the Hermitian inner product is induced on $V^\vee$ from
$\langle -|-\rangle_V$ on $V$; this condition on the unitarity modulo
appropriate rescaling is equivalent to $\lambda_0^T \cdot \Upsilon_0 = \Upsilon_{0^\vee} \cdot \lambda_0^{-1} \; \langle {\bf 1} \ket{\bf 1}^2$, because 
\begin{align}
  \langle \lambda_0^T\Upsilon_0(u),m\rangle = \bra{u} \lambda_0(m)\rangle =?
  \bra{\lambda_0^{-1} u} m \rangle \; \langle {\bf 1} \ket{\bf 1}^2
  = \langle \Upsilon_{0^\vee} \lambda_0^{-1}(u), \; m \rangle \;
  \langle {\bf 1} \ket{\bf 1}^2. 
\end{align}
The property $\lambda_0^T \Upsilon_0 =? \Upsilon_{0^\vee} \lambda_0^{-1} \;
\langle {\bf 1} \ket{\bf 1}^2$ is equivalent to
$(\lambda_0^T \Upsilon_0) \cdot (\lambda_0 \Upsilon_0) =?
\langle {\bf 1} \ket{\bf 1}^2 {\rm id}_V$. 

Back in the context of VOA, the conditions (m)
$\Upsilon^2 = {\rm id}_V$ and (n) $\langle {\bf 1} \ket{\bf 1}=1$ imply,
through the three observations above, that $\Upsilon = \lambda_0 \cdot
\Upsilon_0:V \rightarrow V^\vee \rightarrow V$, $\Upsilon$ is {\it anti-unitary}
$\bra{\Upsilon(u)}\Upsilon(v) \rangle = \bra{v}u\rangle$, 
and $\lambda_0$ is unitary; the property $\Upsilon({\bf 1}) = {\bf 1}$
follows from that of $\lambda_0$ and $\Upsilon_0$. Conversely,
if we impose that $\lambda_0$ of a VOA $(V,Y,{\bf 1}, \omega)$ is unitary
wrpt $\langle -|- \rangle_V$, then it follows, through the three arguments
above, that $\langle {\bf 1} \ket{\bf 1}=1$, $\Upsilon = \lambda_0 \cdot
\Upsilon_0$, $\Upsilon({\bf 1}) = {\bf 1}$, $\Upsilon^2 = \epsilon {\rm id}_V$,
and $\Upsilon$ is anti-unitary, without referring to the sign
$\epsilon \in \{ \pm 1\}$. 
\end{anythng}

\begin{anythng}
  {\bf a VOA of CFT type and a VOA:}
What we introduced as {\it vertex operator algebra} 
in \ref{statmnt:def-VOA} is in fact known as \underline{
  vertex operator algebra of CFT-type} in VOA literature, but we will
not use the term in this article, because we will use the term CFT
in the way we explain in section \ref{sec:CFT} (cf also \S \ref{sec:Intro}). 
There are a few categories of objects where the structure (data set) of
objects has less information or is subject to weaker conditions than
that of vertex operator algebra. 
A {\it vertex operator algebra not necessarily of CFT-type} (known as
\underline{vertex operator algebra} in VOA literature) has the same
defining algebraic data $(V, Y, {\bf 1}, \omega)$ as a VOA does,
but the condition (a) in \ref{statmnt:def-VOA} sis relaxed to the
$\Z$ grading that is lower bounded,
\begin{align}
  V \cong \oplus_{h \in \Z} V_h, \qquad \dim_\C (V_h) < \infty, \quad
  V_h = \{ 0\} \quad {\rm for~} h < {}^\exists h_0,  
\end{align}
and the condition (b) to ${\bf 1} \in V_0$ (instead of $V_0 \cong \C {\bf 1}$)
and $\omega \in V_2$. In a {\it logarithmic VOA}, the condition (h) is relaxed
further;
instead of demanding that $(L_0-h)=0$ on $V_h$, $(L_0-h)$ is required to be
nilpotent on $V_h$.
In this note, however, we will be concerned about VOAs with all of the
conditions (a--$\ell$) imposed, or unitary ones with all of (a--q) imposed. 
\end{anythng}

\begin{anythng}
\label{statmnt:prscr-oprtr-trnstnFcn-coordCh}
{\bf the determinant line bundle associated with change of local coordinates:}
The state-operator correspondence $Y$ is formulated in the defining
algebraic data of a unitary VOA by using a dummy variable $z$ of a
formal power series, but more meanings --- coordinates of a Riemann
surface $\Sigma$---can be assigned to the variable $z$, because the matrix
element
$\bra{s} Y(v,z) \cdot t \rangle$ for a any given choice $s,t,v \in V$ is
in $\C[z,z^{-1}]$ than in $\C[[z,z^{-1}]]$ (the property (a)).
Note that spacetime coordinates are used in a description of a
quantum field theory for three different purposes.
\begin{itemize}
\item (1)  
The operator formalism of a quantum field theory uses
a coordinate system covering a large fraction of spacetime, with which
we assign the vector space of quantum states to an equal-time slice
of spacetime; the time-evolution operator describes how to convert
states between the vector spaces of two time slices.  
\item (2)
Given a quantum field theory, there is an abstract list of {\it fields}
$\{ {\cal O}_i \}_{i\in I}$ a priori, but we use a local coordinate
system and local metric configuration of spacetime when we implement
them as a local operator in the operator formalism at a given point in
a spacetime (e.g., a vector field). 
This is similar to the use of local trivialization in specifying
a section of a vector bundle.
\item (3)
A coordinate system is also used to specify one point---the coordinate
value of the point---in a geometry/spacetime.
\end{itemize}
We can read all the three interpretations
in the dummy variable $z$; an inhomogeneous coordinate $z$ of
$\Sigma \simeq \P^1$ is used for the three different purposes
in a description of a quantum field theory on $\P^1$. 
Here, we review the formulations and notations to deal with those
interpretations (cf \cite[\S2.4]{Polchinski:1998rq},
\cite{MR1448404}, \cite[\S6.3, \S8]{MR2082709}).
Because the three purposes are all different, we adopt a
formulation/notation so that the three roles of the coordinates are separated
from one another.  

Physics intuitions on 2D conformal field theory indicate that, when $p$
is a point in a Riemann surface $\Sigma$, and $(U_\xi, \xi)$ a local
holomorphic coordinate neighborhood of $p$ so $\xi(p)=0$, then
quantum states on the $|\xi|=1$ circle of a given conformal field theory
in the radial quantization constitutes a vector space; this vector space
of states is unique modulo isomorphism (for a given conformal field theory)
for any local coordinate system $(p, U, \xi)$ so long as $\xi(p)=0$ and the
circle $|\xi|=1$ fits within $U$. By applying these intuitions to VOA,
we assign one vector space denoted by $V^{(\xi)}$ to the circle $|\xi|=1$
for any local coordinate system $(p, U_\xi, \xi)$ of a Riemann surface
$\Sigma$, and choose $V^{(\xi)}\cong V$ by using $V$ of the defining
algebraic data of a VOA. 

Let $\xi$ and $\eta$ be a pair of local holomorphic coordinates on $\Sigma$
such that their $|\xi|\leq 1$ disc and $|\eta|\leq 1$ disc are both contained
in a common open set $U \subset \Sigma$. Let $\xi = f(\eta)$ be the
relation between them. We set up the identification between
$V^{(\xi)}$ and $V^{(\eta)}$ by 
\begin{align}
  R_f: V\cong V^{(\xi)} \ni v^{(\xi)} \longmapsto R_f \cdot v^{(\xi)} = v^{(\eta)}
  \in V^{(\eta)} \cong V,
  \label{eq:FBZv-Lemma656-globalTrnsf}  
\end{align}
where
\begin{align}
  R_f := \exp \left[ - \sum_{i=1}^\infty f_i L_i \right] \; f_0^{-L_0} \;
   e^{- f_{-1} L_{-1} }, 
    \label{eq:def-R_f}
\end{align}
when the complex coefficients $f_{-1}$, $f_0$, $f_{1,2,\cdots}$ are determined
by 
\begin{align}
  f(t) = \exp\left[ \sum_{i=1}^\infty f_i t^{i+1}\partial_t \right] \; 
  f_0^{t\partial_t} \; e^{f_{-1} \partial_t} \cdot t
  = f_{-1} + f_0 \left( t + f_1 t^2 + \cdots \right). 
  \label{eq:def-R_f-prepr}
\end{align}
Note that ${\bf 1} \in V \cong V^{(\xi)}$ is mapped by $R_f$
to ${\bf 1} \in V \cong V^{(\eta)}$. 

The identification $R_f : V^{(\xi)}\rightarrow V^{(\eta)}$ described above
is a mathematical declaration (after a few checks on consistency (eg,
$R_{f \circ g} = R_g \cdot R_f$) have been run
(e.g.,\cite[\S6.3]{MR2082709})). As a physics statement, the identification
it is a non-trivial statement that should be verified in conformal field
theories that have weakly coupled path-integral formulation;
here is a physics reasoning: one chooses a circle $\gamma \subset U$ that
encloses both the $|\xi|\leq 1$ disc and the $|\eta|\leq 1$ disc, works out
the evolution of quantum states on the $|\xi|=1$ circle (in $V^{(\xi)}$)
to the states on $\gamma$, and also the evolution of states from
the $|\eta|=1$ circle to those on $\gamma$, and compares to determine
the identification between $V^{(\xi)}$ and $V^{(\eta)}$. The question
is whether this identification agrees with $R_f$. 
In the simplest case $\xi = f_0 \eta$ for some $f_0 \in \C^\times$, 
the identification $R_f: V \rightarrow V$ above agrees with the Euclidean
time evolution from the $|\xi|=1$ circle to the $|\eta|=1$ circle. 
The present author is not knowledgeable about how much this non-trivial physics
statement has been checked. 

Once a coordinate neighborhood $(U,\xi)$ in $\Sigma$ is chosen, then
physics intuitions on 2D conformal field thoery determine how 
the local operator for a given state placed at a given point in $p \in U$
on the vector space of states. This intuition, applied to VOA, is implemented
as the algebraic defining data $Y: V \ni u \mapsto Y(v,\xi(p)) \in
{\rm End}(V)$ of a VOA. For a finite coordinate value
$\xi(p) \in \C$ of the point $p\in \C$, the matrix elements of
$Y(u,\xi(p)) \in {\rm End}(V)$ are well-defined.
Here, we should interpret ${\rm End}(V)$
conceptually as ${\rm End}(V^{(\xi)})$, while $V$ containing $u$
is an abstract list of states/operators (more to be elaborated below)
to be inserted at the point $p$. 

The action of this operator for $u\in V$ at $p\in U$ on the vector
space $V^{(\xi)}$ is given immediately by the $Y : V \rightarrow
{\rm End}(V)[[z,z^{-1}]]$ in the defining algebraic data of a VOA,
but its action on $V^{(\eta)}$ for another local coordinate $\eta$
on $U$ looks different; we need to use the
identification $R_f: V^{(\xi)} \rightarrow V^{(\eta)}$ set up earlier. 
It is known (e.g., \cite[6.5.6 Lemma]{MR2082709}) 
that its action on $V^{(\eta)}$ is as follows:
\begin{align}
  R_f \cdot Y(u,\xi(p)) \cdot R_f^{-1} = Y(R_{f@ \eta(p)} \cdot u,\eta(p))
  \in {\rm End}(V^{(\eta)}),
  \label{eq:FBZv-Lemma656}
\end{align}
where
\begin{align}
 (f@\eta(p))(t) := f(t+\eta(p)) - f(\eta(p)) = f(t+\eta(p)) - \xi(p).
\end{align}
The function $f@\eta(p)$ maps 0 to 0, expressing the local coordinate 
$\xi - \xi(p)$ in terms of $t := \eta - \eta(p)$, both centered at $p$.

Whereas the conversion $R_f: V^{(\xi)}\rightarrow V^{(\eta)}$ reflects the
fact that a spacetime coordinate is used as in (1), the conversion
from $u \in V$ to $R_{f@\eta(p)} \cdot u \in V$ reflects the
fact that a spacetime coordinate is used also as in (2): 
when applied to a VOA, the abstract list of fields $\{{\cal O}_i \}_{i\in I}$
uses $u\in V$ as the label of operators $i\in I$. To split the three
different roles a spacetime coordinate plays, we formulate (while introducing
notations) the {\it implementation
of the local operator for $u \in V$ at a point $p$ in a local coordinate
patch $(U,\zeta)$ represented on the space of states $V^{(\eta)}$ associated
with a coordinate $\eta$ on $U$} as follows: 
\begin{align}
  Y^{(\zeta;\eta)}: V \times U \ni (u,p) \longmapsto Y^{(\zeta;\eta)}(u)_{@p}
  := Y( R_{g @ \zeta(p)}^{-1} \cdot u, \eta(p)) \in {\rm End}(V) \cong
  {\rm End}(V^{(\eta)}).
      \label{eq:state-transf-4-coord-transf} 
\end{align}
Here, $\eta = g(\zeta)$. When we change the coordinate for the
purpose (1) from $\eta$ to $\xi$ (with $\xi = f(\eta)$), while
the same coordinate $\zeta$ is used for the purpose (2), then
the representation changes to the following in ${\rm End}(V^{(\xi)})$, 
\begin{align}
  R_f^{-1} \cdot Y^{(\zeta;\eta)}(u)_{@p} \cdot R_f & \; 
  = R_f^{-1} \cdot Y(R_{g@\zeta(p)}^{-1} \cdot u, \eta(p)) \cdot R_f
  = Y\left( R_{f@\eta(p)}^{-1} \cdot R_{g@\zeta(p)}^{-1} \cdot u, \xi(p) \right) \\
  = & \; Y \left( R_{(f\circ g)@\zeta(p)}^{-1} \cdot u, \xi(p) \right)
  = Y^{(\zeta;\xi)}(u)_{@p},
  \label{eq:vrtxOp-globalTrnsf}
\end{align}
as desired. When we keep the same coordinate $\eta$ for the purpose (1),
but use a new local coordinate $\omega$ for the purpose (2), with
$\zeta = h(\omega)$, then an abstract local operator labeled $u \in V$
is implemented in ${\rm End}(V^{(\eta)})$ as 
\begin{align*}
  Y^{(\omega;\eta)}(u)_{@p} = Y(R_{(g\circ h)@\omega(p)}^{-1} \cdot u, \eta(p))
  = Y(R_{g@\zeta(p)}^{-1} \cdot R_{h@\omega(p)}^{-1} \cdot u, \; \eta(p) ). 
\end{align*}
The same operator implementation is reproduced,
$Y^{(\zeta;\eta)}(u^{(\zeta)})_{@p} = Y^{(\omega;\eta)}(u^{(\omega)})_{@p}$,
when  \cite[\S2.4]{Polchinski:1998rq}
\begin{align}
  V \ni u^{(\zeta)} \longmapsto R_{h@\omega(p)} \cdot u^{(\zeta)}
  = u^{(\omega)} \in V.
  \label{eq:Pol-localTrnsf}
\end{align}
The subscript ${}_{@p}$ of $Y^{(\zeta;\eta)}(u)_{@p}$ referring to the point
$p \in U$ may sometimes be replaced by ${}_{\zeta@\zeta(p)}$ or
${}_{\eta@\eta(p)}$ by using the coordinate values of $p$. 
\end{anythng}

\begin{anythng}
\label{statmnt:coord-at-infty}
The coordinate transformation $\xi = 1/\zeta =: {\rm inv}(\zeta)$ is frequently
used, as we can use two inhomogeneous coordinates in this relation to
cover the whole $\P^1$. For any $p \in \C^\times \subset \P^1$, where
$\xi(p) \neq 0$, $\zeta(p) \neq 0$, 
\begin{align*}
  ({\rm inv}@\zeta(p))(t) = \frac{- t}{\zeta(p) (t+\zeta(p)) }, \qquad
  ({\rm inv}@\zeta(p))_0 = \frac{-1}{\zeta(p)^2} , \quad
  ({\rm inv}@\zeta(p))_1 = \frac{-1}{\zeta(p)}, 
\end{align*}
and $({\rm inv}@\zeta(p))_{i>1} = 0$.  So,
\begin{align}
  R_{{\rm inv}@\zeta(p)} = e^{\frac{1}{\zeta(p)} L_1} (- \zeta(p)^2)^{L_0}, \qquad
  R_{{\rm inv}@\zeta(p)}^{-1} = e^{\zeta(p)L_1} (-1/\zeta(p)^2)^{L_0}. 
\end{align}

The condition (\ref{eq:cond-o-4VOA-onUps}) characterizes the final state
$\bra{\Upsilon(a)} \in {\rm Hom}_\C(V,\C)$ for $a\in V$, when the
condition is used with $u = {\bf 1}$, and in the limit $z \rightarrow 0$.
It does so by inserting an operator 
$Y(R^{-1}_{{\rm inv}@z(p)} \cdot a)_{\xi@1/z(p)} = Y^{(1/\xi;\xi)}(a)_{\xi@1/z(p)}$;
the limit brings the point $p\in \P^1$ to $\xi(p) = \infty$ in the
time coordinate $\xi$ used in the operator formalism.  

The coordinate transformation ${\rm inv}(\zeta) = 1/\zeta$ is also very
special in that it is one of generators of the linear fractional transformation
$[(a,b;c,d)] \in PGL_2\C$, $f(\zeta) = (a\zeta + b)/(c\zeta + d)$, all
of which provide 1-to-1 map from the Riemann surface $\Sigma = \P^1$ to $\P^1$.
For linear fractinal transformation, $R_f$ involves $L_1$, $L_0$ and $L_{-1}$,
as we see in the case $f={\rm inv}$. It follows, for  
any linear fractional transformation $\xi = f(\eta)$, that 
$(R_f)^T: ({\bf 1}^\vee)^{(\eta)} \mapsto ({\bf 1}^\vee)^{(\xi)}$, because
$R_f$ maps $\oplus_{h\geq 1}V_h \subsetneq V^{(\xi)}$ into
$\oplus_{h\geq 1} V_h \subsetneq V^{(\eta)}$ without a spill-over into
$V_{h=0} \cong \C {\bf 1}^{(\eta)} \subset V^{(\eta)}$. In particular,
the expression using $\xi$ as the global coordinate, 
\begin{align}
  \left\langle ({\bf 1}^\vee)^{(\xi)}, \; Y^{(\zeta_1;\xi)}(v_1)_{@p_1} \cdots
  Y^{(\zeta_n;\xi)}(x_n)_{@p_n} \cdot {\bf 1}^{(\xi)} \right\rangle, 
\end{align}
is the same as that of the same expression with $\xi$ replaced by $\eta$,
because we can use (\ref{eq:vrtxOp-globalTrnsf}). For a unitary VOA, 
it is the same as the matrix element of this form
({\it vacuum expectation value}) 
\begin{align}
  \bra{ {\bf 1}^{(\xi)} } \; Y^{(\zeta_1;\xi)}(v_1)_{@p_1} \cdots
  Y^{(\zeta_n;\xi)}(x_n)_{@p_n} \cdot {\bf 1}^{(\xi)} \rangle ; 
\end{align}
instead of the whole $PGL_2\C$ linear fractional transformations
$\xi = f(\eta)$, the transformation $R_f$ is unitary only for the subgroup
$\SU(2)$ of $PGL_2\C$, which is the chiral part of the 2D conformal symmetry
group (the change by a complex phase in Rmk. \ref{rmk:treat-z} is an example). 
Because of this $PGL_2\C$ (or SU(2)) invariance, one may omit explicit
reference to one choice of global coordinate $\xi$ for the overall reference
frame (the purpose (1)) in the notation
(so long as we stay within a family of choices of $\xi$ within the SU(2)
subclass of linear fractional transformations), and the matrix element above
is written as
\begin{align}
   \bra{ {\bf 1} } \; Y^{(\zeta_1)}(v_1)_{@p_1} \cdots
  Y^{(\zeta_n)}(x_n)_{@p_n} \cdot {\bf 1} \rangle . 
\end{align}
We will continue in \ref{statmnt:CFT-noBdry-obsrvbls} reviewing
a slightly broader class of objects to be computed. 
\end{anythng}

\begin{anythng}
\label{statmnt:def-VOA-mod}
{\bf modules, unitarizable modules and unitary modules:} 
A {\it module} of a VOA consists of a set of data $(M, Y_M)$, where $M$ is
a vector space over $\C$ with an $\R$-grading,
\begin{align}
  M \cong \oplus_{h \in \R_{\geq 0}} M_h, \qquad \dim_\C M_h < \infty, 
\end{align}
with each summand being an eigenspace $L_0 M_h = h M_0$ (a-mod, h-mod), and
$Y_M: V \rightarrow {\rm End}(M)[[z,z^{-1}]]$ is a homomorphism that satisfies
the conditions (c-mod) $Y_M({\bf 1}) = {\rm id}_M$, (d, e)
of the VOA with $Y$ just replaced by $Y_M$ ((d,e-mod)), and
(f-mod) $[a]_n$ of a homogeneous state $a \in V_{h_a}$ maps
$M_h$ to $M_{h-n}$. 

The Jacobi identity ($\ell$-mod) to be imposed on $Y_M$ is
\begin{align}
  z_0^{-1} \delta \left( \frac{z_1-z_2}{z_0} \right) & \;
  Y_M(u,z_1) Y_M(v,z_2)
  - z_0^{-1} \delta\left( \frac{z_2-z_1}{-z_0} \right) Y_M(v,z_2) Y_M(u,z_1)
  \nonumber \\
  & - z_2^{-1} \delta \left( \frac{z_1-z_0}{z_2} \right)
  Y_M(Y(u,z_0) \cdot v,z_2) = 0
  \label{eq:Jacobi-module} 
\end{align}
in ${\rm End}(M)[[z_1^{\pm 1}, z_2^{\pm 1}, z_0^{\pm 1}]]$, where
\begin{align}
 \delta(x) := \sum_{n\in \Z} x^n  \qquad \qquad {\rm a~formal~power~series}
\end{align}
in VOA literatures; negative powers of $(z_1-z_2)$, $(z_2-z_1)$ and
$(z_1-z_0)$ in the 1st, 2nd and 3rd terms, respectively, should be
understood as their binomial expansion series with respect to\footnote{
The third term, for example, is a power series in $z_0$ of only finite order
pole, because $[u]_N \cdot v=0$ for large enough $N$, and the binomial
series expansion of $(z_1-z_0)^n$ does not yield a negative power of $z_0$.
The property (j-mod) is obtained by extracting the $z_0^{-(h_u+N)}$ term.
} %
(wrpt)
$z_2/z_1$, $z_1/z_2$ and $z_0/z_1$. The commutativity (j-mod) and associativity
(k-mod) of $Y_M$ follow from the Jacobi identity ($\ell$-mod).
See  \cite[Def.4.1.1]{MR1142494}, 
\cite[Def. 2.7]{MR1338873} 
\cite[Def.1.9]{MR1817614}, 
for example, for more information. 
On the other hand, no conditions analogous to (b, g, i) are imposed on the
data of a unitary module.

As a little digression, useful properties of the power series $\delta(z)$
are noted down here. First,
\begin{align}
  \delta \left( \frac{z_1-z_2}{z_0} \right) (z_0+z_2)^n =
  \delta \left( \frac{z_1-z_2}{z_0} \right) z_1^n, \qquad {}^\forall n \in \Z, 
  \label{eq:rltn-delta-1}
\end{align}
where $(z_0+z_2)^n$ should be treated as its binomial series expansion
wrpt $z_2$; we should not abuse this relation when $n \nin \Z$. Second,
\begin{align}
  \frac{1}{z_0} \delta \left( \frac{z_1-z_2}{z_0} \right) =
  \frac{1}{z_1} \delta \left( \frac{z_0+z_2}{z_1} \right),
  \label{eq:rltn-delta-2}
\end{align}
which can be verified by using properties of the binomial coefficients. 
More generally, 
\begin{align}
  \frac{1}{z_0} \delta \left( \frac{z_1-z_2}{z_0} \right)
  \left( \frac{z_1-z_2}{z_0} \right)^\nu
  = \frac{1}{z_1}\delta \left( \frac{z_0+z_2}{z_1} \right)
  \left(\frac{z_0+z_2}{z_1} \right)^{-\nu} 
\end{align}
for any $\nu \in \R/\Z$ for the same reason (eg., \cite[(7.1)]{MR1233387}).
Finally, although this is not used in this note
(e.g., \cite[(2.1.14)]{MR1142494}), 
\begin{align}
  \frac{1}{z_0} \delta \left( \frac{z_1-z_2}{z_0} \right)
  - \frac{1}{z_0} \delta \left( \frac{z_2-z_1}{-z_0}\right) =
  \frac{1}{z_2} \delta \left( \frac{z_1-z_0}{z_2} \right). 
\end{align}

For a unitary VOA $(V, Y, {\bf 1}, \omega, \Upsilon, \langle -|-\rangle )$,
a set of data $(M, Y_M)$ of a module is accompanied by additional information,
a positive definite Hermitian inner product $\langle -|-\rangle_M$,
to be a {\it unitary module}, when the following condition is
satisfied. Here, in its notation $\langle - | - \rangle_M:
M \times M \ni (m_1,m_2) \longmapsto \langle m_1 | m_2 \rangle \in \C$, 
it is $\C$-linear in $m_2$ and $\C$-anti-linear in $m_1$, as we follow
the convention in physics in this article.
The condition (p) is imposed after replacing $u,v \in V$ by $m_1,m_2\in M$,
while $a \in V$ (the condition (p-mod)). We say that a module $(M, Y_M)$
is {\it unitarizable} when such $\langle - | - \rangle_M$ exists.

When a VOA-module $(M, Y_M)$ is unitarizable, possible choice of
$\langle -|-\rangle_M$ comes at least with a family parametrized by $\R_{>0}$;
we may rescale $\langle -|-\rangle_M$ by any positive real number.
Apart from special cases (such as the defining representation with $(M, Y_M)
=(V, Y)$ where there is a special element such as ${\bf 1}$, $\omega$ in $M$)
none of such choices of $\langle -|-\rangle_M$ in the family is a better
choice than the others. In such non-special cases, a choice of one
Hermitian inner product $\langle -|-\rangle_M$ is not so much as a choice
of additional piece of information on top of $(M, Y_M)$, but as introduction
of a reference frame with which we have more grasp of states in the
representation space $M$ (we can talk of states $m\in M$ with norm
$\langle m | m \rangle =1$, for example). The conditions (m), (o)
and (q) for VOA have their analogues for modules of a VOA, but we will
need a generalization of the CPT isomorphism $\Upsilon$ for modules. 
So, we will come back to this issue in \ref{statmnt:cntrG-mod}. 

A module $(M,Y_M)$ of a VOA $(V, Y, {\bf 1}, \omega)$ is said to be 
{\it simple} if there is no non-trivial submodule $(M_0, Y_{M_0})$ of
$(M, Y_M)$. If a unitary module is a simple module, it does not have
a non-trivial unitary submodule. 

A pair of modules modules $(M, Y_M)$ and $(M'', Y_{M''})$ of a common
VOA $(V,Y,\cdots)$ are regarded {\it equivalent} when there is a $\C$-linear
isomorphism $M \rightarrow M''$ so that the pull-back of $Y_{M''}$ to $M$
is the same as $Y_M$. For a pair of unitary modules
$(M, Y_M, \langle -|-\rangle_M)$, $(M'', Y_{M''}, \langle - | -\rangle_{M''})$
of a common unitary VOA, we further require that the pull-back of
$\langle -|-\rangle_{M''}$ also agrees with $\langle -|-\rangle_M$. 
\end{anythng}

\begin{anythng}
{\bf modules with weaker restrictions:}
A {\it module} $(M, Y_M)$ of a VOA $(V, Y, {\bf 1}, \omega)$ is not required
to admit a Hermitian inner product $\langle - | - \rangle$. The condition
(a-mod, h-mod) for unitary modules of unitary VOAs is relaxed to
\begin{align}
  M \cong \oplus_{h \in \C} M_h, \qquad \dim_\C M_h < \infty, 
\end{align}
with $L_0 M_h = h M_h$, and $M_{h+n} = 0$ for $n < {}^\exists n_0$ for any
given $h \in \C$. Without the condition that $L_0$ is Hermitian,
the conformal weights do not have to be real and positive. 

Any given module [resp. unitary module] is isomorphic to the direct sum
of its submodules each of which is associated with $[h] = h+\Z \in \C/\Z$
[resp. $[h]+\Z \in \R/\Z$]. 

Such objects as an {\it admissible module} of a graded vertex algebra,
or a {\it weak module} of a vertex algebra, $(M, Y_M)$, is with the conditions
on a grading (a-mod, h-mod) relaxed further from the one above for modules.
In this note, however, we will be concerned about unitary modules. 
\end{anythng}

\begin{anythng}
\label{statmnt:cntrG-mod}
{\bf the contragradient module:} Let $(M, Y_M)$ be a module of a VOA
$(V, Y, {\bf 1}, \omega)$. Then another module of the same VOA is constructed
on the dual space 
\begin{align}
  M^\vee := \oplus_{h\in \R_{\geq 0}} M^\vee_h, \qquad M^\vee_h :=
  {\rm Hom}_\C (M_h, \C);  
\end{align}
the natural pairing $M^\vee \times M \rightarrow \C$ is denoted
for $m' \in M^\vee$ and $m \in M$ by $\langle m', m \rangle \in \C$. 
The action of $V$, $Y_{M^\vee}: V \rightarrow {\rm End}(M^\vee)[[z,z^{-1}]]$
is given by 
\begin{align}
  \langle Y_{M^\vee}(v,z) \cdot m', \; m \rangle := \langle m', \;
  Y_M\left( e^{zL_1}(e^{-\pi i} z^{-2})^{L_0} v, 1/z\right) \cdot m \rangle
\end{align}
for ${}^\forall m' \in M'$, ${}^\forall m\in M$ and ${}^\forall v \in V$.
Ref. \cite[\S5.2--5.3]{MR1142494}  
explains why $Y_{M'}$ given above satisfies the Jacobi identity
(\ref{eq:Jacobi-module}). The representation $(M^\vee, Y_{M^\vee})$ 
and any representations equivalent to this are called the
{\it contragradient representation} of $(M, Y_M)$. 

Let $(M_{\Upsilon^{\cal I}}, Y_{\Upsilon^{\cal I}})$ be a module of a VOA
$(V,Y, \cdots)$ that is contragradient to $(M, Y_M)$. There is then
an isomorphism $\lambda_M: M^\vee \rightarrow M_{\Upsilon^{\cal I}}$ of
the $(V,Y)$-modules. The contragradient representation of the $(V,Y)$-module
$(M_{\Upsilon^{\cal I}},Y_{\Upsilon^{\cal I}})$ then contains the original representation
$(M, Y_M)$ in the equivalence class; so there must be an isomorphism of
$(V,Y)$-modules
$\lambda_{M_{\Upsilon^{\cal I}}}: (M_{\Upsilon^{\cal I}})^\vee \rightarrow M$.
It is possible to use $\lambda_{M_{\Upsilon^{\cal I}}} = \lambda_M^T$.
When $(M, Y_M)$ is a simple module and is self-contragradient,
both $\lambda_M$ and $\lambda_M^T$ can be used as an isomorphism
of $(V,Y)$ modules $\lambda: M^\vee \rightarrow M$; $(M,Y_M)$ being
simple means that $\lambda_M^T = \epsilon_M \lambda_M$ for some
non-zero complex number $\epsilon_M$, and $\epsilon_M \in \{ \pm 1\}$
because $((\lambda_M)^T)^T = \lambda_M$ (see \cite{MR1142494}).  

In introducing unitarity structure on a module of a VOA (rather than the VOA itself), we need to pay attention to two extra aspects.
First, an analogue of the CPT isomorphism $\Upsilon: V \rightarrow V$ needs to be formulated for a pair of representations that are contragradient to each other.
Second, we cannot impose the conditions such as (n) $\langle {\bf 1} \ket{\bf 1}=1$ or (o) $\Upsilon({\bf 1}) = {\bf 1}$ because there is no such special states as ${\bf 1}$ in a module $(M, Y_M)$.
One will also find (as reviewed below) that it is not always possible to demand the CPT isomorphism $\Upsilon$ (formulated for modules) to be involutive (it is possible (as in (m)) for VOA).
The refined arguments in \ref{statmnt:unitarity-refined-VOA} \cite[\S1.2]{MR4418715} indicate that the unitary structure, including the CPT isomorphisms for general modules, can be formulated by splitting it into one canonical anti-unitary isomorphism and the other, and by demanding that the other one is unitary, instead of imposing the conditions (m). 

Let us begin by reviewing the canonical anti-unitary isomorphisms.
Let $(M, Y_M, \langle - | - \rangle_M)$ be a unitary module. 
The contragradient representation on the vector space $M^\vee$ is also unitary then; we may introduce a Hermitian inner product on $M^\vee_h \subset M^\vee$ as the orthonormal one for the basis of $M^\vee_h$ dual to an orthonormal basis of $M_h$ wrpt $\langle -|-\rangle_M$.
There is also a $\C$-anti-linear isomorphism $\Upsilon_M: M \rightarrow M^\vee$ given by 
\begin{align}
  \Upsilon_M: M \ni m \longmapsto \langle m | - \rangle_M \in \;\;
  \oplus_{h\in \R_{\geq 0}} {\rm Hom}_\C(M_h, \C) = M^\vee ;
\end{align}
it follows that $\Upsilon_{M^\vee} \cdot \Upsilon_M: M \rightarrow M^\vee
\rightarrow M$ is ${\rm id}_M$, and $\Upsilon_M: M\rightarrow M^\vee$ is anti-unitary, 
$\langle m_1 | m_2 \rangle_M = \langle \Upsilon_{M^\vee}(m_2) | \Upsilon_{M^\vee}(m_1) \rangle_{M^\vee}$. This $\Upsilon_M: M \rightarrow M^\vee$ is the straightforward generalization of $\Upsilon_0: V \rightarrow V^\vee$ in \ref{statmnt:unitarity-refined-VOA}.
We call it the {\it canonical anti-unitary map} for a unitary module $(M,Y_M, \langle - | - \rangle_M)$. 

Let $(M_\gamma, Y_{M_\gamma})$ and $(M_{\Upsilon^{\cal I}(\gamma)}, Y_{M_{\Upsilon^{\cal I}(\gamma)}})$ be a pair of simple unitarizable modules that are contragradient to each other.
We introduce {\it unitary structure} on this pair by choosing
Hermitian inner products $\langle -| - \rangle_{M_\gamma}$ and 
$ \langle -|-\rangle_{M_{\Upsilon^{\cal I}(\gamma)}}$ subject to the condition
(p-mod), and a pair of isomorphisms
$\lambda_\gamma: M_\gamma^\vee \rightarrow M_{\Upsilon^{\cal I}(\gamma)}$ and
$\lambda_{\Upsilon^{\cal I}(\gamma)}: M_{\Upsilon^{\cal I}(\gamma)}^\vee \rightarrow
M_\gamma$ of the VOA modules between equivalent ones, and impose one condition
(m-mod): $\lambda_\gamma$ and $\lambda_{\Upsilon^{\cal I}(\gamma)}$ must be
unitary. Once this condition is imposed, then the {\it CPT isomorphisms}
of VOA modules are set as 
\begin{align}
  \Upsilon_{\Upsilon^{\cal I}(\gamma)\gamma} := \lambda_\gamma \cdot \Upsilon_\gamma:
  M_\gamma \rightarrow M_\gamma^\vee \rightarrow M_{\Upsilon^{\cal I}(\gamma)} 
\end{align}
by combining the canonical anti-unitary $\Upsilon_\gamma$ and unitary
$\lambda_\gamma$. The condition (m-mod) leaves the freedom to change $\lambda_\gamma$ by a complex phase $e^{i\delta} \in S_1$, and also to rescale
$\langle -|-\rangle_\gamma$ by $\Lambda \in \R_{>0}$. 
In the case of the unitary structure of a VOA itself, the conditions
(n) and (o) fixed those two freedoms. The condition (m-mod) reduces the
freedom of $\lambda_\gamma$ (from $\C^\times$ to $S^1$) and also that
of the CPT isomorphism
$\Upsilon_{\Upsilon^{\cal I}(\gamma)\gamma}=\lambda_\gamma \cdot \Upsilon_\gamma$.

For such $\lambda_\gamma$ and $\lambda_{\Upsilon^{\cal I}(\gamma)}$ of
a unitarity structure, there must be a non-zero complex constant $\epsilon_\gamma$
so that $\lambda_{\Upsilon^{\cal I}(\gamma)}^T: M_\gamma^\vee \rightarrow
M_{\Upsilon^{\cal I}(\gamma)}$ is $\epsilon_\gamma \lambda_\gamma$, as we assumed
that both modules are simple. Moreover, it must be a complex phase.
If $(M_\gamma, Y_\gamma)$ and
$(M_{\Upsilon^{\cal I}(\gamma)}, Y_{\Upsilon^{\cal I}(\gamma)})$ are not mutually
equivalent representations, it just follows that 
\begin{align*}
\epsilon_{\Upsilon^{\cal I}(\gamma)}  \;\; \epsilon_\gamma = 1.
\end{align*}
The isomorphism $\lambda_{\Upsilon^{\cal I}(\gamma)}$ being unitary implies
that $\lambda_{\Upsilon^{\cal I}(\gamma)} \cdot \Upsilon_{\Upsilon^{\cal I}(\gamma)}
= \epsilon_\gamma \Upsilon_{M_\gamma^\vee} \cdot \lambda_{\gamma}^{-1}$, so
we arrive at the relation
\begin{align}
  \Upsilon_{\gamma,\Upsilon^{\cal I}(\gamma)} \cdot \Upsilon_{\Upsilon^{\cal I}(\gamma),\gamma} = \epsilon_\gamma {\rm id}_{M_\gamma}. 
\end{align}
By rescaling $\langle - | - \rangle_{\Upsilon^{\cal I}(\gamma)}$ and
changing the phase of $\lambda_{\Upsilon^{\cal I}(\gamma)}$ while keeping
those of $M_\gamma$, it is possible to set
$\epsilon_\gamma = 1 = \epsilon_{\Upsilon^{\cal I}(\gamma)}=1$.
When a simple module $(M_\gamma, Y_\gamma)$ is self-contragradient,
the argument above is repeated by using
$(M_{\Upsilon^{\cal I}(\gamma)},Y_{\Upsilon^{\cal I}(\gamma)})=(M_\gamma, Y_\gamma)$.
In particular, the non-zero complex number $\epsilon_\gamma$ is for
the relation $\lambda_\gamma^T = \epsilon_\gamma \lambda_\gamma$, with
\begin{align*}
  \epsilon_\gamma \in \{ \pm 1\}, \qquad
  \Upsilon_{\gamma,\gamma} \cdot \Upsilon_{\gamma,\gamma} =
  \epsilon_\gamma {\rm id}_{M_\gamma}.
\end{align*}
This generalizes the relation (m) for the defining representation $(V, Y)$.


As a generalization of the condition (q) for a simple unitary module
$(M_\gamma, Y_\gamma,\langle - | - \rangle)$ along with $\lambda_\gamma$,  
we impose (q-mod)
\begin{align}
  \Upsilon_{\Upsilon^{\cal I}(\gamma),\gamma}\left( [a]_n \cdot m \right)
  = \left[ \Upsilon(a) \right]_n \cdot \Upsilon_{\Upsilon^{\cal I}(\gamma),\gamma}(m),
  \qquad a \in V, \qquad m \in M_\gamma, 
 \label{eq:Hconj-VOAmod-wOPE}
\end{align}
where $\Upsilon = \Upsilon_{0,0}$ is that of the unitary VOA, and
$\Upsilon_{\Upsilon^{\cal I}(\gamma),\gamma}$ the anti-unitary map
$M_\gamma \rightarrow M_{\Upsilon^{\cal I}(\gamma)}$. 
\end{anythng}

\begin{anythng}
  {\bf notations:} In this article, we mean by ${\cal I}((V,Y))$
(or ${\cal I}(V)$ or some other similar notations) 
a set of distinct equivalence classes of simple modules\footnote{
As we will deal with VOA modules only of the form $\oplus_{\alpha \in {\cal I}((V, Y))} M_\alpha^{\oplus n_\alpha}$ (and we do not deal with logarithmic VOAs) in this article,
we can use the properties that would follow from semi-simpleness of an algebra even when $(V, Y)$ is not rational. 
 } %
of a VOA whose defining algebraic data is $(V, Y, {\bf 1}, \omega)$;
when we think of a unitary VOA $(V,Y,{\bf 1}, \omega, \Upsilon, \langle -|-\rangle)$, we only include in ${\cal I}((V,Y))$ distinct equivalence classes of simple unitary modules. 
Each simple module (=representation) in the set ${\cal I}((V, Y))$ will be denoted by a Greek letter (e.g., $\alpha$ for $(M_\alpha, Y_{M_\alpha}, \langle - | - \rangle_{M_\alpha})$); the defining representation of the VOA is denoted by 0. For $\alpha \in {\cal I}((V,Y))$, 
$\Upsilon^{\cal I}(\alpha)$ stands for the equivalence class of modules (or unitary modules) contragradient to $\alpha$ (we admit that this is not an ideal choice of notation because the notion of contragradient representation is available without unitarity). See also \ref{statmnt:rationalVOA}.   

For a simple representation/module $\alpha$, all the states $m\in M_\alpha$
share the same $L_0$-eigenvalue ({\it conformal weight})
modulo $+\Z$. Those values are denoted by $h_\alpha + \Z$ or $[h_\alpha]$
in this article. 
\end{anythng}

\begin{anythng}
\label{statmnt:set-I(V)}
In this article, we assume that the set ${\cal I}((V,Y))$
satisfies a couple of conditions stated below.
The set ${\cal I}((V,Y))$ does not necessarily exhaust all the distinct
unitary irreducible representations. We require that (r-1)
$0 \in {\cal I}((V,Y))$, (r-2) $\Upsilon^{\cal I}(\alpha) \in
{\cal I}((V,Y))$ if $\alpha \in {\cal I}((V,Y))$, and (r-3) all the conformal
weights of $\alpha \neq 0 \in {\cal I}((V,Y))$ are strictly positive
(not 0 or negative).

Before stating one more condition (r-4), we introduce
{\it character functional}
\cite[(3.18)]{Moore:1988qv} (cf also \cite[\S4]{MR1317233} 
and \cite[\S2]{MR2387861}) 
of an irreducible representation $\alpha$: 
\begin{align}
  \chi_\alpha(-, \tau) : V \times \mathfrak{H} \ni (v, \tau) \longmapsto
      {\rm Tr}_{M_\alpha} \left[ e^{2\pi i \tau \left( L_0 - \frac{c}{24}\right)}
        \; Y^{(w)}_{M_\alpha}(v)_{w@0}   \right] \in \C,
      \label{eq:def-char-fcnl}
\end{align}
where $\mathfrak{H}$ is the complex upper half plane, $\eta = w$ is the
{\it cylinder coordinate} of a $g=1$ Riemann surface
$\Sigma \cong \C/2\pi(\Z + \tau \Z)$ with the periodicity
$w \sim w+2\pi \sim w+2\pi \tau$, which is related to the projective
coordinate $\xi$ through $\xi = e^{-iw}$. Its {\it character}
$\chi_\alpha({\bf 1},\tau)$ is denoted by $\chi_\alpha(\tau)$. 
Much physics literature talks of enhancing the character
$\chi_\alpha(\tau)$ to a twining characters $\chi_\alpha(\tau;g)$ by
exploiting varieties of $g \in G$ of the symmetry group $G$ of a VOA,
with a motivation to resolve possible coincidence $\chi_\alpha(\tau) = \chi_\beta(\tau)$ for $\alpha \neq \beta$ in ${\cal I}((V,Y))$; in fact,
the character functionals are better suited for resolving them, 
because $\alpha \neq \beta$ implies that the action of some
$v \in V$ on the modules is different. 

The condition (r-4) is that the trace defining the character functional
$\chi_\alpha$ converges (despite infinitely many states in $M_\alpha$)
for all $\alpha \in {\cal I}((V,Y))$. 
Besides the conditions (r-1)--(r-4), or (r) as a whole, we will introduce
a few more conditions on ${\cal I}((V,Y))$ in \ref{statmnt:fusn-alg} and
\ref{statmnt:BnF} that makes a choice of ${\cal I}((V,Y))$ more interesting; 
cf also (r$^{""}$) and (r$^{'""}$) in \ref{statmnt:data-4-CFT}. See also
footnote \ref{fn:ex-I(VY)-torus}.
\end{anythng} 

\begin{anythng}
\label{statmnt:rationalVOA}
{\bf rational VOA:}
There is a subtle variation in the literature in setting the condition for
a VOA to be said {\it rational} (e.g., \cite[\S5]{Gaberdiel:2000qn}). 
It will be safe to demand both semi-simpleness
(e.g., \cite[Def. 5.5.1]{MR1317233}, \cite[p.580]{MR1628239})
and $C_2(V)$ cofiniteness (e.g., \cite{MR1794264})
and one more to give a firm mathematical foundation for what is regarded
as rational VOAs and CFTs in the string theory community.
That one more condition is an assumption (r) that all the modules
of a VOA without a proper submodule are unitary. To the limited knowledge
of the present author, no example of a semi-simple, unitary and $C_2$-cofinite
VOA is known to have such a module that is not unitarizable, but a proof is
not known either. So, the property is counted as an
extra condition to be imposed on rational VOAs discussed in this article. 

Instead of writing down those rigorous definitions and reviewing 
the derivation of various properties of rational VOAs, here we list up 
a few properties that follow from the definition. First, any 
module of a rational VOA $(V, Y)$ is completely reducible
 \cite[Def. 5.5.1]{MR2082709}. Second, there
are only finitely many irreducible modules of
$(V, Y)$ (cf \cite[Thm. 5.2.1]{MR1317233}, \cite[Thm. 8.1]{MR1628239},
\cite[Property 5.5.2(1)]{MR2082709}).
Thirdly, the character functionals for $n$ operator
insertions from $V$---as $(\tau \in \mathfrak{H})$-dependent linear
functionals on $V^{\otimes n}$---for distinct irreducible modules 
are convergent (so the condition (r-3) is satisfied), linearly independent,
and span the genus-1 $n$-insertion conformal block\footnote{
\label{fn:jargon-conf-blck}
The jargon {\it conformal block} has often been used in different meanings
in the VOA community and string theory community. In the string theory community,
a conformal block often refers to the chiral part---a phase ambiguity
remains---of an $n$-point correlation function (as a function of the moduli
of genus-$g$ $n$-pointed Riemann surface)
for a specific choice of $n$ states to be inserted, which should be calculated
by using the intertwining operators whose left-right coupling coefficients
are subject to the condition (t) in \ref{statmnt:cond-Ytot}
(eqs. (\ref{eq:cond-Ytot-repr}--\ref{eq:cond-Ytot-F2})). 
In the VOA community, on
the other hand, conformal block is a vector space of (moduli dependent)
linear functionals on $V^{\otimes n}$ (or $\otimes_{i=1}^n M_{\alpha_i}$
for $\alpha_i \in {\cal I}(V)$ more generally) realized by arbitrary
intertwining operators from the vector space of intertwining operators
${\cal V}_{\beta,\gamma}^\alpha$.
Whereas it is relatively a minor difference whether to focus on a fixed
set of $n$ states to be inserted or to handle a linear functional on
the whole vector space $\otimes_{i=1}^n M_{\alpha_i}$, it is conceptually a big
difference whether to deal with arbitrary intertwining operators or only
those subject to the condition (t). 
We follow the usage in the VOA community in this article.
} %
of $(V,Y)$. Finally, the $g=1$ conformal block is invariant under the
${\rm SL}_2\Z$ action \cite[Thm. 5.3.2]{MR1317233}.  
This property implies, in particular, that there exists an S-transformation
matrix of the character functionals of the untwisted modules of $(V, Y)$;
more review is found in section \ref{ssec:g=1}.

Whenever we deal with a rational VOA, we will choose (r') the set ${\cal I}((V, Y))$ to be that of all the distinct simple representations.
This is partly because the ${\rm SL}_2\Z$ action is closed only when all those representations are included in ${\cal I}((V, Y))$;
the other reason is the observation reviewed in Rmk. \ref{rmk:Moore-Seiberg}. 
\end{anythng}
%

\subsection{Mode Expansion in the Cylinder Coordinate}
\label{ssec:mode-op-cyl}

The way unitarity of a VOA is formulated in section \ref{ssec:review-VOA}
allows the conditions (m--p), especially in (\ref{eq:cond-o-4VOA-onUps}),
to be written down directly in terms of the defining algebraic data of a VOA. 
Physics literature (and also an earlier appearance in VOA literature),
on the other hand, often uses a time coordinate and the Hermitian Hilbert space
to formulate unitarity in a way more natural in quantum field theory;
the physics literature version is often written, however, only for VOAs
that have a Lagrangian / path-integral description.  
Here, we write down a translation of the formulation in
section \ref{ssec:review-VOA} in a style closer in spirit to the one in
physics literature (and also in \cite[\S 12.5 (p.410)]{MR996026}, where
Affine Lie algebras were dealt with),  
so the translated version is applicable to VOAs that do not
necessarily have a weakly coupled Lagrangian and path-integral; advantages
of both sides are there.  
This approach allows a formula (\ref{eq:Hconj-VOA-inCylMdOp}) that
generalizes (\ref{eq:Hconj-VOA-psPrmry}). 
%

\begin{anythng}
  \label{statmnt:cyl-mode-exp}
For a unitary VOA $(V, Y, {\bf 1}, \omega, \Upsilon, \langle - | - \rangle)$
and its representation $(M_\gamma, Y_\gamma, \langle - | - \rangle_M)$,
we introduce the following mode operators using the cylinder coordinate. 
Let $z$ be the dummy variable of the defining algebraic data, as well as
an inhomogeneous coordinate of $\P^1$; a holomorphic coordinate $\theta$
related to $z$ through $z= e^{-i\theta}=: {\rm exp}_-(\theta)$ may well be
used as a local holomorphic coordinate for insertion of vertex operators
at a point in $\P^1$ (cf eq.(\ref{eq:def-char-fcnl})). 
In stead of the mode expansion (\ref{eq:mode-expns-radial})
of the field operator $Y_\gamma(v)_{z@z} = Y^{(z,z)}_\gamma(v,z)$
let us introduce\footnote{
%
Ref. \cite[\S4.2]{MR1317233} also deals with a kind of mode expansion
associated with the coordinate transformation $z = e^{-i \theta}$
[read $z$ here as $w$ there, and $\theta$ here as $-2\pi z$ there].
The field $Y[a,w]$ for $a \in V$ introduced
in \cite[(4.2.2)]{MR1317233} is the same as $Y^{(\theta)}(a)$ here
(modulo the dictionary above); 
%
the mode expansion (\ref{eq:mode-expns-cyl-chi}) in
the power series of $e^{i\theta}$ here, however, is completely
different from the expansion in the power series of $\theta$
in \cite[(4.2.3)]{MR1317233} (or in
\cite[\S2]{MR1794264}, \cite[\S2]{MR1875102}, \cite[\S3]{MR2097833}).
%
%
} %
a mode expansion of the operator implementation
for the local coordinate $\theta$: 
\begin{align}
  Y^{(\theta,z)}_\gamma(v)_{z@z} =: \sum_{n\in \Z} [v]^{\rm cyl.}_n z^{-n},
   \quad 
   Y_\gamma^{{\rm cyl.}}( v) := Y^{(\theta,z)}_\gamma(v)_{z@z(p)} = Y_\gamma
   \left( R_{{\rm exp}_-@\theta(p)}^{-1} \cdot v, e^{-i\theta(p)}\right)
   %
       \label{eq:mode-expns-cyl-chi}
\end{align}
for $v\in V$.
\end{anythng}

%
%

\begin{anythng}
The condition (p-mod) (cf (\ref{eq:cond-o-4VOA-onUps}) for the defining
representation) for unitarity is translated to the following, 
\begin{align}
  \langle [\Upsilon(a)]_n^{\rm cyl.} \cdot m_1 | m_2 \rangle
  = \langle m_1 | [a]_{-n}^{\rm cyl.} \cdot m_2 \rangle , 
  \qquad a \in V_h, \quad m_1,\; m_2 \in M_{\gamma}. 
  \label{eq:Hconj-VOA-inCylMdOp}
\end{align}
This generalizes (\ref{eq:Hconj-VOA-psPrmry}), which was applicable only
when $a \in V$ is quasi-primary. 
%
%
%
\end{anythng}
\proof
Note first that 
\begin{align*}
  R_{{\rm exp}_-@\theta(p)} & \;= \exp
  \left[ \frac{i}{2} L_1 + \frac{i^2}{12}L_2+\frac{i^3}{48}L_3 +\cdots \right]
  (ie^{i\theta(p)})^{L_0}, \\
  R_{{\rm exp}_+@\theta(p)} & \; = \exp
  \left[ \frac{(-i)}{2} L_1 + \frac{(-i)^2}{12}L_2 + \cdots \right]
  (-ie^{-i\theta(p)})^{L_0}, 
\end{align*}
where ${\rm exp}_+(\theta) := e^{+i\theta}$, so
\begin{align*}
  R_{{\rm exp}_-@\theta(p)}^{-1} & \; = (-i e^{-i\theta(p)})^{L_0} \exp
  \left[ - \frac{i}{2}L_1 - \frac{i^2}{12}L_2- \cdots \right]
  =: (-ie^{-i\theta(p)})^{L_0} \exp[L^+ ({\rm exp}_-)], \\
  R_{{\rm exp}_+@\theta(p)}^{-1} & \; = (i e^{i\theta(p)})^{L_0} \exp
  \left[ + \frac{i}{2}L_1 - \frac{i^2}{12}L_2 + \cdots \right]
  =: (ie^{i\theta(p)})^{L_0} \exp[L^+({\rm exp}_+)]. 
\end{align*}

Now, let us use the condition (\ref{eq:cond-o-4VOA-onUps})
with $a = (iz)^{L_0} \exp \left[ L^+({\rm exp}_+) \right] \cdot b$
with $b \in V$. 
In evaluating the LHS (cf Rmk. \ref{rmk:treat-z}),
$\Upsilon(a)= (z/i)^{L_0}
\exp\left[ L^+({\rm exp}_-)\right] \Upsilon(b)$, so the LHS of
(\ref{eq:cond-o-4VOA-onUps}) then turns into 
\begin{align}
  \langle Y_\gamma \left( R_{{\rm exp}_-@\theta(p)}^{-1} \cdot \Upsilon(b), z \right) \cdot m_1 \; | \; m_2 \rangle_\gamma
  = \sum_{n\in \Z} z^{-n} \bra{ [\Upsilon(b)]_n^{\rm cyl.} \cdot m_1 } \; m_2
    \rangle_\gamma.
  \label{eq:temp-unitary-in-cylCoord-LHS}  
\end{align}
The dummy variable $z = e^{-i\theta(p)}$ within the first ($\C$-antilinear)
argument of the inner product is taken outside of $\bra{-}$ as it is (as $z$).
The RHS of (\ref{eq:cond-o-4VOA-onUps}), on the other hand,
is evaluated as follows:
\begin{align}
 & \langle m_1 | Y_\gamma\left( R_{{\rm inv}@e^{-i\theta(p)}}^{-1} \cdot
  (iz)^{L_0} \exp[L^+({\rm exp}_+)] \cdot b,
  \; e^{i\theta(p)}\right) \cdot m_2 \rangle_{M_\gamma} \nonumber \\
 & \;  = \langle m_1 | Y_\gamma \left( R_{{\rm inv}@e^{-i\theta(p)}}^{-1} \cdot
  R_{{\rm exp}_+@-\theta(p)}^{-1} \cdot b, e^{i\theta(p)} \right) \cdot m_2
  \rangle_{M_\gamma}   \label{eq:temp-Hrm-cnj-rltn-trnslt2-cylOp} \\
  & \;  = \langle m_1 | Y_\gamma \left( R_{{\rm exp}_-@ -\theta(p)}^{-1}
  \cdot b, \;  e^{i\theta(p)} \right) \cdot m_2 \rangle_{M_\gamma} 
  = \sum_{n \in \Z} \langle m_1 | [b]^{\rm cyl.}_{-n} \cdot m_2 \rangle_{M_\gamma}
  \; z^{-n},  
 \nonumber 
\end{align}
where we used $R_f \cdot R_g = R_{g\circ f}$ and
${\rm inv} \circ {\rm exp}_+ = {\rm exp}_-$ towards the end;
$(iz)^{-2h_{b}}$ from $R_{{\rm inv}@z}^{-1}$ and
$(iz)^{h_{b}}$ from $R_{{\rm exp}_+@-\theta(p)}^{-1}$ yield
$(z^{-1}/i)^{h_b}$ in $R_{{\rm exp}_-@-\theta(p)}^{-1}$;
we treated $(-z^{-2})^{L_0}$ in (\ref{eq:cond-o-4VOA-onUps}) as
$(iz)^{-2L_0}= (e^{-\pi i}/z^2)^{L_0}$ as in (\ref{eq:cond-o-4VOA-onUps-phys}),
which is a non-trivial choice
of a branch when this calculation is applied beyond VOA (where the
conformal weight of $a$ is not necessarily an integer). 
The condition (\ref{eq:cond-o-4VOA-onUps}) therefore implies
(\ref{eq:Hconj-VOA-inCylMdOp}), and vice versa.

To see that (\ref{eq:Hconj-VOA-inCylMdOp}) generalizes
(\ref{eq:Hconj-VOA-psPrmry}), it is enough to note that
$e^{-\pi i h_a/2}[a]_n = [a]_n^{\rm cyl.}$ for any Virasoro-primary $a \in V_{h_a}$. 
\qed

The property (\ref{eq:Hconj-VOA-wOPE}) is equivalent to this:
\begin{align}
  \Upsilon( [a]_n^{\rm cyl.} \cdot u ) = i^{2h} [\Upsilon(a)]_n^{\rm cyl.} \cdot \Upsilon(u), \qquad a \in V_h, \quad u \in V; 
  \label{eq:Hconj-VOA-wOPE-inCylMdOp}
\end{align}
it is easy to see that (\ref{eq:Hconj-VOA-wOPE-inCylMdOp}) holds true
when $a$ is Virasoro primary. For $a$ that is a Virasoro descendant,
one can use the proof by induction with respect to the level of
excitation. 

\begin{exmpl}
\label{exmpl:Heisenberg-unitarity}
  {\bf the chiral boson (Heisenberg algebra) VOA:} Let us describe data of
  the rank-1 chiral boson VOA that is related to unitarity
  (cf \cite[\S12.5]{MR996026}). 
\begin{align}
  V = {\rm Span}_\C \left\{ \prod_{n=1} (\alpha_{-n})^{N_n} \cdot {\bf 1} \; | \;
  \sum_{n \geq 1} N_n < \infty \right\}, 
\end{align}
where the Hermitian inner product is introduced by
$\langle {\bf 1} | {\bf 1} \rangle =1$, $[\alpha_m, \alpha_n]=m \delta_{m+n}$
and $(\alpha_m)^\dagger = \alpha_{-m}$. The anti-unitary involution
$\Upsilon: V \rightarrow V$ is given by $\Upsilon({\bf 1}) = {\bf 1}$ and
$\Upsilon \cdot \alpha_n = - \alpha_n \cdot \Upsilon$. 

Here are examples of the mode operators using the cylinder
coordinates.\footnote{
\label{fn:CPT-4-boson}
In the eyes of physicists, the sign $-$ in the CPT involution
$\Upsilon (\alpha_{-n}\cdot {\bf 1}) = - \alpha_{-n} \cdot {\bf 1}$ is understood this way; 
\[
X_L(z) := \hat{x}_L -i \alpha_0 \ln(z) + i \sum_{n \neq 0} \alpha_n/(nz^n), \quad
Y(\alpha_{-n} \cdot {\bf 1}) = i (\partial_z^n X_L), \quad
Y^{\rm cyl.}(\alpha_{-n}\cdot {\bf 1}) = i (\partial_\theta^n X_L). 
\]
All the operators $X_L(\theta)$, $\partial_\theta^n X_L$ are Hermitian
($(e^{-in\theta})^\dagger = e^{in\theta}$), so the states
$\alpha_{-n} \cdot {\bf 1} \in V$ are in fact ``{\it real}'' states
multiplied by $i$. The CPT involution maps such a state to the real state
multiplied by $(-i)$. 
} %
\begin{align}
  Y(\alpha_{-1} \cdot {\bf 1}) = \sum_{n\in \Z} \alpha_n/z^{n+1}, \qquad
  Y(\alpha_{-2}\cdot {\bf 1}) = \sum_{n \in \Z} (-n-1) \alpha_n/z^{n+2}.  
\end{align}
Because $Y^{\rm cyl.}(\alpha_{-2} \cdot {\bf 1}) = Y(\alpha_{-2}\cdot {\bf 1}) (z/i)^2  -i/2 \; Y(2\alpha_{-1} \cdot {\bf 1}) (z/i) $,
\begin{align*}
  Y^{\rm cyl.}(\alpha_{-1} \cdot {\bf 1}) = \sum_{n\in \Z} \alpha_n z^{-n} /i,
  \qquad
  Y^{\rm cyl.}(\alpha_{-2} \cdot {\bf 1}) = \sum_{n \in \Z} \alpha_n z^{-n} n,  
\end{align*}
so $[\alpha_{-1}\cdot {\bf 1}]^{\rm cyl.}_n = \alpha_n/i$ and
$[\alpha_{-2} \cdot {\bf 1}]^{\rm cyl.}_n = \alpha_n n$. It is easy to confirm
that the relations (\ref{eq:Hconj-VOA-inCylMdOp},
\ref{eq:Hconj-VOA-wOPE-inCylMdOp}) hold for both
$a= \alpha_{-1} \cdot {\bf 1}$ and $a=\alpha_{-2} \cdot {\bf 1}$. 
\end{exmpl}

\begin{exmpl}
 \label{exmpl:fermion-unitarity}
  {\bf one pair of free chiral fermions:} this is an example
of discussions in \ref{statmnt:def-alg-data-SVOA} in fact.
So, we allow ourselves to use concepts and notations that are introduced
until \ref{statmnt:def-alg-data-SVOA}. One unitary VOA $(V_0, Y_{0,0},\cdots)$
and its unitary module $(V_1, Y_{0,1})$ is constructed with 
\begin{align*}
  V_0 & \; := {\rm Span}_\C \left\{ \psi^{\rm phy.}_{-\mu_1}\cdots
  \psi^{\rm phy.}_{-\mu_m} \bar{\psi}^{\rm phy.}_{-\nu_1}\cdots
  \bar{\psi}^{\rm phy.}_{-\nu_n} \cdot {\bf 1} \; | \;
  0<\mu_1<\cdots, \; 0<\nu_1 < \cdots, \; {\rm even~ }m+n  \right\}, \\
  V_1 & \; := {\rm Span}_\C \left\{ \psi^{\rm phy.}_{-\mu_1}\cdots
  \psi^{\rm phy.}_{-\mu_m} \bar{\psi}^{\rm phy.}_{-\nu_1}\cdots
  \bar{\psi}^{\rm phy.}_{-\nu_n} \cdot {\bf 1} \; | \;
  0<\mu_1<\cdots, \; 0<\nu_1 < \cdots, \; {\rm odd~ }m+n  \right\}; 
\end{align*}
the two Fock spaces generated by $\psi_\mu^{\rm phy.}$ and
$\bar{\psi}_\mu^{\rm phy.}$ with $\mu \in 1/2+\Z$ offer an easiest
construction of an SVOA. Here, we describe only information relevant to
unitarity. 
It is customary to introduce a Hermitian inner product on $V_0$ and $V_1$ by
the relation $\langle {\bf 1} | {\bf 1} \rangle = 1$,
$\{ \psi^{\rm phy.}_\mu, \bar{\psi}^{\rm phy.}_\nu \} = 2 \delta_{\mu+\nu}$,
$\{ \psi^{\rm phy.}_\mu, \psi^{\rm phy.}_\nu\} = 0$,
$\{ \bar{\psi}^{\rm phy.}_\mu, \bar{\psi}^{\rm phy.}_\nu \} = 0$ and
$(\psi^{\rm phy.}_\mu)^\dagger = \bar{\psi}^{\rm phy.}_{-\mu}$,
$(\bar{\psi}^{\rm phy.}_\mu)^\dagger = \psi^{\rm phy.}_{-\mu}$. 

As a unitary $(V_0, Y_{0,0})$-module, however,
there is a freedom in the choice of $\langle -|-\rangle_{V_1}$ parametrized by $\Lambda \in \R_{>0}$,
\begin{align}
  \langle \psi^{\rm phy.}_{-1/2}\cdot {\bf 1} | \psi^{\rm phy.}_{-1/2}\cdot {\bf 1}
  \rangle_{V_1} =
  \langle \bar{\psi}^{\rm phy.}_{-1/2}\cdot {\bf 1} |
  \bar{\psi}^{\rm phy.}_{-1/2}\cdot {\bf 1}  \rangle_{V_1} =  2\Lambda. 
\end{align}
The module $(V_1, Y_{0,1})$ is self-contragradient, so there is
an isometry (norm preserving isomorphism) of $(V_0,Y_{0,0})$ modules,
$\lambda_{V_1}: V_1^\vee \rightarrow V_1$, which comes with a freedom
$e^{i\delta} \in S^1$:
\begin{align}
  \lambda_{V_1}: 2\Lambda (\psi^{\rm phy.}_{-1/2}\cdot {\bf 1})^\vee \longmapsto
  e^{i\delta} \bar{\psi}^{\rm phy.}_{-1/2}\cdot {\bf 1}, \quad
  2\Lambda (\bar{\psi}^{\rm phy.}_{-1/2}\cdot {\bf 1})^\vee \longmapsto
  e^{i\delta} \psi^{\rm phy.}_{-1/2} \cdot {\bf 1}. 
\end{align}
Combined with the anti-unitary $\Upsilon_{V_1}: V_1 \rightarrow (V_1)^\vee$
mapping $\psi_{-1/2}^{\rm phy.} \cdot {\bf 1}$ to $2\Lambda (\psi_{-1/2}^{\rm phy.}
\cdot {\bf 1})^\vee$,
we have an anti-unitary $\Upsilon_{V_1V_1}: V_1 \rightarrow V_1$, mapping
$\psi_{-1/2}^{\rm phy.} \cdot {\bf 1}$ to $e^{i\delta}
\bar{\psi}^{\rm phy.}_{-1/2}\cdot {\bf 1}$.

The condition (p-SVOA) for SVOA constrains the freedom
parametrized by $\Lambda \in \R_{>0}$ and $e^{i\delta} \in S^1$
for a given choice of the component $Y_{1,1}$ of $YY = \{ Y_{s_0,s'_0} \; |
\; s_0,s'_0 \in \Z/2\Z \}$. In this Ex. \ref{exmpl:fermion-unitarity},
however, let us take a perspective that $Y_{1,1}$ is constrained by $\delta$
and $\Lambda$. The choice $Y_{1,0} = \Omega Y_{0,1}$ is mandatory
because of the skew-symmetry condition for SVOA; let us parametrize
$Y_{1,1}:V_1\rightarrow {\rm Hom}(V_1,V_0)[[z,z^{-1}]]$
by $Y_{1,1} = A_r \lambda_{V_0} \cdot (\Theta_r\Omega Y_{0,1}) \cdot
\lambda_{V_1}^{-1}$, with $A_r \in \C^\times$; $r \in \Z$; $\lambda_0: V_0^\vee
\rightarrow V_0$ is the one already discussed in \ref{statmnt:cntrG-mod}. 
By analysing the condition (p-SVOA) with $a \in V_1$, we find
that $A_r = e^{-\pi i r}$, independent of $\delta$ or $\Lambda$
(cf. the relation (\ref{eq:xing-SVOA-Y-vs-Y-vs-lambda-m1})
in \ref{statmnt:xing-S}).
As a result, $Y_{1,1}$ is independent of $r\in \Z$, but depends on $\delta$
and $\Lambda$ through $\lambda_{V_1}^{-1}$. 

Conversely, the information $\langle - |-\rangle_{V_1}$ and $\lambda_{V_1}:
V_1^\vee \rightarrow V_1$ (ie, $\Lambda \in \R_{>0}$ and $e^{i\delta} \in S^1$)
are determined if $Y_{1,1}$ is give a priori; that is the perspective
we choose in \ref{statmnt:def-alg-data-SVOA}. In this example,
it is customary to introduce 
\begin{align}
  \Psi(z) := \sum_{\nu \in 1/2+\Z} \psi^{\rm phy}_\nu z^{-(1/2+\nu)}, \qquad
  \overline{\Psi}(z) := \sum_{\nu \in 1/2+\Z} \bar{\psi}^{\rm phy}_\nu z^{-(1/2+\nu)}
\end{align}
acting on $V_0 \oplus V_1$. 
Because $Y_{1,0}= \Omega Y_{0,1}$, $Y_{1,0}(\psi^{\rm phy.}_{-1/2}\cdot {\bf 1})
= \Psi$ and $Y_{1,0}(\bar{\psi}^{\rm phy.}_{-1/2} \cdot {\bf 1})
= \overline{\Psi}$. When we also set $Y_{1,1}$ so that 
\begin{align}
  Y_{1,1}(\psi^{\rm phy.}_{-1/2}\cdot {\bf 1}) = \Psi, \qquad
  Y_{1,1}(\bar{\psi}^{\rm phy.}_{-1/2} \cdot {\bf 1}) = \overline{\Psi}, 
\end{align}
the condition (p-SVOA) determines $\Lambda = 1$ and $e^{i\delta} = i$. 
The condition (q-SVOA) is satisfied: $\Upsilon \left( [(\psi^{\rm phy.}_{-1/2} \cdot {\bf 1})]_{1/2} \cdot (\bar{\psi}^{\rm phy.}_{-1/2}\cdot {\bf 1}) \right)
= \Upsilon( 2 {\bf 1}) = 2{\bf 1}$, while $(-1) \times [\Upsilon_{1,1}(\psi^{\rm phy.}_{-1/2} \cdot {\bf 1})]_{1/2} \cdot \Upsilon_{1,1}(\bar{\psi}^{\rm phy.}_{-1/2}\cdot {\bf 1}) = (-1) \times i^{2} [(\bar{\psi}^{\rm phy.} \cdot
  {\bf 1})]_{1/2} \cdot (\psi^{\rm phy.}_{-1/2} \cdot {\bf 1}) = 2 {\bf 1}$ in
this independent calculation.
The relation $[\Upsilon(\psi^{\rm phy.}_{-1/2}\cdot {\bf 1})]_\nu^\dagger
= [\psi^{\rm phy.}_{-1/2}\cdot {\bf 1}]_{-\nu} (-i)^{2\cdot 1/2}$
or $([\Upsilon(\psi^{\rm phy.}_{-1/2}\cdot {\bf 1})]_\nu^{\rm cyl.})^\dagger
= [\psi^{\rm phy.}_{-1/2}\cdot {\bf 1}]_{-\nu}^{\rm cyl.}$ with
$e^{i\delta}=i$ governing $\Upsilon$ reproduces the relation
$(\bar{\psi}^{\rm phy.}_{\nu})^\dagger = \psi_{-\nu}^{\rm phy.}$. 
\end{exmpl}
%

\subsection{The Vector Space of Intertwining Operators}
\label{ssec:review-vct-sp-intrtw-op}

The jargon {\it intertwining operators} is used often in the literature
from the VOA community. In the literature from the physics community, the
word closest to intertwining operator will be {\it chiral vertices}.
See also a comment at the end of \ref{statmnt:build-CFT} on the terminology. 

\begin{anythng}
\label{statmnt:def-VOA-inttw-op}
{\bf intertwining operators:}
For irreducible modules
$\alpha, \beta, \gamma$ of a VOA,
the vector space ${\cal V}^\alpha_{\beta,\gamma}$
of {\it intertwining operators} of type $(\alpha;\beta,\gamma)$ consists of
a homomorphism
\[
{\cal Y} = {\cal Y}_{\beta,\gamma}^\alpha : M_\beta \rightarrow {\rm Hom}_\C(M_\gamma,M_\alpha)
[[z,z^{-1}]] z^{-h_\beta-h_\gamma+h_\alpha}, 
\]
satisfying both the condition (e-int.op)
\begin{align}
  {\cal Y}(L_{-1} \cdot m,z) = \partial_z {\cal Y}(m,z), \qquad
  m\in M_\beta, 
\end{align}
and the ($\ell$-int.op) Jacobi identity:
\begin{align}
  z_0^{-1} \delta \left( \frac{z_1-z_2}{z_0} \right) Y_{\alpha}(v,z_1)
  {\cal Y}(m,z_2)
  - z_0^{-1} \delta \left( \frac{z_2-z_1}{-z_0} \right) {\cal Y}(m,z_2)
  Y_\gamma(v,z_1)  \nonumber \\
  - z_2^{-1} \delta \left( \frac{z_1-z_0}{z_2} \right) {\cal Y} \left(
  Y_\beta(v,z_0) \cdot m,z_2 \right) = 0
  \label{eq:def-vct-sp-int.op}
\end{align}
for any $v\in V$ and $m \in M_\beta$. The dimension of this complex
vector space is denoted by $N_{\beta,\gamma}^\alpha$.
See \cite[(2.3a)]{Moore:1988uz} and \cite[(4.1)]{Moore:1988qv}
from physics literature, and \cite[(5.4.4)]{MR1142494}, 
\cite[(11.9)]{MR1233387}, 
\cite[(2.45)]{MR1338873}, 
\cite[Def.1.10]{MR1817614}, 
for example, from math literature.  

The homomorphism $Y$ in the data $(V,Y,{\bf 1},\omega, \cdots)$ of a VOA
is an intertwining operator of type $(0;0,0)$.
The homomorphism
$Y_\alpha=Y_{M_\alpha}$ in the data $(M_\alpha,Y_\alpha, \cdots)$ of an irreducible
module of a VOA is an intertwining operator of type $(\alpha;0,\alpha)$,
where the combination $(-h_\beta-h_\gamma+h_\alpha)+\Z$ is 
$(-\Z -h_\alpha + h_\alpha)+\Z = \Z$. 

We introduce the notation
\begin{align}
 [(\alpha;\beta,\gamma)] := h_\beta + h_\gamma - h_\alpha + \Z \in \R/\Z
\end{align}
for a triplet $\alpha,\beta,\gamma$ of unitary irreducible representations
of a VOA. If a VOA is rational, the value $[(\alpha;\beta,\gamma)]$
is in $\Q/\Z \subset \R/\Z$. 
\end{anythng}

\begin{anythng}
\label{statmnt:fusn-alg}  
{\bf fusion algebra:}  
For a unitary VOA that is rational, its {\it fusion algebra} is denoted\footnote{
When $A$ is a finite group, its group algebra is often denoted by
$\Z[A]$ in the literature. Because one cannot always find a structure
of abelian group on the set ${\cal I}((V,Y))$ to regard a given fusion
algebra as a group ring, it is better to
introduce a separate notation for the fusion algebra. Instead, we chose
not to introduce too many symbols.  
} %
by $\Z[{\cal I}((V,Y))]$ in this note, is the commutative algebra over $\Z$,
\begin{align}
  \Z[{\cal I}((V,Y))] := \left\{ \left. \sum {}_{\alpha \in {\cal I}((V,Y))}
  n_\alpha [e_\alpha] \; \right| \; n_\alpha \in \Z \right\}, \qquad
  [e_\beta] \cdot [e_\gamma] = \sum_\alpha N_{\beta,\gamma}^\alpha [e_\alpha];  
\end{align}
here, $[e_\alpha]$ is a formal element assigned to each element
$\alpha \in {\cal I}((V,Y))$
(e.g., \cite{Verlinde:1988sn}, \cite{Gannon:2007mfj}).

For a unitary VOA that is not rational, it is of interest to think
of a set of---not necessarily all of---unitary and distinct irreducible
representations ${\cal I}((V,Y))$ that satisfies the condition
(r) in \ref{statmnt:set-I(V)}, along with extra conditions
(r'') in \ref{statmnt:BnF} and (s) in \ref{statmnt:data-4-CFT} that we will
discuss later. The set ${\cal I}((V,Y))$
in that case is not necessarily finite then; for a given $\beta,\gamma \in
{\cal I}((V,Y))$, there can be infinitely many $\alpha \in {\cal I}((V,Y))$
where $N_{\beta,\gamma}^\alpha >0$. Such things as a fusion algebra above are
therefore not well-defined; for the product $[e_\beta] \cdot [e_\gamma]$ to
be closed, we should---in general (though not always)---include an infinite
formal sum $\sum_{\alpha \in {\cal I}((V,Y))} n_\alpha [e_\alpha]$ in $\Z[{\cal I}((V,Y))]$,
but the product between two such elements is not well-defined. In other
words, a $g=0$ four-point conformal block is not necessarily of finite
dimensions. 

The Verlinde formula is a relation between the structure constants
$N_{\beta,\gamma}^\alpha$ of the fusion ring and the $S$-transformation
matrix of the character functionals, both available for a rational VOA.
For a VOA that is not rational, it is neither expected that the
fusion algebra is well-defined nor that the $S$-transformation is
closed on the character functionals. 
\end{anythng}

\begin{anythng}
\label{statmnt:skew-xing-iso}
  {\bf skew symmetry isomorphism:} For ${\cal Y}_{\beta,\gamma}^\alpha \in
{\cal V}_{\beta,\gamma}^\alpha$, set
\begin{align}
  (\Omega_r {\cal Y}^\alpha_{\beta,\gamma}): M_\gamma \rightarrow
  {\rm Hom}_\C(M_\beta,M_\alpha) [[z,z^{-1}]]z^{-[(\alpha;\gamma,\beta)]}
\end{align}
for an odd integer $(2r+1) \in 2\Z+1$ by   
\begin{align}
  (\Omega_r{\cal Y}): M_\gamma \times M_\beta \ni (m_\gamma, m_\beta)
  \longmapsto e^{z L_{-1}} \cdot {\cal Y}_{\beta,\gamma}^\alpha
  (m_\beta, e^{\pi i (2r+1)} z) \cdot m_\gamma \in
  M_\alpha [[z,z^{-1}]]z^{-h_\beta-h_\gamma+h_\alpha}.
  \label{eq:def-skew-sym-iso}
\end{align}
The constructed $\Omega_r {\cal Y}^\alpha_{\beta,\gamma}$ is in
${\cal V}^\alpha_{\gamma,\beta}$ in fact    
(Ref. \cite[Prop. 1.11]{MR1817614}, a step-by-step exposition is found
later in Lemma \ref{lemma:skew-sym-iso}),
so we have a family of isomorphisms (the {\it skew-symmetry isomorphisms})
\begin{align}
  \Omega_r: {\cal V}_{\beta,\gamma}^\alpha \longrightarrow
        {\cal V}_{\gamma,\beta}^\alpha, \qquad \quad
        \alpha, \beta,\gamma \in {\cal I}((V,Y)).
\end{align}
The inverse isomorphism
of $\Omega_r: {\cal V}_{\beta,\gamma}^\alpha \rightarrow
{\cal V}_{\gamma,\beta}^\alpha$ is $\Omega_{-1-r} :
{\cal V}_{\gamma,\beta}^\alpha \rightarrow {\cal V}_{\beta,\gamma}^\alpha$.
The isomorphisms $\Omega_r$ are independent of $r\in \Z$
(hence the subscript can be omitted),
if $h_\beta+h_\gamma \equiv h_\alpha + \Z$.

A little more geometric characterization of the isomorphism $\Omega_r$ is 
available; because 
\begin{align*}
  \bra{m_\alpha} {\cal Y}_{\beta,\gamma}(m_\beta, z) \cdot m_\gamma \rangle|_{z \rightarrow e^{\pi i (2r+1)} \underline{z}} 
  & \; = \bra{m_\alpha} e^{z L_{-1}} e^{\underline{z} L_{-1}}
      {\cal Y}_{\beta,\gamma}^{\alpha} (m_\beta,
  e^{\pi i (2r+1)} \underline{z} ) \cdot m_\gamma \rangle \\ 
  & \; = \bra{m_\alpha} e^{zL_{-1}} (\Omega_r{\cal Y})(m_\gamma, \underline{z}) \cdot m_\beta \rangle, 
\end{align*}
the intertwining operator $\Omega_r{\cal Y} \in {\cal V}_{\gamma,\beta}^\alpha$
is the one that reproduces the matrix element on the LHS analytically
continued so that the relative coordinate value $z$ of the two modules
$M_\beta$ and $M_\gamma$ gains complex phase by $+\pi (2r+1)$. 

{\bf crossing symmetry isomorphism:}
For ${\cal Y}_{\beta,\gamma}^\alpha \in {\cal V}_{\beta,\gamma}^\alpha$, set
\begin{align}
  (\Theta_r {\cal Y}_{\beta,\gamma}^\alpha): M_\beta \rightarrow
  {\rm Hom}_\C(M_{\Upsilon^{\cal I}(\alpha)}, M_{\Upsilon^{\cal I}(\gamma)})[[z,z^{-1}]]
  z^{-[(\Upsilon^{\cal I}(\gamma);\beta,\Upsilon^{\cal I}(\alpha))]}
\end{align}
by
\begin{align}
  \langle (\Theta_r {\cal Y}_{\beta,\gamma}^\alpha)(m_\beta,z) \cdot m'_\alpha, \;
  m_\gamma \rangle = \langle m'_\alpha, \; {\cal Y}_{\beta,\gamma}^\alpha (e^{zL_1}
  (e^{\pi i (2r+1)} z^{-2})^{L_0}m_\beta,1/z) \cdot m_\gamma \rangle 
 \label{eq:def-xing-sym-iso}
\end{align}
for $m'_{\alpha} \in M_{\Upsilon^{\cal I}(\alpha)}$ and $m_\gamma \in M_{\gamma}$. 
The constructed $(\Theta_r {\cal Y}_{\beta,\gamma}^\alpha)$
is in ${\cal V}_{\beta,\Upsilon^{\cal I}(\alpha)}^{\Upsilon^{\cal I}(\gamma)}$
(see a step-by-step exposition in Lemma \ref{lemma:xing-sym-iso};
cf \cite[Thm. 5.2.1]{MR1142494} with $\beta =0$ and ${\cal Y} = Y_\gamma$),
so we have a family of isomorphisms (the {\it crossing-symmetry isomorphisms})
\begin{align}
  \Theta_r: {\cal V}_{\beta,\gamma}^\alpha \longrightarrow
        {\cal V}_{\beta,\Upsilon^{\cal I}(\alpha)}^{\Upsilon^{\cal I}(\gamma)},
        \qquad \alpha,\beta,\gamma \in {\cal I}((V,Y)). 
\end{align}
The inverse isomorphism of $\Theta_r$ is $\Theta_{-1-r}:
{\cal V}_{\beta,\Upsilon^{{\cal I}}(\alpha)}^{\Upsilon^{{\cal I}}(\gamma)} \rightarrow
{\cal V}_{\beta,\gamma}^{\alpha}$.
The isomorphisms $\Theta_r$ are independent of $r \in \Z$ (so the
subscript can be omitted) if $[h_\beta] = 0_{+\Z}$.

The intertwining operator $\Theta_r {\cal Y}$ for ${\cal Y} \in
{\cal V}_{\beta,\gamma}^\alpha$ maps $M_\beta$ to the vector space of
formal power series that take values in
${\rm Hom}(M_\alpha^\vee, \; M_\gamma^\vee)$, in fact. To bring it to the
one taking values in
${\rm Hom}(M_{\Upsilon^{\cal I}(\alpha)}, \; M_{\Upsilon^{\cal I}(\gamma)})$,
we need to use
$\lambda_\gamma \cdot (\Theta_r {\cal Y}) \cdot \lambda_\alpha^{-1}$, 
where $\lambda_\alpha$ and $\lambda_\gamma$ are the isomorphisms
of equivalence of representations reviewed in \ref{statmnt:cntrG-mod}. 
This difference matters in section \ref{sec:CFT}. 
\end{anythng}
\begin{anythng}
\label{statmnt:prscr-oprtr-trnstnFcn-coordCh-int.op}  
The formula (\ref{eq:FBZv-Lemma656}) of conversion under the global
coordinate change is not just for the state-operator correspondence $Y$ of
a VOA, and for its representation $Y_M$; the same formula holds
for a general intertwining operator ${\cal Y} \in {\cal V}_{\beta,\gamma}^\alpha$
(if we take care of a subtlety); this is because the conversion formula
reflects the nature of the spaces $V$ and ${\rm End}(V)$ in
$Y: Y \rightarrow {\rm End}(V)$ than details of $Y$. The conversion
operators $R_f$ and $R_{f@p}$ in (\ref{eq:FBZv-Lemma656}) are given
by the mode operators $L_n$ of $\omega$, and we know the action
of $L_n$ on all the modules of the VOA. There are two subtleties, however.
One is that the conversion operator $R_f$ involves $f_0^{-L_0}$, which
is not well-defined when it acts on a module $M_\gamma$ where
$[h_\gamma] \neq 0+\Z$. This subtlety is not an issue, in fact, when we deal
with the representation $Y_\gamma \in {\cal V}_{0,\gamma}^\gamma$ for
any $\gamma$, because
the ambiguity in $f_0^{-L_0}$ in $R_f$ and $f_0^{-L_0}$ in $R_f^{-1}$ cancel.
When it comes to intertwining operators ${\cal Y} \in
{\cal V}_{\beta,\gamma}^\alpha$, we need to face this issue. 
The other subtlety is that the field operator in
${\rm Hom}(M_\gamma,M_\alpha)[[z,z^{-1}]] z^{-h_\beta-h_\gamma+h_\alpha}$
have non-integral powers of $z$; as we hope to deal with $z$ not just as
a formal dummy variable of the algebraic data of a VOA but a local/global
projective coordinate on a Riemann surface, we also need to deal with
the phase ambiguity in $z(p)^{-h_\beta-h_\gamma+h_\alpha}$, where $z(p)\in \C$
is the coordinate value of a point $p$.

Let us restrict our attention to the coordinate transformation $\xi = f(\eta)$
of the form of a linear fractional transformation on an inhomogeneous coordinate
of $\P^1$ here (ie, $f_i=0$ in (\ref{eq:def-R_f}) for $i>1$).
It is enough only to deal with the
case $f_{-1} = f(0) = 0$, because $R_f = R_{f@0} e^{-f(0) L_{-1}}$,
$R_f {\cal Y}(m,f(p)) R_f^{-1} = R_{f@0} {\cal Y}(m,(f@0)(p)) R_{f@0}^{-1}$.
That brings us to think of the coordinate transformation 
\begin{align*}
  \xi = f(\eta) = \frac{\eta A}{(\eta + \eta_0) \eta_0}, \quad A, \eta_0
  \in \C \backslash \{0\}, \quad
  R_f = e^{\frac{1}{\eta_0}  L_1} (A/\eta_0^2)^{-L_0}, \quad
  R_f^{-1} = e^{-\frac{\eta_0}{A}L_1}  (A/\eta_0^2)^{L_0}. 
\end{align*}
Then $R_{f@\eta(p)} = e^{(\eta(p)+\eta_0)^{-1} L_1} [A/(\eta(p)+\eta_0)^2]^{-L_0}$
because 
\begin{align*}
  (f@\eta(p))(t) = \frac{A}{\eta_0} \left[
    \frac{(t+\eta(p))}{(t+\eta(p)+\eta_0)} 
    - \frac{\eta(p)}{(\eta(p)+\eta_0)} \right]
  = \frac{At}{(\eta(p)+\eta(0))[t+\eta(p)+\eta_0]}. 
\end{align*}
Now, we see that 
\begin{align}
  {\rm Arg}\left(\frac{\xi(p)}{\eta(p)} \right) 
  = {\rm Arg}\left( \frac{A}{\eta_0(\eta_0+\eta(p))}\right)
  = {\rm Arg}(A/\eta_0^2) = {\rm Arg}\left( \frac{A}{(\eta_0+\eta(p))^2} \right)
  \label{eq:lin-frct-trnsf-arg-conv}
\end{align}
mod $+2\pi \Z$ for points $p$ in the intervals 
\begin{eqnarray*}
  \eta(p) \in (0,1) \times (-\eta_0) & \leftrightarrow &
  \xi(p) \in (0, +\infty) \times (-A/\eta_0) \quad {\rm or} \\
  \eta(p) \in (0, +\infty) \times \eta_0 & \leftrightarrow &
  \xi(p) \in (0,1) \times (A/\eta_0). 
\end{eqnarray*}
The conversion formula (\ref{eq:FBZv-Lemma656}) still holds true
for intertwining operators ${\cal Y} \in {\cal V}_{\beta,\gamma}^\alpha$
for the points $p$ in the intervals, when we fix the $+2\pi \Z$ ambiguity
in the complex phases of $(A/\eta_0^2)$ in
$f_0 ^{\mp L_0} = (A/\eta_0^2)^{\mp L_0}$ in $R_f^{\pm 1}$,
$(A/\eta_0/(\eta(p)+\eta_0))^{-L_0}$ in $R_{f@\eta(p)}$, and
$(\xi(p)/\eta(p))$ in a way (\ref{eq:lin-frct-trnsf-arg-conv}) holds
literally without a mismatch in $+2\pi \Z$ for the points $p$ in the intervals:
\begin{align*}
  \exp & \; \left[ (-h_\alpha+h_\gamma) {\rm Arg}\left(\frac{A}{\eta_0^2} \right)
    + (-h_\beta-h_\gamma +h_\alpha) {\rm Arg}\left( \frac{\xi(p)}{\eta(p)} \right)
    \right] \\
  & \qquad \qquad \qquad \qquad \qquad \qquad \qquad \qquad \qquad \qquad
  = \exp \left[ -h_\beta {\rm Arg}\left( \frac{A}{\eta_0(\eta(p)+\eta_0)} \right) \right]. 
\end{align*}
So  long as a point $p$ moves away from the line $\eta(p) \in \eta_0 \times
\{ (-1,0)  \cup (0,+\infty)\}$,  but remains not too far away
(i.e., not around $\eta(p) \in -\eta_0 \times (1,\infty)$,
$\xi(p) \in (A/\eta_0) \times (+\infty, 1)$), there is no ambiguity
in the complex phase of $(\xi(p)/\eta(p))^{h_\beta+h_\gamma-h_\alpha}$
$(A/[\eta_0(\eta_p+\eta(p))])^{-h_\beta}$. 
The conversion formula (\ref{eq:FBZv-Lemma656})
for the coordinate transformation for intertwining operators should be
understood with this phase convention for the points
$\eta(p) \nin - \eta_0\times (1,+\infty)$. 
\end{anythng}

\begin{anythng}
\label{statmnt:BnF}
{\bf the braiding and fusion isomorphisms:} First, we review the
{\it commutativity} property (j-int.op), which consists of
(j-a-int.op) and (j-b-int.op). Let us begin with the  
{\it generalized rationality, convergence and extension property
of products} (j-a-int.op) 
in \cite[\S16]{MR1344849}, \cite[p.11]{HUANG1996201}, \cite[\S4]{MR1817614}.
It requires, for any $\alpha, \gamma, \rho,\sigma,\delta \in {\cal I}((V,Y))$,
for any ${\cal Y}_1 \in {\cal V}_{\rho,\delta}^{\alpha}$ and
${\cal Y}_2 \in {\cal V}_{\sigma,\gamma}^\delta$, and for any 
$m_\rho \in M_\rho$, $m_\sigma \in M_\sigma$, $m_\gamma \in M_\gamma$ and
a linear functional $m'_{\alpha'} \in {\rm Hom}_\C(M_\alpha,\C)$,
that there should exist a complex analytic function
$f_{m_\rho,m_\sigma,m_\gamma,m'_{\alpha'}}^{(\alpha;\rho,\sigma;\gamma)}$ on
$(\widetilde{\Sigma}^{(0,2)}_{\rm prd}, {\rm base})$ with the prooperty stated
below; here, $\widetilde{\Sigma}^{(0,2)}_{\rm prd}$ is the universal cover of
$\{ (\zeta_1,\zeta_2) \in (\C^\times)^2 \} \backslash
\{ \zeta_1 = \zeta_2\}$ and base a small contractible open subset of
$\widetilde{\Sigma}^{(0,2)}_{\rm prd}$ that contains one point that is
projected by $\pi: \widetilde{\Sigma}^{(0,2)}_{\rm prd} \rightarrow
(\C^\times)^2$ to $(\zeta_1,\zeta_2) \in (\C^\times)^2$
with $0 < \zeta_2 < \zeta_1 \subset \R_{>0}$; we call $\pi^{-1}(
\{ (\zeta_1,\zeta_2) \in (\R_{>0})^2 \; | \zeta_1 > \zeta_2\})$ the base branch
in $\widetilde{\Sigma}^{(0,2)}_{\rm prd}$. The complex analytic function\footnote{
Here, complex analytic functions include $\tilde{G}(x)$, $\tilde{G}_{\pm}(x)$
that appear in footnotes \ref{fn:Ising4pt-a}--\ref{fn:Ising4pt-c}. When
the VOA in question is not rational, the complex analytic functions may be
regarded as a single-valued function only on the universal cover
$\widetilde{\Sigma}^{(0,2)}_{\rm prd}$ than on a finite cover on
$(\C^\times)^2 \backslash {\rm diagonal}$. 
} %
$f^{(\alpha;\rho,\sigma,\gamma)}_{m_\rho, m_\sigma, m_\gamma,m'_{\alpha'}}$ along the base
branch should have
a Laurent series expansion with respect to $\zeta_2$ and $\zeta_1^{-1}$
(where non-integral powers of $\zeta_2$ and $\zeta_1^{-1}$ are allowed),
and this Laurent series with $\zeta_2 \rightarrow z_2$ and $\zeta_1 \rightarrow z_1$ agrees with the matrix element $\langle m'_{\alpha'}, \;
{\cal Y}_1(m_\rho,z_1) \cdot {\cal Y}_2(m_\sigma,z_2) \cdot m_\gamma \rangle$. 

The condition (j-b-int.op) 
is the existence of a family of the braiding isomorphisms 
${\cal B}_r$ ($r \in \Z$), 
\begin{align}
  {\cal B}_{-1-r}: \oplus_{\delta \in {\cal I}((V,Y))}
  \left( {\cal V}^\alpha_{\beta_1,\delta} \otimes {\cal V}^\delta_{\beta_2,\gamma}
  \right) \rightarrow
  \oplus_{\epsilon \in {\cal I}((V,Y))} \left( {\cal V}^\alpha_{\beta_2,\epsilon}
  \otimes {\cal V}^\epsilon_{\beta_1,\gamma} \right),  
 \label{eq:basic-brid-iso} 
\end{align}
in a way the following properties hold. Let
\begin{align*}
  \left\{ \left. ({\cal Y}\otimes {\cal Y})_i \in
          {\cal V}_{\beta_1,\delta(i)}^\alpha \otimes {\cal V}_{\beta_2,\gamma}^{\delta(i)} \; \right| \; i=1,2,\cdots, \sum_\delta N_{\beta_1,\delta}^\alpha N_{\beta_2,\gamma}^\delta \right\}, \\
  \left\{ \left. ({\cal Y}\otimes {\cal Y})_j \in
          {\cal V}_{\beta_2,\epsilon(j)}^\alpha \otimes {\cal V}_{\beta_1,\gamma}^{\epsilon(j)} \; \right| \; j=1,2,\cdots, \sum_\delta N_{\beta_2,\epsilon}^\alpha N_{\beta_1,\gamma}^\epsilon \right\} 
\end{align*}
be a basis of the two vector spaces, and 
\begin{align}
  {\cal B}_{-1-r}: ({\cal Y}\otimes {\cal Y})_i \longmapsto \sum_{j}
  ({\cal Y}\otimes {\cal Y})_j \;\; {\cal B}_{-1-r}
  (\alpha;\beta_1,\beta_2;\gamma)^i_{\;\; i}   \qquad \qquad {}^\forall i,  
  \label{eq:br-isom-chiral-1}
\end{align}
the matrix representation of the isomorphism ${\cal B}_{-1-r}$ with respect
to this basis; 
the isomorphism ${\cal B}_{-1-r}$ is such that the following relation between
the complex analytic functions holds 
\cite{Moore:1988uz}, 
\cite{Moore:1988qv}, 
\cite[Thm. 3.2]{MR1817614}: 
\begin{align}
 \left( f^{(\alpha;\beta_1,\beta_2;\gamma;i)}
 _{m_{\beta 1}, m_{\beta 2}, m_\gamma, m'_{\alpha'}}(z_1,z_2) \right) & \; 
 _{(z_1-z_2) \rightarrow e^{\pi i(2r+1)} (z_2-z_1)} \nonumber \\
& = \sum_j  
  f^{(\alpha;\beta_2,\beta_1;\gamma;j)}
  _{m_{\beta 2}, m_{\beta 1}, m_\gamma, m'_{\alpha'}}(z_2,z_1) \;\; 
  {\cal B}_{-1-r}(\alpha;\beta_1,\beta_2;\gamma)^j_{\;\; i}
 \label{eq:br-isom-chiral-2}
\end{align}
for any $m_{\beta 2}, m_{\beta 1}, m_{\gamma}$ and $m'_{\alpha'}$ simultaneously. 
Here, the functions $f^{(\alpha;\beta_1,\beta_2;\gamma;i)}
_{m_{\beta 1}, m_{\beta 2}, m_\gamma, m'_{\alpha'}}(z_1,z_2)$ and
$f^{(\alpha;\beta_2,\beta_1;\gamma;j)}
_{m_{\beta 2}, m_{\beta 1}, m_\gamma, m'_{\alpha'}}(z_2,z_1)$ are the ones whose
existence is guaranteed by the condition (i-a-int.op) for
each $({\cal Y}\otimes {\cal Y})_i$ and $({\cal Y}\otimes {\cal Y})_j$, 
and $m_{\beta_1} \in M_{\beta_1}$, $m_{\beta 2} \in M_{\beta_2}$,
$m_\gamma \in M_\gamma$ and $m'_{\alpha'} \in {\rm Hom}_\C(M_\alpha,\C)$. 
To put the characterization of the isomorphisms in (\ref{eq:br-isom-chiral-2})
in a more colloquial expression, the complex analytic function
$f^{(a)}(\zeta_1,\zeta_2)$ on the base branch in
$\widetilde{\Sigma}^{(0,2)}_{\rm prd}$ is regarded as a function of
$0 < z_2 < z_1 \in \R_{>0}$ through $\zeta_1=z_1$, $\zeta_2 = z_2$;
think of its analytic continuation in the way $z_1-z_2$ gains complex phase
by $+\pi(2r+1)$; now $0 < z_1 < z_2 \in \R_{>0}$ and that is what we have on
the LHS of (\ref{eq:br-isom-chiral-2}); on the RHS, the complex
analytic function $\sum_b f^{(b)}(\zeta_1,\zeta_2) ({\cal B}_{-1-r})^j_{\;\; i}$
on $\widetilde{\Sigma}^{(0,2)}_{\rm prd}$ should be chosen so that 
its behavior on the base branch seen as a function of $z_2$ and $z_1$
through $\zeta_1 = z_2$ and $\zeta_2 = z_1$ agrees with the analytically
continued function on the LHS. 

Next, the {\it associativity} condition (k-int.op) consists of two,
(k-a-int.op) and (k-b-int.op). 
The {\it generalized rationality, convergence and extension property
  of iterates} in
\cite[\S16]{MR1344849}, \cite[p.11]{HUANG1996201}, \cite[\S4]{MR1817614})
((k-a-int.op)) is as follows:
for any $\alpha, \gamma, \rho,\sigma,\zeta \in {\cal I}((V,Y))$,
for any ${\cal Y}_o \in {\cal V}_{\zeta,\gamma}^{\alpha}$ and
${\cal Y}_i \in {\cal V}_{\rho,\sigma}^\zeta$, and for any 
$m_\rho \in M_\rho$, $m_\sigma \in M_\sigma$, $m_\gamma \in M_\gamma$ and
a linear functional $m'_{\alpha'} \in {\rm Hom}_\C(M_\alpha,\C)$,
there should exist a complex analytic function
$f_{m_\rho,m_\sigma,m_\gamma,m'_{\alpha'}}^{(\alpha;\rho,\sigma;\gamma)}$ on
$(\widetilde{\Sigma}^{(0,2)}_{\rm itr}, {\rm base})$ with the property stated
below; here, $\widetilde{\Sigma}^{(0,2)}_{\rm itr}$ is the universal cover of
$\{ (\zeta_0,\zeta_2) \in (\C^\times)^2 \} \backslash
\{ \zeta_0+\zeta_2 = 0 \}$ and base a small contractible open subset of
$\widetilde{\Sigma}^{(0,2)}_{\rm itr}$ that contains one point that is
projected by $\pi: \widetilde{\Sigma}^{(0,2)}_{\rm itr} \rightarrow
(\C^\times)^2$ to $(\zeta_0,\zeta_2) \in (\C^\times)^2$
with $\zeta_2, \zeta_0 \subset \R_{>0}$; we call $\pi^{-1}((\R_{>0})^2 )$
the base branch in $\widetilde{\Sigma}^{(0,2)}_{\rm itr}$.
The complex analytic function $f^{(\alpha;\rho,\sigma;\gamma)}_{m_\rho,m_\sigma,m_\gamma,m'_{\alpha'}}$ in the base branch should have a Laurent series expansion with
respect to $\zeta_0$ and $\zeta_2^{-1}$ (non-integral powers of $\zeta_0$ and
$\zeta_2^{-1}$ are allowed), and this Laurent series with $\zeta_0
\rightarrow z_0$ and $\zeta_2 \rightarrow z_2$ agrees with the matrix element
$\langle m'_{\alpha'}, \; {\cal Y}_o\left( 
{\cal Y}_i(m_\rho,z_0) \cdot m_\sigma , z_2 \right) \cdot m_\gamma \rangle$. 

The condition (k-b-int.op) is the existence of a fusion
isomorphism between the finite dimensional vector spaces, 
\begin{align}
  {\cal F}: \oplus_{\delta \in {\cal I}((V,Y))} \left(
  {\cal V}_{\beta_1,\delta}^\alpha \otimes {\cal V}_{\beta_2,\gamma}^\delta \right)
  \rightarrow
  \oplus_{\epsilon \in {\cal I}((V,Y))} \left( {\cal V}_{\epsilon,\gamma}^\alpha
  \otimes {\cal V}_{\beta_1,\beta_2}^\epsilon\right)
  \label{eq:basic-fusn-iso}
\end{align}
in a way the following properties hold. Let 
\[
\left\{
 \left. ({\cal Y}\otimes {\cal Y})_k \in 
  {\cal V}_{\epsilon(k),\gamma}^\alpha \otimes {\cal V}_{\beta_1,\beta_2}^{\epsilon(k)}
  \right| \; k = 1,2,\cdots, \sum_\epsilon N_{\epsilon,\gamma}^\alpha N_{\beta_1,\beta_2}^\epsilon \right\}
\]
be a basis of the latter vector space, and
\begin{align}
  {\cal F}: ({\cal Y}\otimes {\cal Y})_i \longmapsto \sum_k
  ({\cal Y}\otimes {\cal Y})^k \;
  {\cal F}(\alpha;\beta_1, \beta_2;\gamma)^k_{\;\; i}
  \label{eq:fs-isom-chiral-1}
\end{align}
the matrix representation of the isomorphism ${\cal F}$ wrpt this basis;
the isomorphism ${\cal F}$ is such that the following relation between
the functions holds 
 \cite{Moore:1988uz}, 
\cite{Moore:1988qv}, 
\cite[Thm. 3.2]{MR1817614} 
\begin{align}
  \left. 
  f_{m_{\beta 1},m_{\beta 2}, m_\gamma, m'_{\alpha'}}^{(\alpha;\beta_1,\beta_2;\gamma;i)}(z_1,z_2)
  \right|_{z_1 \rightarrow z_0+z_2}
  = \sum_k
  f_{m_{\beta 1},m_{\beta 2}, m_\gamma, m'_{\alpha'}}^{(\alpha;\beta_1,\beta_2;\gamma;k)}(z_2,z_0)
  {\cal F}(\alpha;\beta_1,\beta_2;\gamma)^k_{\;\; i}
  \label{eq:fs-isom-chiral-2}
\end{align}
for any $m_{\beta 2}, m_{\beta 1}, m_{\gamma}$ and $m'_{\alpha'}$ simultaneously. 
Here, the functions 
$f_{m_{\beta 1},m_{\beta 2}, m_\gamma, m'_{\alpha'}}^{(\alpha;\beta_1,\beta_2;\gamma;k)}(z_2,z_0)$
are guaranteed to exist by the condition (k-a-int.op)
for each of the basis element $({\cal Y}\otimes {\cal Y})^c$ and for any
$m_{\beta 1}$, $m_{\beta 2}$, $m_\gamma$ and $m'_{\alpha'}$. 

The conditions (j-int.op) and (k-int.op) are to be imposed on the choice
of the set ${\cal I}((V,Y))$ as well as a VOA $(V,Y, {\bf 1}, \omega, \cdots)$;
we require that the conditions (j-int.op) and (k-int.op) are satisfied
for any of $\oplus_{\delta \in {\cal I}((V,Y))} {\cal V}_{\beta_1,\delta}^\alpha \otimes {\cal V}_{\beta_2,\gamma}^\delta$ for any choice of
$\alpha,\beta_1,\beta_2,\gamma \in {\cal I}((V,Y))$. For the purpose of
emphasizing the aspect that these conditions constrain the choice of the
set ${\cal I}((V,Y))$, we may also call the conditions (j-int.op) and
(k-int.op) combined as the condition (r''). 

\vspace{5mm}

$\Omega(\pm)$ in \cite{Moore:1988qv} correspond to
$\Omega_r$ with $(2r+1)=\mp 1$ here, and ${\cal B}(\pm)$ to
${\cal B}_{s}$ with $(2s+1) = \mp 1$, so the relation ${\cal B}_{-1-r}
= {\cal F}^{-1} \cdot \Omega_r \cdot {\cal F}$ with $(2r+1) = \mp 1$
here 
is ${\cal B}(\mp) = {\cal F}^{-1} \cdot \Omega(\pm) \cdot {\cal F}$
in \cite{Moore:1988qv}. 
In \cite{MR1817614},  
$\Omega_{-1}= \Omega(+)$ is denoted by $\Omega$, and the corresponding
${\cal B}_0 = {\cal B}(-)$ as ${\cal B}$. 

A graphical presentation is assigned to the isomorphisms
$\Omega(-)$, ${\cal F}$ and ${\cal B}(+)$ as in Fig. \ref{fig:OmegaFB}. 
\begin{figure}[tbp]
\begin{center}
    \begin{tabular}{cccc}
      \includegraphics[scale=0.65]{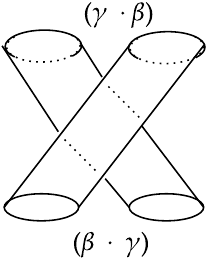} & 
      \includegraphics[scale=0.65]{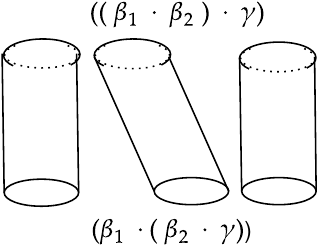} & 
      \includegraphics[scale=0.65]{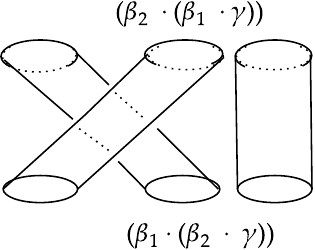} &
      \includegraphics[scale=0.65]{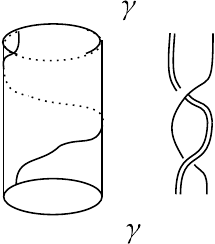} \\      
      (a) & (b) & (c) & (d)
\end{tabular}
\end{center}
\caption{\label{fig:OmegaFB}The graphical presentation of the
 isomorphisms $\Omega(-)=\Omega_{0}$, ${\cal F}$,
 ${\cal B}(+) = {\cal B}_{-1}$ and $\theta$, from (a) to (d). 
Written below those pictures and also above those pictures are
a collection of modules from which states are inserted; parentheses
indicate which operators inserted on $\R_{>0} \subset \C$ are
placed closer; the view point is chosen at $+i\infty \subset \C$. 
The pictures indicate how matrix elements are gradually changed,
where the change is due to the change in the operator insertion points
in $\C \subset \P^1$ in (a--c) starting from the configuration at the bottom
and ending in the configuration at the top, and due to the change in the
choice of local coordinate around the insertion point in (d) [the solid
line on the cylinder or the double line edge of the ribbon indicates
the real and positive direction in the local coordinate]. 
  }
\end{figure}
Those graphs interpolate between a pair of objects; each object
is an ordered collection of irreducible representations from
${\cal I}((V,Y))$, e.g., $\beta_1 \cdot \beta_2 \cdot \cdots \cdot \beta_n$, 
with parentheses indicating proximity between the neighboring letters. 
To this combinatorial information of an object, we assign a modulus of
$g=0$ $(n+1)$-point conformal block where $n$ states $m_1, \cdots, m_n$ from
$M_{\beta_1}, \cdots, M_{\beta_n}$ are inserted at $z(p_i) = z_i \in \C$
with $z_i \in \R_{>0}$, $z_i > z_{i+1}$, while one more state
(whose representation is not specified) is inserted at $z = \infty$.
We adopt a convention that the graphical presentation deals with those
objects by their configuration in the $z$-plane seen from a viewpoint
at $z=+i\infty$ (so the pair of points with $x(p_1) > x(p_2) > 0$
are drawn in a way $p_1$ placed to the left of $p_2$).
The analytic continuation in a way $z(p_1)-z(p_2)$ gains phase by $+\pi$,
as in $\Omega(-)$, is therefore drawn graphically as in
Fig. \ref{fig:OmegaFB} (a).
See section \ref{ssec:ribbon-MTC} for more. 
%
\end{anythng}

\begin{rmk}
\label{rmk:penta-hexa-id}
{\bf the pentagon, hexagon and box identities:}
Think of a rational and unitary VOA whose vector space of intertwining
operators satisfy the (generalized version of the) condition (j', k').
The pentagon identity begins with an observation that there are at least two
isomorphisms
\begin{align*}
  \oplus_{\delta_1,\delta_2}^{\in {\cal I}({\cal A})}
  \left( {\cal V}^\alpha_{\beta_1,\delta_1} \otimes
  \left( {\cal V}^{\delta_1}_{\beta_2,\delta_2} \otimes
       {\cal V}^{\delta_2}_{\beta_3,\beta_4} \right) \right) \rightarrow
  \oplus_{\epsilon_1,\epsilon_2}^{\in {\cal I}({\cal A})}
  \left( \left( {\cal V}_{\beta_1,\beta_2}^{\epsilon_1} \otimes
            {\cal V}_{\epsilon_1,\beta_3}^{\epsilon_2} \right) \otimes
            {\cal V}_{\epsilon_2,\beta_4}^{\epsilon_3} \right) , 
\end{align*}
which are
\begin{align*}
  {\cal F} \cdot {\cal F}: & \;
  {\cal V}_{(\beta_1 \cdot (\beta_2 \cdot (\beta_3 \cdot \beta_4)))}^\alpha \rightarrow
  {\cal V}_{((\beta_1 \cdot \beta_2) \cdot (\beta_3 \cdot \beta_4))}^\alpha \rightarrow
  {\cal V}_{(((\beta_1 \cdot \beta_2)\cdot \beta_3)\cdot \beta_4)}^\alpha, \\
%
%
  {\cal F}\cdot {\cal F} \cdot {\cal F}: & \;
  {\cal V}_{(\beta_1 \cdot (\beta_2 \cdot (\beta_3 \cdot \beta_4)))}^\alpha \rightarrow
  {\cal V}_{(\beta_1 \cdot ((\beta_2 \cdot \beta_3)\cdot \beta_4))}^\alpha \rightarrow
  {\cal V}_{((\beta_1 \cdot (\beta_2\cdot \beta_3))\cdot \beta_4)}^\alpha \rightarrow
  {\cal V}_{(((\beta_1\cdot \beta_2)\cdot \beta_3)\cdot \beta_4)}
%
  \label{eq:2-iso-4-5}
\end{align*}
Here, we introduced a compact notation for the tensor product of the vector
space of intertwining operators with the representations in the intermediate
state summed up. 
The two isomorphisms are always identical, because the two chains of
analytic continuations (for ${\cal F} \cdot {\cal F}$ and
${\cal F}\cdot {\cal F} \cdot {\cal F}$) are the same in the end. 
\begin{figure}[tbp]
  \begin{center}
    \begin{tabular}{ccc}
      \includegraphics[scale=0.7]{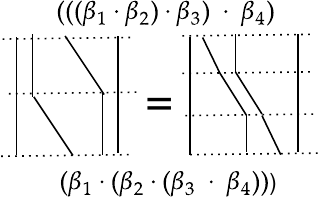} &
      $\qquad \qquad \qquad$ & 
      \includegraphics[scale=0.7]{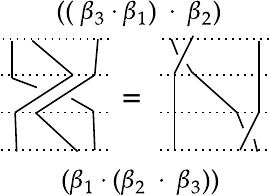}
      \\
      (a) ${\cal F} \cdot {\cal F} = {\cal F} \cdot {\cal F} \cdot {\cal F}$
      & &
      (b) ${\cal F} \cdot \Omega(-) \cdot {\cal F}
      = \Omega(-) \cdot {\cal F} \cdot \Omega(-)$ 
    \end{tabular}
    \end{center}
  \caption{\label{fig:penta-hexa}The pentagon (a) and hexagon (b) identities
    in the graphical presentation of the isomorphisms. }
\end{figure}

The hexagon identity begins with an observation that there are isomorphisms
\begin{align}
  \oplus_\delta  \left( {\cal V}^\alpha_{\beta_1,\delta} \otimes
        {\cal V}^\delta_{\beta_2,\beta_3} \right) \rightarrow
  \oplus_\epsilon \left( {\cal V}^\epsilon_{\beta_3,\beta_1} \otimes {\cal V}_{\epsilon,\beta_2}^\alpha \right). 
\end{align}
They are 
\begin{align*}
  {\cal F}\cdot \Omega(\pm) \cdot {\cal F}: & \;
  {\cal V}_{(\beta_1 \cdot (\beta_2 \cdot \beta_3))}^\alpha \rightarrow
  {\cal V}_{((\beta_1 \cdot \beta_2)\cdot \beta_3)}^\alpha \rightarrow
  {\cal V}_{(\beta_3 \cdot (\beta_1\cdot \beta_2))}^\alpha \rightarrow
  {\cal V}_{((\beta_3 \cdot \beta_1) \cdot \beta_2)}, \\
  \Omega(\pm) \cdot {\cal F} \cdot \Omega(\pm) : & \;
        {\cal V}_{(\beta_1 \cdot (\beta_2 \cdot \beta_3))}^\alpha \rightarrow
        {\cal V}_{(\beta_1 \cdot (\beta_3 \cdot \beta_2))}^\alpha \rightarrow
        {\cal V}_{((\beta_1 \cdot \beta_2) \cdot \beta_3)}^\alpha \rightarrow
        {\cal V}_{(\beta_3 \cdot (\beta_1\cdot \beta_2))}^\alpha .
\end{align*}
The two isomorphisms are identical, because the chain of analytic
continuations for ${\cal F}\Omega(\pm){\cal F}$ is the same as that
of $\Omega(\pm){\cal F}\Omega(\pm)$. This is not true for any $\Omega_r$
with $2r+1 \neq \pm 1$. See Fig. \ref{fig:penta-hexa} for pentagon
and hexagon identities in the graphical presentation. 


Finally, the box identity comes out from an observation that
there are multiple isomorphisms
\begin{align}
  \oplus_\delta \left( {\cal V}_{\beta_1,\delta}^\alpha \otimes {\cal V}_{\beta_2,\gamma}^\delta \right) \rightarrow
  \oplus_{\epsilon} \left( {\cal V}_{\beta_2,\epsilon}^\alpha \otimes {\cal V}_{\beta_1,\gamma}^\epsilon \right); 
\end{align}
examples are
\begin{align*}
  {\cal B}_{-1-r}: &\;
  {\cal V}_{(\beta_1 \cdot (\beta_2 \cdot \gamma))}^\alpha \rightarrow
  {\cal V}_{(\beta_2 \cdot (\beta_1 \cdot \gamma))}^\alpha, \\
  {\cal F}^{-1} \cdot \Omega_r \cdot {\cal F}: & \;
  {\cal V}_{(\beta_1 \cdot (\beta_2\cdot \gamma))}^\alpha \rightarrow
  {\cal V}_{((\beta_1 \cdot \beta_2) \cdot \gamma)}^\alpha \rightarrow
  {\cal V}_{((\beta_2\cdot \beta_1)\cdot \gamma)}^\alpha \rightarrow
  {\cal V}_{(\beta_2 \cdot (\beta_1 \cdot \gamma))}^\alpha .
%
\end{align*}
The two isomorphisms (for any $r\in \Z$) are always the same,
because the two chains of analytic continuations are identical. 
In Moore--Seiberg, the pentagon identity
is \cite[(2.3b)]{Moore:1988uz}=\cite[(4.7)]{Moore:1988qv}, the hexagon identity
\cite[(4.6)]{Moore:1988qv}, 
and the box identity \cite[(4.2)+(4.4)]{Moore:1988qv}. 

Because those pairs of isomorphisms from one common vector space to
another are identically the same by construction, those observations
do not yield any new constraints on the vector space of intertwining
operators of a rational and unitary VOA. Once the conditions
(j') and (k') for the intertwining operators are verified, then
the pentagon, hexagon and box identities among the monodromy
matrices follow automatically.
\end{rmk}

\begin{anythng}
  {\bf the isomorphisms for the local-coordinate twists:} Think of the local
  operator
$(\Omega Y_\gamma)^{(\omega_\theta;\zeta)} ( m_\gamma )_{@p_0}$ of a state in
$m_\gamma \in M_\gamma$
with the choice of a family of local coordinates $\omega_\theta$ that change
relatively to the global coordinate $\zeta$ by 
$(\omega_\theta(p) - \omega_\theta(p_0)) e^{i\theta} = (\zeta(p) - \zeta(p_0))$,
with $\theta \in \R$. As we change the local coordinate from
  $\zeta = \omega_0$ to $\omega_{2\pi}$, the operator changes by
\begin{align}
  \theta: {\cal V}_{\gamma,0}^{\gamma} \ni (\Omega Y_\gamma) =
  (\Omega Y_\gamma^{(\zeta;\zeta)})
  \longmapsto (\Omega Y^{(\omega_{2\pi};\zeta)}_\gamma) =
  (\Omega Y^{(\zeta;\zeta)}_\gamma)(R_{f@p_0}^{-1} -)
  = e^{2\pi i [h_\gamma]} (\Omega Y_\gamma) \in {\cal V}_{\gamma,0}^{\gamma}, 
\end{align}
where $\zeta - \zeta(p_0) = (f@p_0)(\omega_{\theta}-\omega_{\theta}(p_0))
= e^{i\theta} (\omega_\theta - \omega_\theta(p_0))$,
$R_{f@p_0}^{-1} = e^{i\theta L_0}$ with $\theta = 2\pi$. 
The graphical presentation of this isomorphism is shown in
Fig. \ref{fig:OmegaFB} (d); the solid line on the cylinder
(or the double line edge of the ribbon) describes how we change 
the direction in the $\zeta$ plane where $\delta \omega_\theta$ is
real and positive. 
\end{anythng}

\begin{rmk}
\label{rmk:spc-chi-vrtx-1}
The condition (\ref{eq:def-vct-sp-int.op}) on intertwinig operators
is linear on ${\cal Y}$, so it only determines the vector space
of intertwining operators/chiral vertices for each type
$(\alpha;\beta,\gamma)$. There is no intertwining operator 
better motivated than others in the context of VOA
(this situation changes when extra conditions are imposed
in \ref{statmnt:cond-Ytot} in the context of CFT).
When $\beta = 0$ (and $\alpha = \gamma$),
however, there is one special choice ${\cal Y} = Y_\gamma$ for the irreducible
module $(M_\gamma,Y_\gamma, \cdots)$.
The skew-symmetry condition (i) on the defining data $Y =Y_0
\in {\cal V}_{0,0}^0$ of a VOA implies that $\Omega Y = Y$, invariant 
under the skew-symmetry isomorphism from ${\cal V}_{0,0}^0$ to itself.
The crossing-symmetry isomorphism is defined in a way
$\Theta : {\cal V}_{0,\gamma}^\gamma \rightarrow
{\cal V}_{0,\Upsilon^{\cal I}(\gamma)}^{\Upsilon^{\cal I}(\gamma)}$ maps
$Y_\gamma$ to $Y_{\Upsilon^{\cal I}(\gamma)}$ (modulo equivalence).
\end{rmk}

\vspace{5mm}

Comments on special choices of intertwining operators continue
to \ref{statmnt:pair-ann-crtn}. Before doing so, let us digress
for a moment to write down the logic step-by-step in support
of the statements on the skew-symmetry and crossing symmetry isomorphisms
in \ref{statmnt:skew-xing-iso}. 
\begin{lemma}
\label{lemma:skew-sym-iso}
  $(\Omega_r {\cal Y}_{\beta,\gamma}^{\alpha})$ constructed by
  (\ref{eq:def-skew-sym-iso}) from ${\cal Y}_{\beta,\gamma}^\alpha \in
  {\cal V}_{\beta,\gamma}^\alpha$ satisfies the Jacobi identity
  (\ref{eq:def-vct-sp-int.op}) for type $(\alpha;\gamma,\beta)$, and hence
  is in ${\cal V}_{\gamma,\beta}^\alpha$. 
\end{lemma}
\proof The LHS of the condition (\ref{eq:def-vct-sp-int.op})
on ${\cal Y} = (\Omega_r {\cal Y}_{\beta,\gamma}^\alpha)$ of
type $(\alpha;\gamma,\beta)$, acting on $m_\beta \in M_\beta$, is as follows: 
\begin{align}
 & z_0^{-1}\delta \left( \frac{z_1-z_2}{z_0} \right)
  Y_\alpha(v,z_1) \cdot e^{z_2L_{-1}} {\cal Y}_{\beta,\gamma}^\alpha
  \left(m_\beta, e^{\pi i(2r+1)}z_2 \right) \cdot m_\gamma \nonumber \\
 & - z_0^{-1} \delta \left( \frac{z_2-z_1}{-z_0} \right)
  e^{z_2 L_{-1}} {\cal Y}_{\beta,\gamma}^\alpha\left( Y_\beta(v,z_1)\cdot m_\beta, \;
  e^{\pi i (2r+1)}z_2 \right) \cdot m_\gamma \label{eq:temp-skew-iso-inttw-prf} \\
 & - z_2^{-1} \delta \left( \frac{z_1-z_0}{z_2} \right) e^{z_2 L_{-1}}
  {\cal Y}_{\beta,\gamma}^\alpha \left( m_\beta, \; e^{\pi i(2r+1)}z_2 \right) \cdot
  Y_\gamma(v,z_0) \cdot m_\gamma  \nonumber 
\end{align}
for arbitrary elements $v$ and $m_\gamma$ in $V$ and $M_\gamma$,
respectively. 
In the first term, $Y_\alpha(v,z_1) e^{z_2L_{-1}}$ is equal to
$e^{z_2 L_{-1}} Y_\alpha(v,z_1-z_2)$; by rewriting the dummy variables under
the rule
\[
e^{\pi i (2r+1)} z_2 \rightarrow \zeta_2, \quad
z_1 \rightarrow \zeta_0, \quad
z_0 \rightarrow \zeta_1, 
\]
the first term is rewritten as
\begin{align}
 & \;  e^{-\zeta_2 L_{-1}} \frac{1}{\zeta_1} \delta \left( \frac{\zeta_0+\zeta_2}{\zeta_1}\right) Y_\alpha(v,\zeta_0+\zeta_2) {\cal Y}_{\beta,\gamma}^\alpha(m_\beta,\zeta_2),  \nonumber \\
  = & \; e^{-\zeta_2 L_{-1}}
  \frac{1}{\zeta_1} \delta \left( \frac{\zeta_0+\zeta_2}{\zeta_1}\right)
  Y_\alpha(v,\zeta_1) {\cal Y}_{\beta,\gamma}^\alpha(m_\beta,\zeta_2),
  \label{eq:temp-4proof-skewSym-iso-1stTrm} \\
  = & \; e^{-\zeta_2 L_{-1}} \frac{1}{\zeta_0} \delta \left( \frac{\zeta_1-\zeta_2}{\zeta_0}\right) Y_\alpha(v,\zeta_1) {\cal Y}_{\beta,\gamma}^\alpha(m_\beta,\zeta_2);
  \nonumber
\end{align}
we have exploited the $\zeta_1-\zeta_2=\zeta_0$ relation in any integral power
of $(\zeta_0+\zeta_2)$ appearing in $Y_\alpha(v,\zeta_0+\zeta_2)$ in the
first equality, and then used the relation (\ref{eq:rltn-delta-2}).
In the second and third terms, the relation (\ref{eq:rltn-delta-2}) is
used to rewrite the delta functions as
\begin{align}
  \frac{1}{z_0} \delta \left( \frac{z_2-z_1}{-z_0} \right)
  = \frac{1}{\zeta_1}\delta \left( \frac{\zeta_2+\zeta_0}{\zeta_1} \right)
  = \frac{1}{\zeta_2}\delta \left( \frac{\zeta_1-\zeta_0}{\zeta_2} \right),
  \label{eq:temp-4proof-skewSym-iso-2ndTrm} 
\end{align}
and
\begin{align}
  \frac{1}{z_2} \delta \left( \frac{z_1-z_0}{z_2}\right)
  = \frac{1}{-\zeta_2} \delta \left( \frac{\zeta_0-\zeta_1}{-\zeta_2}\right)
  = \frac{1}{\zeta_0} \delta \left( \frac{-\zeta_2+\zeta_1}{\zeta_0} \right)
  = \frac{1}{\zeta_0} \delta \left( \frac{\zeta_2-\zeta_1}{-\zeta_0}\right).
  \label{eq:temp-4proof-skewSym-iso-3rdTrm} 
\end{align}
Now, the three terms in (\ref{eq:temp-skew-iso-inttw-prf}) are the
LHS of the Jacobi identity of type $(\alpha;\beta,\gamma)$
with the dummy variables $\zeta_1,\zeta_2,\zeta_0$ acting on
$m_\gamma \in M_\gamma$, followed by the action of $e^{-\zeta_2L_{-1}}$.
As we have started from ${\cal Y}_{\beta,\gamma}^\alpha \in
{\cal V}_{\beta,\gamma}^\alpha$, the expression (\ref{eq:temp-skew-iso-inttw-prf})
vanishes. \qed

\begin{lemma}
(see \cite[Prop. 3.2]{MR3119224})   
\label{lemma:xing-sym-iso}
$(\Theta_r {\cal Y}_{\beta,\gamma}^\alpha)$ constructed
by (\ref{eq:def-xing-sym-iso})
from ${\cal Y}_{\beta,\gamma}^\alpha \in {\cal V}_{\beta,\gamma}^\alpha$
satisfies the Jacobi identity (\ref{eq:def-vct-sp-int.op}) for the type
$(\Upsilon^{\cal I}(\gamma); \beta,\Upsilon^{\cal I}(\alpha))$, and hence
is in ${\cal V}_{\beta,\Upsilon^{\cal I}(\alpha)}^{\Upsilon^{\cal I}(\gamma)}$. 
\end{lemma}
\proof (The case $\beta =0$ and ${\cal Y} = Y_\gamma$ is covered in
Ref. \cite[Thm. 5.2.1]{MR1142494}.)
Let us evaluate the LHS of the condition (\ref{eq:def-vct-sp-int.op})
with $(\Theta_r {\cal Y}_{\beta,\gamma}^\alpha)$ of type
$(\Upsilon^{\cal I}(\gamma);\beta,\Upsilon^{\cal I}(\alpha))$ as ${\cal Y}$
in (\ref{eq:def-vct-sp-int.op}); the ($z_{0,1,2}$-dependent) operator on the LHS
is converted to a ($z_{0,1,2}$-dependent) number by inserting the operator in
$\langle m_\gamma, \; - \cdot m'_\alpha \rangle$ for 
$m'_\alpha \in M_{\Upsilon^{\cal I}(\alpha)}$ and
$m_\gamma \in {\rm Hom}_\C(M_{\Upsilon^{\cal I}(\gamma)}, \C)$. It is
\begin{align}
  & \frac{1}{z_0} \delta \left( \frac{z_1-z_2}{z_0} \right)
  \left\langle m'_\alpha, \; {\cal Y}_{\beta,\gamma}^\alpha\left(
  e^{z_2L_1}(e^{\pi i (2r+1)}z_2^{-2})^{L_0} \cdot m_\beta, 1/z_2\right)
  \cdot Y_\gamma(R_{f@z_1}^{-1} \cdot v, 1/z_1) \cdot m_\gamma \right\rangle
  \nonumber \\
  & - \frac{1}{z_0} \delta \left( \frac{z_2-z_1}{-z_0}\right)
  \langle m'_\alpha, \; Y_\alpha(R_{f@z_1}^{-1} \cdot v, 1/z_1) \cdot
  {\cal Y}_{\beta,\gamma}^\alpha\left( e^{z_2L_1} (e^{\pi i (2+1)}z_2^{-2})^{L_0}
  \cdot m_\beta, 1/z_2 \right) \cdot m_\gamma \rangle \nonumber \\
 & -\frac{1}{z_2}\delta \left( \frac{z_1-z_0}{z_2} \right)         
  \langle m'_\alpha, \; {\cal Y}_{\beta,\gamma}^\alpha \left(
  e^{z_2L_1}(e^{\pi i (2r+1)}z_2^{-2})^{L_0} \cdot Y_\beta(v,z_0) \cdot
  m_\beta, 1/z_2 \right) \cdot m_\gamma \rangle,
  \label{eq:temp-xing-iso-inttw-prf}
\end{align}
where $R_{f@z}^{-1}(z) = e^{z L_1}(-z^{-2})^{L_0}$. 
We may rewrite the first two terms of this expression
(\ref{eq:temp-xing-iso-inttw-prf}) by using the Jacobi
identity on the intertwining operators of type
$(\alpha;\beta,\gamma)$, with the dummy variables
$\zeta_1 = 1/z_1$, $\zeta_2 = 1/z_2$, $\zeta_0 = -z_0/(z_1z_2)$:
\begin{align*}
  \frac{1}{z_1z_2} \frac{1}{1/z_2}
  \delta \left( \frac{\frac{1}{z_1}-\left(-\frac{z_0}{z_1z_2}\right)}{1/z_2}
  \right)
  \langle m'_\alpha, \; & {\cal Y}_{\beta,\gamma}^\alpha \left(
  Y_\beta ( R_{f@z_1}^{-1}\cdot v,-z_0/(z_1z_2)) \cdot \right.  \\
  & \qquad
  \left. e^{z_2L_1}(e^{\pi i (2r+1)}z_2^{-2})^{L_0} \cdot m_\beta, 1/z_2 \right) \cdot m_\gamma \rangle \\
  =  \frac{1}{z_2}\delta \left( \frac{z_1-z_0}{z_2} \right)
    \langle m'_\alpha, \; & {\cal Y}_{\beta,\gamma}^\alpha \left(
  Y_\beta ( R_{f@z_1}^{-1}\cdot v,-z_0/(z_1z_2)) \cdot \right.  \\
  & \qquad
  \left. e^{z_2L_1}(e^{\pi i (2r+1)}z_2^{-2})^{L_0} \cdot m_\beta, 1/z_2 \right) \cdot m_\gamma \rangle,
\end{align*}
where we used the relation (\ref{eq:rltn-delta-2}) at the equality
in the middle. To prove that $(\Theta_r {\cal Y}_{\beta,\gamma}^\alpha)$ is in
${\cal V}_{\beta,\Upsilon^{\cal I}(\alpha)}^{\Upsilon^{\cal I}(\gamma)}$, therefore, 
all we need to verify is the relation
\begin{align}
  \delta \left( \frac{z_1-z_0}{z_2} \right)
  R_{f@z_2} \cdot Y_\beta (R_{f@z_1}^{-1} \cdot v, -z_0/(z_1z_2)) \cdot
  R_{f@z_2}^{-1} 
  = \delta \left( \frac{z_1-z_0}{z_2} \right)  Y_\beta (v,z_0)
  \label{eq:temp2-xing-iso-inttw-prf}
\end{align}
for all $v \in V$. 

To verify this relation, we use (\ref{eq:FBZv-Lemma656}).
The relation (\ref{eq:FBZv-Lemma656}) 
can be read not just for $Y: V \rightarrow
{\rm End}(V)[[z,z^{-1}]]$ but also for $Y_\beta: V \rightarrow
{\rm End}(M_\beta)[[z,z^{-1}]]$ of any irreducible representation $\beta$.
The phase ambiguity in \ref{statmnt:prscr-oprtr-trnstnFcn-coordCh-int.op}
is not relevant here, because the conformal weight of $v \in V$
is integral. 

The relation (\ref{eq:temp2-xing-iso-inttw-prf}) is almost of the same
form as (\ref{eq:FBZv-Lemma656}), and the difference vanishes when
multiplided by $\delta ( (z_1-z_0)/z_2 )$. To see this,
note first that the function $(f@z_2)@z_0 = f@(z_2+z_0)$ is the same as
$f@z_1$ when multiplied by the delta function; here,
$(f@z)(t) = -t/[(t+z)z]$. So, the relation
(\ref{eq:temp2-xing-iso-inttw-prf}) would follow immediately from
(\ref{eq:FBZv-Lemma656}), if $(f@z_2)(t)|_{t=z_0}$ were $-z_0/(z_1z_2)$. 
Instead,
\begin{align}
  (f@z_2)(t_0) = - \frac{z_0}{z_1z_2} \quad
  {\rm if} \quad t_0 = \frac{z_0z_2}{z_1-z_0}. 
\end{align}
Because $t_0$ can be replaced by $z_0$ when multipled by the delta function,
the relation (\ref{eq:temp2-xing-iso-inttw-prf}) follows from
(\ref{eq:FBZv-Lemma656}) in fact. \qed

\vspace{5mm}

In the rest of this section \ref{ssec:review-vct-sp-intrtw-op},
we will review one more class of well-motivated choices of intertwining
operators (other than $Y_\gamma \in {\cal V}_{0,\gamma}^\gamma$
in \ref{statmnt:pair-ann-crtn}), and also collect
useful formulas that involve either those special choices or
crossing-symmetry isomorphisms. We will pay enough attention to
the complex phases involved in those formulas. 
\begin{anythng}
\label{statmnt:pair-ann-crtn}
{\bf the pair annihilation vertex:} 
There is no {\it unique} choice of an intertwining operator of type
$(0;\gamma, \Upsilon^{\cal I}(\gamma))$ better motivated than others.
A sequence of the isomorphisms
$\Theta_r \cdot \Omega_s: {\cal V}_{0,\gamma}^\gamma \rightarrow
{\cal V}_{\gamma,0}^\gamma \rightarrow {\cal V}_{\gamma,\Upsilon^{\cal I}(\gamma)}^0$
determines ($s$-independent) 
  $\Theta_r \cdot \Omega(Y_\gamma) \in {\cal V}_{\gamma,\Upsilon^{\cal I}(\gamma)}^0$,
%
which we call a {\it chiral pair annihilation vertex}. 
Obviously
\begin{align}
  \Theta_{r+\Delta r}\Omega Y_\gamma =
  e^{2\pi i [h_\gamma] \Delta r} (\Theta_r \Omega Y_\gamma)
  = \Theta_r \left( \theta^{\Delta r}(\Omega Y_\gamma) \right),
  \label{eq:reltn-pairAnnV-var}
\end{align}
so there is a variation of choices---labeled by $r \in \Z$---of
well-motivated intertwining operators in 
${\cal V}_{\gamma,\Upsilon^{{\cal I}}(\gamma)}^0$ unless $[h_\gamma] = 0_{+\Z}$.

It is an option to use $e^{-\pi i (2r+1) h_{\gamma,0}} \Theta_r \Omega Y_\gamma$
as a well-motivated choice from ${\cal V}_{\gamma,\Upsilon^{\cal I}(\gamma)}^0$
(e.g., \cite{MR2387861} and many other references); 
this choice is independent of $r \in \Z$, because of the $r$-dependence
in (\ref{eq:reltn-pairAnnV-var}). We should still keep in mind that
the vertex may change its sign when the conformal weight $h_{\gamma,0}$ of
the ground state of $M_\gamma$ is replaced by the conformal weight
of other excited states in $M_\gamma$.

The skew-symmetry isomorphism maps a chiral pair annihilation vertex
to another in the contragradient representation
%
%
\begin{align}
  \Omega_s: {\cal V}_{\gamma,\Upsilon^{{\cal I}}(\gamma)}^0 \ni
  (\Theta_r \Omega Y_\gamma)  
  \longmapsto (\Theta_{r-1-2s} \Omega Y_{\Upsilon^{\cal I}(\gamma)})
  = \theta^{-(2s+1)} \left( \Theta_r \Omega Y_{\Upsilon^{\cal I}(\gamma)} \right)
  \in {\cal V}_{\Upsilon^{{\cal I}}(\gamma),\gamma}^0. 
 \label{eq:reltn-pairAnnV-conj-repr}
\end{align}
It is possible to use the definition of the crossing symmetry isomorphism
to prove this relation; the easiest way to see this, however, is
to evaluate a pair $(m_\gamma, m'_{\Upsilon(\gamma)}) \in M_\gamma
\times M_{\Upsilon^{\cal I}(\gamma)}$ of Virasoro primary states with a common
conformal weight by both $\Omega_s (\Theta_r\Omega Y_\gamma)$ and
$\Theta_{r-1-2s}\Omega Y_{\Upsilon^{{\cal I}}(\gamma)}$. 

The variations in the chiral pair annihilation vertices ($r \in \Z$) originate
from the phase ambiguity associated with the Jacobian of the coordinate
transformation $\zeta \rightarrow 1/\zeta$. It is useful to assign a
graphical presentation in Fig. \ref{fig:pair-ann} to the pair annihilation
vertices. 
\begin{figure}[tbp]
  \begin{center}
  \includegraphics[scale=0.7]{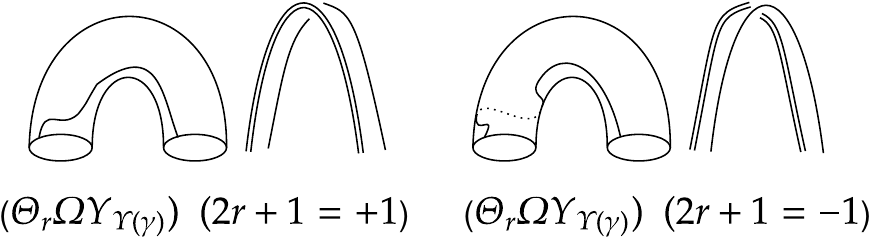}  
  \end{center}
  \caption{ \label{fig:pair-ann}The graphical presentation of chiral pair
    annihilation
vertices: topologically different choices of a solid line on a cylinder or a
double line edge on a ribbon are used to distinguish different chiral pair
annihilation vertices. }
\end{figure}
The relations (\ref{eq:reltn-pairAnnV-var})
and (\ref{eq:reltn-pairAnnV-conj-repr}) are also given a graphical
presentation as in Fig. \ref{fig:}
\begin{figure}[btp]
  \begin{center}
    \begin{tabular}{cc}
      \includegraphics[scale=0.5]{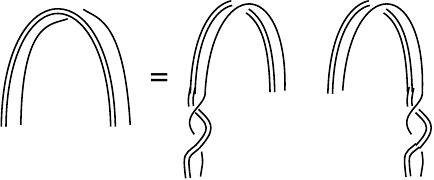} &
      \includegraphics[scale=0.5]{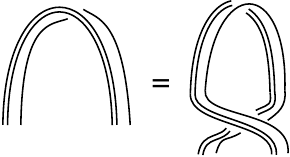} \\
      $\Theta_0\Omega Y_{\Upsilon(\gamma)} = \theta(\Theta_{-1}\Omega Y_{\Upsilon(\gamma),-1})$ &
      $(\Theta_0\Omega Y_{\gamma}) = \Omega_{-1}
      (\Theta_{-1}\Omega Y_{\Upsilon(\gamma)})$
    \end{tabular}
  \end{center}
  \caption{ \label{fig:}The graphical presentation of the relations
    (\ref{eq:reltn-pairAnnV-var}) and (\ref{eq:reltn-pairAnnV-conj-repr}).
}
\end{figure}

The pairing between states in the dual representations is independent
of the choice of local coordinates at the insertion of those states.
The pairing is related to the chiral pair annihilation vertices in this way: 
\begin{align}
  \langle e^{xL_{-1}} \cdot m, \; e^{\xi L_{-1}} \cdot m' \rangle & \; =
  \bra{{\bf 1}} (\Theta_s\Omega Y_\gamma)\left( e^{xL_1}(e^{-\pi  i(2s+1)}x^{-2})^{L_0} \cdot m, 1/x \right) \cdot e^{\xi L_{-1}} \cdot m' \rangle , \nonumber \\
  & \; = \bra{\bf 1} (\Theta_t\Omega Y_{\Upsilon^{{\cal I}}(\gamma)})\left(
  e^{\xi L_1} (e^{-\pi i(2t+1)}\xi^{-2})^{L_0} \cdot m', 1/\xi \right) \cdot
  e^{x L_{-1}} \cdot m \rangle.
  \label{eq:useful-cnvrt-dualParng2oneSide}
\end{align}
\end{anythng}

\begin{anythng}
\label{statmnt:Theta-by-FnOmega}
  {\bf the crossing symmetry isomorphisms $\Theta_r$ by $\Omega_s$ and
    ${\cal F}$:}
The crossing symmetry isomorphisms $\Theta_r: {\cal V}_{\beta,\gamma}^\alpha \ni {\cal Y} \mapsto (\Theta_r {\cal Y}) \in {\cal V}_{\beta,\Upsilon^{\cal I}(\alpha)}^{\Upsilon^{\cal I}(\gamma)}$ are implemented by the fusion and skew-symmetry isomorphisms ${\cal F}$ and $\Omega_s$,
in fact, by using special choices of intertwining operators $Y_\gamma \in {\cal V}_{0,\gamma}^\gamma$, $\Omega Y_\gamma \in {\cal V}_{\gamma,0}^\gamma$ and
$\Theta_t \Omega Y_\gamma \in {\cal V}_{\gamma,\Upsilon^{\cal I}(\gamma)}^0$. 
This claim is reviewed below by refining \cite[(3.8b)=(4.1d)]{Moore:1988qv}.
With this observation, it is then enough to deal only with ${\cal F}$ and
$\Omega$ along with the pentagon and hexagon identities, because the braiding ${\cal B}$ and crossing symmetry $\Theta$ isomorphisms are implemented by ${\cal F}$ and $\Omega$. 

To be more precise, $\Theta_r {\cal Y}$ for ${\cal Y} \in {\cal V}_{\beta,\gamma}^\alpha$ is not of type $(\Upsilon^{\cal I}(\gamma); \beta,\Upsilon^{\cal I}(\alpha))$, but of type $(\gamma^\vee; \beta,\alpha^\vee)$.
We are not concerned about this difference here in \ref{statmnt:Theta-by-FnOmega} and \ref{statmnt:ThetaTheta-vs-fusion}, 
but will raise the level of precision in (\ref{eq:cond-Ytot-skew-N-xing}),
and also in section \ref{sec:Z2}.

To start off, let us digress for a moment to review a similar case where a special choice of intertwining operators simplifies isomorphisms (\cite[(2.3a)]{Moore:1988uz}, \cite[(3.8), (4.1)]{Moore:1988qv}). 
It is an immediate consequence of the defining property (\ref{eq:def-vct-sp-int.op}) of intertwining operators that
\begin{align}
  {\cal F}: & \; 
    {\cal V}_{0,\alpha}^\alpha \otimes {\cal V}_{\beta,\gamma}^\alpha
  \ni Y_\alpha \otimes {\cal Y} \longmapsto {\cal Y} \otimes Y_\beta \in 
      {\cal V}_{\beta,\gamma}^\alpha \otimes {\cal V}_{0,\beta}^\beta ,
      \label{eq:def-vct-sp-int.op-varF1} \\
  {\cal B}: & \; {\cal V}_{0,\alpha}^\alpha \otimes {\cal V}_{\beta,\gamma}^\alpha
  \ni Y_\alpha \otimes {\cal Y} \longmapsto {\cal Y} \otimes Y_\gamma \in       
  {\cal V}_{\beta,\gamma}^\alpha \otimes {\cal V}_{0,\gamma}^\gamma.
\end{align}
The latter may also be translated by using ${\cal B} = {\cal F}^{-1}\cdot \Omega \cdot {\cal F} = \Omega \cdot {\cal F}^{-1} \cdot \Omega^{-1}$ (hexagon identity
in the latter equality): 
\begin{align}
  {\cal F}: {\cal V}_{\beta,\gamma}^\alpha \otimes {\cal V}_{\gamma,0}^\gamma \ni
  {\cal Y} \otimes (\Omega Y_\gamma) \longmapsto
  (\Omega Y_\alpha) \otimes {\cal Y} \in 
  {\cal V}_{\alpha,0}^\alpha \otimes {\cal V}_{\beta,\gamma}^\alpha . 
  \label{eq:def-vct-sp-int.op-varF2}    
\end{align}
By using the box identity along with (\ref{eq:def-vct-sp-int.op-varF2}),
we also have 
\begin{align}
  {\cal B}_{-1-r}: {\cal V}_{\beta,\gamma}^\alpha \otimes {\cal V}_{\gamma,0}^\gamma
  \ni {\cal Y} \otimes (\Omega Y_\gamma) \longmapsto 
  (\Omega_r {\cal Y})\otimes (\Omega Y_\beta) \in 
  {\cal V}_{\gamma,\beta}^\alpha \otimes {\cal V}_{\beta,0}^\beta . 
\end{align}
The braiding isomorphisms can be built from the skewsymmetry isomorphisms
(the box identity), and the skewsymmetry isomorphisms can also be reproduced
from the braiding isomorphisms by using the well-motivated choices of
chiral vertices \cite[(3.8a), (4.1c)]{Moore:1988qv}. 

Now, let us go back to the relation between
${\cal F}$, $\Omega$ and $\Theta$. We claim that 
\begin{align}
  \Omega_s \cdot {\cal F}:
  {\cal V}_{\Upsilon^{\cal I}(\alpha),\alpha}^0 \otimes {\cal V}_{\beta,\gamma}^\alpha
  & \; \ni (\Theta_s\Omega Y_{\Upsilon^{\cal I}(\alpha)})\otimes {\cal Y}
  \label{eq:xing-iso-by-FnOmega-a} \\ & \; \longmapsto
  (\Theta_s\Omega Y_{\Upsilon^{\cal I}(\gamma)}) \otimes (\Theta_{-1-s}{\cal Y}) \in 
       {\cal V}_{\Upsilon^{\cal I}(\gamma),\gamma}^0 \otimes
       {\cal V}_{\beta,\Upsilon^{\cal I}(\alpha)}^{\Upsilon^{\cal I}(\gamma)} \nonumber
\end{align}
for any $s \in \Z$. The crossing symmetry isomorphism $\Theta$ is therefore
reproduced from the fusion isomorphism along with a well-motivated choice
of a chiral vertex (cf \cite[(3.8b), (4.1d)]{Moore:1988qv}), and in particular,
${\cal B}$ and $\Theta$ can be derived from ${\cal F}$ and $\Omega$. 
Note that the formula above only describes the image of an element
labeled by $s \in \Z$ by the map labeled by the same $s$, but this 
is enough because $\Omega_{s+\Delta s} =
\Omega_{s} e^{2\pi i \Delta s([h_\gamma]-[h_\beta]-[h_\alpha])}$, and
${\cal V}_{\Upsilon^{\cal I}(\alpha),\alpha}^0 \cong
\C (\Theta_s \Omega Y_{\Upsilon^{\cal I}(\alpha)})$. 
\proof In eq. (\ref{eq:def-xing-sym-iso}) defining a crossing symmetry
isomorphism, let us substitute $m'_\alpha \rightarrow e^{x L_{-1}} m'_\alpha
\in M_{\Upsilon^{\cal I}(\alpha)}[[x]][x^{-1}]$
and $m_\gamma \rightarrow e^{\xi L_{-1}} \cdot m_\gamma \in
M_\gamma[[\xi]][\xi^{-1}]$. Then LHS = RHS, where
\begin{align*}
  LHS & \; = \langle e^{\xi L_{-1}} \cdot m_\gamma, \; e^{xL_{-1}} \cdot
  (\Theta_r {\cal Y})(m_\beta, z-x) \cdot m'_\alpha \rangle,  \\
  & \; = \langle {\bf 1}, (\Theta_h \Omega Y_{\Upsilon^{\cal I}(\gamma)})\left[
  e^{x L_1} (e^{-\pi i (2h+1)}x^{-2})^{L_0} \cdot
  (\Theta_r{\cal Y})( m_\beta, z-x) \cdot m'_\alpha, \; 1/x \right] \cdot
  e^{\xi L_{-1}} \cdot m_\gamma \rangle 
\end{align*}
for any $h \in \Z$ as we may use (\ref{eq:useful-cnvrt-dualParng2oneSide}),
and
\begin{align*}
  RHS & \; = \langle e^{x L_{-1}} \cdot m'_\alpha,\ ; e^{\xi L_{-1}} \cdot
  {\cal Y} \left( e^{z L_1} (e^{\pi i (2r+1)}z^{-2})^{L_0} \cdot m_\beta, z^{-1}-\xi
  \right) \cdot m_\gamma \rangle \\
  & \; = \langle {\bf 1}, \; (\Theta_s\Omega Y_{\Upsilon^{\cal I}(\alpha)})
  \left( e^{xL_1}(e^{-\pi i(2s+1)}x^{-2})^{L_0} \cdot m'_\alpha, 1/x \right) \\
  & \qquad \qquad \qquad \qquad \qquad \qquad 
  \cdot
  {\cal Y}\left( e^{zL_1} (e^{\pi i(2r+1)}z^{-2})^{L_0} \cdot m\beta, 1/z \right)
  \cdot e^{\xi L_{-1}} \cdot m_\gamma \rangle 
\end{align*}
for any $s\in \Z$. As we can parametrize the image of
$(\Theta_s \Omega Y_{\Upsilon^{\cal I}(\alpha)}) \otimes {\cal Y} \in
{\cal V}_{\Upsilon^{\cal I}(\alpha),\alpha}^0 \otimes {\cal V}^{\alpha}_{\beta,\gamma}$ by
${\cal F}: {\cal V}_{\Upsilon^{\cal I}(\alpha),\alpha}^0 \otimes
{\cal V}_{\beta,\gamma}^\alpha \rightarrow
{\cal V}_{\Upsilon^{\cal I}(\gamma),\gamma}^0 \otimes
{\cal V}_{\Upsilon^{\cal I}(\alpha),\beta}^{\Upsilon^{\cal I}(\gamma)}$ by
$(\Theta_h \Omega Y_{\Upsilon^{\cal I}(\gamma)}) \otimes {\cal Y}_1$
(because $\dim_\C {\cal V}_{\Upsilon^{\cal I}(\gamma),\gamma}^0=1$), 
\begin{align*}
  RHS = \langle {\bf 1}, \; (\Theta_h\Omega Y_{\Upsilon^{\cal I}(\gamma)}) & \;
  \left[
    {\cal Y}_1\left( e^{xL_1}(e^{-\pi i(2s+1)}x^{-2})^{L_0} \cdot m'_\alpha, \;
    x^{-1}-z^{-1} \right) \right.  \\
    & \qquad \qquad \qquad
    \left. \cdot e^{zL_1}(e^{\pi i(2r+1)}z^{-2})^{L_0} \cdot
    m_\beta, \; 1/z \right] \cdot e^{\xi L_{-1}} \cdot m_\gamma \rangle. 
\end{align*}
For the equality LHS = RHS to hold for arbitrary $\xi$ and $m_\gamma$,
we should have $(*)_L = (*)_R$, where
\begin{align*}
  (*)_L := e^{\frac{1}{x} L_{-1}} \cdot e^{xL_1} (e^{-\pi i(2h+1)}x^{-2})^{L_0} \cdot
  (\Theta_r{\cal Y})\left( m_\beta, z-x \right) \cdot m'_\alpha, 
\end{align*}
while
\begin{align*}
 (*)_R := e^{\frac{1}{z}L_{-1}} \cdot {\cal Y}_1\left( e^{x L_1} (e^{-\pi i(2s+1)}x^{-2})^{L_0} \cdot m'_\alpha, \; 1/x - 1/z\right) \cdot e^{zL_1} (e^{\pi i (2r+1)}z^{-2})^{L_0} \cdot m_\beta. 
\end{align*}
Remember that ${\cal Y}_1$ depends implicitly on $s, h\in \Z$;
we will focus on the choice $h=s$ as we will not lose anything by doing so
after this, but still ${\cal Y}_1$ depends on $s =h$ because
$[h_\alpha]$ is not necessarily the same as $[h_\gamma]$.

Let us rewrite $(*)_R$ in a way that it looks closer to $(*)_L$. First, 
\begin{align*}
  (*)_R = e^{x^{-1} L_{-1}} (\Omega_t {\cal Y}_1) \left[ e^{zL_1}(e^{\pi i(2r+1)}z^{-2})^{L_0} \cdot m_\beta, e^{-\pi i(2t+1)}\left( \frac{1}{x}-\frac{1}{z}\right) \right]
  \\
  \cdot e^{xL_1}(e^{-\pi i(2s+1)}x^{-2})^{L_0} \cdot m'_\alpha 
\end{align*}
for any $t \in \Z$. This expression can be rewritten by using the
transformation law (\ref{eq:FBZv-Lemma656}) under the coordinate change
on $\P^1$,
so long as $z \nin x \R_{>1}$ as we can fix the phase ambiguity properly
then as discussed in \ref{statmnt:prscr-oprtr-trnstnFcn-coordCh-int.op});
the present situation corresponds to $\eta_0 = x$, $\eta(p) = z-x$,
$\xi(p) = e^{-\pi i(2t+1)} (x^{-1}-z^{-1})$, and $A = e^{-\pi i(2s+1)}$. 
So, we need to choose $t=s$. Now,
\begin{align*}
  (*)_R = e^{x^{-1}L_{-1}} e^{xL_1}(e^{-\pi i (2s+1)}x^{-2})^{L_0} \cdot
  (\Omega_s {\cal Y})\left[ e^{\pi i (2r+1+2s+1) L_0} \cdot m_\beta, \; z-x \right]
  \cdot m'_\alpha. 
\end{align*}
The relation $(*)_L = (*)_R$ characterizing the crossing symmetry isomorphism
is therefore translated to the relation $\Theta_r {\cal Y} = \Omega_s{\cal Y}_1
\cdot e^{\pi i (2r+1+2s+1) [h_\beta]}$. To summarize, $\Omega_s \cdot {\cal F}$
maps $(\Theta_s \Omega Y_{\Upsilon^{\cal I}(\alpha)}) \otimes {\cal Y} \in
{\cal V}_{\Upsilon^{\cal I}(\alpha),\alpha}^0 \otimes {\cal V}_{\beta,\gamma}^\alpha$
to $(\Theta_s\Omega Y_{\Upsilon^{\cal I}(\gamma)}) \otimes {\cal Y}_1 \in
{\cal V}_{\Upsilon^{\cal I}(\gamma),\gamma}^0 \otimes
{\cal V}_{\Upsilon^{\cal I}(\alpha),\beta}^{\Upsilon^{\cal I}(\gamma)}$ and then to
$(\Theta_s \Omega Y_{\Upsilon^{\cal I}(\gamma)}) \otimes (\Theta_r{\cal Y})
\cdot e^{\pi i (2r+1+2s+1) h_\beta}  \in {\cal V}_{\Upsilon^{\cal I}(\gamma),\gamma}^0
\otimes {\cal V}_{\beta,\Upsilon^{\cal I}(\alpha)}^{\Upsilon^{\cal I}(\gamma)}$.  \qed 

There are a few more variations to (\ref{eq:xing-iso-by-FnOmega-a}).
One version follows immediately from (\ref{eq:xing-iso-by-FnOmega-a}):
\begin{eqnarray}
  (\Omega_s \otimes \Omega_s) \cdot {\cal F}: {\cal V}_{\Upsilon^{\cal I}(\alpha),\alpha}^0 \otimes {\cal V}_{\beta,\gamma}^{\alpha} & \rightarrow 
  {\cal V}_{\Upsilon^{\cal I}(\alpha),\beta}^{\Upsilon^{\cal I}(\gamma)} \otimes
  {\cal V}_{\Upsilon^{\cal I}(\gamma),\gamma}^0 \rightarrow & 
  {\cal V}_{\gamma, \Upsilon^{\cal I}(\gamma)}^0 \otimes
  {\cal V}_{\beta, \Upsilon^{\cal I}(\alpha)}^{\Upsilon^{\cal I}(\gamma)} ,
  \label{eq:xing-iso-by-FnOmega-b}  \\
  (\Theta_s \Omega Y_{\Upsilon^{\cal I}(\alpha)}) \otimes {\cal Y} & \longmapsto &
  (\Theta_{-1-s}\Omega Y_\gamma) \otimes (\Theta_{-1-s}{\cal Y}) ;
  \nonumber 
 \end{eqnarray}
we just used (\ref{eq:reltn-pairAnnV-conj-repr}) on top
of (\ref{eq:xing-iso-by-FnOmega-a}), and another \cite[(4.7)]{MR2387861}
is ($\Omega_{-1-s} = \Omega_s^{-1}$)
\begin{align}
  {\cal F}: {\cal V}_{\Upsilon^{\cal I}(\alpha),\alpha}^0 \otimes
  {\cal V}_{\beta,\gamma}^\alpha & \; \longrightarrow 
          {\cal V}_{\Upsilon^{\cal I}(\gamma),\gamma}^0 \otimes
          {\cal V}_{\Upsilon^{\cal I}(\alpha),\beta}^{\Upsilon^{\cal I}(\gamma)} ,
      \label{eq:xing-iso-by-FnOmega-d}    \\
  (\Theta_s \Omega Y_{\Upsilon^{\cal I}(\alpha)}) \otimes {\cal Y}
  & \; \longmapsto (\Theta_s \Omega Y_{\Upsilon^{\cal I}(\gamma)}) \otimes
  (\Omega_{-1-s} \Theta_{-1-s} {\cal Y}).   \nonumber
\end{align}

Another version is to use the isomorphism 
\begin{align*}
  {\cal F}\cdot \Omega_r \cdot {\cal F}^{-1}:
  {\cal V}_{((\beta \cdot \gamma) \cdot \Upsilon^{\cal I}(\alpha))}^0
\rightarrow
  {\cal V}_{(\beta \cdot (\gamma \cdot \Upsilon^{\cal I}(\alpha)))}^0 
\rightarrow
   {\cal V}_{(\beta \cdot (\Upsilon^{\cal I}(\alpha) \cdot \gamma))}^0 
   \rightarrow
   {\cal V}_{((\beta \cdot \Upsilon^{\cal I}(\alpha))\cdot \gamma)}^0 ; 
\end{align*}
this is an attempt\footnote{
The relation between $\Theta(\pm)$ in \cite{Moore:1988qv} and $\Theta_{r}$
or $\Theta[\pm]$ here is quite likely not very simple (such as
$2r+1 \in \{\pm 1\}$), to say the least; in Ref. \cite{Moore:1988qv},
$\Theta(+) = e^{2\pi i(h_\beta+h_\gamma-h_\alpha)} \Theta(-)$, while
$\Theta_0 = e^{2\pi i h_\beta} \Theta_{-1}$ here.
Ref. \cite{Moore:1988qv} also has a view that there is a unique
natural choice of chiral pair annihilation vertex from
${\cal V}_{\Upsilon^{\cal I}(\alpha),\alpha}^0$ and also a unique and
natural one from ${\cal V}_{\Upsilon^{\cal I}(\gamma),\gamma}^0$, but
we are not sure which one it is (cf. \ref{statmnt:pair-ann-crtn}).
} %
to make sense of the
statement \cite[(3.8b)=(4.1d)]{Moore:1988qv}. It should be noted
that this looks similar to the braiding isomorphism ${\cal B}_{-1-r}$
but it is not. We claim that
\begin{align}
   {\cal F}\cdot \Omega_r \cdot {\cal F}^{-1}: & \; 
   {\cal V}_{\beta,\gamma}^{\alpha}\otimes {\cal V}_{\alpha,\Upsilon^{\cal I}(\alpha)}^0 
   \ni 
      {\cal Y} \otimes (\Theta_s\Omega Y_{\alpha}) \longmapsto
       \label{eq:xing-iso-by-FnOmega-c} \\
 &  (\Theta_{t}{\cal Y}) \otimes (\Theta_u\Omega Y_\gamma) \cdot
 e^{2\pi i h_\alpha(s-r)}
 e^{2\pi i h_\beta(r-t)}
 e^{2\pi i h_\gamma (-r-u-1)} \in
 {\cal V}_{\beta,\Upsilon^{\cal I}(\alpha)}^{\Upsilon^{\cal I}(\gamma)}
  \otimes {\cal V}_{\Upsilon^{\cal I}(\gamma),\gamma}^0.  \nonumber 
\end{align}
The phase factors at the end vanish if $r=s=t$ and $u = -1-r$. 
\proof We prove this (\ref{eq:xing-iso-by-FnOmega-c}) for the case
$2r+1=2s+1=2t+1=-\epsilon$, $2u+1= \epsilon$ ($\epsilon \in \{ \pm 1\}$),
first, by using the hexagon identities and (\ref{eq:xing-iso-by-FnOmega-b}).
Generalizations to arbitrary $r,s,t,u \in \Z$ are straightforward. 

Note, first, that the isomorphism
$(\Omega(\epsilon)\otimes \Omega(\epsilon))\cdot {\cal F}$
in (\ref{eq:xing-iso-by-FnOmega-b}) is sandwiched by $\Omega(-\epsilon)$
to yield another isomorphism
\[
\Omega(-\epsilon) \cdot \left( (\Omega(\epsilon)\otimes \Omega(\epsilon))
\cdot {\cal F} \right) \cdot \Omega(-\epsilon):
      {\cal V}_{\beta,\gamma}^\alpha \otimes {\cal V}_{\alpha,\Upsilon^{\cal I}(\alpha)}^0
      \rightarrow {\cal V}_{\beta,\Upsilon^{\cal I}(\alpha)}^{\Upsilon^{\cal I}(\gamma)}
      \otimes {\cal V}_{\Upsilon^{\cal I}(\gamma),\gamma}^0.
\]
The result (\ref{eq:xing-iso-by-FnOmega-b}) implies, by
using (\ref{eq:reltn-pairAnnV-conj-repr}), that this isomorphism maps
\begin{align}
  {\cal Y} \otimes (\Theta[-\epsilon]\Omega Y_{\alpha}) \longmapsto
  (\Theta[-\epsilon]\Omega {\cal Y}) \otimes (\Theta[\epsilon]\Omega Y_\gamma); 
\end{align}
here, $\Theta[\epsilon] := \Theta_s|_{2s+1= -\epsilon}$ for
$\epsilon \in \{ \pm 1\}$. It is possible to use the hexagon identity
to see that the isomorphism
$\Omega(-\epsilon) \cdot \left( (\Omega(\epsilon)\otimes \Omega(\epsilon))
\cdot {\cal F} \right) \cdot \Omega(-\epsilon)$ is the same as
${\cal F} \cdot \Omega(-\epsilon) \cdot {\cal F}^{-1}$; in fact,
it is enough to compare the braid group elements corresponding
to the two isomorphisms. \qed 
\end{anythng}

\begin{anythng}
\label{statmnt:ThetaTheta-vs-fusion}
{\bf the crossing symmetry isomorphism before and after fusion:}
There are two ways in applying the crossing symmetry isomorphisms to
the chain of two intertwining operators, ${\cal Y}_i \otimes {\cal Y}_j \in 
{\cal V}_{\beta_1,\delta}^\alpha \otimes {\cal V}_{\beta_2,\gamma}^\delta$.
One is to apply the crossing symmetry isomorphism on ${\cal Y}_i$ and
${\cal Y}_j$ separately, bringing it to an operator in
${\cal V}_{\beta_1,\Upsilon^{\cal I}(\alpha)}^{\Upsilon^{\cal I}(\delta)}
\otimes {\cal V}_{\beta_2, \Upsilon^{\cal I}(\delta)}^{\Upsilon^{\cal I}(\gamma)}$.
The other is to apply the fusion isomorphism first, then the crossing symmetry
isomorphism, and the inverse of the fusion isomorphism at the end.
Here we describe the way the two chains of isomorphisms are precisely equal;
see (\ref{eq:ThetaTheta-after-fusion}).

Think of a couple of isomorhisms ${\cal V}_{(\Upsilon^{\cal I}(\alpha) \cdot (\beta_1 \cdot (\beta_2 \cdot \gamma)))}^0 \rightarrow
{\cal V}_{(\gamma \cdot (\beta_2 \cdot (\beta_1 \cdot \Upsilon^{\cal I}(\alpha))))}^0$.
One is via the composition of 
\begin{align}
  (\Omega(\epsilon) \cdot {\cal F}): & \; 
  {\cal V}_{\Upsilon^{\cal I}(\alpha),\alpha}^0 \otimes
  {\cal V}_{\beta_1,\delta}^\alpha \rightarrow
  {\cal V}_{\beta_1,\Upsilon^{\cal I}(\alpha)}^{\Upsilon^{\cal I}(\delta)} \otimes
  {\cal V}_{\Upsilon^{\cal I}(\delta),\delta}^0, \nonumber \\
  (\Omega(\epsilon)\otimes \Omega(\epsilon)) \cdot {\cal F}:  & \;
  {\cal V}_{\Upsilon^{\cal I}(\delta),\delta}^0 \otimes {\cal V}^\delta_{\beta_2,\gamma}
  \rightarrow {\cal V}_{\beta_2,\Upsilon^{\cal I}(\delta)}^{\Upsilon^{\cal I}(\gamma)}
  \otimes {\cal V}^0_{\gamma,\Upsilon^{\cal I}(\gamma)},
  \label{eq:Theta-Theta-a}
\end{align}
which maps 
\begin{align}
  (\Theta[\epsilon]\Omega Y_{\Upsilon^{\cal I}(\alpha)}) \otimes
  {\cal Y}_i \otimes {\cal Y}_j \longmapsto
  (\Theta[-\epsilon]\Omega Y_\gamma) \otimes
  (\Theta[-\epsilon] {\cal Y}_j) \otimes (\Theta[-\epsilon]{\cal Y}_i); 
  \label{eq:Theta-Theta-a-result}
\end{align}
we have applied (\ref{eq:xing-iso-by-FnOmega-a},
\ref{eq:xing-iso-by-FnOmega-b}) twice. An alternative route to
obtain the result (\ref{eq:Theta-Theta-a-result}) is to use
the composition of the isomorphisms 
\begin{align}
  (\Omega(\epsilon)\otimes \Omega(\epsilon)) \cdot {\cal F}:
  & \;   {\cal V}_{\Upsilon^{\cal I}(\alpha),\alpha}^0 \otimes
  {\cal V}_{\beta_1,\delta}^\alpha \rightarrow
  {\cal V}_{\beta_1,\Upsilon^{\cal I}(\alpha)}^{\Upsilon^{\cal I}(\delta)} \otimes
  {\cal V}_{\delta, \Upsilon^{\cal I}(\delta)}^0, \nonumber \\
  (\Omega(-\epsilon)\cdot {\cal F})^{-1} = ({\cal F}^{-1}\cdot \Omega(\epsilon)):
  & \; {\cal V}_{\beta_2,\gamma}^\delta \otimes {\cal V}_{\delta,\Upsilon^{\cal I}(\delta)}^0
  \rightarrow 
  {\cal V}_{\gamma,\Upsilon^{\cal I}(\gamma)}^0
  \otimes {\cal V}_{\beta_2,\Upsilon^{\cal I}(\delta)}^{\Upsilon^{\cal I}(\gamma)}.
  \label{eq:Theta-Theta-b}
\end{align}
That the two routes (\ref{eq:Theta-Theta-a}) and (\ref{eq:Theta-Theta-b})
yield the same isomorphism follows from the fact that the braid group elements
for the two routes are the same.

Another isomorphism is the composition of
\begin{align}
  {\cal F}: & \; \oplus_\delta \left( {\cal V}_{\beta_1,\delta}^\alpha \otimes
  {\cal V}_{\beta_2,\gamma}^\delta \right) \rightarrow
  \oplus_\epsilon \left( {\cal V}_{\epsilon,\gamma}^\alpha \otimes
         {\cal V}_{\beta_1,\beta_2}^\epsilon \right),
  \nonumber \\
  (\Omega(\epsilon)\otimes \Omega(\epsilon)) \cdot {\cal F}: & \;
  {\cal V}_{\Upsilon^{\cal I}(\alpha),\alpha}^0 \otimes
  {\cal V}_{\epsilon,\gamma}^\alpha \rightarrow
  {\cal V}_{\gamma,\Upsilon^{\cal I}(\gamma)}^0 
  {\cal V}_{\epsilon,\Upsilon^{\cal I}(\alpha)}^{\Upsilon^{\cal I}(\gamma)} , \nonumber \\
  (\Omega(-\epsilon)\cdot {\cal F})^{-1} =
  ({\cal F}^{-1}\cdot \Omega(\epsilon)): & \; \oplus_{\epsilon} \left(
  {\cal V}_{\beta_1,\beta_2}^\epsilon \otimes
  {\cal V}_{\epsilon,\Upsilon^{\cal I}(\alpha)}^{\Upsilon^{\cal I}(\gamma)}
  \right) \rightarrow \oplus_{\delta} \left(
        {\cal V}_{\beta_2,\Upsilon^{\cal I}(\delta)}^{\Upsilon^{\cal I}(\gamma)} \otimes
        {\cal V}_{\beta_1,\Upsilon^{\cal I}(\alpha)}^{\Upsilon^{\cal I}(\delta)}
        \right). 
\label{eq:Theta-Theta-c}        
\end{align}
This chain of isomorphisms is determined so that the braid group element
of (\ref{eq:Theta-Theta-c}) is the same as that of (\ref{eq:Theta-Theta-a},
\ref{eq:Theta-Theta-b}). So, the isomorphism (\ref{eq:Theta-Theta-c})
must also yield the result (\ref{eq:Theta-Theta-a-result}).
Therefore, by stripping off the chiral pair annihilation vertices
$\Theta[\epsilon]\Omega Y_{\Upsilon^{\cal I}(\alpha)}$ and
$\Theta[-\epsilon]\Omega Y_\gamma$ that are common to the two routes,
we have (cf \cite[(4.9a)]{Moore:1988qv})
\begin{eqnarray}
  {\cal F} & ({\rm id}\otimes \Theta[-\epsilon]) &
  \left(\Omega(-\epsilon) \cdot {\cal F} \right)^{-1}
  \label{eq:ThetaTheta-after-fusion} \\
  {\cal V}_{\beta_1,\delta}^\alpha \otimes {\cal V}_{\beta_2,\gamma}^\delta \rightarrow
  & \; \oplus_\epsilon \left( {\cal V}_{\beta_1,\beta_2}^\epsilon \otimes
  {\cal V}_{\epsilon,\gamma}^\alpha \right) \rightarrow \oplus_\epsilon \left(
  {\cal V}_{\beta_1,\beta_2}^\epsilon \otimes
  {\cal V}_{\epsilon,\Upsilon^{\cal I}(\alpha)}^{\Upsilon^{\cal I}(\gamma)} \right)
  & \; \rightarrow
  {\cal V}_{\beta_2, \Upsilon^{\cal I}(\delta)}^{\Upsilon^{\cal I}(\gamma)} \otimes
  {\cal V}_{\beta_1,\Upsilon^{\cal I}(\alpha)}^{\Upsilon^{\cal I}(\delta)}, \nonumber \\
  {\cal Y}_i \otimes {\cal Y}_j \quad & \longmapsto & 
  (\Theta[-\epsilon] {\cal Y}_j) \otimes (\Theta[-\epsilon]{\cal Y}_i).
  \nonumber 
\end{eqnarray}
To put it in colloquial language, applying the crossing symmetry isomorphism
$\Theta[-\epsilon]$ on both ${\cal Y}_i$ and ${\cal Y}_j$ is the same as
applying $\Theta[-\epsilon]$ on the joined vertex in
${\cal V}_{\epsilon,\gamma}^{\alpha}$ in the fused
description, if that is acompanied by $\Omega(\epsilon)$
on the vertex in ${\cal V}_{\beta_1,\beta_2}^\epsilon$ that fuses the two
inserted states.  
\end{anythng}

\subsection{Algorithms to Compute Intertwining Operators}
\label{ssec:algrthm-int.op}

Here, we address the question of how to compute the vector space
${\cal V}_{\beta,\gamma}^\alpha$ of intertwining operators explicitly,
given a VOA $(V,Y, \cdots)$ and a set ${\cal I}$ of its unitary
representations. The definition of general elements of
${\cal V}_{\beta,\gamma}^\alpha$
is given already in (\ref{eq:def-vct-sp-int.op}); what is reviewed
below is a practical algorithm to determine the elements of
${\cal V}_{\beta,\gamma}^\alpha$ for a general VOA that is not necessarily
rational. 

\begin{anythng}
The {\it Verma module} $V(h,c)$ of a Virasoro algebra of central charge
$c \in \R_{\geq 0}$ and $h\in \R_{\geq 0}$ is spanned by a basis
\begin{align}
  V(h,c) := {\rm Span}_\C \left\{ L_{-1}^{n_1}L_{-2}^{n_2}\cdots L_{-d}^{n_d} \cdot
  m^{h,c}_0 \; | \; d \in \Z_{\geq 0}, \;\;
  n_{1,2,\cdots} \geq 0 \right\}, 
\end{align}
where the state $m^{h,c}_0$ is Virasoro primary, $L_n \cdot m^{h,c}_0 = 0$
for all $n\in \Z_{>0}$.
The action of the Virasoro algebra on $V(h,c)$ can be computed by using
the property of $m^{h,c}_0$ above and the commutation relation
of $L_n$'s. The vector space $V(0,c)/V(1,c)$ is given a structure of a VOA, 
and $V(h,c)$ is a module of the VOA.
We will sometimes omit referring to $c$ in $V(h,c)$ or $m_0^{h,c}$ in the text
below. 
For three Verma modules, $V(h_1,c)$, $V(h_2,c)$ and $V(h_3,c)$, the vector
space of intertwining operators
${\cal V}_{V(h_1),V(h_2)}^{V(h_3)}$ is determined in a three-step algorithm
reviewed below (cf \cite{Gaberdiel:1996kx}). 

Step 1: For ${\cal Y} \in {\cal V}_{V(h_1),V(h_2)}^{V(h_3)}$, find constraints
on 
\begin{align}
  {\cal Y}(m^{h_1}_0,z) \cdot m^{h_2}_0 =:
  c_{h_1,h_2}^{h_3} z^{h_3-h_1-h_2} \left( 1 + a_{1} L_{-1} z
  + (a_{2} L_{-1}^2 + a_{0,1}L_{-2}^2)z^2+ \cdots m\right) m^{h_3}_0 
\end{align}
in $V(h_3,c)[[z]][z^{-1}]$ for the Virasoro primary states
$m^{h_1} \in V(h_1,c)$ and $m^{h_2}_0 \in V(h_2,c)$.
To do so, we may use the property
\begin{align}
  [L_m, {\cal Y}(m^{h_1}_0,z)] = z^m \left( z \partial_z + (m+1) h_1 \right)
  {\cal Y}(m^{h_1}_0,z), \qquad {}^\forall m \in \Z
  \label{eq:TY-OPE-4-primary}
\end{align}
and $L_n m^{h_2}_0 = 0$ for $n>0$, because both
$m^{h_1}_0$ and $m^{h_2}_0$ are Virasoro primary. 
The relation
\begin{align}
  L_N \cdot {\cal Y}(m^{h_1}_0,z) \cdot m^{h_2}_0 =
  \left( z^{N+1} \partial_z + (N+1) h_1 z^n \right) {\cal Y}(m^{h_1}_0,z)
  \cdot m^{h_2}_0, \qquad N=1,2,\cdots, 
\end{align}
yields infinitely many linear relations (labeled by $m$) that constrain
$a_{n_1,n_2,\cdots}$'s with the excitation level $\sum_d n_d d = m+N$ on the LHS
relatively to those with the level-$m$ on the RHS. Practically, it is enough
to impose this relation only for $N=1,2$ (cf, $L_{N+1} = (N-1)^{-1} [L_N,L_1]$
for $N\geq 2$). 

Those infinitely many linear relations can be used as a recursion relation,
determining $a_{n_1,n_2,\cdots}$'s for the level-$m$ excitations by
those for the level-$(m-N)$ excitations ($N=1,2$), and eventually by
the coefficient $a_0 =1$ of the ground state. The overall coefficient
$c_{h_1,h_2}^{h_3}$ is not constrained by the linear relations, except 
when there are too many linear
relations at some level-$m$ than the coefficients to be determined.  
When $V(h_3)$ has a non-trivial {\it singular vector} 
(a descendant state of $m_0^{h_3}$ that is also Virasoro primary),
the coefficient $a_*$ of a singular vector is not constrained, because it
does not show up in the LHS of the recursion relations. 
 
Step 2: determine ${\cal Y}(m^{h_1}_0,z) \cdot m_2 \in V(h_3,c)[[z]][z^{-1}]$,
where $m_2 \in V(h_2,c)$ may not be Virasoro primary, although we still
assume that $m^{h_1}_0$ is Virasoro primary. Suppose that
$m_2 = L_{-k} \cdot m'_2$ for some $k>0$. Then
\begin{align}
  {\cal Y}(m^{h_1}_0,z) \cdot m_2 & \;
  = L_{-k} \cdot {\cal Y}(m^{h_1}_0,z) \cdot m'_2
  - [L_{-k}, {\cal Y}(m^{h_1}_0,z)] \cdot m'_2 \nonumber \\
  & \; =  \left( L_{-k} -z^{-k}(z\partial_z + (1-k)h_1)\right)
  {\cal Y}(m^{h_1}_0,z) \cdot m'_2.
\end{align}
So, ${\cal Y}(m^{h_1}_0,z) \cdot m_2$ can be computed from
${\cal Y}(m^{h_1}_0,z)\cdot m^{h_2}_0$. 

Step 3: determine ${\cal Y}(m_1,z) \cdot m_2$ for a general
$m_1 \in V(h_1,c)$ and $m_2 \in V(h_2,c)$. Suppose that
$m_1 = L_{-k}\cdot m'_1$. One may then relate ${\cal Y}(m_1,z)$
to ${\cal Y}(m'_1,z)$ by using the Jacobi identity for
two states $\omega \in V(0,c)$ and $m'_1 \in V(h_1,c)$; because
\begin{align*}
  {\rm Res}_{z_1} {\rm Res}_{z_0}\left[\frac{z_0^{-k+1}}{z_2}
    \delta \left( \frac{z_1-z_0}{z_2} \right)
           {\cal Y}\left( Y(\omega,z_0) \cdot m'_1, z_2 \right) \right]
  = {\cal Y}\left( L_{-k} \cdot m'_1,z_2\right)
\end{align*}
the Jacobi identity (\ref{eq:def-vct-sp-int.op}) reduces the action
of ${\cal Y}(m_1,z_2)$ on $V(h_2,c)$ to
\begin{align}
  {\rm Res}_{z_1} & \;
  {\rm Res}_{z_0}\left[ z_0^{-k} \delta \left( \frac{z_1-z_2}{z_0}
    \right) Y(\omega,z_1) {\cal Y}(m'_1,z_2)-z_0^{-k}\delta \left( \frac{z_2-z_1}{-z_0}\right){\cal Y}(m'_1,z_2) Y(\omega,z_1) \right] \nonumber \\
 & \; = {\rm Res}_{z_1}\left[(z_1-z_2)^{1-k} T(z_1) {\cal Y}(m'_1,z_2)-(-1)^k
    (z_2-z_1)^{1-k} {\cal Y}(m'_1,z_2) T(z_1) \right]; 
\end{align}
as a reminder, $(z_1-z_2)^{-k}$ in the first term [resp.
  $(z_2-z_1)^{-k}$ in the second term] should be treated as its
binomial expansion with respect to $z_2/z_1$ [resp. $z_1/z_2$]. 
\end{anythng}

\begin{anythng}
A simple module of the Virasoro algebra of central charge $c$ is obtained
by taking a quotient of the module $V(h,c)$,
by the maximal proper submodules of $V(h,c)$ generated by the action of
the Virasoro algebra on singular vectors.
This quotient module is denoted by $M(h,c)$. 
The vector space ${\cal V}_{h_1,h_2}^{h_3} = {\cal V}_{M(h_1),M(h_2)}^{M(h_3)}$
of three irreducible unitary modules $M(h_1,c)$, $M(h_2,c)$, $M(h_3,c)$
of the Virasoro algebra
(the VOA with $V = M(0,c)$) is of practical interest. 

Suppose that $V(h_3)$ contains a singular vector. The step 1 procedure
leaves the coefficient $a_*$ of the singular vector undetermined for
${\cal Y} \in {\cal V}_{V(h_1),V(h_2)}^{V(h_3)}$; all of those free coefficients
are projected out when ${\cal Y}$ is cast to ${\cal V}_{V(h_1),V(h_2)}^{M(h_3)}$,
however, as we take quotient $V(h_3) \rightarrow M(h_3)$.  

Suppose that $V(h_2)$ has a singular vector $m_*^{h'_2}$. For ${\cal Y} \in
{\cal V}_{V(h_1),V(h_2)}^{M(h_3)}$ to define an intertwining operator
in ${\cal V}_{V(h_1),M(h_2)}^{M(h_3)}$, a condition 
${\cal Y}(m_1,z) \cdot m_*^{h'_2} \subset J_3[[z,z^{-1}]]z^{h_3-h_1-h_2}$
has to be imposed for ${}^\forall m_1 \in V(h_1)$; $J_3$ is the kernel of
$V(h_3)\rightarrow M(h_3)$. Once this condition is imposed, then
${\cal Y}(m_1,z)$ also map all the descendants of the singular vector
$m_*^{h'_2}$ to $J_3$. The same is true when $V(h_1)$ contains a singular
vector. 

It sounds a little odd: on one hand, existence of a singular vector in
$V(h_2)$ or $V(h_1)$ may rule out $c_{h_1,h_2}^{h_3} \neq 0$ in
${\cal V}_{h_1,h_2}^{h_3}$ even when any $c_{h_1,h_2}^{h_3}$ is allowed 
for ${\cal Y} \in {\cal V}_{V(h_1),V(h_2)}^{V(h_3)}$; on the other hand, 
existence of a singular vector in $V(h_3)$ results in only projecting
out the coefficients $a_*$ of the states in the proper submodules
undetermined in ${\cal Y} \in {\cal V}_{V(h_1),V(h_2)}^{V(h_3)}$, with
$c_{h_1,h_2}^{h_3}$ untouched in the projection to
${\cal V}_{V(h_1),V(h_2)}^{M(h_3)}$. That does not sound democratic among the
three Verma modules. Presumably
what is happening here is that the existence of a singular vector in $V(h_3)$ has
already taken a toll in Step 1, not in the process after Step 3.
Suppose that a singular vector $m_*^{h'_3} \in V(h_3)$ is at level-$m$
excitation, and think of the recursion relation for $N=1,2$ for the states
at level-$(m-N)$. The fact that $a_*$ fails to show up in the LHS
of the relation certainly leaves $a_*$ undetermined, but that also
makes it harder to solve the recursion relation with a non-zero
$c_{h_1,h_2}^{h_3}$.

Another resolution of this democracy puzzle is also provided
by following the argument of \cite[\S3]{Belavin:1984vu}. There,
one considers a chiral three-point function on $\P^1$ of
three Verma modules $V(h_{1,2,3})$ instead of a $V(h_3)$-valued intertwining
operators from $V(h_1)$ and $V(h_2)$.  
The three-point conformal block is at most of 1-dimension, parametrized by
the overall coefficient $c_{h_1,h_2}^{h_3}$ of the three point function of
the ground states of the three Verma modules; the three-point functions
of any descendant (excited) states above the ground states can be determined
from that of the ground states by using the Ward identity of the energy
momentum tensor multiple times. When we use the chiral three-point conformal
block as the language, the existence of a singular vector in any one of the
three Verma modules imposes a constraint on the conformal block of
the three simple modules $M(h_{1,2,3})$. When $V(h_3)$ has a singular vector,
we should also note that the bilinear form on $V(h_3) \times V(h_3)$ given 
by the chiral two-point function is not invertible, however. This is why
the $g=0$ conformal block of the three modules $V(h_{1,2,3})$ being at
most dimension-1 is consistent with having more than one parameter
for ${\cal V}_{V(h_1),V(h_2)}^{V(h_3)}$. Upon projection to
${\cal V}_{V(h_1),V(h_2)}^{M(h_3)}$, where $M(h_3)$ is a simple module, there
can be at most one free parameter $c_{h_1,h_2}^{h_3}$. 
\end{anythng}

\begin{anythng}
\label{statmnt:algthm-vct-int.op-bynd-Virasoro-1}
  When a given unitary VOA is not purely the Virasoro algebra $V = M(0,c)$,
we still have a decomposition
\begin{align}
  V \cong \oplus_{h_V \in A_V} M(h_V,c),  \qquad
  M_\alpha \cong \oplus_{h_\alpha \in A_\alpha} M(h_\alpha,c) \quad
  {}^\forall \alpha \in {\cal I}(V). 
\end{align}
for each of its unitary modules. Given a VOA and its representations, 
we are given a choice of $\{ c_{h_V,h_\alpha}^{h'_\alpha} \; | \; h_V \in A_V, \;
h_\alpha, h'_\alpha \in A_\alpha \}$ for $\alpha \in {\cal I}(V)$.  
The vector space
${\cal V}_{\beta,\gamma}^\alpha$ of intertwining operators of three
unitary modules $M_\beta$, $M_\gamma$ and $M_\alpha$ is within
(ignore subtleties associated with infinite $\#(A_{\alpha,\beta,\gamma})$)
\begin{align}
  \oplus_{h_\beta \in A_\beta} \oplus_{h_\gamma \in A_\gamma} \oplus_{h_\alpha \in A_\alpha}
  {\cal V}_{h_\beta,h_\gamma}^{h_\alpha}
\end{align}
parametrized by $\{ c_{h_\beta,h_\gamma}^{h_\alpha} \; | h_\alpha \in A_\alpha, \;
h_\beta \in A_\beta, \; h_\gamma \in A_\gamma \}$'s. Here is a practical procedure
to narrow down where ${\cal V}_{\beta,\gamma}^\alpha$ is.  

Pick homogeneous states $m_\alpha \in M(h_\alpha) \subset M_\alpha$,
$m_\beta \in M(h_\beta) \subset M_\beta$, $m_\gamma \in M(h_\gamma)
\subset M_\gamma$ and $m_V \in M(h_V) \subset V$; their
conformal weights are $h(m_\alpha)$, $h(m_\beta)$, $h(m_\gamma)$ and $h(m_V)$,
respectively. Think of their chiral four-point function.
The Jacobi identity (\ref{eq:def-vct-sp-int.op})
implies that\footnote{
cf \cite[\S2]{MR1363971}.
Also use (\ref{eq:rltn-delta-2}) for (\ref{eq:ass-rltn-4-int.op}). 
} %
\begin{align}
  (*A) \cdot (z_1 - z_2)^{N_{m_V,m_\beta}} & \; = (*B) \cdot (z_1-z_2)^{N_{m_V,m_\beta}},
   \label{eq:comm-rltn-4-int.op}
  \\
  (*A)_{z_1 \rightarrow (z_0 + z_2)} \cdot (z_0+z_2)^{N_{m_V,m_\gamma}} & \; =
  (*C) \cdot (z_0+z_2)^{N_{m_V,m_\gamma}}; \label{eq:ass-rltn-4-int.op}
\end{align}
here, 
\begin{align*}
  (*A) & :=
  \bra{m_\alpha} Y_\alpha (m_V,z_1) {\cal Y}(m_\beta,z_2) m_\gamma \rangle \in \C[z_2^{-1},z_1][[z_2,z_1^{-1}]] z_2^{h_\alpha-h_\beta-h_\gamma}, \\
  (*B) & \; :=\bra{m_\alpha} {\cal Y}(m_\beta,z_2) Y_\gamma (m_V,z_1) m_\gamma \rangle \in \C[z_2,z_1^{-1}][[z_2^{-1},z_1]] z_2^{h_\alpha-h_\beta-h_\gamma}, \\
  (*C) & \; := \bra{m_\alpha} {\cal Y}\left( Y_\beta(m_V,z_0) \cdot m_\beta, \; z_2\right) m_\gamma \rangle \in \C[z_0^{-1}, z_2][[z_0, z_2^{-1}]]
  z_2^{h_\alpha-h_\beta-h_\gamma},  
\end{align*}
and
\begin{align*}
  N_{m_V,m_\beta} := h_V + (h(m_\beta) - h_{\beta,0}), \qquad
  N_{m_V,m_\gamma} := h_V + (h(m_\gamma) - h_{\gamma,0}) 
\end{align*}
are the order of the severest pole in the $m_V$--$m_\beta$ OPE and
$m_V$--$m_\gamma$ OPE, respectively; $h_{\beta,0}$ and $h_{\gamma,0}$ are
the smallest conformal weight in $M_\beta$ and $M_\gamma$. On the LHS of
(\ref{eq:ass-rltn-4-int.op}), we should replace any integral power of
$z_1$ by the binomial expansion series of the same power of $(z_0+z_2)$
wrpt $z_2/z_0$. The commutativity relation (\ref{eq:comm-rltn-4-int.op}) 
[resp. associativity relation (\ref{eq:ass-rltn-4-int.op})]
implies that both sides are not formal power series anymore, but
are in $\C[z_1,z_1^{-1},z_2,z_2^{-1}] z_2^{h_\alpha - h_\beta - h_\gamma}$
[resp. in $\C[z_0, z_0^{-1},z_2,z_2^{-1}] z_2^{h_\alpha-h_\beta -h_\gamma}$], 
and both sides agree.

Suppose that we include only the intermediate states in
$\oplus_{h'_\alpha \in A_\alpha}^{(h'_\alpha \leq h(m_\alpha) + \delta N_\alpha)}
M(h'_\alpha)$ in (*A), $\oplus_{h'_\beta \in A_\beta}^{(h'_\beta \leq h(m_\beta)
  + \delta N_\beta)}M(h'_\beta)$ in (*C) and $\oplus_{h'_\gamma \in A_\gamma}^{(h'_\gamma \leq h(m_\gamma) + \delta N_\gamma)} M(h'_\gamma)$ in (*B); here, $\delta N_\alpha$, $\delta N_\beta$, $\delta N_\gamma$ are some non-negative integers. Then we cannot
compute all of the terms of the form
$z_1^{\mu_1}z_2^{\mu_2}$ in (\ref{eq:comm-rltn-4-int.op}) and
those of the form $z_0^{\mu_0} z_2^{\mu_2}$ in (\ref{eq:ass-rltn-4-int.op}),
but we still have reliable results on both sides for some of those terms.
Each term of the commutativity relation (\ref{eq:comm-rltn-4-int.op})
in the range 
\begin{align}
  \mu_2 - (h_\alpha-h_\beta-h_\gamma) \in [-\delta N_\gamma + N_{m_V,m_\beta}, \delta N_\alpha], \quad 
  \mu_1 + h_V \in [-\delta N_\alpha + N_{m_V,m_\beta} , \delta N_\gamma] 
\end{align}
yields a linear relation among $\{ c_{h_\beta,h_\gamma}^{h'_\alpha} \; | \;
h'_\alpha \leq h(m_\alpha) + \delta N_\alpha\}$ and $\{ c_{h_\beta,h'_\gamma}^{h_\alpha}
\; | \; h'_\gamma \leq h(m_\gamma) + \delta N_\gamma \}$. Each term of the
associativity relation (\ref{eq:ass-rltn-4-int.op}) in the range
\begin{align}
  \mu_2 - (h_\alpha-h_\beta-h_\gamma) \in [-\delta N_\beta + N_{m_V,m_\gamma}, \delta N_\alpha], \quad
  \mu_0 + h_V \in [-\delta N_\alpha + N_{m_V,m_\gamma}, \delta N_\beta]
\end{align}
also gives rise to a linear relation among 
$\{ c_{h_\beta,h_\gamma}^{h'_\alpha} \; | \; h'_\alpha \leq h(m_\alpha)
+ \delta N_\alpha\}$ and $\{ c_{h'_\beta,h_\gamma}^{h_\alpha} \; | \; h'_\beta
\leq h(m_\beta) + \delta N_\beta \}$. As we increase $\delta N_\alpha, \delta N_\beta, \delta N_\gamma$, the number of parameters to be constrained---$c^{h'_\alpha}_{h_\beta,h_\gamma}$,
$c^{h_\alpha}_{h'_\beta,h_\gamma}$ and $c^{h_\alpha}_{h_\beta,h'_\gamma}$---increases
linearly, while the number of constraints (the number of $(\mu_1,\mu_2)$ and
$(\mu_0,\mu_2)$) increases quadratically, so this is a highly constrained
system. One may repeat the same process for different choices of the
homogeneous states $m_\alpha$, $m_\beta$, $m_\gamma$ and $m_V$, to constrain
$\{ c^{h'_\alpha}_{h'_\beta,h'_\gamma} \}$ with arbitrary $h'_\alpha \in A_\alpha$,
$h'_\beta \in A_\beta$ and $h'_\gamma \in A_\gamma$. 
\end{anythng}

\begin{anythng}
An algorithm alternative to (and not essentially different from)
the one in \ref{statmnt:algthm-vct-int.op-bynd-Virasoro-1} is to use
the Ward identity. For any homogeneous state $m_V \in V$ with $h_V > 0$,
and for any three homogeneous states $m_{\alpha} \in
M_{\alpha}$, $m_\beta \in M_\beta$ and $m_\gamma \in M_\gamma$,
it is of the form\footnote{
The present author thanks K. Hori for explaining the
derivation \cite{Hori:2008CFT}:
the difference between both sides is regular as a function
of $\zeta \in \C \subset \P^1$ because the RHS is chosen that way; the
RHS also vanishes in the limit $\zeta \rightarrow \infty$, so the rest
is to show that the LHS also vanishes at $\zeta \rightarrow \infty$.
A regular function of $\zeta \in \P^1$ that vanishes at one point is zero
everywhere. Because the LHS behaves as $\zeta^{-h_V}$ ($\zeta^{-2h_V}$ if
$m_V$ is quasi-primary), $h_V >0$ is enough. 
} %
\begin{align}
  \langle Y(m_v,\zeta) & \; {\cal Y}\left( \Upsilon(m_\alpha),z_0 \right)
          {\cal Y}(m_\beta, z) m_\gamma \rangle  \\
    = & \;    \sum_{n=1}^{N_{m_V,\Upsilon(m_\alpha)}} (\zeta - z_0)^{-n} \langle
              {\cal Y}([m_V]_{-h_V+n} \cdot \Upsilon(m_\alpha), z_0)
              {\cal Y}(m_\beta,z) m_\gamma \rangle \nonumber \\
    & \;    +    \sum_{n=1}^{N_{m_V, m_\beta}} (\zeta - z)^{-n}
          \langle {\cal Y}(\Upsilon(m_\alpha),w_0)
                  {\cal Y}([m_V]_{-h_V+n} \cdot m_\beta, z)m_\gamma \rangle
                  \nonumber \\
    & \;   + \sum_{n=1}^{N_{m_V,m_\gamma}} \zeta^{-n}
              \langle {\cal Y}(\Upsilon(m_\alpha),w_0)
                      {\cal Y}(m_\beta,z) [m_V]_{-h_V+n} m_\gamma \rangle .
                      \nonumber 
\end{align}
An easy way to extract a practically useful consequence is
to focus on a case $m_\alpha$ is Virasoro primary, and
$h_\alpha - h_{\alpha,0} < m_V$. Then we can multiply $z_0^{2h_\alpha}$
to the Ward identity and take the limit $z_0 \rightarrow \infty$, to
have 
\begin{align*}
  \bra{m_\alpha } Y(m_V,\zeta) \; \cdot & \;
      {\cal Y}(m_\beta,z) \cdot m_\gamma \rangle = \\
  = & \; \sum_{n=1}^{N_{m_V,m_\beta}} (\zeta -z)^{-n}
  \langle m_\alpha | {\cal Y} \left( [m_V]_{-h_V+n} m_\beta, z \right) \cdot
  m_\gamma \rangle \\
  & \; + \sum_{n=1}^{N_{m_V,m_\gamma}} \zeta^{-n}
  \langle m_\alpha | {\cal Y}\left( m_\beta, z\right) \cdot
      [m_V]_{-h_V+n} \cdot m_\gamma \rangle .
\end{align*}
The first term on the RHS of the Ward identity has dropped out in the limit
$z_0 \rightarrow \infty$. There are only terms with a negative power in $\zeta$,
as we expand $(\zeta -z)^{-n}$ in the binomial series wrpt $z/\zeta$. 
For the coefficient of $\zeta^{-k}$ of each $k =1,2,\cdots$, we have a
linear relation among a finite subset of 
$\{ c_{h'_\beta,h'_\gamma}^{h'_\alpha} \}_{h'_\alpha \in A_\alpha, h'_\beta \in A_\beta, h'_\gamma \in A_\gamma}$. The linear relation for $k=h_V$, in particular, provides a
constraint that involves the zero-mode charge operator $[m_V]_0$ of the
state $m_V$. 
\end{anythng}

\begin{rmk}
Here, we chose to review a few algorithms that can handle the intertwining
operators directly and that can apply to any unitary VOA. 
There can be easier approaches, when one is interested only in chiral
three/four point functions but not in details of the intertwining operators,
or for a specific family of VOAs (such as Affine Lie algebra with Sugawara
construction for the Virasoro generators). 
\end{rmk}
%

\subsection{Genus-1 Equations of Rational VOAs: an Example}
\label{ssec:g=1}

All of the skewsymmetry, crossing-symmetry, braiding and fusion
isomorphisms deal with state/operator insertions of a VOA on a complex
plane.
There are a few more---$S$ and $T$-transformations, and the $a$-move
and $b$-move---when we deal with state/operator insertions of a rational
VOA on a Riemann surface of genus 1. Just like there are identities
among ${\cal F}$, $\Omega$ and ${\cal B}$'s, there are also identities
(\ref{eq:g=1-proj-repr-SL2Z-rltn}) and
(\ref{eq:g=1-MSeq-3}, \ref{eq:g=1-MSeq-3a}, \ref{eq:g=1-MSeq-3b}) 
among those extra isomorphisms ({\it genus-1 equations of a rational VOA}). 
This section \ref{ssec:g=1} does not add anything to what has been
presented in such references as \cite{Moore:1988uz}, \cite{Moore:1988qv}
(cf also \cite{MR1317233}, \cite{MR1875102}, \cite{MR2175093})
and \cite{MR2387861}, but
presents one example of the genus-1 equation, in the Ising VOA
(the unitary Virasoro minimal VOA with $c=1/2$). We will also add
a little more explanation on a few subtle points in the genus-1
equations when Refs. \cite{Moore:1988uz}, \cite{Moore:1988qv} and
\cite{MR2387861} offer very brief explanations. 

\begin{anythng}
Ref. \cite{Moore:1988uz} and \cite[\S3]{Moore:1988qv} introduce two
isomorphisms $T(\beta)$ and $S(\beta)$ for a given rational and unitary VOA
${\cal A}$ and $\beta \in {\cal I}({\cal A})$; $T(\beta)$ and $S(\beta)$
constitute a set of generators the projective ${\rm SL}_2\Z$ action on
the vector space
\begin{align}
  {\cal V}_\beta^{(g=1)} := \oplus_{\gamma \in {\cal I}({\cal A})} {\cal V}_{\beta,\gamma}^\gamma. 
\end{align}
For $g = (a,b; c,d) \in {\rm SL}_2\Z$ and $j_g(\tau) := c\tau+d$, where 
$\tau \in \mathfrak{H}$, this condition
\begin{align}
  \chi \left({\cal Y}; (j_g(\tau))^{-L_0} \; m, \; g \cdot \tau \right) =:
  \chi \left( {\cal Y}|[g]; m, \tau \right), \qquad
      {}^\forall m \in M_\beta, \quad {}^\forall \tau \in \mathfrak{H}
   \label{eq:cond-SL2Z-transf-on-int.tw.op}
\end{align}
determines an isomorphism ${\cal V}_{\beta}^{(g=1)} \ni {\cal Y} \mapsto
{\cal Y}|[g] \in {\cal V}_\beta^{(g=1)}$; here, 
\begin{align}
  \chi_\gamma({\cal Y}; m,\tau) := {\rm Tr}_{M_\gamma}\left[
    e^{2\pi i \tau \left( L_0-\frac{c}{24}\right)}
    ({\cal Y}_{\beta,\gamma}^\gamma)^{(w)}(m)_{w@0} \right], \qquad
      {\cal Y} \in {\cal V}_{\beta}^{(g=1)},
      \label{eq:def-int-tw-op-ins-char-fcnl}
\end{align}
$\chi({\cal Y};m,\tau) = \{ \chi_\gamma({\cal Y};m,\tau) \; | \;
\gamma \in {\cal I}({\cal A})\}$ and the
condition (\ref{eq:cond-SL2Z-transf-on-int.tw.op}) should be read as
an equality for each of ${}^\forall \gamma \in {\cal I}({\cal A})$. 
$T(\beta)$ and $S(\beta)$ are for $g=(1,1;0,1)$ and $(0,-1;1,0)$,
respectively. So, $T(\beta)$ acts as multiplication of
$e^{2\pi i ([h_\gamma]-c/24)}$ on ${\cal V}_{\beta,\gamma}^\gamma \subset
{\cal V}_{\beta}^{(g=1)}$. 
\end{anythng}

\begin{rmk}
By starting off purely in algebraic perspective, as in
section \ref{ssec:review-VOA}, it is a non-trivial question for a given
unitary rational VOA whether the modular transformations on the objects
$\chi({\cal Y};-,-)$ in
(\ref{eq:def-int-tw-op-ins-char-fcnl}) are facilitated by isomorphisms
from ${\cal V}_\beta^{(g=0)}$ to itself. Indeed this is not expected to be
true for a general unitary VOA. 
Ref. \cite[\S5]{MR1317233} 
proved that they are for a rational VOA, when $\beta$ is the defining
representation of the VOA. The case with a general $\beta$ was established
in \cite[Thm. 5.1]{MR1875102} and \cite[Thm. 7.2+7.3]{MR2175093}.
\end{rmk}

\begin{rmk}
\label{rmk:SL2Z-act-frmR-onVrtxOp}
By using a basis $\{ {\cal Y}_i \}_{i=1,2,\cdots, \sum_\gamma N_{\beta,\gamma}^\gamma}$
of ${\cal V}_{\beta}^{(g=1)}$, we may have a matrix projective representation
of $g \in {\rm SL}_2\Z$ on ${\cal V}_{\beta}^{(g=1)}$:
\begin{align}
  {\cal Y}_i|[g] = \sum_j (g)_i^{\;\; j} \; {\cal Y}_j; \qquad
  \sum_i c^i {\cal Y}_i \mapsto \left(\sum_i c^i {\cal Y}_i|[g] \right)
  = \sum_{i,j} c^i \; (g)_i^{\;\; j} \; {\cal Y}_j.
 \label{eq:SL2Z-transf-matrix-repr-on-g=1-optr}
\end{align}
The notation is chosen this way, because the way ${\cal Y}|[g]$ is introduced
indicates that 
\begin{align}
 {\cal Y}|[g] = \left( {\cal Y}|[g_1] \right)|[g_2]
\end{align}
for $g = g_1 \cdot g_2 \in {\rm SL}_2\Z$. 
%
%
%
\end{rmk}
\begin{anythng}    
The action of $g \in {\rm SL}_2\Z$ on ${\cal V}_\beta^{(g=1)}$ is only a
projective representation \cite[(4.2a), (4.2b)]{Moore:1988uz}
and \cite[(4.11), (4.12)]{Moore:1988qv}:
\begin{align}
  (S(\beta))^2 = \Theta_{-1}, \qquad
   (T(\beta) S(\beta))^3 = (S(\beta))^2;
 \label{eq:g=1-proj-repr-SL2Z-rltn}
\end{align}
here,
\begin{align}
  \Theta_{-1}: {\cal V}_{\beta}^{(g=1)} \ni {\cal Y} = \left\{
        {\cal Y}_{\beta,\gamma}^\gamma \right\}_{\gamma \in {\cal I}({\cal A})}
        \mapsto
        \Theta_{-1}({\cal Y}) :=
        \left\{ \Theta_{-1}\left({\cal Y}_{\beta,\Upsilon^{\cal I}(\delta)}^{\Upsilon^{\cal I}(\delta)}\right) \right\}_{\delta \in {\cal I}({\cal A})} \in {\cal V}_{\beta}^{(g=1)}, 
\end{align}
and $T(\beta) \cdot \Theta_{-1} = \Theta_{-1} \cdot T(\beta)$
because $[h_\gamma] = [h_{\Upsilon^{\cal I}(\gamma)}]$.
\end{anythng}

\begin{rmk}
It is intuitively obvious for those with physics background that
$(S(\beta))^2$ is somehow related to a crossing symmetry isomorphism $\Theta$;
the $S$-transformation once changes the torus modulus parameter $\tau$ and 
the cylinder coordinate $w$ (where $z=e^{-iw}$, $z \sim z e^{-2\pi i \tau}$) to
$-1/\tau$ and $w^{(1)} := w/\tau$, respectively, and the second
$S$-transformation to $\tau$ and $w^{(2)} := w^{(1)}/(-1/\tau)=-w$,
respectively; the Euclidean time ${\rm Im}(w^{(2)})$ after $S(\beta)
\cdot S(\beta)$ flows in the direction opposite to that of ${\rm Im}(w)$. 
It is less obvious, however, that we should use $\Theta_{-1}$ among the crossing
symmetry isomorphisms $\Theta_r$ parametrized by $r \in \Z$. 

Here is how we see this. The $S$-transformation twice brings
$\chi_\gamma({\cal Y};m,\tau)$ in (\ref{eq:def-int-tw-op-ins-char-fcnl})
to
\begin{align}
  & {\rm Tr}_{M_\gamma} \left[ e^{2\pi i \tau \left( L_0-\frac{c}{24}\right)}
    ({\cal Y}_{\beta,\gamma}^{\gamma})\left( R_{{\rm exp}_-@w(p)}^{-1}
    \cdot \tau^{-L_0} \cdot (-1/\tau)^{-L_0} \cdot m \right)
    \right] \nonumber\\
  & \; = {\rm Tr}_{M_\gamma} \left[ e^{2\pi i \tau \left( L_0-\frac{c}{24}\right)}
    ({\cal Y}_{\beta,\gamma}^{\gamma})\left( R_{{\rm exp}_-@w(p)}^{-1} \cdot
    e^{-\pi i L_0} \cdot m \right) \right];
\end{align}
we have combined $(\tau \cdot (-1/\tau))$ into $e^{ + \pi i}$ because we deal
with
both $\tau$ and $-1/\tau$ in the upper half plane as a complex number whose
argument is within $(0, +\pi)$; this is in fact the only subtlety, and the rest
is a manipulation. The trace over $M_\gamma$ is rewritten by using an orthonormal
basis $\{ \tilde{m}\}$; we will focus on only the essential part of computation;
\begin{align*}
  \bra{\tilde{m}} ({\cal Y}_{\beta,\gamma}^{\gamma})\left( R^{-1}_{{\rm exp}_-@w(p)}
  \cdot e^{-\pi i L_0} \cdot m \right) \cdot \tilde{m} \rangle 
  = \langle \Upsilon(\tilde{m}), \; ({\cal Y}_{\beta,\gamma}^{\gamma})\left(
  R^{-1}_{ {\rm exp}_- \circ e^{-\pi i } @-w(p)} \cdot m \right) \cdot
  \tilde{m} \rangle  \\
  = \langle \Upsilon(\tilde{m}), \; ({\cal Y}_{\beta,\gamma}^{\gamma})\left(
  R^{-1}_{{\rm inv}@e^{iw(p)}} \cdot R_{{\rm exp}_-@-w(p)}^{-1} \cdot m \right)
  \cdot \tilde{m} \rangle =
  \langle (\Theta {\cal Y})\left( R_{{\rm exp}_-@-w(p)}^{-1} \cdot m \right)
  \cdot \Upsilon(\tilde{m}), \tilde{m} \rangle, 
\end{align*}
where we used ${\rm exp}_-( -w) = 1/{\rm exp}_-(w) = {\rm inv} \circ
{\rm exp}_-(w)$ in the middle, so $(S(\beta))^2$ is indeed given by a
crossing symmetry isomorphism $\Theta$. To read out the subscript of
$\Theta_r$, it is fine to focus on the case $m$ is a Virasoro primary, when
all the $L_i$'s with $i>0$ in $R_{\rm inv}^{-1}$ and $R_{{\rm exp}_-}^{-1}$ can
be ignored. Now, $(e^{-iw(p)}/i)^{L_0} \cdot e^{-\pi i L_0}$ in the first line
should be equal to $(e^{\pi i (2r+1)}/e^{2iw(p)})^{L_0}\cdot (e^{iw(p)}/i)^{L_0}$
in the second line. For this to be the case regardless of the conformal weight
of $m$, we should have $e^{\pi i (2r+1)} = e^{-\pi i}$. 
\end{rmk}

\begin{exmpl}
The Virasoro minimal model VOA with $c=1/2$, also known as the Ising VOA,
has three irreducible modules labeled by ${\cal I} = \{ 0, \epsilon, \sigma\}$.
Each of the modules has a non-degenerate ground state with the conformal
weight $h_0=0$, $h_\epsilon = 1/2$ and $h_\sigma = 1/16$, respectively. The
multiplication law of the fusion algebra $\Z[{\cal I}] = {\rm Span}_\Z\{
[e_0], [e_\epsilon], [e_\sigma] \}$ is
\[
[e_{\epsilon}] \cdot [e_{\epsilon}] = [e_0], \quad
[e_\epsilon] \cdot [e_\sigma] = [e_\sigma], \quad
[e_\sigma]  \cdot [e_\sigma] = [e_0] + [e_\epsilon].
\]

On the vector space ${\cal V}_{0}^{(g=1)} = {\cal V}_{0,0}^0 \oplus
{\cal V}_{0,\epsilon}^\epsilon \oplus {\cal V}_{0,\sigma}^\sigma$, we may
use $\{ Y, Y_\epsilon, Y_\sigma\}$ as the basis. 
\begin{align}
  S(0) = \frac{1}{2} \left( \begin{array}{ccc} 1 & 1 & \sqrt{2} \\
    1 & 1 & - \sqrt{2} \\ \sqrt{2} & -\sqrt{2} & 0 \end{array} \right)
  \label{eq:Smtrx-Ising-0}
\end{align}
when the basis above is used. The relation (\ref{eq:g=1-proj-repr-SL2Z-rltn})
is satisfied. 

Whereas ${\cal V}_\sigma^{(g=1)}$ is empty, the vector space
${\cal V}_\epsilon^{(g=1)} \cong {\cal V}_{\epsilon,\sigma}^\sigma$ is of
1-dimension. So, there must be a scalar multiplication of $S(\epsilon)$ and also
$T(\epsilon)$ on this 1-dimensional vector space. To determine them,
it is enough to study the ${\rm SL}_2\Z$ action on one generator ${\cal Y}_0$
of the one-dimensional ${\cal V}_\epsilon^{(g=1)}$, and even to study
${\cal Y}_0(m)$ for just one state $m \in M_\epsilon$. We choose $m$ to be
the Virasoro primary ground state of $M_\epsilon$, which is also denoted by
$\epsilon$ here. The
Ward identity in \cite[eq.(6)]{Eguchi:1987qd}
for the one-point function of this state $\epsilon$
on a $g=1$ Riemann surface immediately indicates, by using the fact that
the one-point function is independent of the point the operator is inserted,
that $\chi_\sigma({\cal Y}_0;\epsilon,\tau) \propto \eta(\tau)$,
the Dedekind eta-function. So, 
\begin{align}
 S(\epsilon) = e^{-\pi i/4}, \qquad T(\epsilon) = e^{2\pi i \frac{1}{24}}.
 \label{eq:Smtrx-Ising-e}
\end{align}
One also finds $\Theta_{r}: {\cal Y}_0 \mapsto {\cal Y}_0 e^{\pi i(2r+1)/2}$
by looking at ${\cal Y}_0(\epsilon) \cdot \sigma \propto z^{-1/2}\sigma
+ \cdots$ and $h_\epsilon = 1/2$. Therefore, both of the relations
in (\ref{eq:g=1-proj-repr-SL2Z-rltn}) are satisfied.


\end{exmpl}

\begin{anythng}
There is a relation that involves $a$-{\it move} and $b$-{\it move} as well
as $S(\beta)$'s; it is to say that two isomorphisms from
$\oplus_{\delta} ({\cal V}_{\delta}^{(g=1)} \otimes {\cal V}_{\beta_1,\beta_2}^\delta)$
to itself are equal \cite[(4.2c)]{Moore:1988uz}, \cite[(4.16)]{Moore:1988qv}. 
The idea is to start from a configuration of inserting 
a state $m_1 \in M_{\beta_1}$ at $w= w_1 + 2\pi (\Z \oplus \tau \Z)$ and
$m_2 \in M_{\beta_2}$ at $w=w_2 + 2\pi (\Z \oplus \tau \Z)$, where
$w_1, w_2 + 2\pi (\Z \oplus \tau \Z) \in \C/2\pi(\Z \oplus \tau \Z)$
specify two points in a torus, placed relatively to each other in a way
described in Figure \ref{fig:a-move-N-b-move}~(a).
\begin{figure}[tbp]
\begin{center}
  \begin{tabular}{ccc}
      \includegraphics[scale=0.8]{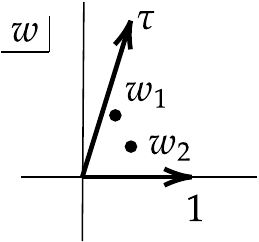} &
      \includegraphics[scale=0.8]{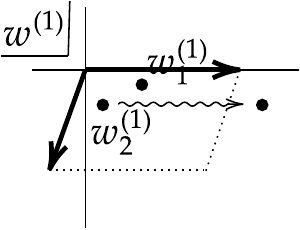} &
      \includegraphics[scale=0.8]{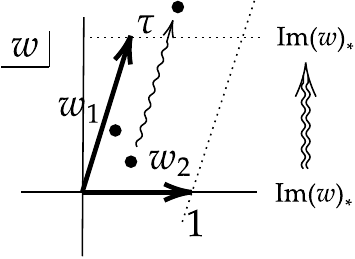}  \\
      (a) initial configuration & (b) the $a$-move in $w^{(1)}$ &
      (c) the $b$-move in $w$
  \end{tabular}
\caption{\label{fig:a-move-N-b-move}The $b$-move in the cylinder coordinate
$w$ is the $a$-move in the coordinate $w^{(1)}:= w/\tau$. As we choose the
initial configuration so that ${\rm Im}(w_1)>{\rm Im}(w_2)$ and
${\rm }(w^{(1)}_1) > {\rm Im}(w^{(1)}_2)$, the relative coordinate $w_1 - w(p_2)$
gains---does not loses---complex phase under the $b$-move of $p_2$. }
 \end{center}
\end{figure}
The initial configuration
is set so that ${\rm Im}(w_1) > {\rm Im}(w_2)$, and
${\rm Im}(w_1/\tau) > {\rm Im}(w_2/\tau)$. 

On the moduli $(\tau, w_i)$ of $g=1$ two point functions, $g = (p,q;r,s)
\in {\rm SL}_2\Z$ and $a^\mu b^\nu := (\mu,\nu) \in \Z^{\oplus 2}$ act through
\begin{align}
  g: \left( \tau, w \right) \mapsto \left( \frac{p \tau + q}{r \tau + s},
  \; \frac{w}{r \tau + s} \right), \qquad 
  a^\mu b^\nu: \left( \tau, w \right) \mapsto
  \left( \tau, w + \mu + \nu \tau \right). 
\end{align}
For $S = (0,-1;1,0) \in {\rm SL}_2\Z$, $a = (1,0) \in \Z^{\oplus 2}$
and $b = (0,1) \in \Z^{\oplus 2}$,
\begin{align}
 S^{-1} \cdot a \cdot S = b 
\end{align}
as an action on the moduli; it says that the $a$-move
$w^{(1)} \mapsto w^{(1)} +1$ after the $S$-transformation (where
$w^{(1)} := w/\tau$) is the same as the $b$-move
$w \mapsto w + \tau$ in the coordinate $w$ before the
$S$-transformation (see Fig. \ref{fig:a-move-N-b-move}~(b, c)). 
Noting that the projective action of ${\rm SL}_2\Z$ on the intertwining
operators is opposite
(cf Rmk. \ref{rmk:SL2Z-act-frmR-onVrtxOp}), its matrix representation
on the vector space 
$\oplus_\delta \left( {\cal V}_\delta^{(g=1)}\otimes
{\cal V}_{\beta_1,\beta_2}^\delta \right)$ has a relation 
\begin{align}
  S \cdot a \cdot S^{-1} = b;
  \label{eq:g=1-MSeq-3}
\end{align}
the matrix $S$ on $\oplus_{\delta} \left( {\cal V}_\delta^{(g=1)} \otimes
{\cal V}_{\beta_1,\beta_2}^\delta\right)$ here should be $((s^T) \otimes {\rm id})$
using the transpose matrix $(s^T)$ of the matrix representation $s$ described
in (\ref{eq:SL2Z-transf-matrix-repr-on-g=1-optr}).  

The $a$-move in the description using the cylinder coordinate $w^{(1)}=w/\tau$
is described best by exploiting the isomorphism 
${\cal F}: \oplus_{\lambda,\kappa \in {\cal I}} ({\cal V}_{\beta_1,\lambda}^\kappa
\otimes {\cal V}_{\beta_2,\kappa}^\lambda) \rightarrow
\oplus_{\delta,\kappa \in {\cal I}} ({\cal V}_{\delta,\kappa}^{\kappa} \otimes
  {\cal V}_{\beta_1,\beta_2}^{\delta})$; 
as we assumed ${\rm Im}(w^{(1)}_1) > {\rm Im}(w^{(1)}_2)$, the operator
from ${\cal V}_{\beta_2,\kappa}^\lambda$ acts on the vector space
of chiral states $\oplus_\kappa M_\kappa$ before an operator from
${\cal V}_{\beta_1,\lambda}^\kappa$ does. The change $w^{(1)}_2 \rightarrow
w^{(1)}_2+1$ (Fig. \ref{fig:a-move-N-b-move} (b)), and hence the change
$e^{-iw^{(1)}_2} \rightarrow e^{-iw^{(1)}_2} e^{-2\pi i}$ is reproduced by changing
an operator from ${\cal V}_{\beta_2,\kappa}^\lambda$ by
its multiple by $e^{2\pi i (h_\kappa-h_\lambda)}$; 
\begin{align}
  a = {\cal F} \cdot \left( {\rm id}\otimes e^{2\pi i (h_\kappa - h_\lambda)} \right)
  \cdot {\cal F}^{-1}: & \; 
  \oplus_{\delta} \left( {\cal V}_\delta^{(g=1)} \otimes
        {\cal V}_{\beta_1,\beta_2}^\delta \right)  \rightarrow 
  \oplus_{\kappa,\lambda} \left(
        {\cal V}_{\beta_1,\lambda}^\kappa \otimes {\cal V}_{\beta_2,\kappa}^\lambda
        \right)  \label{eq:g=1-MSeq-3a} \\
   & \qquad    \rightarrow
    \oplus_{\kappa,\lambda} \left(
        {\cal V}_{\beta_1,\lambda}^\kappa \otimes {\cal V}_{\beta_2,\kappa}^\lambda
        \right) \rightarrow  \oplus_{\delta} \left( {\cal V}_\delta^{(g=1)} \otimes
        {\cal V}_{\beta_1,\beta_2}^\delta \right).  \nonumber        
\end{align}
The $b$-move in the description using the cylinder coordinate $w$
changes the configuration with ${\rm Im}(w_1) > {\rm Im}(w_2)$
to the configuration where ${\rm Im}(w_2+\tau) > {\rm Im}(w_1)$
(Fig. \ref{fig:a-move-N-b-move} (c)). 
The two-point function after the change is reproduced by replacing the
operator in $\oplus_{\delta} \left( {\cal V}_{\delta}^{(g=1)}
\otimes {\cal V}_{\beta_1,\beta_2}^\delta \right)$ by its image by 
$({\rm id}\otimes \Omega_0)$; we should use the specific skewsymmetry
isomorphism $\Omega_0$ (not $\Omega_r$ with any other $r \in \Z$),
because the  relative coordinate $w_1-w_2$ changes in its
complex phase by $+\pi$ (see Fig. \ref{fig:a-move-N-b-move} (c)). 
A combination of operators in $\oplus_{\delta} ({\cal V}_{\delta}^{(g=1)}\otimes
{\cal V}_{\beta_2,\beta_1}^\delta)$ that determines a chiral $g=1$ two-point function is translated by an isomorphism 
\begin{align}
  \left({\cal F} \cdot {\rm exch} \cdot {\cal F}^{-1} \right):  
 & \; 
  \oplus_{\delta \in {\cal I}({\cal A})} \left(
        {\cal V}_{\delta}^{(g=1)} \otimes {\cal V}_{\beta_2,\beta_1}^\delta \right)
        \rightarrow
        \oplus_{\kappa,\lambda \in {\cal I}({\cal A})} \left(
              {\cal V}_{\beta_2,\lambda}^{\kappa}
  \otimes {\cal V}_{\beta_1,\kappa}^\lambda \right) \\
 & \qquad  \rightarrow
  \oplus_{\lambda,\kappa} \left( {\cal V}_{\beta_1,\kappa}^\lambda \otimes
        {\cal V}_{\beta_2,\lambda}^{\kappa} \right) \rightarrow \oplus_{\epsilon}
        \left( {\cal V}_\epsilon^{(g=1)} \otimes
  {\cal V}_{\beta_1,\beta_2}^\epsilon \right) ;  \nonumber 
\end{align}
the isomorphism ${\rm exch}$ corresponds to changing the time slice---at
${\rm Im}(w) = {\rm Im}(w)_*$---of the
vector space of states in the operator formalism from the one
with ${\rm Im}(w)_* < {\rm Im}(w_1)$, ${\rm Im}(w_2+\tau)$
to the one with ${\rm Im}(w_1) < {\rm Im}(w)_* <
{\rm Im}(w_2+\tau)$ (Fig. \ref{fig:a-move-N-b-move} (c)); in math language,
${\rm Tr}[A B] = {\rm Tr}[B A]$ for anly linear operations $A$, $B$. 
To summarize,
\begin{align}
  b= {\cal F} \cdot {\rm exch} \cdot {\cal F}^{-1} \cdot ({\rm id} \otimes \Omega_0): \oplus_\delta
  ({\cal V}_\delta^{(g=1)}\otimes {\cal V}_{\beta_1,\beta_2}^\delta) \longrightarrow
  \oplus_\delta
({\cal V}_\delta^{(g=1)}\otimes {\cal V}_{\beta_1,\beta_2}^\delta). 
  \label{eq:g=1-MSeq-3b}
\end{align}
See Ref. \cite[pp.200--201]{Moore:1988qv}. 
\end{anythng}

\begin{exmpl}
In the Ising VOA, the $S \cdot a \cdot S^{-1} = b$ relation on
$\oplus_\delta ({\cal V}_\delta^{(g=1)} \otimes {\cal V}_{\beta_1,\beta_2}^\delta)$
must hold for each of the $3 \times 3$ choices of $\beta_1,\beta_2
\in {\cal I}$. For the cases $(\beta_1,\beta_2) = (1,\sigma)$, $(\sigma,1)$,
$(\epsilon, \sigma)$ and $(\sigma,\epsilon)$, the vector space
$\oplus_\delta ({\cal V}_\delta^{(g=1)} \otimes {\cal V}_{\beta_1,\beta_2}^\delta)$
is empty in the first place. For the cases $(\beta_1,\beta_2) = (1,\epsilon)$, 
$(\epsilon,1)$, where the vector space is of 1-dimension, and
$(\beta_1,\beta_2) = (1,1)$, where the vector space is of 3-dimensions,
both $S \cdot a \cdot S^{-1}$ and $b$ on $\oplus_\delta ({\cal V}_\delta^{(g=1)}
\otimes {\cal V}_{\beta_1,\beta_2}^\delta)$ are the identity map, so the relation
holds. In the two remaining cases, both $S \cdot a \cdot S^{-1}$ and $b$
are non-trivial, and agree, as we see explicitly below. 

Think of the case $(\beta_1,\beta_2) = (\epsilon, \epsilon)$, where
${\cal V}_{\beta_1,\beta_2}^\delta$ is non-empty when $\delta = 0 \in
{\cal I}({\cal A})$. The vector
space $\oplus_\delta ({\cal V}_\delta^{(g=1)} \otimes
{\cal V}_{\beta_1,\beta_2}^\delta)$ is of 3-dimensions. By using
$\{ Y \otimes (\Theta_r \Omega Y_\epsilon), \; Y_\epsilon \otimes (\Theta_r\Omega Y_\epsilon), \; Y_\sigma \otimes (\Theta_r \Omega Y_\epsilon) \}$ as a basis,
the matrix representation of $S \cdot a \cdot S^{-1}$ is given by 
\begin{align}
 S \cdot a \cdot S^{-1} & \; =  \frac{1}{2} \left( \begin{array}{ccc} 1 & 1 & \sqrt{2} \\ 1 & 1 & -\sqrt{2} \\ \sqrt{2} & - \sqrt{2} & 0 \end{array} \right) \cdot
  \left( \begin{array}{ccc} -1 & & \\ & -1 & \\ & & 1 \end{array} \right)
  \cdot 
  \left( \begin{array}{ccc} 1 & 1 & \sqrt{2} \\ 1 & 1 & -\sqrt{2} \\ \sqrt{2} & - \sqrt{2} & 0 \end{array} \right) \frac{1}{2} \nonumber \\
 & \; = \left( \begin{array}{ccc} & -1 & \\ -1 & & \\ & & -1 \end{array}
  \right);
  \label{eq:Ising-g1-SaS-eeCase}
\end{align}
here, the matrix in the middle, $\diag(-1,-1,1)$, is the representation of
$({\rm id}\otimes e^{2\pi i (h_\kappa - h_\lambda)})$ on a basis
of $({\cal V}_{\epsilon,\epsilon}^0\otimes {\cal V}_{\epsilon,0}^\epsilon)$,
$({\cal V}_{\epsilon,0}^\epsilon \otimes {\cal V}_{\epsilon,\epsilon}^0)$
and $({\cal V}_{\epsilon,\sigma}^\sigma \otimes {\cal V}_{\epsilon,\sigma}^\sigma)$;
details of the choice of a generator in each of the three 1-dimensional spaces do not matter in the calculation
$a = {\cal F} \cdot ({\rm id}\otimes e^{2\pi i (h_\kappa-h_\lambda)}) \cdot
{\cal F}^{-1}$ above, because the isomorphism ${\cal F}$ provides a
one-to-one correspondence between the three 1-dimensional spaces in the middle
with the three 1-dimensional spaces $\{ {\cal V}_{0,\kappa}^\kappa \otimes {\cal V}_{\epsilon, \epsilon}^0 \; | \; \kappa \in {\cal I}({\cal A})\}$
at the beginning/end. 

In computing the matrix representation of $b$, a little more detail
is necessary. The fusion isomorphism
${\cal F}: \oplus_{\lambda \in {\cal I}}
({\cal V}_{\epsilon,\lambda}^\kappa \otimes {\cal V}_{\epsilon,\kappa}^\lambda)
\rightarrow \oplus_{\delta \in {\cal I}}
({\cal V}_{\delta,\kappa}^\kappa \otimes {\cal V}_{\epsilon,\epsilon}^\delta)$
is of the form 
\begin{align}
  {\cal V}_{\epsilon,\epsilon}^0 \otimes {\cal V}_{\epsilon,0}^\epsilon
  \ni (\Theta_r \Omega Y_\epsilon)\otimes (\Omega Y_\epsilon)
  & \mapsto Y \otimes (\Theta_r \Omega Y_\epsilon) \in
      {\cal V}_{0,0}^0 \otimes {\cal V}_{\epsilon,\epsilon}^0, \nonumber \\
 {\cal V}_{\epsilon,0}^\epsilon \otimes {\cal V}_{\epsilon,\epsilon}^0
 \ni (\Omega Y_\epsilon) \otimes (\Theta_r\Omega Y_\epsilon)
 & \mapsto Y_\epsilon \otimes (\Theta_r\Omega Y_\epsilon)
 \in {\cal V}_{0,\epsilon}^\epsilon \otimes {\cal V}_{\epsilon,\epsilon}^0,
 \label{eq:fusionI-Ising-a} \\
     {\cal V}_{\epsilon,\sigma}^\sigma \otimes {\cal V}_{\epsilon,\sigma}^\sigma
     \ni {\cal Y}_0 \otimes {\cal Y}_0 & \mapsto \propto 
     Y_\sigma \otimes (\Theta_r \Omega Y_\epsilon) \in 
   {\cal V}_{0,\sigma}^\sigma \otimes {\cal V}_{\epsilon,\epsilon}^0,    
  \nonumber 
\end{align}
for $\kappa = 0,\epsilon, \sigma$, respectively.
We will be more specific about a generator ${\cal Y}_0$ of
${\cal V}_{\epsilon,\sigma}^\sigma$ later. The result in the first line
is a special case of (\ref{eq:def-vct-sp-int.op-varF2}). 
The result in the second line is non-trivial.\footnote{
\label{fn:Ising4pt-a}
Think of the $g=0$ chiral correlation function
$\tilde{G}(x) := \bra{\sigma_3} \sigma_4(1) \sigma_1(x) \ket{\sigma_2}$,
where $x \in \C \backslash \{0,1\}$ and $\sigma_{i=1,2,3,4}$ are Virasoro
primary states of conformal weight $h_{i=1,2,3,4}$. When the state
$\sigma_1$ has a null vector at its level-2 excitation (ie, at $h=h_1+2$)
(cf. \cite[\S5]{Belavin:1984vu}, \cite[\S4.1, \S5.2]{Ginsparg:1988ui})
%
\begin{align}
  \left[ \frac{3}{2(2h_1+1)} \frac{d^2}{dx^2}
    + \left( \frac{1}{x} + \frac{1}{x-1}\right) \frac{d}{dx}
    - \left( \frac{h_2}{x^2} + \frac{h_4}{(x-1)^2}
    - \frac{h_1+h_2+h_3-h_4}{x(x-1)} \right) \right]\tilde{G}(x) = 0.
 \label{eq:g0n4-nullEq}
\end{align}

In the second line of (\ref{eq:fusionI-Ising-a}), we use the fact that
a solution to this (\ref{eq:g0n4-nullEq}) with $h_{i=1,2,3,4}=1/2$ is
$\tilde{G}(x) \propto e^{\pi i (2r+1)/2}(1-x+x^2)/[x(1-x)]$. Its expansion
$e^{\pi i(2r+1)/2}x^{-1}(1+{\cal O}(x))$ is interpreted as that of
$(\Omega Y_\epsilon)\otimes (\Theta_r\Omega Y_\epsilon)$ before the fusion, and
the other expansion $e^{\pi i (2r+1)/2}(1-x)^{-1}\cdot [1+{\cal O}(1-x)]$
interpreted as $Y_\epsilon \otimes (\Theta_r \Omega Y_\epsilon)$ after
the fusion. 
} %
Using the result above, we see that the translation ${\cal F} \cdot {\rm exch}
\cdot {\cal F}^{-1}$ is represented by
\begin{align}
  \left( \begin{array}{ccc} & 1 & \\ 1 & & \\ & & 1 \end{array} \right)
\end{align}
on the basis $\{ Y\otimes (\Theta_r \Omega Y_\epsilon), \;
Y_\epsilon \otimes (\Theta_r \Omega Y_\epsilon), \;
Y_\sigma \otimes (\Theta_r \Omega Y_\epsilon) \}$. Because the skewsymmetry
isomorphisms on ${\cal V}_{\epsilon,\epsilon}^0$ just multiply $(-1)$,
the representation matrix of $b$ agrees with (\ref{eq:Ising-g1-SaS-eeCase}). 

Think of the remaining case, $(\beta_1,\beta_2) = (\sigma, \sigma)$. 
The vector space $\oplus_\delta ({\cal V}_\delta^{(g=1)} \otimes
{\cal V}_{\sigma,\sigma}^\delta)$ is of 4-dimensions; we may use
\begin{align}
  \{ Y \otimes (\Theta_r \Omega Y_\sigma), \;
Y_\epsilon \otimes (\Theta_r \Omega Y_\sigma), \;
Y_\sigma \otimes (\Theta_r \Omega Y_\sigma), \;
{\cal Y}_0 \otimes (e^{\pi i(2s+1)/2} \Theta_r \Omega_s {\cal Y}_0) \},
 \label{eq:basis-g1n2Ising-ssCase}
\end{align}
as a basis of this vector space, where
${\cal Y}_0$ is the generator of ${\cal V}_{\epsilon,\sigma}^\sigma$
characterized by the property ${\cal Y}_0(\epsilon) \cdot \sigma
= z^{-1/2} \sigma + {\cal O}(z^{1/2})$. The fusion isomorphism
${\cal F}: \oplus_\lambda ({\cal V}_{\sigma,\lambda}^\kappa \otimes
{\cal V}_{\sigma,\kappa}^\lambda) \rightarrow \oplus_\delta
({\cal V}_{\delta,\kappa}^\kappa \otimes {\cal V}_{\sigma,\sigma}^\delta)$
is given by
\begin{align}
  (\Theta_r\Omega Y_\sigma) \otimes (\Omega Y_\sigma) & \; \mapsto
  Y \otimes (\Theta_r \Omega Y_\sigma),  \nonumber \\
  (\Theta_r \Omega_s {\cal Y}_0) \otimes (\Omega_{-1-s} {\cal Y}_0) & \; \mapsto
  Y_\epsilon \otimes (\Theta_r \Omega Y_\sigma) 2, 
  \label{eq:fusionI-Ising-b} \\
  \left( (\Omega Y_\sigma) \otimes (\Theta_r \Omega Y_\sigma), \;
  (\Omega_{-1-s} {\cal Y}_0) \otimes (\Theta_r \Omega_s {\cal Y}_0) \right) &
  \nonumber \\
  & \!\!\!\!\!\!\!\!\!\!\!\!\!\!\!\!\!\!\!\!\!\!\!\!
  \!\!\!\!\!\!\!\!\!\!\!\!\!\!\!\!\!\!\!\!\!\!\!\! \mapsto 
  \left( Y_\sigma \otimes (\Theta_r\Omega Y_\sigma),
  {\cal Y}_0 \otimes (e^{\pi i (2s+1)/2} \Theta_r \Omega_s {\cal Y}_0)  \right)
  \left( \begin{array}{cc} 1 & 2 \\ 1/2 & -1 \end{array} \right)
  \frac{1}{\sqrt{2}}   \nonumber 
\end{align}
for $\kappa = 0, \epsilon$ and $\sigma$, respectively.
The result in the first line of (\ref{eq:fusionI-Ising-b}) is a special case
of (\ref{eq:def-vct-sp-int.op-varF2}), while those in the last two lines of
(\ref{eq:fusionI-Ising-b}) are non-trivial.\footnote{
\label{fn:Ising4pt-b}
In the second line of (\ref{eq:fusionI-Ising-b}), we use the fact that
$\tilde{G}(x) = e^{\pi i(2r+1)/16}(1+x)/[(1-x)^{1/8}x^{1/2}]$ is a solution to
(\ref{eq:g0n4-nullEq}) with $h_{1,4}=1/16$ and $h_{2,3}=1/2$.
Its expansion $e^{\pi i(2s+1)/2}x^{-1/2}[e^{\pi i\left(\frac{(2r+1)}{16}-\frac{(2s+1)}{2}\right)} +{\cal O}(x)]$ is interpreted as
that of $(\Theta_r\Omega_s {\cal Y}_0) \otimes (\Omega_{-1-s} {\cal Y}_0)$
before
the fusion, and the other expansion $2e^{\pi i(2r+1)/16}(1-x)^{-1/8} [1 + {\cal O}(1-x)]$
interpreted as that of $2 Y_\epsilon \otimes (\Theta_r\Omega Y_\sigma)$ after
the fusion. 
}\raisebox{5pt}{,}\footnote{
\label{fn:Ising4pt-c}
In the third line of (\ref{eq:fusionI-Ising-b}), we use the fact that
$\tilde{G}_{\pm} (x) = e^{\pi i (2r+1)/16} \sqrt{1 \pm \sqrt{1-x}} /
[x^{1/8}(1-x)^{1/8}]$ are
two independent solutions to (\ref{eq:g0n4-nullEq}) with
$h_{1,2,3,4}=1/16$. Before the fusion, the two solutions
$2^{-1/2}\tilde{G}_+(x)$ and $2^{1/2}\tilde{G}_-(x)$ are for
$(\Omega Y_\sigma) \otimes (\Theta_r \Omega Y_\sigma)$ and
$(\Omega_{-1-s} {\cal Y}_0)\otimes (\Theta_r\Omega_s {\cal Y}_0)$, respectively.
After the fusion, $(\tilde{G}_+(x) + \tilde{G}_-(x))/2$ and
 $(\tilde{G}_+(x)-\tilde{G}_-(x))$ are for
 $Y_\sigma \otimes (\Theta_r \Omega Y_\sigma)$ and
 ${\cal Y}_0 \otimes (\Theta_r\Omega {\cal Y}_0)$, respectively.
}%

In computing the matrix representation of the $b$-move, therefore,
the matrix representation of ${\cal F} \cdot {\rm exch} \cdot {\cal F}^{-1}$
on the basis (\ref{eq:basis-g1n2Ising-ssCase}) is 
\begin{align*}
  \left( \begin{array}{cccc} 1 & & & \\ & 2 & & \\ & & 2^{-1/2} & 2^{1/2}\\
    & & 2^{-3/2} & -2^{-1/2} \end{array} \right) \left(
  \begin{array}{cccc} & & 1 & \\ & & & 1 \\ 1 & & & \\ & 1 & &
  \end{array} \right)
  \left( \begin{array}{cccc} 1 & & & \\ & 1/2 & & \\ & & 2^{-1/2} & 2^{1/2} \\
    & & 2^{-3/2} & -2^{-1/2} \end{array} \right). 
\end{align*}
The skewsymmetry isomorphism $({\rm id} \otimes \Omega_0)$ is
$e^{-2\pi i/16} \diag(1,1,1,e^{\pi i/2})$ on this basis. So, the $b$-move
is represented on the basis (\ref{eq:basis-g1n2Ising-ssCase}) by the matrix
\begin{align}
  b = \frac{e^{-2\pi i/16}}{\sqrt{2}}\left( \begin{array}{cccc}
    & & 1 & 2 e^{\pi i/2} \\
    & & 1 & -2 e^{\pi i/2} \\
    1 & 1 & & \\
    1/2 & -1/2 & & \end{array} \right).
  \label{eq:b-repr-g1n2-Ising-ssCase} 
\end{align}
The $a$-move is represented on the same basis by
\begin{align*}
  e^{-2\pi i/16}
  \left( \begin{array}{cccc} 1 & & & \\ & 2 & & \\ & & 2^{-1/2} & 2^{1/2}\\
    & & 2^{-3/2} & -2^{-1/2} \end{array} \right) \left(
 \begin{array}{cccc} 1 & & & \\ & - 1 & & \\ & &
   e^{2\pi i/8} & \\ & & & - e^{2\pi i/8}
  \end{array} \right)
  \left( \begin{array}{cccc} 1 & & & \\ & 1/2 & & \\ & & 2^{-1/2} & 2^{1/2} \\
    & & 2^{-3/2} & -2^{-1/2} \end{array} \right) \\
  = e^{-2\pi i/16} \left( \begin{array}{cccc} 1 & & & \\ & -1 & & \\
    & & & e^{2\pi i/8}2 \\
    & & e^{2\pi i/8}/2 & \end{array} \right).
\end{align*}
It is now a straightforward computation by using
$S = S(0) \oplus S(\epsilon)$, with (\ref{eq:Smtrx-Ising-0},
\ref{eq:Smtrx-Ising-e}),
to see that the 4 x 4 matrix $S \cdot a \cdot S^{-1}$ agrees with
that of the $b$-move (\ref{eq:b-repr-g1n2-Ising-ssCase}). 
\end{exmpl}

\begin{anythng}
Ref. \cite{Moore:1988uz} also derives a corollary of the relation
$S \cdot a \cdot S^{-1} = b$ on
$\oplus_{\delta \in {\cal I}({\cal A})} ({\cal V}_\delta^{(g=1)}
\otimes {\cal V}_{\beta_1,\beta_2}^\delta)$ in its version with 
$\beta_1 = \Upsilon^{\cal I}(\beta_2)$;  it is in the form of
\begin{align*}
  e^{2\pi i h_\beta}
  \left.\left( S \cdot a \cdot S^{-1} \right)\right|_{{\cal V}_0^{(g=1)} \otimes
    {\cal V}_{\Upsilon^{\cal I}(\beta),\beta}^0}
  = \left. \left( {\cal F} \cdot {\rm exch} \cdot {\cal F}^{-1} \right)
  \right|_{{\cal V}_0^{(g=1)} \otimes {\cal V}_{\Upsilon^{\cal I}(\beta),\beta}^0}. 
\end{align*}
The skewsymmetry isomorphism $({\rm id}\otimes \Omega_0)$ as a part of
the $b$-move acts on the subspace ${\cal V}_0^{(g=1)} \otimes
{\cal V}_{\Upsilon^{\cal I}(\beta),\beta}^0$ as a scalar multiplication of
$e^{-2\pi i h_\beta}$, and this scalar multiplication has been moved to the
LHS in the relation above. The complex phase
$e^{2\pi i (h_\kappa -h_\lambda)}$ within the implementation of the $a$-move and
the scalar multiplication $e^{2\pi i h_\beta} = e^{2\pi i h_{\beta_2}}$ in the
LHS above are combined to become $e^{2\pi i (h_\kappa+h_{\beta_2}-h_\lambda)}$, which
is the response of ${\cal Y} \in {\cal V}_{\beta_2,\kappa}^\lambda$ under
the change $e^{-iw_2} \rightarrow e^{-i(w_2+2\pi)}$. 

Verlinde formula of a rational VOA follows from the relation above.
Ref. \cite[p.456]{Moore:1988uz} provides a sketch of how to rewrite
the RHS and the combination $e^{2\pi i h_\beta} a$ on the LHS in the relation
above to do so. The present author found Ref. \cite[\S3--\S4]{MR2387861}
(and also \cite[\S2]{Moore:1988ss}) easier to follow.
\end{anythng}

\subsection{Braided/Modular Tensor Category of a VOA}
\label{ssec:ribbon-MTC}

Such a sentence as ``a rational VOA is a braided/modular tensor category'' often appears in physics literature. Of course, this sentence itself has too little information to be either true or false. The present author has struggled while going through physics literature to find out a minimum version of such statements; a minimum version should include a description of at least the set of objects, the sets of morphisms, and what the $\otimes$ functor does in practice, but individual papers from the physics community often lack some of that information. Math literature, on the other hand, poses other challenges to physicists; it is often a culture of the math community to keep distance from intuitions, and also to avoid talking---in publications---about motivations, or what the essence of a proof is.

The following notes (\S \ref{ssec:ribbon-MTC}) are intended primarily as an exposition with which the present author is satisfied, but these notes might also ease the challenge of communication a little for some other physicists. The exposition below focuses on writing down a main story; it is not self-contained or intended as a lecture note.  It is more like literature notes while the present author was reading a part of \cite{MR1691794}, \cite[\S\S 1--3]{MR1797619}, \cite{MR2468370}.

Given a VOA, $(V, Y, {\bf 1}, \omega)$, there are at least two different
ways to handle information associated with the VOA in terms of category
theory. One is ${\cal C}_{\rm VOA}^{\rm tree}$, the {\it category of trees}; 
it is a way to extract a set of data from the original VOA so that 
the set of data fits into the definition of braided tensor category, and
even of ribbon category. If the VOA is rational, 
with additional data extracted from the original VOA, the category of trees
has almost all the materials to be regarded as a modular tensor category,
but it still stops short of being a modular tensor category
(see Rmk. \ref{rmk:treeVSmod}). An advantage of
this approach is that all the data in this approach are explained/described
by using the materials and logic physicists are familiar with (those reviewed
in \S \ref{ssec:review-vct-sp-intrtw-op}). A disadvantage
is that the category of trees is only meant as a way of packaging some
of the information of the original VOA, while other detailed information
fades into the background. 

The other is ${\cal C}_{\rm VOA}^{\rm mod}$, the {\it category of $(V,Y)$-modules}.
The data set of ${\cal C}_{\rm VOA}^{\rm mod}$ retains all the information of the VOA, and---if the VOA is rational---also fits into the definition of modular tensor category.
One disadvantage is that we need to introduce a construction---the $\boxtimes$-product (reviewed in Ex. \ref{exmpl:BTC-for-VOA-1})---that is often avoided in physics literature in offering a description of the data set of ${\cal C}_{\rm VOA}^{\rm mod}$.  

In the description of both ${\cal C}_{\rm VOA}^{\rm tree}$ and
${\cal C}_{\rm VOA}^{\rm mod}$, we deal with vector spaces over
the field $k= \C$. 
In the case the VOA is rational, however, it may be possible to replace
$k=\C$ by $\overline{\Q}$, the field of all the algebraic numbers. 

\begin{defn}
\label{defn:BTC-1}
  A {\it braided tensor category} (BTC) consists of a set of data
\[  
\left( {\cal C}, \otimes, {\bf 1}_{\cal C}, \{ \lambda_U \}, \{ \rho_U \},
\{ \alpha_{U,V,W} \}, \{ c_{U,V} \} \right)
\]
that is explained in Def. \ref{defn:BTC-1} (on $\otimes$, ${\bf 1}_{\cal C}$,
$\lambda_U$ and $\rho_U$) and Def. \ref{defn:BTC-2} (on $\alpha_{U,V,W}$
and $c_{U,V}$). ${\cal C}$ is a category, which has its set of objects ${\rm Obj}({\cal C})$,
and the set of morphisms ${\rm Mor}_{\cal C}(U, V)$ for any pair of
objects $U, V \in {\rm Obj}({\cal C})$. $\otimes$ is a bifunctor
$\otimes: {\cal C} \times {\cal C} \rightarrow {\cal C}$, where the category
${\cal C} \times {\cal C}$ is with ${\rm Obj}({\cal C} \times {\cal C}) =
{\rm Obj}({\cal C}) \times {\rm Obj}({\cal C})$, and
${\rm Mor}_{{\cal C}\times {\cal C}}((U_1,U_2), \; (V_1,V_2)) =
{\rm Mor}_{\cal C}(U_1,V_1) \times {\rm Mor}_{\cal C}(U_2,V_2)$.
For $(U_1,U_2) \in {\rm Obj}({\cal C}\times
 {\cal C})$, $\otimes (U,V) \in {\rm Obj}({\cal C})$ may also be denoted
 by $U_1 \otimes U_2$; for $(f_1,f_2) \in {\rm Mor}_{{\cal C}\times {\cal C}}(
 (U_1,U_2), \; (V_1,V_2))$, the image of $(f_1, f_2)$ under the functor
 $\otimes$, $\otimes(f_1,f_2)$ may also be denoted by $f_1\otimes f_2$. 

For one special object of ${\cal C}$, called {\it the unit object} and
denoted by ${\bf 1}_{\cal C}$, $\{ \lambda_U \; | \;
U \in {\rm Obj}({\cal C}) \}$
and $\{ \rho_U \;  | \; U \in {\rm Obj}({\cal C})\}$ are a collection
of invertible morphisms (isomorphisms) $\lambda_U \in
{\rm Mor}_{\cal C}({\bf 1}_{\cal C} \otimes U , U )$ and $\rho_U \in
{\rm Mor}_{\cal C}(U \otimes {\bf 1}_{\cal C}, U)$, respectively.
The collection of morphisms $\{ \lambda_U\}$ and $\{ \rho_U\}$ should satisfy
\begin{align}
  f \cdot \lambda_U & \; = \lambda_V \cdot ({\rm id}_{{\bf 1}_{\cal C}} \otimes f) \in {\rm Mor}_{\cal C}({\bf 1}_{\cal C} \otimes U, V), \label{eq:cond-BTC-LR1-natural} \\
  f \cdot \rho_U & \; = \rho_V \cdot (f \otimes {\rm id}_{{\bf 1}_{\cal C}})
  \in {\rm Mor}{\cal C}(U \otimes {\bf 1}_{\cal C}, V), \qquad
  {}^\forall f \in {\rm Mor}_{\cal C}(U, V); \nonumber
\end{align}
this condition is a colloquial translation of a condition that
$\{ \lambda_U \}$ [resp. $\{ \rho_U \}$] be a {\it natural transformation}
({\it natural isomorphism} if all the morphisms are invertible)
from the functor $({\bf 1}_{\cal C}, {\rm id}_{{\bf 1}_{\cal C}}) \otimes -:
{\cal C} \rightarrow {\cal C}$ [resp. $- \otimes ({\bf 1}_{\cal C},
  {\rm id}_{{\bf 1}_{\cal C}}): {\cal C} \rightarrow
  {\cal C}$]
to the functor ${\rm id}_{\cal C}: {\cal C} \rightarrow {\cal C}$. 
$\qquad$ This Def. \ref{defn:BTC-1} continues to Def. \ref{defn:BTC-2}. 
\end{defn}

\begin{anythng}
Under a widely accepted definition of BTC (as reviewed in \ref{defn:BTC-1}),
  ${\rm Mor}_{\cal C}(U,V)$ is
required only to be a set, where the composition law 
${\rm Mor}_{\cal C}(U,V) \times {\rm Mor}_{\cal C}(V, W) \rightarrow
{\rm Mor}_{\cal C}(U,W)$ is a set-theoretical map such that 
the composition law is associative, and each ${\rm Mor}_{\cal C}(U,U)$
has one identity element ${\rm id}_U$. In the categories
${\cal C}_{\rm VOA}^{\rm tree}$ and ${\cal C}_{\rm VOA}^{\rm mod}$ reviewed
below, the sets of morphisms are vector spaces over a field $k$. 
\end{anythng}

\begin{exmpl}
\label{exmpl:BTC-for-VOA-1}
Given a VOA, $(V, Y, {\bf 1}, \omega)$ and a set ${\cal I}((V,Y))$ of its distinct simple modules, we set ${\rm Obj}({\cal C}_{\rm VOA}^{\rm tree})$ as the collection of all the length-$n$ words with letters from ${\cal I}((V,Y))$ decorated by $(n-1)$ parentheses that describe distance between neighboring letters; the length $n$ is an arbitrary positive integer. An example is $U = (\beta_1 \cdot (\beta_2 \cdot \beta_3))$, where $\beta_{1,2,3}\in {\cal I}((V,Y))$ (cf \ref{statmnt:BnF} and Rmk. \ref{rmk:penta-hexa-id}). We call them {\it parenthesized words}. The parenthesized words that are the same as a word but are different in the decoration by the parentheses are treated as different objects in the category ${\cal C}_{\rm VOA}^{\rm tree}$. Alternatively, one may assign to a length-$n$ object of ${\cal C}_{\rm VOA}^{\rm tree}$ a tree trivalent graph with $n$ branches and one trunk, along with one flower ($\beta_i \in {\cal I}((V, Y))$) at the end of each branch, but not at the root of the trunk. The information carried by the parentheses can be replaced by the way the branches of the graph merge (see \cite{MR1691794}). On the other hand, the category ${\cal C}_{\rm VOA}^{\rm mod}$ is that of modules of the VOA made out as the direct sum of simple modules in ${\cal I}((V, Y))$; if the VOA $(V, Y)$ is rational, any module of the VOA is in ${\rm Obj}({\cal C}_{\rm VOA}^{\rm mod})$. 

The set of morphisms in the category ${\cal C}_{\rm VOA}^{\rm mod}$ is, for a pair of $(V,Y)$-modules $U, W \in {\rm Obj}({\cal C}_{\rm VOA}^{\rm mod})$, the set of homomorphisms as $(V,Y)$-modules; ${\rm Mor}_{{\cal C}_{\rm VOA}^{\rm mod}}(U, W) = {\rm Hom}_{(V,Y)}(U,W)$.
For any object $U$ in this category,\footnote{
To accommodate the case that $(V, Y)$ is not rational, it is better not to
enforce $N_U^\alpha < \infty$ strictly. 
} %
there is a unique choice of $\{ (N^\alpha_U)_{\alpha \in {\cal I}((V,Y))} \; | \; N^\alpha_U \in \Z_{\geq 0}\}$ so that there is an invertible morphism $U \cong \oplus_{\alpha \in {\cal I}((V,Y))} M_\alpha^{\oplus N_U^\alpha}$, where $M_\alpha$ is an irreducible module of $(V,Y)$.
It then follows that the set of morphisms ${\rm Mor}_{{\cal C}_{\rm VOA}^{\rm mod}}(U, W)$ is isomorphic to the vector space of matrices $\oplus_{\alpha \in {\cal I}((V,Y))} M_{N^\alpha_W, N^\alpha_U}(k)$. 

The set of morphisms ${\rm Mor}_{{\cal C}_{\rm VOA}^{\rm tree}}(U, W)$ in the category ${\cal C}_{\rm VOA}^{\rm tree}$, on the other hand, needs to be specified (contrary to the case in ${\cal C}_{\rm VOA}^{\rm mod}$).
We set it to be
\begin{align}
  {\rm Mor}_{{\cal C}_{\rm VOA}^{\rm tree}}(U, W) := \oplus_{\alpha \in {\cal I}((V,Y))}
  {\rm Hom}_{k}\left( ({\cal V}_U^\alpha)^\vee, \; ({\cal V}_{W}^{\alpha})^\vee
  \right);
\end{align}
here, for a length-$n$ parenthesized word $U \in {\rm Obj} ({\cal C}_{\rm VOA}^{\rm tree})$ (and its tree trivalent graph), ${\cal V}^\alpha_U$ denotes the genus-0 conformal block with $n$ incoming states from the simple modules designated by the $n$ letters appearing in $U$, and one outgoing state from the simple module $\alpha$ (this notation has appeared in Rmk. \ref{rmk:penta-hexa-id}).
There is a natural construction for the vector space ${\cal V}_{U}^\alpha$ that follows from the branching pattern of the tree graph (parenthesis information) of $U \in {\rm Obj}({\cal C}_{\rm VOA}^{\rm tree})$. 
When $U = (\beta_1 \cdot (\beta_2 \cdot \beta_3))$, for example, we have a construction ${\cal V}_U^\alpha := \oplus_{\gamma \in {\cal I}((V,Y))} {\cal V}_{\beta_1,\gamma}^\alpha \otimes_k {\cal V}_{\beta_2,\beta_3}^{\gamma}$.
For a lengh-1 word $U=\beta$, we set ${\cal V}_{\beta}^\alpha$ to be the
vector space $\delta_{\beta}^{\;\; \alpha} k$. 
We may call it the conformal block of type $\left( \alpha; U \right)$ instead of type $(\alpha;\beta_1,\gamma)\otimes (\gamma;\beta_2,\beta_3)$.
We also introduce the notation $N^\alpha_U := \dim_k {\cal V}^\alpha_U$, which is $\sum_{\gamma} N^\alpha_{\beta_1,\gamma} N^\gamma_{\beta_2,\beta_3}$ in the example above. We will explain shortly why we will deal with the vector-space homomorphisms between the dual vector spaces $({\cal V}_U^\alpha)^\vee$ and $({\cal V}_W^\alpha)^\vee$. 
For now, it is enough that ${\rm Mor}_{{\cal C}_{\rm VOA}^{\rm tree}}(U,W)$ is calculable for any given VOA and $U, W \in {\rm Obj} ({\cal C}_{\rm VOA}^{\rm tree})$. 

The functor $\otimes: {\cal C}_{\rm VOA}^{\rm tree} \times {\cal C}_{\rm VOA}^{\rm tree} \rightarrow {\cal C}_{\rm VOA}^{\rm tree}$ maps a pair of parenthesized words $U_1, U_2$ to $(U_1 \cdot U_2)$, the concatenation of the parenthesized words with
an extra parenthesis outside. A pair $(f_1,f_2)$ with $f_1 \in \oplus_\alpha {\rm Hom}_k(({\cal V}^\alpha_{U_1})^\vee, ({\cal V}^\alpha_{W_1})^\vee)$ and
$f_2 \in \oplus_\alpha {\rm Hom}_k(({\cal V}^\alpha_{U_2})^\vee, ({\cal V}^\alpha_{W_2})^\vee )$ is mapped by $\otimes$ to
\begin{align}
  \otimes(f_1,f_2) :=
  \left( f_1\otimes_k f_2 \otimes_k {\rm id}\right):
  & \; \oplus_{\delta,\epsilon \in {\cal I}}
  \left( ({\cal V}_{U_1}^\delta)^\vee \otimes_k ({\cal V}_{U_2}^\epsilon)^\vee
  \otimes_k ({\cal V}_{\delta,\epsilon}^{\alpha})^\vee \right) \\
  & \; \rightarrow
  \oplus_{\delta,\epsilon \in {\cal I}}
  \left( ({\cal V}_{W_1}^\delta)^\vee \otimes_k ({\cal V}_{W_2}^\epsilon)^\vee
  \otimes_k ({\cal V}_{\delta,\epsilon}^{\alpha})^\vee \right).  \nonumber 
\end{align}
Once again, this is how we set the $\otimes$-functor on the set of morphisms in the category ${\cal C}_{\rm VOA}^{\rm tree}$, rather than a consequence that follows from some principle. This is a natural choice, however, as we will see shortly. 

To describe the functor $\otimes: {\cal C}_{\rm VOA}^{\rm mod} \times {\cal C}_{\rm VOA}^{\rm mod} \rightarrow {\cal C}_{\rm VOA}^{\rm mod}$, on the other hand, we need a construction that is often avoided in physics literature (cf \cite[(280)]{Gaberdiel:1999mc} and \cite[(5.3.3)]{Gannon:2007mfj}).
As a first step, one may generalize the definition of the vector space of intertwining operators in \ref{statmnt:def-VOA-inttw-op} for three simple representations of a VOA to that for three representations each of which is a direct sum of simple representations;
no subtlety arises in doing so because $z^{-h_\beta-h_\gamma + h_\alpha}$ is well-defined on each of the direct summand.  For a pair of $(V,Y)$-modules $U_1 \cong \oplus_{\delta \in {\cal I}((V,Y))} (M_\delta)^{\oplus N^\delta_{U_1}}$ and $U_2 \cong \oplus_{\delta \in {\cal I}((V,Y))} (M_\delta)^{\oplus N^\delta_{U_2}}$, the object $\otimes (U_1, U_2)$ in ${\cal C}_{\rm VOA}^{\rm mod}$ is set as
\begin{align}
  \otimes (U_1, U_2) = U_1 \boxtimes U_2 := \oplus_{\alpha \in {\cal I}((V,Y))}
  M_\alpha \otimes_k ({\cal V}_{U_1,U_2}^\alpha)^\vee.
  \label{eq:def-BTC-byMod-boxtimes}
\end{align}
Intuition behind this construction is that states in $M_{U_1}$ and $M_{U_2}$ are mapped into $M_\alpha$ by all the $\dim_{k} {\cal V}_{U_1,U_2}^\alpha$ independent different ways.
There are a couple of different ways\footnote{
In math literature, the $(V,Y)$-module $U_1 \boxtimes U_2$ is characterized by the universal property that any intertwining operator ${\cal Y} \in {\cal V}_{U_1,U_2}^\alpha$ is realized as $f \cdot \widetilde{\cal Y}$ where $f: U_1 \boxtimes U_2 \rightarrow M_\alpha$ is a homomorphism of $(V,Y)$-modules appropriate for ${\cal Y}$, and $\widetilde{\cal Y}$ an intertwining operator that takes values in
a module $U_1 \boxtimes U_2$ (see \cite[\S5]{MR2691082}, \cite{MR1338873}).
Characterizing $U_1 \boxtimes U_2$ as in (\ref{eq:def-BTC-byMod-boxtimes}) is a physicist-friendly translation of this. 
Indeed, suppose that $\{ {\cal Y}_{i=1,\cdots, N_{U_1, U_2}^\alpha}\}$ is a basis of the vector space ${\cal V}_{U_1,U_2}^\alpha$; then $U_1 \boxtimes U_2 = \oplus_{\alpha \in {\cal I}((V,Y))} \left( \oplus_{i=1}^{N_{U_1,U_2}^\alpha} M_\alpha \eta^i \right)$ according to (\ref{eq:def-BTC-byMod-boxtimes}),  
where $\{ \eta^{i=1,\cdots, N_{U_1,U_2}^\alpha} \}$ is the basis dual to $\{ {\cal Y}_i \}$. For ${\cal Y} = \sum_i c^i {\cal Y}_i \in {\cal V}_{U_1,U_2}^\alpha$, where $c^i \in k$, we may choose the homomorphism $f: U_1 \boxtimes U_2 \rightarrow M_\alpha$ as $\eta^i \mapsto c^i$. 
} %
 to see that we should have $\otimes_k ({\cal V}_{U_1,U_2}^\alpha)^\vee$ than $\otimes_k {\cal V}_{U_1,U_2}^\alpha$ in (\ref{eq:def-BTC-byMod-boxtimes});
 the most intuitive one is to note that we hope $U_1 \boxtimes U_2$ to be something like $U_1 \otimes_k U_2$ rather than their dual space;
 if we had used $\otimes_k {\cal V}_{U_1,U_2}^\alpha$ in (\ref{eq:def-BTC-byMod-boxtimes}), $U_1 \boxtimes U_2$ would have become something like $M_\alpha^{\otimes 2} \otimes (U_1^\vee \otimes U_2^\vee)$ instead. 
For the object $\otimes (U_1 , U_2) = U_1 \boxtimes U_2$, $N^\alpha_{\otimes (U_1,U_2)} = \sum_{\delta,\epsilon \in {\cal I}((V,Y))} N^\delta_{U_1} N^\epsilon_{U_2} N_{\delta,\epsilon}^\alpha = N_{U_1,U_2}^\alpha$.
It follows that the set of morphisms is
\begin{align*}
  {\rm Mor}_{{\cal C}_{\rm VOA}^{\rm mod}} & \; \left( U_1 \boxtimes U_2, \; W_1 \boxtimes W_2\right) \cong \\
  & \; \oplus_{\alpha \in {\cal I}((V,Y))} {\rm Hom}_k \left(
  \oplus_{\delta,\epsilon} (({\cal V}_{\delta,\epsilon}^\alpha)^\vee)^{\oplus N^\delta_{U_1}N^\epsilon_{U_2}} , \; \oplus_{\delta,\epsilon} (({\cal V}_{\delta,\epsilon}^\alpha)^\vee)^{\oplus N^\delta_{W_1}N^\epsilon_{W_2}} \right).
\end{align*}

For a pair $(f_1,f_2)$ of homomorphisms of $(V,Y)$-modules,
$f_1 \in {\rm Mor}_{{\cal C}_{\rm VOA}^{\rm mod}}(U_1, W_1)$ and
$f_2 \in {\rm Mor}_{{\cal C}_{\rm VOA}^{\rm mod}}(U_2, W_2)$, the vector space
of intertwining oeprators ${\cal V}_{W_1,W_2}^\alpha$ is pulled back by
$f_1^*$ and $f_2^*$ to ${\cal V}_{U_1,U_2}^\alpha$. 
The morphism $\otimes (f_1,f_2)$ is the dual of this $(f_1^* \otimes_k f_2^*)$.

It must now be obvious why the set of morphisms in the category
${\cal C}_{\rm VOA}^{\rm tree}$ was chosen as above. There is a functor
$\imath: {\cal C}_{\rm VOA}^{\rm tree} \rightarrow {\cal C}_{\rm VOA}^{\rm mod}$; 
to $U \in {\rm Obj}({\cal C}_{\rm VOA}^{\rm tree})$ as a parenthesized word,
such as $U = (\beta_1 \cdot (\beta_2 \cdot \beta_3))$, the functor $\imath$
assigns the $\boxtimes$-product of the irreducible modules in the word,
such as $M_U := M_{\beta_1} \boxtimes (M_{\beta_2} \boxtimes M_{\beta_3})$. 
The sets of morphisms in ${\cal C}_{\rm VOA}^{\rm tree}$ are chosen so that  
${\rm Mor}_{{\cal C}_{\rm VOA}^{\rm tree}}( \otimes (\beta_1, \beta_2), \;
\otimes (\gamma_1,\gamma_2) )$ for
$\beta_1,\beta_2,\gamma_1,\gamma_2 \in {\cal I}((V,Y))$ (as well as
the length-1 objects in ${\rm Obj}({\cal C}_{\rm VOA}^{\rm tree})$) is
the same as ${\rm Mor}_{{\cal C}_{\rm VOA}^{\rm mod}}( \otimes(M_{\beta_1} , M_{\beta_2}),
\; \otimes(M_{\gamma_1},M_{\gamma_2}))$, i.e., so that $\imath$ is a functor.
Note that the map $\imath: {\rm Obj}({\cal C}_{\rm VOA}^{\rm tree}) \rightarrow
{\rm Obj}({\cal C}_{\rm VOA}^{\rm mod})$ is not surjective. 
The zero-module $\{0\}$ as an object of ${\cal C}_{\rm VOA}^{\rm mod}$
is never obtained as $M_U$ of a parenthesized word $U$ of length $n\geq 1$;
when the VOA in question has a fusion ring that has a structure of a group
ring, $M_U$ has $N^\alpha_U \neq 0$ only for just one $\alpha \in
{\cal I}((V,Y))$, and $N^\alpha_U = 1$ for that $\alpha$, regardless of
length of the parenthesized word $U \in {\rm Obj}({\cal C}_{\rm VOA}^{\rm tree})$. 

The unit object of ${\cal C}_{\rm VOA}^{\rm tree}$ is the length-1 word $0$,
where $0 \in {\cal I}$ is the defining representation of the VOA $(V, Y)$. 
The unit object of ${\cal C}_{\rm VOA}^{\rm mod}$ is $V = M_0$ (i.e.,
$N^\alpha_0 = \delta^\alpha_0$). 

In both ${\cal C}_{\rm VOA}^{\rm tree}$ and
${\cal C}_{\rm VOA}^{\rm mod}$, $\lambda_U$
 $\in \oplus_{\alpha \in {\cal I}}{\rm Hom}_k
 (({\cal V}_{(0\cdot U)}^\alpha)^\vee , \; ({\cal V}_U^\alpha)^\vee )$
is set as the dual of  
\begin{align}
  \lambda_U^*: {\cal V}_U^\alpha \ni {\cal Y} \mapsto \left(
  1 \otimes_k {\cal Y} \otimes_k Y_\alpha\right)_{\delta=0,\epsilon=\alpha}
  \in \oplus_{\delta,\epsilon} \left( {\cal V}_{0}^\delta \otimes_k {\cal V}_U^\epsilon \otimes_k {\cal V}_{\delta,\epsilon}^\alpha \right). 
\end{align}
Similarly, $\rho_U \in \oplus_{\alpha \in {\cal I}}{\rm Hom}_k(
({\cal V}_{(U \cdot 0)}^\alpha)^\vee , ({\cal V}_U^\alpha)^\vee )$ is set
as the dual of 
\begin{align}
  \rho_U^*: {\cal V}_U^\alpha \ni {\cal Y} \mapsto \left(
  {\cal Y} \otimes_k 1 \otimes_k  \Omega Y_\alpha \right)_{\epsilon=0,\delta=\alpha}
  \in \oplus_{\delta,\epsilon} \left( {\cal V}_U^\delta \otimes_k
      {\cal V}_{0}^\epsilon \otimes_k {\cal V}_{\delta,\epsilon}^\alpha \right). 
\end{align}
This is indeed a way we can construct a basis of
${\cal V}_{ (0 \cdot U)}^\alpha$ and ${\cal V}_{(U \cdot 0)}^\alpha$ for
$U \in {\rm Obj}({\cal C}_{\rm VOA}^{\rm tree})$, and of 
${\cal V}_{0,U}^\alpha$ and ${\cal V}_{U,0}^\alpha$ for
$U \in {\rm Obj}({\cal C}_{\rm VOA}^{\rm mod})$. 
That this choice of $\{ \lambda_U \}$ and $\{ \rho_U \}$
satisfies (\ref{eq:cond-BTC-LR1-natural}) follows from the way
$\otimes(f_1,f_2)$ was set, as above.
\end{exmpl}

\begin{defn}
\label{defn:BTC-2} This is a continuation of Def. \ref{defn:BTC-1}. 
The collection of invertible morphisms $\{ \alpha_{U,V,W} \}$
for any triple of $U,V,W \in {\rm Obj}({\cal C})$ should give a natural
isomorphism $\alpha: \otimes \cdot ({\rm id}_{\cal C}, \otimes) \Rightarrow
\otimes \cdot (\otimes, {\rm id}_{\cal C})$, where both of the functors
are from ${\cal C}\times {\cal C}\times {\cal C}$ to ${\cal C}$:
for any triple $U_{i}, W_{i} \in {\rm Obj}({\cal C})$
and any $f_{i} \in {\rm Mor}_{\cal C}(U_{i}, W_{i})$ with $i=1,2,3$, 
\begin{align}
  ((f_1 \otimes f_2) \otimes f_3) \cdot \alpha_{U_1,U_2,U_3}
  = \alpha_{W_1,W_2,W_3} \cdot (f_1 \otimes (f_2 \otimes f_3))   
 \label{eq:cond-BTC-alpha-natural}
\end{align}
in ${\rm Mor}_{\cal C}(U_1 \otimes (U_2 \otimes U_3), \;
(W_1 \otimes W_2) \otimes W_3 )$. The collection of invertible
morphisms $\{ c_{U, W} \}$ for any pair of $U, W \in {\rm Obj}({\cal C})$
should give a natural isomorphism $c: \otimes \Rightarrow \otimes
\cdot (1\leftrightarrow 2)$, where both of the functors are
from ${\cal C} \times {\cal C} \rightarrow {\cal C}$; the functor
$(1\leftrightarrow 2)$ exchanges the first and second ${\cal C}$'s:
for any pair $U_i, W_i \in {\rm Obj}({\cal C})$ and for any $f_i \in {\rm Mor}_{\cal C}(U_i, W_i)$ with $i=1,2$,
\begin{align}
 (f_2 \otimes f_1) \cdot c_{U_1,U_2} = c_{W_1,W_2} \cdot (f_1 \otimes f_2)
 \label{eq:cond-BTC-c-natural}
\end{align}
in ${\rm Mor}_{\cal C}(U_1 \otimes U_2, \; W_2 \otimes W_1)$.

Those isomorphisms in ${\cal C}$ for the natural isomorphisms
must satisfy the pentagon identity
for any quartet of objects in ${\cal C}$, the hexagon identities
for any triple of objects in ${\cal C}$, and the triangle identity
$(\rho_U \otimes {\rm id}_W) \cdot \alpha_{U,{\bf 1}_{\cal C}, W} =
{\rm id}_U \otimes \lambda_W$ in ${\rm Mor}_{\cal C}( U \otimes
({\bf 1}_{\cal C} \otimes W), U \otimes W)$ for any pair $U, W \in {\rm Obj}({\cal C})$. 
\end{defn}

\begin{exmpl}
In the category ${\cal C}_{\rm VOA}^{\rm tree}$, the isomorphism
$\alpha_{U,V,W}$ is chosen as the dual of 
\begin{align}
  {\cal F}(\alpha;\alpha_U,\alpha_V;\alpha_W)^{-1} &
  \; \otimes_k {\rm id} \otimes_k {\rm id}\otimes_k {\rm id} :
  \label{eq:choice-BTC-alpha-4Cpre} \\
  {\cal V}_{((U\cdot V)\cdot W)}^\alpha & \; =
  \left( \oplus_{\epsilon \in {\cal I}}
 {\cal V}_{\alpha_U,\alpha_V}^{\epsilon} \otimes_k {\cal V}_{\epsilon, \alpha_W}^\alpha
 \right)\otimes_k 
 {\cal V}^{\alpha_U}_U \otimes_k {\cal V}^{\alpha_V}_V \otimes_k {\cal V}^{\alpha_W}_W
 \nonumber \\
 & \longrightarrow
 \left( \oplus_{\delta \in {\cal I}} {\cal V}^\alpha_{\alpha_U, \delta} \otimes_k
 {\cal V}^\delta_{\alpha_V, \alpha_W} \right)  \otimes_k 
 {\cal V}^{\alpha_U}_U \otimes_k {\cal V}^{\alpha_V}_V \otimes_k
 {\cal V}^{\alpha_W}_W = {\cal V}_{(U \cdot (V \cdot W))}^\alpha,  \nonumber 
\end{align}
and the isomorphism $c_{U,V}$ as the dual of\footnote{
Although we could have used $\Omega(-)^{-1} = \Omega(+)$ instead of
$\Omega(+)^{-1}$ above for the associativity morphisms and commutativity
morphisms $\{ \alpha \}$ and $\{ c\}$ to satisfy the hexagon identity, we use
$\Omega(+)^{-1}$ so that the condition (\ref{eq:cond-ribbon-theta-wTnsr})
is satisfied when we adopt the convention (\ref{eq:theta-in-ribbonC}). 
} %
\begin{align}
  \Omega(+)^{-1} \otimes_k {\rm id} \otimes_k {\rm id}:  
  {\cal V}^\alpha_{(V \cdot U)} & \; = {\cal V}^\alpha_{\alpha_V, \alpha_U}
  \otimes_k {\cal V}^{\alpha_V}_V \otimes_k {\cal V}^{\alpha_U}_U
  \label{eq:choice-BTC-c-4Cpre} \\
  & \; \longrightarrow {\cal V}^\alpha_{\alpha_U,\alpha_V} \otimes_k
  {\cal V}^{\alpha_V}_V \otimes_k {\cal V}^{\alpha_U}_U
    = {\cal V}^\alpha_{(U \cdot V)};   \nonumber 
\end{align}
The triangle identity in Def. \ref{defn:BTC-2} follows from the
condition (\ref{eq:def-vct-sp-int.op}) and the way $\lambda_V$ and
$\rho_U$ are set in ${\cal C}_{\rm VOA}^{\rm tree}$. 
In the category ${\cal C}_{\rm VOA}^{\rm mod}$,
the isomorphism $\alpha_{U,V,W}$ for
$U, V, W \in {\rm Obj}({\cal C}_{\rm VOA}^{\rm mod})$ is the dual of 
\begin{align}
  {\cal F}(\alpha; \alpha_U, \alpha_V; \alpha_W)^{-1} & \; \otimes_k
  {\rm id} \otimes_k {\rm id} \otimes_k {\rm id}:
  \label{eq:choice-BTC-alpha-4C} \\
 & \; \left( \oplus_{\epsilon \in {\cal I}} {\cal V}_{\alpha_U,\alpha_V}^\epsilon \otimes_k {\cal V}_{\epsilon,\alpha}^\alpha \right) \otimes_k k^{N^{\alpha_U}_U} \otimes_k
  k^{N^{\alpha_V}_V} \otimes_k k^{N^{\alpha_W}_W} \nonumber \\
 & \;  \longrightarrow
  \left( \oplus_{\delta \in {\cal I}} {\cal V}_{\alpha_U,\delta}^\alpha \otimes_k
       {\cal V}_{\alpha_V,\alpha_W}^\delta \right) \otimes_k
        \otimes_k k^{N^{\alpha_U}_U} \otimes_k
  k^{N^{\alpha_V}_V} \otimes_k k^{N^{\alpha_W}_W} \nonumber 
\end{align}
The isomorphism $c_{U,V}$ in the category ${\cal C}_{\rm VOA}^{\rm mod}$ is
also the dual of
\begin{align}
  \Omega(+)^{-1} \otimes_k {\rm id} \otimes_k :
  {\cal V}_{\alpha_V,\alpha_U}^\alpha \otimes_k k^{N^{\alpha_V}_V} \otimes_k
  k^{N^{\alpha_U}_U} \rightarrow {\cal V}_{\alpha_U,\alpha_V}^\alpha \otimes_k
  k^{N^{\alpha_V}_V} \otimes_k k^{N^{\alpha_U}_U} . 
\end{align}
The condition (\ref{eq:cond-BTC-alpha-natural}) follows obviously, because
${\cal F}[\alpha;\alpha_U,\alpha_V;\alpha_W]$ acts on the first two tensored
factors in (\ref{eq:choice-BTC-alpha-4Cpre}, \ref{eq:choice-BTC-alpha-4C}),
whereas $f_{1,2,3}$ act on the last three tensored factors in
(\ref{eq:choice-BTC-alpha-4Cpre}, \ref{eq:choice-BTC-alpha-4C}).
The condition (\ref{eq:cond-BTC-c-natural}) is also satisfied for the
same reason. 
\end{exmpl}

\begin{defn}
\label{defn:ribbon}  
A braided tensor category $({\cal C}, \otimes, {\bf 1}_{\cal C}, \{ \lambda_U \},
\{ \rho_U\}, \{ \alpha_{U,V,W}\}, \{ c_{U,V}\})$ accompanied by
an extra set of data (to be explained below)
\begin{align}
    \left( *, \{ i_U\}, \{ e_U \}, \{\theta_U\}  \right)
\end{align}
is called a {\it ribbon category}.
Here, $*$ is a map ${\rm Obj}({\cal C}) \rightarrow {\rm Obj}({\cal C})$; $*(U)$ for $U \in {\rm Obj}({\cal C})$ is also denoted by $U^*$.
All the other extra data are collections of morphisms for all $U \in {\rm Obj}({\cal C})$; 
the morphisms $i_U \in {\rm Mor}_{\cal C}({\bf 1}_{\cal C}, \; U \otimes U^*)$, 
$e_U \in {\rm Mor}_{\cal C}(U^* \otimes U, {\bf 1}_{\cal C})$ for each object $U$ {\it should} satisfy the {\it right-rigidity condition} ({\it right-zig-zag-equation})
\begin{align}
  ({\rm id}_U \otimes e_U) \cdot (i_U \otimes {\rm id}_U) = {\rm id}_U, \qquad
  (e_U \otimes {\rm id}_{U^*}) \cdot ({\rm id}_{U^*} \otimes i_U) = {\rm id}_{U^*}.
  \label{eq:cond-ribbon-Rrigid}
\end{align}
Under the assignment
\begin{align}
 f^* := (e_W \otimes {\rm id}_{U^*}) \cdot ({\rm id}_{W^*}\otimes f \otimes {\rm id}_{U^*}) \cdot ({\rm id}_{W^*} \otimes i_U) \in {\rm Mor}_{\cal C}(W^*,U^*)
\end{align}
for $f \in {\rm Mor}_{\cal C}(U, W)$, we then have a contravariant functor $(-)^*: {\cal C}\rightarrow {\cal C}$, and an ordinary functor $(-)^{**}: {\cal C} \rightarrow {\cal C}$. 
The object $U^*$ is called the {\it right-dual} of $U$, while $e_U$ and $i_U$ the {\it evaluation morphism} and {\it coevaluation morphism}, 
respectively. 

A collection of endomorphisms $\{ \theta_U \}$ {\it should} be a natural isomorphism $\theta: {\rm id}_{\cal C} \Rightarrow {\rm id}_{\cal C}$, meaning that
\begin{align}
  \theta_V \cdot f = f \cdot \theta_U, \qquad \qquad
        {}^\forall f \in {\rm Mor}_{\cal C}(U,V),
  \label{eq:cond-ribbon-theta-nat}
\end{align}
such that
\begin{align}
  \theta_{U \otimes W} = \left( \theta_{U} \otimes \theta_W \right) \cdot
  \left( c_{W, U} \cdot c_{U,W} \right) .
  \label{eq:cond-ribbon-theta-wTnsr}
\end{align}
Moreover, they {\it should} satisfy
\begin{align}
  (\theta_U)^* = \theta_{U^*}
  \label{eq:cond-ribbon-theta-wRgd}
\end{align}
in ${\rm Mor}_{\cal C}(U^*, U^*)$ for all $U \in {\rm Obj}({\cal C})$.
%
\end{defn}
\begin{anythng}
A set of data only with $({\cal C}, {\bf 1}_{\cal C}, \otimes, \{\lambda \}, \{\rho\}, \{\alpha\})$ without $\{ c\}$ is called a {\it monoidal category},
without the conditions involving $\{c\}$. Because the conditions
on $(*, \{i_U\}, \{e_U\})$ in Def. \ref{defn:ribbon} do not involve the
morphisms $\{c_{U,W}\}$, it is possible to think of adding enhanced structure
step by step on top of the underlying monoidal category. By adding 
$(*, \{ i_U\}, \{ e_U\})$ to the set of data 
$({\cal C}, \otimes, {\bf 1}_{\cal C}, \{ \lambda \}, \{ \rho \}, \{ \alpha\})$
of a monoidal category, along with the rigidity condition
(\ref{eq:cond-ribbon-Rrigid}), we have a {\it right-rigid monoidal
  category}.

Note that we have started to omit $\lambda_U$, $\rho_U$
and $\alpha_{U,V,W}$ from equations, eg. as in (\ref{eq:cond-ribbon-Rrigid}). 
Once we begin to talk about a right-rigid monoidal category,
isomorphisms between a pair of categories that are built from $\{ \lambda\}$,
$\{ \rho\}$, $\{\alpha\}$, $\{ e\}$ and $\{i\}$ are also often omitted from
equations. Examples include $\lambda_{({\bf 1}_{\cal C})^*} \cdot i_{{\bf 1}_{\cal C}} 
\in {\rm Mor}_{\cal C}({\bf 1}_{\cal C}, ({\bf 1}_{\cal C})^*)$, and
$(e_We_U \otimes {\rm id}_{(U\otimes W)^*}) \cdot [\alpha{\rm 's}] \cdot
({\rm id}_{(W^*\otimes U^*)}\otimes i_{(U\otimes W)}) \cdot
\rho_{(W^*\otimes U^*)}^{-1} \in {\rm Mor}_{\cal C}(W^* \otimes U^*,
(U\otimes W)^*)$; the expressions in (\ref{eq:cond-ribbon-delta-Tnsr}) already
exploit this convention. Those isomorphisms are called {\it canonical
  isomorphisms}.

In a right-rigid monoidal category,
the morphism $(e_U)^*$ is equal to $i_{U^*}$ modulo the canonical isomorphisms,
and so is $(i_U)^*$ to $e_{U^*}$ modulo the canonical isomorphisms. It follows
that $(e_U)^{**} = e_{U^{**}}$ and $(i_U)^{**} = i_{U^{**}}$. One can also derive
\begin{align}
  e_W \cdot ({\rm id}_{W^*} \otimes f) = e_U \cdot (f^* \otimes {\rm id}_U) \in
  {\rm Mor}_{\cal C}(W^* \otimes U, {\bf 1}_{\cal C}),
  \label{eq:prprty-right-rigid-e} \\
  ({\rm id}_W \otimes f^*) \cdot i_W =(f \otimes {\rm id}_{U^*}) \cdot i_U
  \in {\rm Mor}_{\cal C}({\bf 1}_{\cal C}, W \otimes U^*).
  \label{eq:prprty-right-rigid-i}
\end{align}
\end{anythng}
\begin{defn}
\label{defn:pivot-sphr}
Let $({\cal C}, \otimes, {\bf 1}_{\cal C}, \{ \lambda_U\}, \{\rho_U\},
\{\alpha_{U,V,W}\}, *, \{e_U\}, \{i_U\})$ 
be a monoidal category that is right-rigid. 
A set of morphisms $\{ \delta_U \}$ implementing a natural isomorphism
$\delta: {\rm id}_{\cal C} \Rightarrow (-)^{**}$ is called a {\it pivotal structure}, i.e.,
\begin{align}
  \delta_W \cdot f = f^{**} \cdot \delta_U \in {\rm Mor}_{\cal C}(U, W^{**}),
  \qquad \qquad {}^\forall f \in {\rm Mor}_{\cal C}(U,W), 
  \label{eq:cond-ribbon-delta-natC}
\end{align}
if it is compatible not just with the structure of the category of ${\cal C}$,
but also with that of $({\cal C}, \otimes, {\bf 1}_{\cal C}, \{\lambda\}, \{\rho\},
\{\alpha\})$; that is to {\it require} $\delta$ to be {\it monoidal},
\begin{align}
  \delta_{U \otimes V} = \delta_U \otimes \delta_V, \quad
  \delta_{{\bf 1}_{\cal C}} = {\rm id}_{{\bf 1}_{\cal C}}. 
  \label{eq:cond-ribbon-delta-Tnsr}
\end{align}
Note that the property
\begin{align}
  (\delta_{U^*})^{-1} = (\delta_U)^* \in {\rm Mor}_{\cal C}(W^*, U^*)
  \label{eq:prprty-ribbon-delta-*}
\end{align}
for $f \in {\rm Mor}_{\cal C}(U, W)$ follows from combining
(\ref{eq:cond-ribbon-delta-natC}),
(\ref{eq:cond-ribbon-delta-Tnsr}), (\ref{eq:prprty-right-rigid-e}) and
(\ref{eq:cond-ribbon-Rrigid}). 

A pivotal structure is said to be {\it spherical}, if 
\begin{align}
  {\rm tr}_R[f] = {\rm tr}_L[f], \qquad \qquad
  {}^\forall f \in {\rm Mor}_{\cal C}(U,U),
  \label{eq:cond-ribbon-sphere}
\end{align}
where 
\begin{align}
  {\rm tr}_R[f] & \; :=
  e_{U^*} \cdot ((\delta_U \cdot f) \otimes {\rm id}_{U^*}) \cdot i_U ,
  \label{eq:def-trace-RL} \\
{\rm tr}_L[f] & \; :=
e_{U} \cdot ({\rm id}_{U^*} \otimes (f \cdot \delta_U^{-1}) ) \cdot i_{U^*}
= e_{U^{**}} \cdot ((\delta_{U^*} \cdot f^*) \otimes {\rm id}_{U^{**}} ) \cdot i_{U^*}
\nonumber \\
 & \; = e_U \cdot (\delta_{U_*} \otimes f) \cdot i_{U_*}, 
\end{align}
where $U_*$ is the object whose right-dual is $U$.


%
\end{defn}

\begin{anythng}
\label{statmnt:thetaFirst-vs-deltaFirst}
  Suppose that a monoidal category is equipped both with a right-rigidity
and braiding structure; there are two next steps in enriching structure; one
is to add twists $\theta$ to make it a ribbon category (as in
Def. \ref{defn:ribbon}), and the other to
add a spherical pivotal structure $\delta$ (as in Def. \ref{defn:pivot-sphr}).
The two enrichments are, in fact, almost the same in the sense explained below. 

Suppose, first, that we are given $\{ \theta_U \}$, so we have a ribbon
category as in Def. \ref{defn:ribbon}.  Then there is a canonical way
to find a spherical pivotal structure: 
\begin{align}
  \delta_U & \; := u_U \cdot \theta_U = \theta_{U^{**}} \cdot u_U, \\
  u_U & \; := 
  (e_U \otimes {\rm id}_{U^{**}}) \cdot ({\rm id}_{U^*} \otimes c_{U,U^{**}}^{-1}) \cdot (i_{U^*} \otimes {\rm id}_U) \in {\rm Mor}_{\cal C}(U, U^{**}).
  \label{eq:def-Drinfeld-morph}
\end{align}
There is also an alternative expression 
\begin{align}
  u_U = (e_U \otimes {\rm id}_{U^{**}}) \cdot (c_{U,U^*} \otimes {\rm id}_{U^{**}})
  \cdot ({\rm id}_U \otimes i_{U^*}). 
\end{align}
%
%
All the desired properties (\ref{eq:cond-ribbon-delta-natC},
\ref{eq:cond-ribbon-delta-Tnsr}, \ref{eq:cond-ribbon-sphere}) of 
$\delta_U$ given above follow from the defining conditions
(\ref{eq:cond-ribbon-theta-nat}, \ref{eq:cond-ribbon-theta-wTnsr},
\ref{eq:cond-ribbon-theta-wRgd}) on $\theta$ and those of the underlying
right-rigid braided tensor category. 

Conversely, suppose that we are given a natural isomorphism $\delta : {\rm id}_{\cal C} \Rightarrow (-)^{**}$ subject to the conditions (\ref{eq:cond-ribbon-delta-natC}, \ref{eq:cond-ribbon-delta-Tnsr}, \ref{eq:cond-ribbon-sphere})
in addition to a right-rigid braided tensor category, as in
Def. \ref{defn:pivot-sphr}. By maintaining
the definition of $u_U$ in (\ref{eq:def-Drinfeld-morph}), and setting
\begin{align}
 \theta_U := u_U^{-1} \cdot \delta_U \in {\rm Mor}_{\cal C}(U,U),  
\end{align}
almost all we need for a ribbon category is obtained; the properties
(\ref{eq:cond-ribbon-theta-nat}, \ref{eq:cond-ribbon-theta-wTnsr}) on $\theta$
follow from the conditions (\ref{eq:cond-ribbon-delta-natC},
\ref{eq:cond-ribbon-delta-Tnsr}) on $\delta$. The condition
(\ref{eq:cond-ribbon-sphere}) on $\delta$ can be translated to that of
$\theta$ as follows:
\begin{align}
  e_U \cdot ((f^* \cdot (\theta_U)^*) \otimes {\rm id}_U) \cdot c_{U,U^*} \cdot
  i_U = 
  e_U \cdot ((f^* \cdot \theta_{U^*}) \otimes {\rm id}_U) \cdot c_{U,U^*} \cdot
  i_U 
\end{align}
in ${\rm Mor}_{\cal C}({\bf 1}_{\cal C}, {\bf 1}_{\cal C})$ for
${}^\forall f \in {\rm Mor}_{\cal C}(U,U)$. So, the property
(\ref{eq:cond-ribbon-theta-wRgd}) follows if we make an extra assumption
that any difference between $(\theta_U)^*$ and $\theta_{U^*}$ would be
detected by varying choices of $f \in {\rm Mor}_{\cal C}(U,U)$. 
\end{anythng}

\begin{exmpl}
In the category ${\cal C}_{\rm VOA}^{\rm tree}$, the map $*:
{\rm Obj}({\cal C}_{\rm VOA}^{\rm tree}) \rightarrow
{\rm Obj}({\cal C}_{\rm VOA}^{\rm tree})$ inverts $U \in {\rm Obj}
({\cal C}_{\rm VOA}^{\rm tree})$ as a parenthesized word of length $n$
from left to right, with the letters from ${\cal I}$ labeling
representations replaced by those of the contragradient
representations; for example, if $U = (\beta_1 \cdot (\beta_2 \cdot \beta_3))$,
then $U^* = ((\Upsilon^{\cal I}(\beta_3) \cdot \Upsilon^{\cal I}(\beta_2))
\cdot \Upsilon^{\cal I}(\beta_1))$. In the category ${\cal C}_{\rm VOA}^{\rm mod}$,
$U^*$ is the contragradient module of $U \in {\cal C}_{\rm VOA}^{\rm mod}$. 
So, $({\bf 1}_{\cal C})^* = {\bf 1}_{\cal C}$ in both of the categories. 
Now, $U^{**} = U$ for all the objects $U$ in the two categories. We choose
$\delta_U$ to be ${\rm id}_U$. 

For an arbitrary $r \in \Z$, we may specify the evaluation and coevaluation
morphisms $\{ e_U \}$ and $\{ i_U \}$ of ${\cal C}_{\rm VOA}^{\rm tree}$ and
${\cal C}_{\rm VOA}^{\rm mod}$ in the following way. In describing those
morphisms, we begin with the case $U \in {\rm Obj}({\cal C}_{\rm VOA}^{\rm tree})$
is a word of length-1 ($U= \beta$), and the case
$U = M_\beta \in {\rm Obj}({\cal C}_{\rm VOA}^{\rm mod})$ is a simple module. 
Note that $e_U \in {\rm Mor}_{\cal C}(U^* \otimes U, {\bf 1}_{\cal C}) \cong
{\rm Hom}_k(({\cal V}_{U^* \otimes U}^0)^\vee, k)$ is specified by
an element in ${\cal V}_{U^* \otimes U}^0$, and $i_U \in {\rm Mor}_{\cal C}({\bf 1}_{\cal C}, U \otimes U^*) \cong {\rm Hom}_k(k, ({\cal V}_{U \otimes U^*}^0)^\vee)$ by
an element in $({\cal V}_{U \otimes U^*}^0)^\vee$. The prescription
in \cite[p. 886]{MR2468370} in the case $U=\beta$ is to set 
\begin{align}
  e_\beta = F_\beta^{-1} (\Theta_r \Omega Y_{\Upsilon^{\cal I}(\beta)})
  \in {\cal V}_{\Upsilon^{\cal I}(\beta),\beta}^0, \qquad
  i_\beta : {\cal V}_{\beta, \Upsilon^{\cal I}(\beta)}^0 \ni
  (\Theta_r \Omega Y_\beta) \longmapsto 1;
  \label{eq:evNcoev-in-VOA}
\end{align}
here, $F_\beta$ is the $r$-independent coefficient in the $\epsilon = 0$
component in 
\begin{align}
  {\cal F} : \quad 
  \oplus_\delta ({\cal V}_{\beta,\delta}^\beta \otimes {\cal V}^\delta_{\Upsilon^{\cal I}(\beta),\beta} ) & \; \longrightarrow
  \oplus_{\epsilon} ({\cal V}_{\epsilon,\beta}^\beta \otimes {\cal V}_{\beta,\Upsilon^{\cal I}(\beta)}^\epsilon),   \label{eq:def-F_alpha} \\
  (\Omega Y_\beta) \otimes (\Theta_r \Omega Y_{\Upsilon^{\cal I}(\beta)}) & \;
  \longmapsto
  F_\beta \cdot Y_\beta \otimes (\Theta_r \Omega Y_\beta) + (\epsilon \neq 0 \;
  {\rm components}).  \nonumber 
\end{align}
The composition of the morphisms $({\rm id}_\beta \otimes e_\beta) \cdot
\alpha_{\beta, \beta^*, \beta} \cdot (i_\beta \otimes {\rm id}_\beta) \in {\rm Mor}_{\cal C}(\beta \otimes {\bf 1}_{\cal C}, {\bf 1}_{\cal C} \otimes \beta)$
in the right-rigidity condition
(\ref{eq:cond-ribbon-Rrigid}) is evaluated in the dual space, 
\begin{eqnarray*}
  {\cal V}_{\beta, 0}^\beta \rightarrow {\cal V}_{(\beta \cdot (\Upsilon^{\cal I}(\beta) \cdot \beta))}^\beta & \rightarrow & {\cal V}_{((\beta \cdot \Upsilon^{\cal I}(\beta))\cdot \beta)}^\beta \rightarrow {\cal V}_{0,\beta}^\beta \\
  \Omega Y_\beta \longmapsto F_\beta^{-1} (\Omega Y_\beta) \otimes (\Theta_r \Omega Y_{\Upsilon^{\cal I}(\beta)}) & \longmapsto & (\Theta_r \Omega Y_\beta) \otimes Y_\beta
  \longmapsto Y_\beta, 
\end{eqnarray*}
where the three maps are the dual of $e_\beta$, ${\cal F}$, and the dual of
$i_\beta$ from the left to right. So, this choice of $\{e_\beta \}$ and
$\{ i_\beta\}$ satisfies the first condition in (\ref{eq:cond-ribbon-Rrigid});
the second one is also verified by the same calculation 
\cite[Props. 3.3 and 3.7, Thm. 3.8]{MR2468370}.  

For ${\cal C}_{\rm VOA}^{\rm mod}$, the morphisms $e_U$ and $i_U$ for
a module $U = M_U \in {\rm Obj}({\cal C}_{\rm VOA}^{\rm mod})$ that is not
simple can be determined by decomposing $M_U$ into a direct sum of
simple modules, and apply the prescription above to each of the simple
components. For ${\cal C}_{\rm VOA}^{\rm tree}$, however, we still have to
state how to set $e_U$ and $i_U$ for a parenthesized word $U \in
{\rm Obj}({\cal C}_{\rm VOA}^{\rm tree})$ of length $n > 1$.  We set
the element of ${\cal V}_{(U^* \cdot U)}^0$ for $e_U$ as the
image of
\begin{align*}
  F_{\beta_1}^{-1} \cdots F_{\beta_n}^{-1} \cdot
  (\Theta_r \Omega Y_{\Upsilon^{\cal I}(\beta_1)}) \otimes \cdots
  \otimes (\Theta_r \Omega Y_{\Upsilon^{\cal I}(\beta_n)}) \otimes
  Y_{\beta_2} \otimes \cdots \otimes Y_{\beta_n} \in
  {\cal V}_{(U^* \cdot U)_{\rm inside~out}}^0
\end{align*}
by the fusion isomorphisms from there to ${\cal V}_{(U^* \cdot U)}^0$. 
Here, when the parenthesized word $U$ is $\beta_1\beta_2\cdots \beta_n$
as a word, $(U^* \cdot U)_{\rm inside~out}$ is a parenthesized word that
shares the same word as $(U^* \cdot U)$, but with parentheses placed
as in $\cdots (\Upsilon^{\cal I}(\beta_2) \cdot ((\Upsilon^{\cal I}(\beta_1)
\cdot \beta_1) \cdot \beta_2)) \cdots$, from inside to outside. The element of
${\rm Hom}_k({\cal V}_{(U \cdot U^*)}^0, k)$ for $i_U$ should then be
a chain of fusion isomorphisms ${\cal V}_{(U \cdot U^*)}^0 \rightarrow
{\cal V}_{(U \cdot U^*)_{\rm inside~out}}^0$ followed by extracting the
coefficient of the generator
\begin{align*}
  (\Theta_r\Omega Y_{\beta_n}) \otimes \cdots & \;
  \otimes (\Theta_r\Omega Y_{\beta_1})
  \otimes (\Omega Y_{\beta_{n-1}}) \otimes \cdots \otimes (\Omega Y_{\beta_1})
   \\
& \in {\cal V}_{\beta_n,\Upsilon^{\cal I}(\beta_n)}^0 \otimes
\cdots \otimes {\cal V}_{\beta_1, \Upsilon^{\cal I}(\beta_1)}^0 \otimes
       {\cal V}_{\beta_{n-1},0}^{\beta_{n-1}} \otimes \cdots \otimes
       {\cal V}_{\beta_1,0}^{\beta_1} \subset
       {\cal V}_{(U \cdot U^*)_{\rm inside~out}}^0. 
\end{align*}

The natural isomorphisms $\{\theta_U \}$ is chosen for $U \in
{\rm Obj}({\cal C})$ as
\begin{align}
  \left( e^{2\pi i h_\alpha} \times \right) \in \oplus_{\alpha \in {\cal I}((V,Y))}
       {\rm Hom}_k \left( ({\cal V}_U^\alpha)^\vee, \; ({\cal V}_U^\alpha)^\vee \right) = {\rm Mor}_{\cal C}(U, U),
  \label{eq:theta-in-ribbonC}
\end{align}
for both ${\cal C} = {\cal C}_{\rm VOA}^{\rm mod}$ and
${\cal C}_{\rm VOA}^{\rm tree}$. The conditions (\ref{eq:cond-ribbon-theta-nat},
\ref{eq:cond-ribbon-theta-wRgd}) are satisfied obviously. The condition
(\ref{eq:cond-ribbon-theta-wTnsr}) is verified when $U = \alpha_U$ and
$W = \alpha_W$ are simple, by noting that
$\Omega(-) \cdot \Omega(-): {\cal V}_{\beta,\gamma}^\alpha \rightarrow
    {\cal V}_{\beta,\gamma}^\alpha$ multiplies $e^{-2\pi i (h_\beta + h_\gamma - h_\alpha)}$.
This is enough even when $U, W \in {\rm Obj}({\cal C}_{\rm VOA}^{\rm tree})$
are words of length more than 1, because the morphism $c_{U, W}$
in (\ref{eq:choice-BTC-c-4Cpre}) is non-trivial only on one tensor factor
in (\ref{eq:choice-BTC-c-4Cpre}). 
\end{exmpl}

\vspace{5mm}

Having reviewed ribbon categories associated with a VOA $(V,Y,\cdots)$ that is not necessarily rational and a set ${\cal I}((V,Y))$ of its simple modules,
we now talk about additional data beyond those in the ribbon categories. 
\begin{anythng}
For any object $U$ in a ribbon category ${\cal C}$ that has a canonical isomorphism from the set ${\rm Mor}_{\cal C}({\bf 1}_{\cal C}, {\bf 1}_{\cal C})$ to a field $k \subset \C$, one may assign a number $d_U \in k$ by
\begin{align}
  d_U := {\rm tr} [ {\rm id}_U ] \in {\rm Mor}_{\cal C}
   ({\bf 1}_{\cal C}, {\bf 1}_{\cal C}) \cong k; 
\end{align}
as the sphericity (\ref{eq:cond-ribbon-sphere}) is guaranteed in a ribbon category (cf \ref{statmnt:thetaFirst-vs-deltaFirst}), the subscript of ${\rm tr}$ is omitted.  For a pair $U, W$ of objects of such a ribbon category ${\cal C}$, one may assign a number 
\begin{align}
  \tilde{s}_{U,W} & \; := (e_{U^*} \otimes e_{W^*}) \cdot
  (\delta_U \otimes {\rm id}_{U^*} \otimes \delta_{W^*} \otimes {\rm id}_{W^*})
  \cdot ({\rm id}_U \otimes (c_{W,U^*} \cdot c_{U^*,W}) \otimes {\rm id}_{W^*})
  \cdot (i_U \otimes i_W) \nonumber \\
  & \qquad \qquad \in {\rm Mor}_{\cal C}({\bf 1}_{\cal C}, {\bf 1}_{\cal C})
  \cong k.
\end{align}
\end{anythng}

\begin{defn}
A category ${\cal C}$ is said to be a {\it pre-additive category},
if the set ${\rm Mor}_{\cal C}(U,V)$ for any $U, V \in {\rm Obj}({\cal C})$
is an abelian group, and the composition law ${\rm Mor}_{\cal C}(U,V) \times
{\rm Mor}_{\cal C}(V, W)  \rightarrow {\rm Mor}_{\cal C}(U,W)$ is bilinear. 

For a pre-additive category to be an {\it additive category}, two more
conditions need to be satisfied: there should exist ${\bf 0}_{\cal C} \in
{\rm Obj}({\cal C})$ where ${\rm Mor}_{\cal C}({\bf 0}_{\cal C}, U) = \{ 0\}$ and
${\rm Mor}_{\cal C}(U, {\bf 0}_{\cal C}) = \{ 0\}$ for any $U \in
{\rm Obj}({\cal C})$,  and there should exist an object $U \oplus V$
characterized by the universal property for any $U, V \in {\rm Obj}({\cal C})$.
An isomorphism class of an object $S$ of an additive category is said to be
{\it simple}, when its subobjects consist of the isomorphism class of $S$
itself and that of ${\bf 0}_{\cal C}$. 

An additive category is called an {\it abelian category} after a few more
conditions are imposed, and is called a {\it semi-simple category} after
a few more conditions are imposed; we do not copy those conditions
and write them down here. 
\end{defn}

\begin{exmpl}
The category ${\cal C}_{\rm VOA}^{\rm tree}$ is pre-additive, but it is not
additive; the object $0 \in {\rm Obj}({\cal C}_{\rm VOA}^{\rm tree})$
(the parenthesized word of length-1, the letter $0 \in {\cal I}((V,Y))$)
does not have the property required for ${\bf 0}_{\cal C}$, while we do
not include a parenthesized word of length 0 in
${\rm Obj}({\cal C}_{\rm VOA}^{\rm tree})$. An object $U \oplus W$ does not
necessarily exist in ${\cal C}_{\rm VOA}^{\rm tree}$, either.
To see this, think of a rational VOA where the fusion ring has the structure of a group ring.
Then $N^\alpha_U = 1$ for just one $\alpha \in {\cal I}((V,Y))$ and $N^\alpha_U = 0$ for all other $\alpha \in {\cal I}((V,Y))$, for any $U \in {\rm Obj}({\cal C}_{\rm VOA}^{\rm tree})$.
So, for $U, V \in {\rm Obj}({\cal C}_{\rm VOA}^{\rm tree})$, there is no $W \in {\rm Obj}({\cal C}_{\rm VOA}^{\rm tree})$ where $N_W^\alpha = N_U^\alpha + N_V^\alpha$ for all $\alpha \in {\cal I}((V,Y))$.
The category ${\cal C}_{\rm VOA}^{\rm mod}$, on the other hand, is semi-simple.
The simple objects of ${\cal C}_{\rm VOA}^{\rm mod}$ modulo isomorphisms are labeled by ${\cal I}((V,Y))$. 

Although the notion of a simple object is defined (above) only for an additive category (because ${\bf 0}_{\cal C} \in {\rm Obj}({\cal C})$ is used in the definition), one may look for isomorphism classes of objects in a category whose subobjects (mod isomorphisms) are either itself or ${\bf 0}_{\cal C}$ if exists, and call them ``simple''.  Under this definition, the ``simple'' objects of ${\cal C}_{\rm VOA}^{\rm tree}$ are still labeled by ${\cal I}((V,Y))$;
each of the ``simple'' objects corresponds to $U \in {\rm Obj}({\cal C}_{\rm VOA}^{\rm tree})$ such that $N_U^\alpha = 1$ for one $\alpha \in {\cal I}((V,Y))$ and $N_U^\beta = 1$ for all other $\beta \in {\cal I}((V,Y))$.
This ``simple'' object is not a conventional terminology. 
\end{exmpl}

\begin{defn}
\label{defn:MTC}  
A ribbon category $({\cal C}, {\bf 1}_{\cal C}, \otimes, \cdots )$ that has a canonical isomorphism ${\rm Mor}_{\cal C}({\bf 1}_{\cal C}, {\bf 1}_{\cal C}) \cong k$ is said to be a {\it modular tensor category} (MTC), if it is semi-simple, 
its set of simple objects ${\cal I}$ is finite, and the finite matrix $(\tilde{s}_{U,V})_{U,V \in {\cal I}}$ is invertible.   
\end{defn}

\begin{rmk}
\label{rmk:ratVSnonrat}
The condition that the set ${\cal I}$ is finite is to be able to
argue about whether a matrix is invertible without a subtlety on
convergence in computing the product of matrices. If this
subtlety is resolved in some way or another, the condition $\#{\cal I}<\infty$
is not strictly necessary for the definition above to be well-defined.
It is also a separate question whether the matrix $(\tilde{s}_{U,V}
)_{U,V \in {\cal I}}$ is of some interest in the case $\#{\cal I}$ is not finite. 
\end{rmk}
\begin{rmk}
\label{rmk:treeVSmod}  
The process of restricting our attention from the matrix
$(\tilde{s}_{U,V})_{U,V \in {\rm Obj}({\cal C}_{\rm VOA}^{\rm mod})}$ to
$(\tilde{s}_{U,V})_{U,V \in {\cal I}}$ is structurally transparent
in the semi-simple category ${\cal C}_{\rm VOA}^{\rm mod}$; 
it is easy to see in ${\cal C}_{\rm VOA}^{\rm mod}$ that the matrix $(\tilde{s}_{U,V})_{U,V \in {\rm Obj}({\cal C}_{\rm VOA}^{\rm mod})}$ is not invertible at all,
because $\tilde{s}_{U_1 \oplus U_2,V} = \tilde{s}_{U_1,V} + \tilde{s}_{U_2,V} \in k$.
This additivity in the value $\tilde{s}_{U, V} \in k$ is in line with the additivity ($\oplus$ operation) in the underlying additive category of ${\cal C}_{\rm VOA}^{\rm mod}$, while there is no such operation among the objects of ${\cal C}_{\rm VOA}^{\rm tree}$. 
For this reason, ${\cal C}_{\rm VOA}^{\rm tree}$ does not fit the definition of modular tensor category (even when the VOA is rational),
although the matrix $(\tilde{s}_{U,V})$ with ``simple'' $U$ and $V$ is well-defined and is the same as that of ${\cal C}_{\rm VOA}^{\rm mod}$. 
\end{rmk}

\begin{anythng}
Given an MTC, it is possible to define matrices
\begin{align}
  S_{\beta,\gamma} := \frac{\tilde{s}_{\beta,\gamma}}{\sqrt{p^+ p^-}}, \qquad
  T_{\beta,\gamma} := \delta_{\beta,\gamma} \theta_\beta (p^-/p^+)^{1/6}, 
\end{align}
where 
\begin{align}
  p^{\pm} := \sum_{\beta \in {\cal I}} \theta_\beta^{\pm 1}(d_{\beta})^2 \in k
\end{align}
The two matrices $(S_{U,V})_{U,V \in {\cal I}}$ and $(T_{U,V})_{U,V\in {\cal I}}$ satisfy the relation (\ref{eq:g=1-proj-repr-SL2Z-rltn}) for $\beta = 0 \in {\cal I}((V,Y))$. The Verlinde formula and also the relation $\tilde{s}_{{\bf 1}_{\cal C},U} = d_U$ for $U \in {\cal I}$ also hold true. See \cite[Thm. 3.17]{MR1797619}
for $S(\beta)$ and $T(\beta)$ for $\beta \in {\cal I}$ other than ${\bf 1}_{\cal C}$.

The reverse dictionary is given by
\begin{align}
  d_\beta = \frac{S_{\beta 0}}{S_{00}}, \qquad
  \sqrt{p^+p^-} = \frac{1}{S_{00}}, \qquad
  (p^+/p^-)^{1/6} = e^{2\pi i c/24}, \qquad
  \tilde{s}_{\beta,\gamma} = \frac{S_{\beta,\gamma}}{S_{00}}. 
\end{align}
\end{anythng}

\section{2D Conformal Field Theory}
\label{sec:CFT}

\subsection{Defining Algebraic Data}
\label{ssec:CFT-data}
%
\begin{anythng}
  \label{statmnt:r-mover}
{\bf right-moving VOA:} As a preparation for introducing the defining algebraic
data of a 2D conformal field theory, we explain what we mean by a
{\it right-moving} (or {\it anti-holomorphic}) {\it unitary vertex operator
  algebra} and its {\it modules and intertwining operators}.
The set of defining data of a right-moving unitary VOA is the same as that
of a unitary VOA reviewed in section \ref{ssec:review-VOA}, except that
we use $\bar{z}$ as the dummy variable in a formal power series instead
of $z$, subject to the conditions (a--p) and (r). The defining algebraic data
of a right-moving unitary VOA is often denoted by
$(\widetilde{V}, \widetilde{Y}, \tilde{\bf 1}, \tilde{\omega},
\widetilde{\Upsilon}, \langle - | - \rangle)$ in this article. 
We will often put $\sim$ on a symbol to indicate that the object in question
is for a right-moving VOA; this is to follow the convention in a
widely accepted textbook \cite{Polchinski:1998rq}, \cite{Polchinski:1998rr}
in the string-theory community. 

The matrix elements of an intertwining operator in an ordinary unitary VOA 
depend holomorphically on the local coordinates of a Riemann surface $\Sigma$
(cf \ref{statmnt:prscr-oprtr-trnstnFcn-coordCh}).
Those in a right-moving unitary VOA, on the other hand, are set to depend
anti-holomorphically on the local coordinates of a Riemann surface.
This is done by interpreting the dummy variable $\bar{z}$ of a
right-moving unitary VOA as the anti-holomorphic coordinate value
$\bar{z}(p) = (z(p))^{cc}$ of a point $p \in \Sigma$, where
$cc$ stands for complex conjugation, and by replacing the relation
(\ref{eq:state-transf-4-coord-transf}) 
between the coordinate-dependent operator implementations as
\begin{align}
  \widetilde{Y}^{(\eta)}(\tilde{a})_{@p} =
  \widetilde{Y}^{(\eta)}(\tilde{a})_{\bar{\eta}@(\eta(p))^{cc}} =
  \widetilde{Y}^{(\xi)}(\tilde{R}_{f@\eta(p)}^{-1} \cdot \tilde{a}
  )_{\bar{\xi}@[f(\eta(p))]^{cc}}, \quad {}^\forall \tilde{a} \in \widetilde{V},
  \quad {}^\forall p \in U. 
\end{align}
where $\xi = f(\eta)$ is the change of local holomorphic coordinate, and 
\begin{align}
  \tilde{R}_{f@\eta(p)} := \exp \left[ - \sum_{i=1}^\infty (f@\eta(p))^{cc}_i
    \tilde{L}_i
    \right] (f@\eta(p))_0^{cc, -\tilde{L}_0}, 
\end{align}
replacing $L_n$ by $\tilde{L}_n$ and $f_i$'s with $(f_i)^{cc}$'s
from the left-moving one (\ref{eq:def-R_f}). The character functional
of an irreducible representation $\tilde{\alpha}$ is
\begin{align}
  \tilde{\chi}_{\tilde{\alpha}}(-,\tau): \widetilde{V} \times \mathfrak{H} \ni
  (\tilde{v}, \tau) \longmapsto {\rm Tr}_{\widetilde{M}_\alpha}\left[
    e^{-2\pi i \tau^{cc} \left( \tilde{L}_0 - \frac{\tilde{c}}{24} \right)} \; 
    \widetilde{Y}^{(w)}_{\tilde{\alpha}}(\tilde{v})_{\bar{w}@0} \right] \in \C. 
\end{align}
%
When it is necessary to make a clear contrast between an ordinary
VOA and a right-moving VOA, we refer to the ordinary one as
a left-moving VOA (or holomorphic VOA) in this article.  

The vector space of intertwining operators of type
$(\tilde{\alpha};\tilde{\beta},\tilde{\gamma})$, where
$\tilde{\alpha}$, $\tilde{\beta}$, $\tilde{\gamma}$
label unitary irreducible modules of a right-moving unitary VOA,
 is denoted by $\widetilde{\cal V}_{\tilde{\beta},\tilde{\gamma}}^{\tilde{\alpha}}$. 
 One can introduce the isomorphisms $\widetilde{\Omega}_r$,
 $\widetilde{\Theta}_s$, $\widetilde{\cal F}(\tilde{\alpha};\tilde{\beta}_1,\tilde{\beta}_2;\tilde{\gamma})$ and
 $\widetilde{\cal B}_r(\tilde{\alpha};\tilde{\beta}_1,\tilde{\beta}_2;\tilde{\gamma})$ by using the same manipulations on the dummy variables as in
 introducing the isomorphisms $\Omega_r$, $\Theta_r$, ${\cal F}$ and
 ${\cal B}_r$. To take the skew-symmetry isomorphism
 $\widetilde{\Omega}_r$ as an example, we should replace
 $e^{zL_{-1}} {\cal Y}^{\alpha}_{\beta,\gamma}(m_\beta,e^{\pi i (2r+1)}z)$ in
 (\ref{eq:def-skew-sym-iso}) by
 $e^{\bar{z}\tilde{L}_{-1}} \widetilde{\cal Y}^{\tilde{\alpha}}_{\tilde{\beta},\tilde{\gamma}}(\tilde{m}_{\tilde{\beta}}, e^{\pi i (2r+1)}\bar{z})$ in introducing
 $\widetilde{\Omega}_r$; this is just a matter of convention, however.   
\end{anythng}
\begin{anythng}
\label{statmnt:data-4-CFT}
  {\bf defining data set of a 2D modular invariant and unitary CFT:}
The class of {\it 2D modular invariant and unitary conformal field
  theory} (CFT) to be considered in this article has a defining
set of algebraic data
(cf \cite{Moore:1988uz}, \cite{Moore:1988qv},
\cite{Kapustin:2000aa}, \cite{MR1933066}, \cite{MR2130840}, for example) 
%
\begin{align}
&  \left( V,Y, {\bf 1}, \omega, \Upsilon, \langle - | - \rangle \right), \quad
  \left( \widetilde{V},\widetilde{Y}, \tilde{\bf 1}, \tilde{\omega}, \widetilde{\Upsilon}, \langle - | - \rangle \right), \quad
  L, \;\; 
  \{ Y_{\rm tot}^{(\alpha;\beta,\gamma)(\tilde{\alpha};\tilde{\beta},\tilde{\gamma})}\},
   \label{eq:data-set-alg-4-CFT-viaChiral}  \\
&  {\cal I}(V), \;\; \tilde{I}(\widetilde{V}), \quad
  \left\{ (M_\alpha, Y_\alpha, \langle -| - \rangle_\alpha, \lambda_\alpha)
  \; | \; \alpha \in {\cal I}(V) \right\}, \quad
  \left\{ (\widetilde{M}_{\tilde{\alpha}}, \widetilde{Y}_{\tilde{\alpha}},
  \langle -| - \rangle_{\tilde{\alpha}}, \tilde{\lambda}_{\tilde{\alpha}})
  \; | \; \tilde{\alpha} \in \tilde{\cal I}(\widetilde{V}) \right\}; 
  \nonumber 
\end{align}
it begins with one left-moving unitary VOA and one right-moving unitary VOA; 
their central charges $c$ and $\tilde{c}$, respectively, are not necessarily
identical. 
%
The first [resp. second] item in the second line, $({\cal I}((V,Y))$
[resp. $\tilde{\cal I}((\widetilde{V},\widetilde{Y}))$],  
is a set of distinct equivalence classes of unitary and simple
representations of the left-moving [resp. right-moving] VOA that satisfies
the conditions 
(r, r', r'') in \ref{statmnt:set-I(V)} and \ref{statmnt:BnF}.
For each equivalence class, one implementation (e.g., $(M_\alpha, Y_\alpha,
\langle - | -\rangle_\alpha)$ is chosen, along with a norm-preserving
isomorphism
$\lambda_\alpha: M_\alpha^\vee \rightarrow M_{\Upsilon^{\cal I}(\alpha)}$ of
the VOA modules. 


A $\Z_{\geq 0}$-valued matrix
$L = (L^{\alpha,\tilde{\alpha}})_{\alpha \in {\cal I}(V), \;
  \tilde{\alpha} \in \tilde{\cal I}(\widetilde{V})}$ called 
the {\it multiplicity matrix} should satisfy (s) the following transformation
law under the modular transformation; the linear functional
$Z: V \otimes \widetilde{V} \rightarrow \C$ is given by
\begin{align}
  Z(\tau) : v\otimes \tilde{v} \mapsto
  \sum_{\alpha \in {\cal I}((V,Y)), \;
    \tilde{\alpha} \in \tilde{\cal I}((\widetilde{V},\widetilde{Y}))}
  L^{\alpha,\tilde{\alpha}}
  \chi_\alpha(v,\tau) \tilde{\chi}_{\tilde{\alpha}}(\tilde{v},\bar{\tau});
\end{align}
$g = (a,b; c,d) \in {\rm SL}_2\Z$ acts on this $Z$ by
$Z \rightarrow Z|[g]$, where 
\begin{align}
  (Z|[g])(v\otimes \tilde{v},\tau) := Z\left( (c \tau+d)^{-L_0} v \otimes
  (c\bar{\tau}+d)^{-\tilde{L}_0} \tilde{v}, (a\tau+b)/(c\tau+d)\right); 
\end{align}
the required transformation law is \cite{Alvarez-Gaume:1983ihn}
\begin{align}
  Z|[g] = \chi_\eta(g)^{-(c-\tilde{c})} \; Z,
   \label{eq:cond-mod-inv}
\end{align}
where $\chi_\eta(T) = e^{2\pi i/24}$ and $\chi_\eta(S) = e^{2\pi i (-3/24)}$.
In particular, $[h_\alpha] = [\tilde{h}_{\tilde{\alpha}}]$ (mod $\Z$)
when $L^{\alpha \tilde{\alpha}} > 0$. 
%
%
%
We also require that (b-CFT) $L^{0,\tilde{0}} = 1$, and that  ($r^{'""}$)
there is ${}^\exists \tilde{\beta} \in \tilde{\cal I}((\widetilde{V},\widetilde{Y}))$ for each $\alpha \in {\cal I}((V,Y))$
[resp. ${}^\exists \beta \in {\cal I}((V,Y))$ for each $\tilde{\alpha} \in \tilde{\cal I}((\widetilde{V},\widetilde{Y}))$] such that 
$L^{\alpha, \tilde{\beta}} >0$ [resp. $L^{\beta,\tilde{\alpha}} > 0$]. 
The multiplicity matrix should also have the property that 
$L^{\Upsilon^{\cal I}(\alpha), \widetilde{\Upsilon}^{\tilde{\cal I}}(\tilde{\alpha})}
= L^{\alpha,\tilde{\alpha}}$.
%
The condition (s) is imposed on both the sets
${\cal I}((V,Y))$, $\tilde{\cal I}((\widetilde{V}, \widetilde{Y}))$ and the
matrix $L$; it is not easy to find exhaustive solutions to this condition (s)
if $(V,Y)$ and $(\widetilde{V}, \widetilde{Y})$ are not
rational.\footnote{
\label{fn:ex-I(VY)-torus}
When both $(V,Y)$ and $(\widetilde{V},\widetilde{Y})$ are the rank-1
Heisenberg algebra (one chiral boson), as in
Ex. \ref{exmpl:Heisenberg-unitarity}, the answer is
known \cite{Dijkgraaf:1989hb}.  
While unitary and simple representations of $(V,Y)$ are labeled by
the eigenvalues of $\alpha_0$ (notations in
Ex. \ref{exmpl:Heisenberg-unitarity}) taking values in $\R$, we may choose
${\cal I}((V,Y))$ as a subset
$\Z^{\oplus 2} \cong \{ \alpha_0 = (n/R + wR/2) \in \R \; | \; n,w\in \Z\}
\subset \R$, and ${\cal I}((\widetilde{V},\widetilde{Y}))$
as $\{ \tilde{\alpha}_0 = (\tilde{n}/R - \tilde{w}R/2) \; | \; \tilde{n},\tilde{w}\in \Z\}$, with $L^{(n,w), (\tilde{n},\tilde{w})} = \delta^{n,\tilde{n}}
\delta^{w,\tilde{w}}$.
Solutions to the condition (s) are parametrized by $R \in \R_{>0}$. 
} %

The last piece in the set of data
$Y_{\rm tot} = \{ Y_{\rm tot}^{(\alpha;\beta,\gamma)(\tilde{\alpha};\tilde{\beta},\tilde{\gamma})}\}$ consists of $L^{\beta,\tilde{\beta}} L^{\gamma,\tilde{\gamma}}
L^{\alpha,\tilde{\alpha}}$ choices  
\begin{align}
  Y_{\rm tot}^{(\alpha;\beta,\gamma)(\tilde{\alpha};\tilde{\beta},\tilde{\gamma})}(r) \in
{\cal V}_{\beta,\gamma}^{\alpha} \otimes
\widetilde{\cal V}_{\tilde{\beta},\tilde{\gamma}}^{\tilde{\alpha}},
\qquad 1\leq r_1 \leq L^{\beta\tilde{\beta}},
\quad 1\leq r_2 \leq L^{\gamma,\tilde{\gamma}}, \quad
1 \leq r_3 \leq L^{\alpha,\tilde{\alpha}}
\end{align}
for all the triples $\alpha$, $\beta$, $\gamma$ from ${\cal I}((V,Y))$
with a non-zero $N_{\beta,\gamma}^\alpha$,
and all the triples $\tilde{\alpha}$, $\tilde{\beta}$, $\tilde{\gamma}$
from $\tilde{\cal I}((\widetilde{V},\widetilde{Y}))$ with a non-zero
$\widetilde{N}_{\tilde{\beta},\tilde{\gamma}}^{\tilde{\alpha}}$ and non-zero
$L_{\alpha,\tilde{\alpha}}$, $L_{\beta,\tilde{\beta}}$ and $L_{\gamma,\tilde{\gamma}}$.
An extra set of conditions (t) reviewed below (\ref{statmnt:cond-Ytot})
is imposed on the choice of $Y_{\rm tot}$. The condition (t) includes (t-1)
a special choice for $\beta = \tilde{\beta} = 0$; the
$L^{\alpha,\tilde{\alpha}} L^{\beta,\tilde{\beta}} L^{\gamma,\tilde{\gamma}} =
(L^{\alpha,\tilde{\alpha}})^2$ choices from
$({\cal V}_{0,\alpha}^\alpha \otimes
\widetilde{\cal V}_{\tilde{0},\tilde{\alpha}}^{\tilde{\alpha}})$ should be
$Y_{\rm tot}^{(\alpha;0,\alpha)(\tilde{\alpha};\tilde{0},\tilde{\alpha})}(r_2,r_3) =
(Y_\alpha \otimes \widetilde{Y}_{\tilde{\alpha}}) \delta^{r_3}_{\;\; r_2}$
where $r_3 = 1,\cdots, L^{\alpha,\tilde{\alpha}}$ and
$r_2 = 1,\cdots, L^{\gamma,\tilde{\gamma}} = L^{\alpha,\tilde{\alpha}}$. 

An extreme case, $\tilde{c}=0$, is called a {\it holomorphic CFT}
or {\it chiral CFT}. Then the right-moving VOA $(\widetilde{V},
\widetilde{Y}, \cdots)$ is trivial (cf comments at the end
of \ref{statmnt:def-VOA}), and the set
$\tilde{\cal I}((\widetilde{V},\widetilde{Y}))$ is trivial,
$\{ \tilde{0} \}$.

We will omit explicit mention of 2D and/or modular invariant in the rest
of this article, but both remain implicit whenever we say unitary
CFT hereafter. 
\end{anythng}  

\begin{rmk}
\label{rmk:nonUnitary-CFT?}
It must be possible to formulate 2D modular invariant CFT without unitarity;
that is done by throwing away Hermitian inner products $\langle - |
- \rangle_{\alpha}$ from the defining algebraic data, ignoring the conditions
(m-q), while retaining $\lambda_\alpha$. In defining the normalized correlation
functions in (\ref{eq:algrthm-multi-prhrl}), we can replace
$\bra{\tilde{\bf 1}{\bf 1}} - \rangle$ using the Hermitian inner product
by $\langle ({\bf 1}\tilde{\bf 1})^\vee, -\rangle$ using the natural paring
between dual vector spaces. 

Sections \ref{sec:CFT} and \ref{ssec:fCFT} of this note only deal with
unitary CFTs, primarily because of lack of experience in non-unitary
theories of the present author. 
\end{rmk}

\begin{anythng}
\label{statmnt:build-CFT}
Before stating the rest of the condition (t) on the choices $Y_{\rm tot}$,
we review a few constructions that do not have to refer to details of
$Y_{\rm tot}$, introduce notations, and have a remark.
The {\it Hilbert space of a unitary CFT} (in the class of CFTs considered
in this article)\footnote{
We will also ignore the difference between
the direct sum of infinitely many vector spaces and its completion
using the norm associated with the Hermitian inner product.
} %
is
\begin{align}
  {\cal H}_{\rm tot} = \oplus_{\alpha \in I((V,Y))}
  \oplus_{\tilde{\alpha} \in \tilde{\cal I}((\widetilde{V},\widetilde{Y}))}
  (M_\alpha \otimes \widetilde{M}_{\tilde{\alpha}})^{\oplus L_{\alpha,\tilde{\alpha}}};  
\label{eq:str-Htot-bos-byVOA}
\end{align}
the states ${\bf 1}\otimes \tilde{\bf 1}$, $\omega \otimes \tilde{\bf 1}$
and ${\bf 1} \otimes \tilde{\omega}$ in
$M_{0} \otimes \widetilde{M}_{\tilde{0}} \subset {\cal H}_{\rm tot}$ are
also denoted by ${\bf 1}\tilde{\bf 1}$, $\omega$ and $\tilde{\omega}$.
The Hermitian positive definite inner product $\langle - | - \rangle_{\rm tot}$
on each summand, and hence on ${\cal H}_{\rm tot}$, follows from
the Hermitian positive definite inner products $\langle - | - \rangle_{M_\alpha}$
and $\langle - | - \rangle_{\widetilde{M}_{\tilde{\alpha}}}$ of the
unitary modules $M_\alpha$ and $\widetilde{M}_{\tilde{\alpha}}$. 

A $\C$-anti-linear (anti-unitary) involution $\Upsilon_{\rm tot}:
{\cal H}_{\rm tot} \rightarrow {\cal H}_{\rm tot}$ is determined by
pairing up $L^{\alpha,\tilde{\alpha}} =
L^{\Upsilon^{\cal I}(\alpha),\widetilde{\Upsilon}^{\tilde{\cal I}}(\tilde{\alpha})}$ copies of
$(M_\alpha \otimes \widetilde{M}_{\tilde{\alpha}})$ and
$(M_{\Upsilon^{\cal I}(\alpha)} \otimes
\widetilde{M}_{\widetilde{\Upsilon}^{\tilde{\cal I}}(\tilde{\alpha})})$, and using 
$\Upsilon_{\Upsilon^{\cal I}(\alpha)\alpha} \otimes \widetilde{\Upsilon}_{\widetilde{\Upsilon}^{\cal I}(\tilde{\alpha})\tilde{\alpha}}: M_\alpha
\otimes \widetilde{M}_{\tilde{\alpha}} \rightarrow M_{\Upsilon^{\cal I}(\alpha)}
\otimes \widetilde{M}_{\Upsilon^{\cal I}(\tilde{\alpha})}$
between those $L^{\alpha,\tilde{\alpha}}$ pairs. As a reminder,
$\Upsilon_{\Upsilon^{\cal I}(\alpha) \alpha} = \lambda_\alpha \cdot \Upsilon_\alpha$,
as reviewed in \ref{statmnt:cntrG-mod}. 

The $L^{\beta,\tilde{\beta}} L^{\gamma,\tilde{\gamma}} L^{\alpha,\tilde{\alpha}}$ choices
of $Y_{\rm tot}^{(\alpha;\beta,\gamma)(\tilde{\alpha};\tilde{\beta},\tilde{\gamma})}$
determine
\begin{align}
  Y_{\rm tot}: & \; (M_\beta \otimes \widetilde{M}_{\tilde{\beta}})^{\oplus L^{\beta,\tilde{\beta}}} \longrightarrow \\
  & {\rm Hom}_\C\left( (M_\gamma\otimes \widetilde{M}_{\tilde{\gamma}})^{\oplus L^{\gamma,\tilde{\gamma}}}, \;
  (M_\alpha\otimes \widetilde{M}_{\tilde{\alpha}})^{\oplus L^{\alpha,\tilde{\alpha}}} \right)
  [[z,\bar{z},z^{-1},\bar{z}^{-1}]] z^{-[(\alpha;\beta,\gamma)]}\bar{z}^{-[(\tilde{\alpha};\tilde{\beta},\tilde{\gamma})]}, \nonumber 
\end{align}
and hence
\begin{align}
  Y_{\rm tot}: {\cal H}_{\rm tot} \longrightarrow {\rm End}({\cal H}_{\rm tot})
  [[z,\bar{z},z^{-1},\bar{z}^{-1},(z\bar{z})^{\R}]]. 
\end{align}
One way to specify a choice of $Y_{\rm tot}$ is by introducing a label
$p = 1,2,\cdots,$ over the $\sum_\alpha \sum_{\tilde{\alpha}}
L^{\alpha,\tilde{\alpha}}$ irreducible blocks in (\ref{eq:str-Htot-bos-byVOA}),
and also a basis of the vector space ${\cal V}_{\beta,\gamma}^\alpha$
and $\widetilde{\cal V}_{\tilde{\beta},\tilde{\gamma}}^{\tilde{\alpha}}$, and 
\begin{align}
  Y_{\rm tot}(p_3;p_1,p_2) = \sum_{i=1}^{N_{\beta(p_1),\gamma(p_2)}^{\alpha(p_3)}}
  \sum_{\tilde{\imath}=1}^{N_{\tilde{\beta}(p_1),\tilde{\gamma}(p_2)}^{\tilde{\alpha}(p_3)}}
    ({\cal Y}_{\beta(p_1),\gamma(p_2)}^{\alpha(p_3)})_{i} \otimes
  (\widetilde{\cal Y}_{\tilde{\beta}(p_2),\tilde{\gamma}(p_2)}^{\tilde{\alpha}(p_3)}
  )_{\tilde{\imath}}
  \; \; \; d(p_3;p_1,p_2)^{i,\tilde{\imath}};
   \label{eq:def-LR-coupl-matrix-byVOA}
\end{align}
the coefficients $d(p_3;p_1,p_2)^{i,\tilde{\imath}}$
({\it left-right-coupling matrix})
along with the chosen basis specify an element in the vector space
${\cal V}_{\beta,\gamma}^\alpha \otimes
\widetilde{\cal V}_{\tilde{\beta},\tilde{\gamma}}^{\tilde{\alpha}}$.

The $\R$-grading on the modules $M_\alpha$ and $\widetilde{M}_{\tilde{\alpha}}$
introduces the bi-$\R$-grading
\begin{align}
  (M_\alpha \otimes \widetilde{M}_{\tilde{\alpha}}) \cong \oplus_{h \in \R_{\geq0}, \tilde{h} \in \R_{\geq 0}}^{(h\in [h_\alpha], \tilde{h} \in [\tilde{h}_{\tilde{\alpha}}])}
  M_{\alpha,h} \otimes \widetilde{M}_{\tilde{\alpha},\tilde{h}}
  =: \oplus_{h \in \R_{\geq0}, \tilde{h} \in \R_{\geq 0}}^{(h\in [h_\alpha], \tilde{h} \in [\tilde{h}_{\tilde{\alpha}}])} 
  [M_\alpha \otimes \widetilde{M}_{\tilde{\alpha}}]_{h,\tilde{h}}. 
\end{align}
The mode expansion of the operator $Y_{\rm tot}(p_3;p_1,p_2)(x)$ for 
$x \in [M_\beta\otimes \widetilde{M}_{\tilde{\beta}}]_{h,\tilde{h}}$ is
denoted by 
\begin{align}
  Y_{\rm tot}^{(\alpha;\beta,\gamma)(\tilde{\alpha};\tilde{\beta},\tilde{\gamma})}
  (p_3;p_1,p_2) & \; 
  : [M_\beta \otimes \widetilde{M}_{\tilde{\beta}}]_{h,\tilde{h}} \ni x
  \longmapsto   \label{eq:mode-expns-radial-CFT} \\
 & Y_{\rm tot}^{(\alpha;\beta,\gamma)(\tilde{\alpha};\tilde{\beta},\tilde{\gamma})}(p_3;p_1,p_2)(x)
  = \sum_{\delta h, \delta \tilde{h}}^{(h-\delta h \in [(\alpha;\beta,\gamma)], \;
    \tilde{h}-\delta \tilde{h}\in [(\tilde{\alpha};\tilde{\beta},\tilde{\gamma})])}
  \frac{[x]_{-\delta h, -\delta \tilde{h}}}{z^{h-\delta h}\bar{z}^{\tilde{h}-\delta \tilde{h}}} \nonumber 
\end{align}
in this article; $\delta h - \delta \tilde{h} \in \Z$ because of
the first one of the condition (s).

In physics literature, $Y_{\rm tot}(x)$ for any $x \in {\cal H}_{\rm tot}$
is called a vertex operator; the VOA community, on the other hand,
prefers to reserve the word only for $Y(v)$ for $v \in V$, and
general elements ${\cal Y} \in {\cal V}_{\beta,\gamma}^\alpha$ are referred to as intertwining operators. 
In order to avoid confusion, we will refer to the choice $Y_{\rm tot}$ subject
to the condition (t) as a {\it CFT vertex operator} in this article. 
\end{anythng}

\begin{rmk}
\label{rmk:extrct-chi-alg}
Two unitary CFTs are regarded {\it identical} when there is an isomorphism
between their data $({\cal H}_{\rm tot}, Y_{\rm tot}, {\bf 1}\tilde{\bf 1}, \omega,
\tilde{\omega}, \Upsilon_{\rm tot}, \langle - | - \rangle_{\rm tot})$. 
For one unitary CFT, therefore, there can be multiple constructions
whose data sets (\ref{eq:data-set-alg-4-CFT-viaChiral}) are not isomorphic. 
It is therefore desirable to write down the conditions directly in terms
of $({\cal H}_{\rm tot}, Y_{\rm tot}, {\bf 1}\tilde{\bf 1}, \cdots)$
so that they are equivalent to the conditions (a--t); we do not try
to do so in this article, however.
cf \cite{Kapustin:2000aa}, \cite{MR1933066}, \cite{MR2130840}.

For a given unitary CFT $({\cal H}_{\rm tot}, Y_{\rm tot}, {\bf 1}\tilde{\bf 1},
\cdots)$, it is possible to extract one left-moving VOA ${\cal A}$ and
one right-moving VOA $\widetilde{\cal A}$ independently from how it is
constructed. The unique left-moving and right-moving VOAs are built on
the $\tilde{L}_0=0$ vector subspace $V' \subset {\cal H}_{\rm tot}$ and
the $L_0=0$ vector subspace $\widetilde{V}' \subset {\cal H}_{\rm tot}$. 
Because of the condition (r-3), $V' \cong \oplus_{\alpha \in {\cal I}((V,Y))}
M_\alpha^{\oplus L_{\alpha,\tilde{0}}}$ and
$\widetilde{V}' \cong \oplus_{\tilde{\alpha} \in \tilde{\cal I}((\widetilde{V},\widetilde{Y}))} \widetilde{M}_{\tilde{\alpha}}^{\oplus L_{0,\tilde{\alpha}}}$.
We call ${\cal A} = (V', Y_{\rm tot}|_{V'}, {\bf 1}\tilde{\bf 1}, \omega, \Upsilon_{\rm tot}|_{V'}, \langle -|-\rangle_{\rm tot}|_{V'})$ the {\it left-moving chiral algebra of the CFT}, and $\widetilde{\cal A} = (\widetilde{V}',
Y_{\rm tot}|_{\widetilde{V}'}, {\bf 1}\tilde{\bf 1}, \tilde{\omega},
\Upsilon_{\rm tot}|_{\widetilde{V}'},
\langle -|-\rangle_{\rm tot}|_{\widetilde{V}'})$
the {\it right-moving chiral algebra of the CFT}. 
The original VOA $(V,Y,{\bf 1}, \cdots )$ is a vertex operator subalgebra
of ${\cal A}$; there are both cases, $V = V'$ and $V \subsetneq V'$. 
Under the maximally extended choice of $V$ and $\widetilde{V}$,
$L_{0,\tilde{\alpha}} = 0$ for $\tilde{\alpha} \in
\tilde{\cal I}(\widetilde{\cal A}) \backslash \{ \tilde{0} \}$,
and $L_{\alpha,\tilde{0}} = 0$ for $\alpha \in {\cal I}({\cal A}) \backslash
\{ 0\}$.

In this article, the words CFT, VOA and chiral algebra will be used
in the sense explained above. For example, the level-$k$ $\mathfrak{su}(2)$
Affine Lie algebra is a VOA, and the level-$k$ $\SU(2)$ Wess--Zumino--Witten
model is a CFT whose left-moving and right-moving chiral algebras are both
the level-$k$ $\mathfrak{su}(2)$ Affine Lie algebra. 
\end{rmk}

\begin{rmk}
\label{rmk:Moore-Seiberg}
(cf Ref. \cite[\S4]{Moore:1988ss})  
The representation-theory structure (\ref{eq:str-Htot-bos-byVOA})
of ${\cal H}_{\rm tot}$ wrpt the VOAs $(V,Y)$ and $(\widetilde{V},\widetilde{Y})$
simplifies considerably when we choose them to be the chiral algebras
${\cal A}$ and $\widetilde{\cal A}$. 
Ref. \cite[\S4]{Moore:1988ss} points out two things on the multiplicity
matrix $L^{\alpha,\tilde{\alpha}}$ when the VOAs on both sides are rational.

The first observation is that
$\sum_{\tilde{\alpha} \in \tilde{\cal I}(\widetilde{\cal A})}L^{\alpha,\tilde{\alpha}}
\leq 1$ and $\sum_{\alpha \in {\cal I}({\cal A})} L^{\alpha,\tilde{\alpha}} \leq 1$;
as we have assumed that (r$^{'""}$) there is at least one entry in each row
and at least one entry in each column of the matrix that is positive,
this means that there is just one entry in each row, and just one entry in each
column where $L^{\alpha,\tilde{\alpha}}=1$, and all the other entries of the
multiplicity matrix are zero. Ref. \cite[\S4]{Moore:1988ss} did so
in a proof by contradiction; suppose that there were two irreducible
modules of $(V,Y,\cdots) = {\cal A}$ and $(\widetilde{V},\widetilde{Y},\cdots)
= \widetilde{\cal A}$ in (\ref{eq:str-Htot-bos-byVOA}) that are both
in one unitary irreducible representation---denoted by $\alpha$---of the
left-mover chiral algebra ${\cal A}$. Let the two irreducible modules
of ${\cal A} \times \widetilde{\cal A}$ be
$(M_\alpha \otimes \widetilde{M}_{\tilde{\beta}_1})$ and
$(M_\alpha \otimes \widetilde{M}_{\tilde{\beta}_2})$; $\tilde{\beta}_1$ and
$\tilde{\beta}_2$ in $\tilde{\cal I}(\widetilde{\cal A})$ may be equal 
($L^{\alpha,\tilde{\beta}}\geq 2$), or different
($L^{\alpha,\tilde{\beta}_1} \geq 1$ and $L^{\alpha,\tilde{\beta}_2}\geq 1$). 
Then there are two different ways to write down the chiral four-point
correlation function, with three incoming states from
$(M_\alpha \otimes \widetilde{M}_{\tilde{\beta}_1})$,
$(M_{\Upsilon^{\cal I}(\alpha)} \otimes \widetilde{M}_{\widetilde{\Upsilon}^{\tilde{\cal I}}(\tilde{\beta}_1)})$ and
$(M_\alpha \otimes \widetilde{M}_{\tilde{\beta}_2})$ and one outgoing state
from $(M_\alpha \otimes \widetilde{M}_{\tilde{\beta}_1})$; 
one is to use the matrix elements of
\begin{align}
  \sum_{p} & \; \sum_{i,\tilde{\imath},j,\tilde{\jmath}}
  d(1;1,p)^{i,\tilde{\imath}} ({\cal Y}_{\alpha,\gamma(p)}^\alpha)_i
  (\widetilde{\cal Y}_{\tilde{\beta}_1,\tilde{\gamma}(p)}^{\tilde{\beta}_1})_{\tilde{\imath}} \;\; 
  d(p;2^\vee,2)^{j,\tilde{\jmath}} ({\cal Y}_{\Upsilon^{\cal I}(\alpha),\alpha}^{\gamma(p)})_j
  (\widetilde{\cal Y}_{\widetilde{\Upsilon}^{\tilde{\cal I}}(\tilde{\beta}_2),\tilde{\beta}_2}^{\tilde{\gamma}(p)})_{\tilde{\jmath}} \nonumber \\
 & \;  = \sum_p \sum_{i,j}\sum_{\gamma' \in {\cal I}({\cal A})}
  \sum_{k_{\gamma'} ,\ell_{\gamma'}, \tilde{\imath},\tilde{\jmath}}
  ({\cal Y}_{\gamma',\alpha}^\alpha )_{k_{\gamma'}}
  ({\cal Y}_{\alpha,\Upsilon^{\cal I}(\alpha)}^{\gamma'})_{\ell_{\gamma'}} \; \;
  (\widetilde{\cal Y}_{\tilde{\beta}_1,\tilde{\gamma}(p)}^{\tilde{\beta}_1}
  )_{\tilde{\imath}} (\widetilde{\cal Y}_{\widetilde{\Upsilon}^{\tilde{\cal I}}(\tilde{\beta}_2),\tilde{\beta}_2}^{\tilde{\gamma}(p)}
  )_{\tilde{\jmath}} \label{eq:temp-MSnatural-S4-bF-chann} \\
  & \qquad \qquad \qquad  d(1;1,p)^{i,\tilde{\imath}}
  d(p;2^\vee,2)^{j,\tilde{\jmath}} \;\; 
  {\cal F}(\alpha;\alpha,\Upsilon^{\cal I}(\alpha);\alpha
  )^{k_{\gamma'},\ell_{\gamma'}}_{\qquad i,j},
  \nonumber
\end{align}
and the other to use the matrix elements of
\begin{align}
  \sum_{p'} & \; \sum_{k,\tilde{k},\ell,\tilde{\ell}} d(1;p',2)^{k,\tilde{k}}
  ({\cal Y}^\alpha_{\gamma'(p'),\alpha})_k (\widetilde{\cal Y}^{\tilde{\beta}_1}
  _{\tilde{\gamma}'(p'),\tilde{\beta}_2})_{\tilde{k}} \;\;
  d(p';1,2^\vee)^{\ell,\tilde{\ell}}
  ({\cal Y}^{\gamma'(p')}_{\alpha,\Upsilon^{\cal I}(\alpha)})_\ell (\widetilde{\cal Y}^{\tilde{\gamma}'(p')}_{\tilde{\beta}_1,\Upsilon^{\tilde{\cal I}}(\tilde{\beta}_2)})_{\tilde{\ell}},
  \nonumber \\
  & \; = \sum_{p'} \sum_{k,\ell} \sum_{\tilde{\gamma} \in
    \tilde{\cal I}(\widetilde{\cal A})}
  \sum_{\tilde{\imath}_{\tilde{\gamma}},\tilde{\jmath}_{\tilde{\gamma}}, \tilde{k},\tilde{\ell}}
  ({\cal Y}^\alpha_{\gamma'(p'),\alpha})_k
  ({\cal Y}^{\gamma'(p')}_{\alpha,\Upsilon^{\cal I}(\alpha)})_\ell \;\; 
  (\widetilde{\cal Y}_{\tilde{\beta}_1,\tilde{\gamma}(p)}^{\tilde{\beta}_1})_{\tilde{\imath}} (\widetilde{\cal Y}_{\widetilde{\Upsilon}^{\tilde{\cal I}}(\tilde{\beta}_2),\tilde{\beta}_2}^{\tilde{\gamma}(p)})_{\tilde{\jmath}} \label{eq:temp-MSnatural-S4-aF-chann} \\
  & \qquad \qquad \qquad \qquad
  d(1;p',2)^{k,\tilde{k}} d(p';1,2^\vee)^{\ell,\tilde{\ell}} \;\; 
  [\widetilde{\cal F}(\tilde{\beta}_1;\tilde{\beta}_1,\widetilde{\Upsilon}^{\tilde{\cal I}}(\tilde{\beta}_2); \tilde{\beta}_2)^{-1}]^{\tilde{\imath}_{\tilde{\gamma}},\tilde{\jmath}_{\tilde{\gamma}}}_{\qquad  \tilde{k},\tilde{\ell}}.   \nonumber
\end{align}
The matrix elements computed in the two different ways should agree because
we have explicitly assumed that $(V,Y, \cdots )$ and ${\cal I}((V,Y))$ in the
defining algebraic data (\ref{eq:data-set-alg-4-CFT-viaChiral}) of a CFT
satisfies the condition (k-int.op), and so does $(\widetilde{V},
\widetilde{Y},\cdots)$ and $\tilde{\cal I}((\widetilde{V},\widetilde{Y}))$. 
Ref. \cite[\S4]{Moore:1988ss} focuses on the coefficient of
the intertwining operator
$({\cal Y}^\alpha_{\gamma',\alpha})_k({\cal Y}^{\gamma'}_{\alpha, \Upsilon^{\cal I}(\alpha)})_\ell \cdot (\widetilde{\cal Y}^{\tilde{\beta}_1}_{\tilde{\beta}_1,\tilde{\gamma}})_{\tilde{\imath}} (\widetilde{\cal Y}^{\tilde{\gamma}}_{\widetilde{\Upsilon}^{\tilde{\cal I}}(\tilde{\beta}_2),\tilde{\beta}_2})_{\tilde{\jmath}}$ with
$\gamma' = 0 \in {\cal I}({\cal A})$ and $\tilde{\gamma} =
\tilde{0} \in \tilde{\cal I}(\widetilde{\cal A})$. Then there is a unique $k$,
$\ell$, $\tilde{\imath}$ and $\tilde{\jmath}$ (as those vector spaces are of
1-dimensions),
so we drop those super/subscripts. The fact that we deal with the
irreducible decomposition (\ref{eq:str-Htot-bos-byVOA}) under the maximally
extended VOAs (${\cal A} \times \widetilde{\cal A}$) implies that there
is just one $({\cal A} \times \widetilde{\cal A})$-irreducible block $p'$
with $\gamma'(p')=0$, and there is just one irreducible block $p$
with $\tilde{\gamma}(p) = \tilde{0}$. The coefficient calculated in the
two different ways are therefore
\begin{align}
  d(1;1,0\tilde{0}) & \; d(0\tilde{0};2^\vee,2) \;
  {\cal F}(\alpha;\alpha,\Upsilon^{\cal I}(\alpha);\alpha) \nonumber \\
  & = 
  d(1;0\tilde{0},2) \; d(0\tilde{0};1,2^\vee) \;
  [\widetilde{\cal F}(\tilde{\beta}_1;\tilde{\beta}_1,\widetilde{\Upsilon}^{\tilde{\cal I}}(\tilde{\beta}_2); \tilde{\beta}_2)^{-1}];
  \label{eq:MS-natural-S4-contradict}
\end{align}
here, the $0\tilde{0}$ is the label $p'$ and $p$ of the irreducible
block with $\gamma'(p) = 0$ and $\tilde{\gamma}(p) = \tilde{0}$;
the sub/superscripts $i$, $j$ are also unique and are therefore omitted 
because the relevant vector space of intertwining operators is of 1-dimension.  
Ref. \cite[\S4]{Moore:1988ss} further points out that $d(1;0\tilde{0},2) = 0$
because of the condition (t-1), whereas $d(1;1,0\tilde{0})$ and
$d(0\tilde{0};2^\vee,2)$ are not because of (t-2). This is a contradiction. 

The observation above was stated in Ref. \cite[\S4]{Moore:1988ss}
with a case both ${\cal A}$ and $\widetilde{\cal A}$ are rational in mind.
The argument repeated above, however, does not seem to rely very much
on a property specific to rational VOAs; the condition (k-b-int.op) is
the key element in the argument, which we have already
imposed on the input data (\ref{eq:data-set-alg-4-CFT-viaChiral}), 
not the rationality or ${\rm SL}_2\Z$ invariance. 
Two subtleties in extending the argument above to the cases with non-rational
${\cal A}$ and/or $\widetilde{\cal A}$ are the followings.

First, the argument of \cite[\S4]{Moore:1988ss} (reviewed above) implicitly
uses linear independence of the matrix elements of the intertwining operators
\[
  {\cal Y}\otimes {\cal Y} \otimes \widetilde{\cal Y}\otimes \widetilde{\cal Y} \in \oplus_{\gamma',\tilde{\gamma}} \left( {\cal V}^\alpha_{\gamma',\alpha}\otimes {\cal V}^{\gamma'}_{\alpha,\Upsilon^{\cal I}(\alpha)} \otimes
  \widetilde{\cal V}^{\tilde{\beta}_1}_{\tilde{\beta}_1,\tilde{\gamma}}
  \otimes \widetilde{\cal V}^{\tilde{\gamma}}_{\widetilde{\Upsilon}^{\tilde{\cal I}}(\tilde{\beta}_2),\tilde{\beta}_2} \right)
\]
when extracting the coefficients (\ref{eq:MS-natural-S4-contradict}) from
(\ref{eq:temp-MSnatural-S4-bF-chann}, \ref{eq:temp-MSnatural-S4-aF-chann});
although we have chosen ${\cal Y}$'s and
$\widetilde{\cal Y}$'s as bases of the vector spaces ${\cal V}$ and
$\widetilde{\cal V}$, we take a trace over the states in
$M_{\gamma'}$ and $\widetilde{M}_{\tilde{\gamma}}$, and have a sum
over $\gamma' \in {\cal I}({\cal A})$ and $\tilde{\gamma} \in
\tilde{\cal I}(\widetilde{\cal A})$.

Second, we need ${\cal F}(\alpha;\alpha,
\Upsilon^{\cal I}(\alpha);\alpha) \neq 0$ to claim that there is a
contradiction in (\ref{eq:MS-natural-S4-contradict}).  This number,
equal to $F_\alpha$ introduced in (\ref{eq:def-F_alpha}), 
is indeed non-zero when the left-mover chiral algebra ${\cal A}$ is
rational; in the case ${\cal A}$ is rational, 
this number $F_\alpha$ is associated with the quantum dimension of the
representation $\alpha$, and \cite[Prop.5.1]{MR2387861} also uses
the Verlinde formula /genus-1 equations (cf \S \ref{ssec:g=1}) to give
a mathematical proof that $F_\alpha \neq 0$.
When ${\cal A}$ is not rational, the Verlinde formula is not available,
so the proof cannot be recycled, on one hand. On the other hand,
$F_\alpha=0$ for some $\alpha \in {\cal I}({\cal A})$ would make it
impossible to find a rigidity structure in (\ref{eq:evNcoev-in-VOA}). 
It may not be a very bad idea to impose an extra condition that
$F_\alpha \neq 0$ for all $\alpha \in {\cal I}((V,Y))$ as (k-c-int.op). 

Another observation in \cite[\S4]{Moore:1988ss} is that all the representations
the maximally extended left-mover chiral algebra ${\cal A}$ shows up in
(\ref{eq:str-Htot-bos-byVOA}), if ${\cal A}$ is rational. This is the other
reason that we always choose ${\cal I}((V,Y))$ of a rational VOA
$(V,Y, \cdots)$ as the set of {\it all} the distinct irreducible modules
(cf (r$^{'""}$)). 
Ref. \cite[\S4]{Moore:1988ss} uses the Verlinde formula and existence of
the $S$-transformation matrix among the character functionals of ${\cal A}$
to show that all the representations should be
in (\ref{eq:str-Htot-bos-byVOA}). This argument does not extend to
non-rational cases. The $S^1$-target CFTs with a generic radius are the
easiest examples (footnote \ref{fn:ex-I(VY)-torus});
${\cal A}$ and $\widetilde{\cal A}$ are the rank-1
Heisenberg algebra (the chiral boson VOA), and
${\cal I}({\cal A}) \cong \Z^{\oplus 2}$ in this family of CFTs, while
the set of all the distinct unitary irreducible representations of ${\cal A}$
is $\R$. 

The argument of \cite[\S4]{Moore:1988ss} extended to the cases ${\cal A}$
and $\widetilde{\cal A}$ are  not necessarily rational, with the two
subtleties ignored, indicates that there is a set-theoretical
isomorphism $\pi: {\cal I}({\cal A}) \rightarrow
\tilde{\cal I}(\widetilde{\cal A})$ with which 
\begin{align}
  {\cal H}_{\rm tot} \cong \oplus_{\alpha \in {\cal I}({\cal A})}
  M_\alpha \otimes \widetilde{M}_{\pi(\alpha)}.  
\label{eq:str-Htot-bos-byChiAlg}
\end{align}
There is a relation 
$\pi \cdot \Upsilon^{\cal I} = \widetilde{\Upsilon}^{\tilde{\cal I}} \cdot \pi$
as a map ${\cal I}({\cal A}) \rightarrow \tilde{\cal I}(\widetilde{\cal A})$. 

When ${\cal A}$ and $\widetilde{\cal A}$ are both rational, 
the set-theoretical isomorphism  
$\pi: {\cal I}({\cal A}) \rightarrow \tilde{\cal I}(\widetilde{\cal A})$
also induces an isomorphism between the fusion algebras; to see this,
note that the second condition in (\ref{eq:cond-mod-inv}) implies
$S_{\pi(\beta),\pi(\alpha)} = (S_{\beta,\alpha})^{cc}$. Using the Verlinde
formula, we see that $\widetilde{N}_{\pi(\beta),\pi(\gamma)}^{\pi(\alpha)}
= (N_{\beta,\gamma}^\alpha)^{cc} = N_{\beta,\gamma}^\alpha$ \cite{Moore:1988ss}.
The isomorphism between the fusion algebras is also denoted here by 
 $\pi: \Z[{\cal I}({\cal A})] \rightarrow
\Z[\tilde{\cal I}(\widetilde{\cal A})]$ in this article.
\end{rmk}

\begin{anythng}
\label{statmnt:cond-Ytot}
{\bf continuation from \ref{statmnt:data-4-CFT}:} We have yet to
 describe the condition (t) on the data $Y_{\rm tot}$
 ({\it CFT vertex operator}) to complete \ref{statmnt:data-4-CFT}, so we get
 it done here.  Here, we use the description available when $(V,Y, \cdots)$ and
$(\widetilde{V}, \widetilde{Y},\cdots)$ agree with the chiral
algebras ${\cal A}$ and $\widetilde{\cal A}$, as reviewed
in Rmk. \ref{rmk:Moore-Seiberg}. The data $Y_{\rm tot}$ then
consists of one choice
\begin{align}
Y_{\rm tot}^{(\alpha;\beta,\gamma)} :=
Y_{\rm tot}^{(\alpha;\beta,\gamma)(\pi(\alpha);\pi(\beta),\pi(\gamma))}
\in {\cal V}_{\beta,\gamma}^\alpha \otimes
\widetilde{\cal V}_{\pi(\beta),\pi(\gamma)}^{\pi(\alpha)}
\end{align}
for each triplet $\alpha$, $\beta$, $\gamma$ from ${\cal I}({\cal A})$ such
that $N_{\beta,\gamma}^\alpha \neq 0$. A choice of $Y_{\rm tot}^{(\alpha;\beta,\gamma)}$
can be parametrized by a square matrix $d(\alpha;\beta,\gamma)$ of size
$N_{\beta,\gamma}^\alpha$ ({\it left-right-coupling matrices}),
\begin{align}
  Y_{\rm tot}^{(\alpha;\beta,\gamma)} = \sum_{i=1}^{N_{\beta,\gamma}^\alpha}
  \sum_{\tilde{\imath}=1}^{N_{\pi(\beta),\pi(\gamma)}^{\pi(\alpha)}} 
  ({\cal Y}_{\beta,\gamma}^{\alpha})_{i}
  \otimes (\widetilde{\cal Y}_{\pi(\beta),\pi(\gamma)}^{\pi(\alpha)})_{\tilde{\imath}}
  \; d(\alpha;\beta,\gamma)^{i,\tilde{\imath}}
  \label{eq:def-LR-coupl-matrix}
\end{align}
The condition (t) on $Y_{\rm tot}$ (hence on the matrices
$d(\alpha;\beta,\gamma)$) consists of all the followings.

First, as stated already as the condition (t-1), $Y_{\rm tot}$ must be
compatible with the representation of the chiral algebras
${\cal A}$ and $\widetilde{\cal A}$; that is, for $\beta=0$ (and hence
$\alpha = \gamma$), 
\begin{align}
  Y_{\rm tot}^{(\gamma;0,\gamma)} = Y_\gamma \otimes \widetilde{Y}_{\pi(\gamma)}.
  \label{eq:cond-Ytot-repr}
\end{align}
Second (t-2)=(t-2sk, t-2cr), $Y_{\rm tot}$ has a skew symmetry and
crossing symmetry,
\begin{align}
 & (\Omega_r\otimes \widetilde{\Omega}_{-1-r})(Y_{\rm tot}^{(\alpha;\beta,\gamma)})
  = Y_{\rm tot}^{(\alpha;\gamma,\beta)},   \label{eq:cond-Ytot-skew-N-xing} \\
 & (\lambda_\gamma \otimes \tilde{\lambda}_{\pi(\gamma)}) \cdot
  (\Theta_r\otimes \widetilde{\Theta}_{-1-r})(Y_{\rm tot}^{(\alpha;\beta,\gamma)})
  \cdot (\lambda_\alpha^{-1} \otimes \tilde{\lambda}_{\pi(\alpha)}^{-1})
  = Y_{\rm tot}^{(\Upsilon^{\cal I}(\gamma);\beta,\Upsilon^{\cal I}(\alpha))} 
 \nonumber
\end{align}
for all the triplets $\alpha,\beta,\gamma \in {\cal I}({\cal A})$ with
$N_{\beta,\gamma}^\alpha \neq 0$ and $r \in \Z$; this generalizes the
condition (h) in the defining algebraic data of a VOA. One may also
write this condition as $\Omega_{\rm tot}(Y_{\rm tot}) = Y_{\rm tot}$ and
$(\lambda\tilde{\lambda}) (\Theta_{\rm tot}(Y_{\rm tot}))
(\lambda \tilde{\lambda})^{-1} = Y_{\rm tot}$. The crossing symmetry condition
is both on the choice of a CFT vertex operator $Y_{\rm tot}$ and 
$\{ \lambda_\alpha \}$ and $\{\tilde{\lambda}_{\tilde{\alpha}}\}$; without
referring to $\lambda \tilde{\lambda}$, we cannot even compare
$\Theta_{\rm tot}Y_{\rm tot}$ with $Y_{\rm tot}$. Finally, $Y_{\rm tot}$ has
the associativity \cite[(2.3e)]{Moore:1988uz}; that is, the isomorphism 
\begin{align}
  ({\cal F}\otimes \widetilde{\cal F}):
  \oplus_{\delta \in {\cal I}({\cal A})} & \;
  \left( {\cal V}_{\beta_1,\delta}^\alpha \otimes
  \widetilde{\cal V}_{\pi(\beta_1),\pi(\delta)}^{\pi(\alpha)} \right) \otimes 
  \left( {\cal V}_{\beta_2,\gamma}^\delta \otimes
  \widetilde{\cal V}_{\pi(\beta_2),\pi(\gamma)}^{\pi(\delta)} \right)
  \label{eq:cond-Ytot-F1} \\
  & \qquad \longrightarrow 
  \oplus_{\epsilon \in {\cal I}({\cal A})} \left( {\cal V}_{\beta_1,\beta_2}^\epsilon
  \otimes \widetilde{\cal V}_{\pi(\beta_1),\pi(\beta_2)}^{\pi(\epsilon)} \right)
  \otimes \left( {\cal V}_{\epsilon,\gamma}^\alpha \otimes
  \widetilde{\cal V}_{\pi(\epsilon),\pi(\gamma)}^{\pi(\alpha)} \right)  \nonumber 
\end{align}
maps $\{ (Y_{\rm tot}^{(\alpha;\beta_1,\delta)}\otimes
Y_{\rm tot}^{(\delta;\beta_2,\gamma)})_{\delta \in {\cal I}({\cal A})} \}$ to
$\{ (Y_{\rm tot}^{(\epsilon;\beta_1,\beta_2)} \otimes
Y_{\rm tot}^{(\alpha;\epsilon,\gamma)})_{\epsilon \in {\cal I}({\cal A})} \}$.
In terms of the left-right-coupling
matrices $d(\alpha;\beta,\gamma)$ and the matrix representations of the fusion
isomorphisms, this associativity of $Y_{\rm tot}$ is equivalent to 
\begin{align}
  d(\alpha;\epsilon,\gamma)^{k,\tilde{k}} 
  d(\epsilon;\beta_1,\beta_2)^{\ell,\tilde{\ell}}  & \; = 
    \sum_{\delta \in {\cal I}({\cal A})} 
    \sum_{i_\delta,j_\delta} \sum_{\tilde{\imath}_\delta, \tilde{\jmath}_\delta}
  d(\alpha;\beta_1,\delta)^{i_\delta,\tilde{\imath}_\delta}
  d(\delta;\beta_2,\gamma)^{j_\delta,\tilde{\jmath}_\delta}
  \label{eq:cond-Ytot-F2}  \\
  &     
      {\cal F}(\alpha;\beta_1,\beta_2;\gamma)_{\quad i_\delta,j_\delta}^{k,\ell} \;
  \widetilde{\cal F}(\pi(\alpha);\pi(\beta_1),\pi(\beta_2);\pi(\gamma)
  )^{\tilde{k},\tilde{\ell}}_{\quad \tilde{\imath}_\delta, \tilde{\jmath}_\delta} \;  
 \nonumber 
\end{align}
Although this equation is
homogeneous with respect to the left-right-coupling matrices $d$,
some of them are already determined by
(\ref{eq:cond-Ytot-repr}, \ref{eq:cond-Ytot-skew-N-xing}),
so this is a quadratic polynomial equation on the residual freedom in $d$. 
We may also write this condition symbolically as
${\cal F}_{\rm tot}(Y_{\rm tot} \otimes Y_{\rm tot}) = Y_{\rm tot} \otimes
Y_{\rm tot}$.  Because of the relation ${\cal B}_{-1-r} = {\cal F}^{-1}
\cdot \Omega_r \cdot {\cal F}$ and the counter part in the right-mover
chiral algebra, one relation follows automatically from the condition
(t) that is best written symbolically
as ${\cal B}_{\rm tot}(Y_{\rm tot} \otimes Y_{\rm tot}) =
Y_{\rm tot} \otimes Y_{\rm tot}$.

A large fraction of the crossing symmetry condition (t-2cr) is also
redundant, in fact. The condition (t-2cr) {\it is} non-trivial, when
applied to $Y_{\rm tot}^{(0;\Upsilon^{\cal I}(\gamma),\gamma)}$, and its consequence
(along with (t-1) and (t-2sk)) is that we should choose it to be the chiral
pair annihilation vertex (cf \ref{statmnt:pair-ann-crtn}) of the left-mover
and right-mover combined. For all other types, however, the condition (t-2cr)
is satisfied automatically as the consequence of (t-3) and (t-2sk),
because the crossing symmetry isomorphisms are implemented by appropriate
combinations of the skew-symmetry and fusion isomorphisms
(cf \ref{statmnt:Theta-by-FnOmega}) when $Y_{\rm tot}$ of the type
$(0;\Upsilon^{\cal I}(\gamma),\gamma)$ is given by the pair annihilation
vertices. 
\end{anythng}

\begin{anythng}
A unitary CFT is said to be {\it rational} when both of the left-moving and right-moving chiral algebras ${\cal A}$ and $\widetilde{\cal A}$ are rational as reviewed in \ref{statmnt:rationalVOA}.
\end{anythng}

\begin{anythng}
A holomorphic 2D unitary modular invariance CFT is always a unitary VOA whose underlying vector space $V$ is that of the left-mover chiral algebra.
Indeed, the right-mover unitary VOA $(\widetilde{V}, \widetilde{Y},\cdots)$ with $\tilde{c}=0$ must be the trivial one, so there is just one irreducible representation $\tilde{0} \in \tilde{\cal I}((\widetilde{V},\widetilde{Y}))$ (cf \ref{statmnt:def-VOA}).
The structure of the irreducible decomposition (\ref{eq:str-Htot-bos-byChiAlg}) indicates then that just one representation $0$ of the maximally extended left-mover VOA is involved in ${\cal H}_{\rm tot}$;
${\cal I}({\cal A}) = \{ 0\}$.  $Y_{\rm tot}=Y$ for the maximally extended left-mover defining data ${\cal A}= (V,Y, \cdots)$;
the condition $\Omega_{\rm tot}(Y_{\rm tot}) = Y_{\rm tot}$ is nothing more than the condition (i) of ${\cal A}$, and the condition ${\cal F}_{\rm tot}(Y_{\rm tot} \otimes Y_{\rm tot}) = Y_{\rm tot} \otimes Y_{\rm tot}$ is nothing more than a consequence of the Jacobi identity of the VOA ${\cal A}$;
they are not an extra condition to be imposed on ${\cal A}$.
When we see a 2D unitary holomorphic CFT as a VOA, therefore, the invariance of the character functional of the defining representation under the $S$-transformation is the only one non-trivial constraint on a VOA. 
\end{anythng}

\subsection{Some Observables of a 2D CFT}
\label{ssec:CFT-obs}

\begin{anythng}
\label{statmnt:CFT-noBdry-obsrvbls}  
Here, we introduce two classes of {\it observables} for any unitary CFT satisfying the conditions (a--t): {\it normalized correlation functions on a closed Riemann surface} and {\it unnormalized correlation functions on a closed 2-dimensional Riemannian manifold}.
The jargon {\it observables} does not mean more than well-defined things in this article. 

The first class of observables is for a closed Riemann surface $\Sigma$
of genus $g$, $n$ ordered points $\{p_1,\cdots, p_n\}$ in $\Sigma$,
and a choice of local holomorphic coordinates $\{(U_i, z_i)_{i=1,\cdots,n}\}$
where $U_i\subset \Sigma$ is an open neighborhood of $p_i$ and
$z_i: U_i \rightarrow \C$ is a holomorphic coordinate.
The {\it normalized correlation functions} assign linearly 
\begin{align}
  ({\cal H}_{\rm tot})^{\otimes n} \ni  x_1 \otimes \cdots \otimes x_n \longmapsto
  \langle Y_{\rm tot}^{(z_1)}(x_1)_{@p_1} Y_{\rm tot}^{(z_2)}(x_2)_{@p_2} \cdots
  Y_{\rm tot}^{(z_n)}(x_n)_{@p_n} \rangle^{\rm nrm}_{\Sigma} \in \C ;
  \label{eq:notn-CFT-closed-crrltn-fcn-nrm}
\end{align}
we have just introduced the notation for this class of observables here,
and review how to compute the assigned values shortly; they are called
functions as we may see them for a given choice of
$x_1 \otimes \cdots \otimes x_n$
a function of the moduli of $(\Sigma, \{ (p_i, U_i, z_i) \})$. 

When a 2D modular invariant and unitary CFT has a weakly coupled
description with a path integral formulation, the normalized
correlation functions are meant---in physics terms---to be
the correlation functions on $\Sigma$, with the vertex operators 
of the states $x_1, \cdots, x_n$ inserted at the points $p_1,\cdots, p_n$
in the operator implementation for the local coordinates $(U_i, z_i)$,
normalized by the partition function on $\Sigma$. The Riemannian metric
on $\Sigma$ is assumed to be Cartesian wrpt the coordinates $z_i$ in
a small neighborhood of $p_i$ in $U_i$ and is otherwise arbitrary. 
Then the physics reasonings indicate (sewing algorithm \cite[\S9.4]{Polchinski:1998rq})
how to compute those observables
denoted as in (\ref{eq:notn-CFT-closed-crrltn-fcn-nrm}) in terms of
the defining algebraic data $({\cal H}_{\rm tot}, Y_{\rm tot}, {\bf 1}\tilde{\bf 1},
\omega, \tilde{\omega}, \Upsilon_{\rm tot}, \langle - | - \rangle_{\rm tot})$
of the CFT. When a CFT does not have a weakly coupled path integral
formulation, or if one intends to lay theoretical foundation without
relying on physics reasonings, that computation algorithm is taken as the
definition of the {\it normalized correlation functions} denoted as
in (\ref{eq:notn-CFT-closed-crrltn-fcn-nrm}).  

Think of the simplest case where $\Sigma = \P^1$ with an inhomogeneous
coordinate $z$ covering $U := \Sigma \backslash \infty$, and $(U,z)$ is
used as the local coordinate neighborhood of all the $n$ points
(i.e., $z_i = z$ for all $i=1,\cdots, n$).
In this case, one of the computation algorithms (called
{\it multi-peripheral} in \cite{Moore:1988qv}) is to set
\begin{align}  
&  
  \langle Y^{(z)}_{\rm tot}(x_1)_{@p_1} \cdots Y^{(z)}_{\rm tot}(x_n)_{@p_n}
  \rangle_{\Sigma}^{\rm nrm} =   \label{eq:algrthm-multi-prhrl} \\
& \qquad \quad
  \sum_{\delta_1,\cdots,\delta_{n-1}\in {\cal I}({\cal A})}
  \bra{\tilde{\bf 1}{\bf 1} } \; Y_{\rm tot}^{(0;\beta_1,\delta_1)}(x_1)_{@p_1}
  \; Y_{\rm tot}^{(\delta_1;\beta_2,\delta_2)}(x_2)_{@p_2} \cdots
  Y_{\rm tot}^{(\delta_{n-1};\beta_n,0)}(x_n)_{@p_n} \cdot {\bf 1}\tilde{\bf 1}
  \rangle   \nonumber 
\end{align}
when $x_i \in M_{\beta_i} \otimes \widetilde{M}_{\pi(\beta_i)}
\subset {\cal H}_{\rm tot}$. There are multiple other computation
algorithms\footnote{
The algorithm called {\it staircase} in \cite{Moore:1988qv} uses
$Y_{\rm tot}^{(\epsilon_1;\beta_1,\beta_2)}$,
$Y_{\rm tot}^{(\epsilon_2;\epsilon_1,\beta_3)}$, ... and
$Y_{\rm tot}^{(0;\epsilon_{n-1},\beta_n)}$ instead. 
} %
that yield the same result, because of the associativity and
commutativity 
${\cal F}_{\rm tot}(Y_{\rm tot} \otimes Y_{\rm tot}) =
Y_{\rm tot}\otimes Y_{\rm tot}$ and
${\cal B}_{\rm tot}(Y_{\rm tot} \otimes Y_{\rm tot}) =
Y_{\rm tot} \otimes Y_{\rm tot}$ imposed on the defining algebraic
data of the CFT in question. 
For a more general choice of local coordinates, 
\begin{align}
&  
  \langle Y^{(z_1)}_{\rm tot}(x_1)_{@p_1} \cdots Y^{(z_n)}_{\rm tot}(x_n)_{@p_n}
  \rangle_{\Sigma}^{\rm nrm} = \\
& \qquad \quad
  \sum_{\delta_1,\cdots,\delta_{n-1}\in {\cal I}({\cal A})}
  \bra{{}^{(z)}\tilde{\bf 1}{\bf 1} } \; Y_{\rm tot}^{(0;\beta_1,\delta_1)(z;z)}
  (R_{f_1@z_1(p_1)}^{-1}\tilde{R}_{f_1@z_1(p_1)}^{-1} x_1)_{z@z(p_1)} \cdots
  \nonumber \\
  & \qquad \qquad \qquad \qquad \qquad 
  \cdots Y_{\rm tot}^{(\delta_{n-1};\beta_n,0)(z;z)}(
  R_{f_n@z_n(p_n)}^{-1}\tilde{R}_{f_n@z_n(p_n)}^{-1} x_n)_{@p_n} \cdot
  {\bf 1}\tilde{\bf 1}^{(z)} \rangle ,   \nonumber 
\end{align}
where $z=f_i(z_i)$ is the dictionary between the coordinates,
and we read $L_i$'s and $\tilde{L}_i$'s in $R_{f_i@z_i(p_i)}$
and $\tilde{R}_{f_i@z_i(p_i)}$ as their representations in
$(M_{\beta_i} \otimes \widetilde{M}_{\pi(\beta_i)})$
In the expression above,
${\bf 1}\tilde{\bf 1}^{(z)}$ stands for the image of
${\bf 1}\tilde{\bf 1} \in {\cal H}_{\rm tot}$ under
${\cal H}_{\rm tot} \cong {\cal H}_{\rm tot}^{(z)}$,
using the global coordinate $z$ of $\Sigma = \P^1$ for
the frame choice and trivialization of ${\rm End}({\cal H}_{\rm tot})$
(as reviewed in \ref{statmnt:prscr-oprtr-trnstnFcn-coordCh} with
VOA in mind). The action of $R_{f_i@z_i(p_i)} \tilde{R}_{f_i@z_i(p_i)}$
on $(M_{\beta_i} \otimes \widetilde{M}_{\pi(\beta_i)})$ is well-defined;
although the $L_0$ and
$\tilde{L}_0$ eigenvalues $h$ and $\tilde{h}$ are not necessarily
integers on $M_\beta \otimes \widetilde{M}_{\pi(\beta)}$,
we deal with $f_0^{-h_\beta} (f_0^{cc})^{-\tilde{h}_{\pi(\beta)}}$ as
$|f_0|^{-2\tilde{h}_{\pi(\beta)}} f_0^{\tilde{h}_{\pi(\beta)}-h_\beta}$, which is
well-defined because $\tilde{h}_{\pi(\beta)}-h_\beta \in \Z$ in a modular
invariant CFT. 

For a closed Riemann surface $\Sigma$ with $g\geq 1$, the algorithm
begins with finding a sewing construction/pants decomposition
\cite[\S9.3]{Polchinski:1998rq}, \cite[\S1.1--\S1.2]{MR1448404}
%
%
of $(\Sigma, \{ p_i \})$, where $\Sigma$ is regarded as multiple
$\P^1$'s glued together, each $\P^1$ is covered by coordinate charts
whose coordinates are linear fractional transformations of the
homogeneous coordinates of the $\P^1$, and the coordinate
transformations between the charts glued (sewn) together are
also linear fractional transformations (${\rm PGL}_2\C$)
(this additional set of data on top of $\Sigma$ is called the
{\it projective structure}). Once this structure is given,
then one may pick one of the coordinate charts, and use
the coordinate there to provide a frame and trivialization of 
${\cal H}_{\rm tot}$ (like ${\cal H}_{\rm tot}^{(z)}$ in the case $g=0$). 
The normalized correlation
function (\ref{eq:notn-CFT-closed-crrltn-fcn-nrm}) is computed
by using the pants decomposition and the defining algebraic
data of the given CFT (e.g., $Y_{\rm tot}^{(\alpha;\beta,\gamma)}$'s);
see \cite[\S9.4]{Polchinski:1998rq} for more information; one may 
call this computation algorithm of the normalized correlation functions 
for closed Riemann surfaces as the {\it sewing algorithm}. 

The second class of observables is for a two-dimensional closed
Riemannian
manifold $(M, ds^2)$ of genus $g$, which specifies a Riemann surface
$\Sigma$, with $n$ ordered points with local holomorphic coordinates,
$\{ (p_i, U_i, z_i) \}$. In the rest of this section \ref{ssec:CFT-obs},
we only deal with 2D unitary and modular invariant CFTs with $c=\tilde{c}$.
The {\it unnormalized correlation functions} 
assign linearly
\begin{align}
  ({\cal H}_{\rm tot})^{\otimes n} \ni x_1\otimes \cdots \otimes x_n
  \longmapsto
  Z\left( M,ds^2; Y_{\rm tot}^{(z_1)}(x_1)_{@p_1} \cdots Y_{\rm tot}^{(z_n)}(x_n)_{@p_n} \right) \in \C;
  \label{eq:notn-CFT-closed-crrltn-fcn-UNnrm}
\end{align}
once again, we have just introduced the notation for this class of
observables. When a unitary CFT has a weakly coupled path integral
formulation (where the integration is over the configuration space of
the elementary fields {\it and} their canonical conjugation fields),
and if the metric $ds^2$ is Cartesian $dz^i \otimes d\bar{z}^{i}$
in a small neighborhood of $p_i$ within $U_i$ ({\it fortunate cases}),
then this class of observables is meant---in physics terms---to
be the correlation functions without dividing them by the partition
function. Physics reasoning available in such a CFT determines 
the algorithms to compute this class of observables, which
we review below. This computation algorithm is taken as the math
definition of the {\it unnormalized correlation functions on $(M,ds^2)$};
they are meant as functions on the moduli of $(M,ds^2, \{ p_i,U_i,z_i\})$
for individual choices of $x_1\otimes \cdots \otimes x_n$. 

The {\it unnormalized correlation function} in the case $M$ is closed
is
\begin{align}
  Z\left( M, ds^2; Y_{\rm tot}^{(z_1)}(x_1)_{@p_1} \cdots
  Y_{\rm tot}^{(z_n)}(x_n)_{@p_n} \right) := Z\left( M, ds^2 \right)
  \langle Y_{\rm tot}^{(z_1)}(x_1)_{@p_1} \cdots
  Y_{\rm tot}^{(z_n)}(x_n)_{@p_n} \rangle^{\rm nrm}_\Sigma, 
\end{align}
where $Z(M,ds^2)$ is the unnormalized correlation function in the
case $n=0$, and is called the {\it partition function} on a genus-$g$
Riemannian manifold $(M,ds^2)$. The equality above is a consequence
of physics reasonings in fortunate cases, but is regarded as a definition
otherwise. This definition implies, when we see $(M,ds^2)$ as a Riemann
surface $\Sigma$ with a choice of K\"{a}hler form compatible with
the complex structure of $\Sigma$, that the unnormalized correlation
functions depend on the choice of a K\"{a}hler form only through
the partition function. 

The partition functions on a Riemann surface $\Sigma$ and a pair of
its compatible K\"{a}hler metrics, $ds^2_*$ and $ds^2 = e^{2\omega} ds^2_*$,
where $\omega: \Sigma \rightarrow \R$ is a smooth function
({\it conformal transformation parameter}), are related by
\cite[\S4]{Polchinski:1998rq} ($c=\tilde{c}$ here)
\begin{align}
  Z(M,ds^2_* e^{2\omega} ) & \; := Z(M,ds^2_*) \; W[M,ds_*^2, e^{2\omega}], \\
  W[M,ds_*^2,e^{2\omega}ds_*^2] & \; := \exp \left[ \frac{c}{24\pi}
    \int_M {\rm vol}_* \; \left\{
    h_*^{ab} (\partial_a \omega)(\partial_b \omega) + R_* \omega \right\}
    \right],
  \label{eq:def-fctr-Liouville}
\end{align}
where $h_{*ab}d\sigma^a \otimes d\sigma^b$ is the description of the
metric $ds_*^2$ using a local coordinate $\{ \sigma^a \; | \; a=1,2\}$
on $M$, ${\rm vol}_*$ and $R_*$ the volume form and the Ricci scalar
curvature of the metric $ds_*^2$; $R_* = +2$ for the unit-radius round
sphere metric on $\P^1$. We still need an algorithm to compute the
partition function for a Riemann surface $\Sigma$ and at least
one choice of a K\"{a}hler metric $ds_*^2$. 

There is a computation algorithm for the partition function that 
uses a sewing construction of two-dimensional Riemannian manifolds
$(M,ds^2)$. The sewing construction for $(M,ds^2)$ goes as follows
\cite[\S9.3]{Polchinski:1998rq}. 
Suppose, first, that $(M_1, ds_1^2)$ and $(M_2,ds_2^2)$ are two-dimensional
Riemannian manifolds, $p_1 \in M_1$, $p_2 \in M_2$ and $(U_1,z_1)$ and
$(U_2,z_2)$ are holomorphic coordinate neighbourhoods of $p_1$ and $p_2$,
respectively, such that $z_i(p_i)=0$ and the unit disc fits within $z_i(U_i)$ 
for both $i=1,2$. 
The discs $\{ p \in U_i \; | \; z_i(p)\leq 1\}$ are denoted by $B_i$.
Suppose further that $ds_i^2 = dz_i\otimes d\bar{z}_i/|z_i|^2$ within
the annulus region $1-\epsilon \leq |z_i(p)| \leq 1 + \epsilon$ for some
small $\epsilon \in \R_{>0}$ inside $U_i$ ($i=1,2$). 
Then the pair of the two-dimensional Riemannian manifolds are sewn together
to form a manifold $M := M_1\infty M_2((p_1,z_1),(p_2,z_2),1)$ by
cutting out the interior of the discs $B_i$, 
and identifying the coordinates by $z_1z_2 =1$ around the gluing loci
$|z_i(p)|=1$, when the metric $ds_1^2$ smoothly continues
to $ds_2^2$ to form a metric $ds_M^2$ on $M$.
When $M_1$ and $M_2$ are of genus $g_1$ and $g_2$, respectively,
then $M$ is of genus $g_1+g_2$. 
Suppose, next, that $(M_0, ds_0^2)$ is a two-dimensional Riemannian
manifold, $p_1, p_2 \in M_0$ and $(U_1,z_1)$ and
$(U_2,z_2)$ are holomorphic coordinate neighborhoods of $p_1$ and $p_2$,
respectively, such that $z_i(p_i)=0$ and the unit disc fits within
$z_i(U_i)$  for both $i=1,2$. 
Suppose further that $ds_0^2 = dz_i\otimes d\bar{z}_i/|z_i|^2$ within
the annulus region $1-\epsilon \leq |z_i(p)| \leq 1 + \epsilon$ for some
small $\epsilon \in \R_{>0}$ inside $U_i$ ($i=1,2$). 
Then the neighborhood of the boundary of the two discs $B_i$ ($i=1,2$) in $M_0$ are sewn together to form a manifold $M := M_08((p_1,z_1),(p_2,z_2),1)$
by cutting out the interior of $B_i$, 
and identifying the coordinates by $z_1z_2 =1$ around the gluing loci
$|z_i(p)|=1$, when the metric $ds_0^2$ on both sides of the
gluing locus smoothly continues to the other side of the gluing locus,  
to form a metric $ds_M^2$ on $M$. When $M_0$ is of genus $g_0$, $M$
is of genus $g_0+1$. By combining those two methods, it is possible
to construct any Riemann surface $\Sigma$ from $\P^1$'s, along with
at least one K\"{a}hler metric on $\Sigma$.  

Physics reasoning relying on path-integral formulation indicates
that 
\begin{align}
  Z(M_1\infty M_2, ds_M^2) = & \; \sum_{x,y}
  \frac{Z(M_1,ds_1^2; Y_{\rm tot}^{(z_1)}(x)_{@p_1} )}
       {Z(\P^1,{\rm sph}_1)^{1/2} \; W[B_1,{\rm sph}_1|_{B_1},ds_1^2|_{B_1}] }
       \\
 & \qquad   \underline{\cal G}^{xy} \quad 
  \frac{Z(M_2,ds_2^2; Y_{\rm tot}^{(z_2)}(y)_{@p_2} )}
       {Z(\P^1,{\rm sph}_1)^{1/2} \; W[B_2,{\rm sph}_1|_{B_2},ds_2^2|_{B_2}] },
         \nonumber 
\end{align}
where each of $\{x\}$ and $\{y\}$ is a basis of ${\cal H}_{\rm tot}$,
and $\underline{\cal G}^{xy}$ is the inverse of the the matrix
\begin{align}
  \underline{\cal G}_{xy} := \langle Y_{\rm tot}^{(z)}(x)_{z@0} \;
  Y_{\rm tot}^{(1/z)}(y)_{1/z@0} \rangle_{\P^1}^{\rm nrm}. 
\end{align}
The unit-radius round sphere metric on $\P^1$ is denoted by ${\rm sph}_1$,
and a unit disc (such as $B_1$ and $B_2$) is identified with
$\P^1$ cut by its equator through the stereographic projection;
so ${\rm sph}_1|_{B_i}$ makes sense. 
Although the factor (\ref{eq:def-fctr-Liouville}) was introduced
for a closed Riemannian manifold, we use the same formula
(\ref{eq:def-fctr-Liouville}) here for $B_i$'s with boundaries,
because the equator of $\P^1$ has vanishing geodesic curvature
wrpt the round sphere metric. Physics reasoning also indicates that
\begin{align}
  Z(M_08, ds_M^2) = \sum_{x,y}
  \frac{Z(M_0,ds_0^2; Y_{\rm tot}^{(z_1)}(x)_{@p_1} Y_{\rm tot}^{(z_2)}(y)_{@p_2})}
       {Z(\P^1,{\rm sph}_1) \; W[B_1,{\rm sph}_1|_{B_1},ds_0^2|_{B_1}]
         W[B_2,{\rm sph}_1|_{B_2}, ds_0^2|_{B_2}]} \quad 
   \underline{\cal G}^{xy}. 
\end{align}
Therefore, those two formulas reduce the computation of
normalized correlation functions on closed two-dimensional Riemannian
manifolds to computation of those on closed Riemannian manifolds of
lower genera, normalized correlation functions, the
factor (\ref{eq:def-fctr-Liouville}) and the value of $Z(\P^1,{\rm sph}_1)$.  
So, we have an algorithm to compute the partition functions
and unnormalized correlation functions on closed two-dimensional
Riemannian manifolds by using the defining algebraic data and
the value of $Z(\P^1,{\rm sph}_1)$ of a CFT in question. 

The present authors are not aware of an algorithm to compute
$Z(\P^1, {\rm sph}_1)$ of a CFT from its defining algebraic data.
It is a widely accepted belief with lots of evidence, when a CFT is a non-linear sigma model
whose target space metric\footnote{
The Polyakov action involves the target space metric of mass-dimension $-2$
divided by $\alpha'$; what matters here is this combination, a metric
that is dimensionless. 
 } %
has small curvature, that its value is the volume of the target space,
possibly accompanied by a constant of order unity: 
\begin{align}
  Z(\P^1, {\rm sph}_1) \sim {\rm vol.}({\rm target~space}).
  \label{eq:sph-Z-prpt-vol}
\end{align}

\end{anythng}

\newpage


\section{With a $\Z/2\Z$-grading (with fermionic excitations)}
\label{sec:Z2}

\subsection{Vertex Operator Superalgebra}
\label{ssec:SVOA}

\begin{anythng}
 \label{statmnt:def-alg-data-SVOA}
  {\bf algebraic data of a vertex operator superalgebra:}
The defining set of data of a unitary {\it vertex operator superalgebra}
(SVOA or VOSA) consists of $(V, YY, {\bf 1}, \omega, \Upsilon,
\langle - | - \rangle)$
satisfying the conditions below. $V$ is a vector space over $\C$ with a
${\cal G}:= \Z/2\Z$ grading,
\begin{align}
 V = V_0 \oplus V_1 = \oplus_{s_0 \in {\cal G}} V_{s_0},  
\end{align}
and $YY: V \rightarrow {\rm End}(V)[[z,z^{-1}]]$ a homomorphism;
$YY$ is described by a quartet of homomorphisms $\{ Y_{s_0, \; s'_0} \; |\;
s_0, s'_0 \in {\cal G} \}$, where $Y_{s_0,s'_0}: V_{s_0} \rightarrow
{\rm Hom}(V_{s'_0}, V_{s_0 + s'_0})[[z,z^{-1}]]$, so $YY$ maps
$V_0$ to the even part of ${\rm End}(V)$ and $V_1$ to the odd part
of ${\rm End}(V)$.  Both ${\bf 1}$ and $\omega$ are in $V_0$. 
A part of the data $(V_0, Y_{0,0}, {\bf 1}, \omega)$ should satisfy
the conditions (a--$\ell$) of a set of data of a VOA. The conditions
(a, i--$\ell$) on the entire set of data $(V, YY, \cdots)$ are generalized
as follows, however. 

First (a-SVOA), with an isomorphism $\hat{\phi}: {\cal G}=\Z/2\Z \ni s_0 \mapsto
s_0/2 +\Z \in 2^{-1}\Z/\Z \subset \R/\Z$, 
\begin{align}
  {\rm for~}s_0 \in {\cal G} = \Z/2\Z, \qquad 
  V_{s_0} \cong \oplus_{h \in \hat{\phi}(s_0)}^{(h\geq 0)} V_{s_0,h} , \qquad
  \dim_\C(V_{s_0,h}) < \infty.
\end{align}
The mode operators of a homogeneous state $u \in V_{s_0,h}$ are of the form
\begin{align}
 YY(u,z) =: \sum_{\nu \in - \hat{\phi}(s_0)} [u]_\nu z^{-h-\nu}
\end{align}
so that all the powers $-(h+\nu)$ are integral.

The super-Jacobi identity ($\ell$-SVOA) is, for arbitrary
$u_1 \in V_{s_{0,1}}$ and $u_2 \in V_{s_{0,2}}$, 
\begin{align}
  \frac{1}{z_0} \delta \left( \frac{z_1-z_2}{z_0}\right) & \; YY(u_1, z_1)
  \cdot YY(u_2, z_2)
  - \frac{c(s_{0,1}, s_{0,2})}{z_0}
  \delta \left( \frac{z_2-z_1}{-z_0}\right)
  YY(u_2, z_2) \cdot YY(u_1,z_1)  \nonumber  \\
  & \; = \frac{1}{z_2}
    \delta \left( \frac{z_1-z_0}{z_2}\right)
    YY \left( YY(u_1,z_0) \cdot u_2, z_2\right),
    \label{eq:super-Jacobi-SVOA}
\end{align}
where $c: {\cal G} \times {\cal G} \rightarrow S^1$ in the second term
should have the property (\ref{eq:cond-bimultipl}, \ref{eq:cond-skewsym-C})
and is set 
\begin{align}
  {\cal G} \times {\cal G} \ni (s_{0,1}, s_{0,2}) \mapsto
  c(s_{0,1}, \; s_{0,2}) = (-1)^{+s_{0,1}s_{0,2}} 
  \label{eq:fermion-exch-phase}
\end{align}
for SVOA. 
The skew-symmetry condition (i-SVOA) is of the
form (e.g., \cite[(2.41)]{MR1841356})
%
\begin{align}
  Y_{s_{0,1},s_{0,2}} (u_1,z) \cdot u_2 & \; = (-1)^{- s_{0,1}s_{0,2}} e^{z L_{-1}}
  Y_{s_{0,2},s_{0,1}} (u_2,-z) \cdot u_1 , \nonumber \\
  & \; = c(s_{0,1}, \; s_{0,2})^{-1}  e^{z L_{-1}}
  Y_{s_{0,2},s_{0,1}} (u_2,-z) \cdot u_1 .
  \label{eq:skewsym-SVOA}
\end{align}

An SVOA always has a group of automorphisms $\hat{\cal G}\cong \{ \hat{0},
\hat{g}\}$ where $\hat{0} = {\rm id}_V$, and the generator $\hat{g}$
multiplies $e^{2\pi i s_0/2}$ on $V_{s_0} \subset V$. 
The automorphism group $\hat{\cal G}$ is dual to ${\cal G}$, so we may
regard $\hat{\cal G} \cong {\rm Hom}({\cal G}, S^1) \cong
{\rm Hom}({\cal G}, \R/\Z)$.

The set of data also involves (cf Rmk. \ref{rmk:nonUnitary-CFT?})
a Hermitian positive definite inner product
$<-|->$ in which $V_{s_0}$ with different $s_0 \in {\cal G}$ are orthogonal,
and also a $\C$-anti-unitary $\Upsilon_{-s_0,s_0}: V_{s_0}
\rightarrow V_{-s_0}$. This $\Upsilon_{-s_0,s_0}$ is split into
$\lambda_{s_0} \cdot \Upsilon_{s_0}$, where $\lambda_{s_0}: V_{s_0}^\vee
\rightarrow V_{-s_0}$ is a unitary isomorphism of $(V_0, Y_{0,0})$
modules (cf \cite[\S1.2]{MR4418715}, also \ref{statmnt:unitarity-refined-VOA} and \ref{statmnt:cntrG-mod} in this note).    
With ${\cal G} \cong \Z/2\Z$, all $(V_{s_0}, Y_{0,s_0})$ are self-contragradient;
if $(V_{s_0}, Y_{0,s_0})$ is a simple module of the VOA $(V_0, Y_{0,0})$, this means that $\lambda_{s_0}^T = \epsilon_{s_0} \lambda_{s_0}$ with $\epsilon_{s_0} \in \{ \pm \}$, and it follows that $(\Upsilon_{s_0,s_0})^2 = \epsilon_{s_0} {\rm id}_{V_{s_0}}$.
The conditions (m-o, p) remain the same as for a VOA, but we should have one word about the condition (p): eq. (\ref{eq:cond-o-4VOA-onUps}) is imposed for $a, u, v \in V= V_0 \oplus V_1$ as it is, without inserting any sign.
To be more explicit,\footnote{
\label{fn:right-mover-opp-sign-4-z/i}
We need to take extra care when $a \nin V_0$
in (\ref{eq:cond-o-4VOA-onUps}) because the conformal weights are not
necessarily integers; one specific branch $(e^{- \pi i}/x^2)^{L_0}$ is chosen 
for the expression $(-x^{-2})^{L_0}$ (as in (\ref{eq:cond-o-4VOA-onUps-phys},
\ref{eq:temp-Hrm-cnj-rltn-trnslt2-cylOp}));
when we deal with a right-mover SVOA,
the same expression $(-\bar{x}^{-2})^{\tilde{L}_0}$ is treated
as $(e^{+\pi i}/\bar{x}^2)^{\tilde{L}_0}$.
} %
(cf Rmk. \ref{rmk:treat-z})
\begin{align}
  \langle Y_{-s_{0,1}, \; s_{0,1}+s_{0,2}} (\Upsilon(a), z) \cdot u | v \rangle_{s_{0,2}}
  = \langle u | Y_{s_{0,1}, \; s_{0,2}}\left( e^{z L_1}(e^{-\pi i}/z^2)^{L_0} \cdot
  a , 1/z\right) \cdot v \rangle_{s_{0,1}+s_{0,2}}
  \label{eq:cond-o-4VOA-onUps-S}
\end{align}
for $s_{0,1}, s_{0,2} \in {\cal G}$, $v\in V_{s_{0,2}}$, $a\in V_{s_{0,1}}$,
and $u \in V_{s_{0,1}+s_{0,2}}$. 
The condition (p) only with $a \in V_0$ would leave the normalization of
$\langle -|-\rangle_{V_1}$ undetermined;
the condition (p) with $a \in V_1$ fixes the normalization of
$\langle - | - \rangle_{V_1}$ relatively to that of $V_0$, and
also the unitary isomorphism $\lambda_1$ by using $Y_{1,1}$ of the SVOA.
So, we refer to this version of the condition (p) as (p-SVOA).

The condition (q) is modified to (q-SVOA)
\begin{align}
  \Upsilon_{(-s_{0,1}-s_{0,2}),(s_{0,1}+s_{0,2})} \left( [u_{s_{0,1}}]_\nu \cdot v_{s_{0,2}}
  \right) & \; =
  (-1)^{- s_{0,1} s_{0,2}} \left[ \Upsilon_{-s_{0,1},s_{0,1}}(u_{s_{0,1}})
 \right]_\nu \cdot \Upsilon_{-s_{0,2},s_{0,2}}(v_{s_{0,2}}), \nonumber \\
  = & \; 
  c(s_{0,1}, \;  s_{0,2})^{-1} \;  \left[ \Upsilon_{-s_{0,1},s_{0,1}}(u_{s_{0,1}})
    \right]_\nu \cdot \Upsilon_{-s_{0,2},s_{0,2}}(v_{s_{0,2}}),
  \label{eq:Hconj-VOA-wOPE-S}
\end{align}
for $u_{s_{0,1}} \in V_{s_{0,1}}$ and $v_{s_{0,2}} \in V_{s_{0,2}}$.
\end{anythng}

\begin{rmk}
The choice of conditions (p-SVOA, q-SVOA) for SVOA above is equivalent to
that of Ref. \cite{MR3671191}.  Indeed, the non-trivial factor
$(-1)^{s_{0,1}s_{0,2}}$ for $\Upsilon_{-s_{0,1},s_{0,1}}$ in (q-SVOA) is gone
for a modified choice
$\Upsilon'_{1,1} := i \Upsilon_{1,1} $ ($\lambda'_1:= i \lambda_1$);
at the same time, (p-SVOA) for the modified choice $\Upsilon'$
in SVOA has $(e^{-\pi i})^{L_0 + 2 (L_0)^2}$ on the RHS instead of
$(e^{-\pi i})^{L_0}$, which is how Ref. \cite{MR3671191} formulated unitarity
structure. There are no
such things as THE right choice of conditions of (p-SVOA) or (q-SVOA)
separately, because some complex phases can be traded off from one to
the other. The example \ref{exmpl:fermion-unitarity} is compatible with
the choice of the conditions (p-SVOA, q-SVOA) above, and of course with
that of \cite{MR3671191}. 

With a view toward generalization for a class of structures containing
SVOA, where ${\cal G}$ is a finite abelian group ${\cal G}$ and $c$ has
the property (\ref{eq:cond-bimultipl}, \ref{eq:cond-skewsym-C})
(see \ref{statmnt:genG-w2cond}), it may not always be possible to
eliminate the phase factor $c$ in (q-SVOA) above by modifying
$\{ \lambda_{s_0}\}$ (or $Y_{s_{0,1}, s_{0,2}}$) at the cost of modifying (p-SVOA). 
So, there are two separate routes of generalization: whether to modify the
condition (p) or the condition (q). Which route is better is determined
ultimately by whether there are good examples, and also by whether nice
structure develops. The following discussion in section \ref{ssec:SVOA}
is written so that one can read it for a general ${\cal G}$ and $c$
in the route the condition (q) is modified by $c$ (but not the condition (p))
as we did above. 
\end{rmk}

\begin{rmk}
\label{rmk:SVOA-fermionic}
The definition of SVOA described above implies that $(V_1, Y_{0,1})$ is
a module of the underlying VOA $(V_0, Y_{0,0}, {\bf 1}, \omega, \cdots)$.
Even when an SVOA is {\it simple}, in that there is no non-trivial
$\Z/2\Z$-graded ideal of the SVOA, the $(V_0, Y_{0,0})$-module
$(V_1, Y_{0,1})$ is not necessarily simple. Even if we assume that
the $(V_0, Y_{0,0})$-module $(V_1, Y_{0,1})$ is also simple,
the vector space ${\cal V}_{1,1}^1$ may not be empty. There are such
examples of SVOA indeed. 

One may sometimes be interested in a smaller class of SVOA by assuming
that the SVOA is simple, $N_{0,1}^1 = N_{1,1}^0=1$ (so $(V_1, Y_{0,1})$ is
an irreducible module), and $N_{1,1}^1=0$, and furthermore that the fusion
of $V_1$ on a simple module $s \in {\cal I}((V_0, Y_{0,0}))$ yields
just one simple module $1 \cdot s \in {\cal I}((V_0, Y_{0,0}))$, and
the fusion of $V_1$ on $1 \cdot s$ once more yields $s$ back again.
This extra set of assumptions
captures and generalizes the properties of fermionic excitations in physics. 
In the language of VOA, a structure $(V, YY, {\bf 1}, \cdots)$ with this extra
set of assumptions is regarded as a simple-current extension of the VOA
$(V_0, Y_{0,0}, {\bf 1}, \cdots)$; cf \ref{statmnt:genG-w2cond}.
We call an SVOA with this set of extra properties {\it fermionic SVOA}
in this article.  This extra assumption is implicit in
\ref{statmnt:int.op-S-NSonly}, \ref{statmnt:int.op-S-wR} and beyond.
\end{rmk}
\begin{rmk}
The present author is tempted to impose one more condition:
invariance under the crossing symmetry.
We postpone this argument until \ref{statmnt:xing-S}. 
\end{rmk}

\begin{anythng}
An {\it $N\geq 1$ SVOA} (also referred to as {\it Neveu--Schwarz SVOA}
in some literature) consists
of a set of data $(V, YY, {\bf 1}, \omega, g)$
where the part of its data $(V, YY, {\bf 1}, \omega)$ is an SOVA and
$g \in V_{1,3/2} \subset V_1$ whose mode operators $G_\nu = [g]_\nu$ with
$\nu \in 1/2+\Z$ satisfy the following relations: 
\begin{align}
  [L_m, G_\nu] = \left( \frac{m}{2} - \nu \right) G_{m+\nu}, \qquad
  \{ G_\mu, G_\nu \} = 2L_{\mu + \nu} + \frac{c}{3}\left(\mu^2 - \frac{1}{4}
  \right) \delta_{\mu + \nu}.
\end{align}
\end{anythng}

\begin{anythng}
\label{statmnt:NS-repr}
  {\bf a representation of an SVOA:} A module of an SVOA $(V, YY, {\bf 1},
\omega, \cdots)$ consists of a set of data $(S_M, M, YY_M)$ where
$S_M$ is a set of two elements on which the group ${\cal G} = \Z/2\Z$ acts
non-trivially,
\[
f[M;0,M]: {\cal G} \times S \rightarrow S,
\]
$M \cong \oplus_{s \in S_M} M_s$ a vector space made of the direct sum of
the representation spaces of those in ${\cal I}((V_0, Y_{0,0}))$, and
$YY_M = \{ Y_{s_0, s} \; | \;
s_0 \in {\cal G}, \; s \in S_M\}$ a quartet of homomorphisms 
$V_{s_0} \rightarrow \oplus_{s \in S} {\rm Hom}(M_s, M_{f(s_0, s)}) [[z,z^{-1}]]$;
we may also write $f(s_0, s)$ as $s_0 \cdot s$ because
$f=f[M;0,M]$ is a group action on the set $S$.
The homomorphism $YY_M$ is subject to the super-Jacobi identity
($\ell$-SVOA-mod) 
\begin{align}
  \frac{1}{z_0} \delta & \; \left( \frac{z_1-z_2}{z_0} \right) YY_M(u,z_1)
  \cdot YY_M(v, z_2) - \frac{c( s_{0,1}, s_{0,2})}{z_0} \delta \left(
  \frac{z_2-z_1}{-z_0} \right) YY_M(v,z_2) \cdot YY_M(u, z_1)  \nonumber \\
  & \; = \frac{1}{z_2} \delta \left( \frac{z_1-z_0}{z_2} \right) YY_M\left(
  YY(u,z_0) \cdot v, z_2 \right) , \qquad
  u \in V_{s_{0,1}}, \quad v \in V_{s_{0,2}}.
  \label{eq:Jacobi-module-S}
\end{align}
For each of $s \in S_M$, $(M_s, YY_{0,s})$ is a module of the VOA $V_0$;
$[h_{f(1,s)}]  = [h_s] + 1/2$ (mod $\Z$). So, $M_s$ and $M_{f(1,s)}$ are
different modules of the VOA $(V_0,Y_{0,0})$. 

Here, we do not talk about twisted representations
until \ref{statmnt:Ramond-repr}.
The triple $({\cal G}, V, YY)$ is a module of the SVOA $(V, YY, {\bf 1},
\cdots)$, where $f[0;0,0]: {\cal G} \times {\cal G} \rightarrow {\cal G}$
is the group law of ${\cal G} \cong \Z/2\Z$. We call it the {\it defining
representation} of the SVOA. 
\end{anythng}

\begin{anythng}
\label{statmnt:int.op-S-NSonly}
{\bf intertwining operators of an SVOA:}
Let ${\cal II}$ be a set of simple representations\footnote{
For an SVOA module of the form introduced above, it is always
possible to decompose each of $M_s$ ($s\in S_M)$ into simple modules
of the VOA $(V_0, Y_{0,0})$. When a module $(M, YY_M)$ has a decomposition
into a direct sum of modules $(N_1, YY_{N_1})$ and $(N_2, YY_{N_2})$, it does
so also as a $\Z/2\Z$-graded module. 
} %
  of an SVOA
$(V, YY, \cdots )$ (the notation ${\cal II}^{\hat{0}}$ may also be used to
emphasize that this is a set of untwisted ones). Before we review the notion of intertwining operators of
the SVOA of type $(\alpha;\beta,\gamma)$, where
$\alpha, \beta,\gamma \in {\cal I}{\cal I}$, we need a preparation. 

Suppose, by extending the condition for a fermionic SVOA in
Rmk. \ref{rmk:SVOA-fermionic}, that the fusion of $V_1$ acts as permutation
on each of the sets $S_\alpha$, $S_\beta$ and $S_\gamma$.
When ${\cal V}^{\alpha s_\alpha}_{\beta s_\beta, \gamma s_\gamma}$ of the
underlying VOA $(V_0, Y_{0,0},\cdots)$ is not empty for ${}^\exists s_\alpha
\in S_\alpha$, there is a set-theoretical map\footnote{
More generally, in the context of \ref{statmnt:genG-w2cond} beyond SVOA,
we should replace a map by a correspondence $f \subset (S_\beta \times S_\gamma)
\times S_\alpha$ if the action of ${\cal G}$ on $S_\alpha$, $S_\beta$ or $S_\gamma$
is not faithful; the compatibility with the ${\cal G}$ action is of
the form $(s_0 \cdot s_\alpha) \in f(s_0 \cdot s_\beta,s_\gamma)$ and
$(s_0 \cdot s_\alpha) \in f(s_\beta, s_0 \cdot s_\gamma)$ if $s_\alpha \in
f(s_\beta,s_\gamma)$. 
} %
\begin{align}
 f[\alpha;\beta,\gamma]:  S_\beta \times S_\gamma \rightarrow S_\alpha 
\end{align}
that is compatible with the action of ${\cal G}$, 
\begin{align}
  f[\alpha;\beta,\gamma](s_0 \cdot s_\beta, s_\gamma) = s_0 \cdot f[\alpha;\beta,\gamma](s_\beta, s_\gamma) = f[\alpha; \beta,\gamma](s_\beta, s_0 \cdot s_\gamma)
\end{align}
for any $s_\beta \in S_\beta$, $s_\gamma \in S_\gamma$ and $s_0 \in S_0 =
{\cal G}$, so that $N_{\beta s_\beta, \gamma s_\gamma}^{\alpha f(s_\beta, s_\gamma)} > 0$.
Reference to the type $[\alpha;\beta,\gamma]$ may be omitted ($f$ instead of
$f[\alpha;\beta,\gamma]$) when
the relevant type is obvious from the context (or unimportant).

We define a vector space of {\it intertwining operators} of the SVOA of
type $(\alpha;\beta,\gamma)$ as follows, by choosing one
set-theoretical identification $i_\beta: S_\beta \cong \Z/2\Z$ and
fixing the choice, if some of
${\cal V}_{\beta s_\beta, \gamma s_\gamma}^{\alpha s_\alpha}$ are non-empty.  
It consists of a quartet ${\cal YY} = \{
{\cal Y}_{s_\beta,s_\gamma} \} \in \oplus_{s_\beta, s_\gamma}
{\cal V}_{\beta s_\beta, \gamma s_\gamma}^{\alpha f(s_\beta,s_\gamma)}$, where
\begin{align}
 {\cal Y}_{s_\beta,s_\gamma}: M_{\beta s_\beta} \rightarrow
{\rm Hom}(M_{\gamma s_\gamma}, M_{\alpha f(s_\beta,s_\gamma)})[[z,z^{-1}]]
z^{-(h_{\beta s_\beta}+h_{\gamma s_\gamma}-h_{\alpha f(s_\beta, s_\gamma)})}
\label{eq:int.op-S-formal-expr}
\end{align}
with a common non-integral part of the power
$h_{\beta s_\beta} + h_{\gamma s_\gamma} -
h_{\alpha f(s_\beta,s_\gamma)} +\Z \in \R/\Z$, satisfying
\begin{align}
 & \frac{1}{z_0} \delta \left( \frac{z_1-z_2}{z_0} \right)
  Y_{s_0, \; \; \alpha f(s_\beta,s_\gamma)}(u, z_1) \cdot
  {\cal Y}_{s_\beta, \;\; s_\gamma}(m_\beta, z_2) \cdot m_\gamma \nonumber \\
 & -
  \frac{c(s_0, i_\beta(s_\beta))}{z_0} 
  \delta \left( \frac{z_2-z_1}{-z_0}
  \right) {\cal Y}_{s_\beta, \;\; s_0 \cdot s_\gamma}(m_\beta, z_2) \cdot
  Y_{s_0, \;\; \gamma s_\gamma}(u, z_1)  \cdot m_\gamma \label{eq:def-vct-sp-int.op-S} \\
  & = \frac{1}{z_2} \delta \left( \frac{z_1-z_0}{z_2} \right)
  {\cal Y}_{s_0 \cdot s_\beta, \;\; s_\gamma}\left( Y_{s_0, \;\; \beta s_\beta}(u, z_0)
  \cdot m_\beta, z_2 \right) \cdot m_\gamma \nonumber 
\end{align}
for any $u \in V_{s_0}$, $m_\beta \in M_{\beta s_\beta}$ and $m_\gamma \in M_{\gamma s_\gamma}$; here, $s_0 \in S_0 = {\cal G}$, $s_\beta \in S_\beta$ and
$s_\gamma \in S_\gamma$. The vector space of such ${\cal YY}$ is denoted
by ${\cal VV}_{\beta,\gamma}^\alpha$. This is a way to keep track of choices
of intertwining operators ${\cal Y}_{s_\beta,s_\gamma}$ of the underlying
VOA $(V_0, Y_{0,0})$ that are related
(together with the representations $YY_M= \{ Y_{s_{0,1},s_M} \}$)
by the fusion/braiding isomorphisms. We could introduce a term such as
intertwining {\it super}operator, or {\it supermultiplet} of intertwining
operators, we will just use the plain word intertwining operators (of SVOA)
in this article, and the distinction is made by notations
(${\cal YY}$ vs ${\cal Y}_{s_\beta,\;s_\gamma}$, ${\cal VV}_{\beta, \gamma}^\alpha$ vs
${\cal V}_{\beta s_\beta, \gamma s_\gamma}^{\alpha s_\alpha}$).
(cf \cite{MR1233387})

Here is a physics reasoning behind setting the definition as above;
implicit assumptions introduced in the name of intuitions are made explicit
in \ref{statmnt:genG-w2cond}, as we arrive at the same conclusion in a VOA
language there by setting ${\cal G} \cong \Z/2\Z$. 
The RHS of this super-Jacobi identity remains the same as ($\ell$-int.op)
for a VOA, because nothing should happen (in the physics intuition based on
radial quantization) so long as we insert the state $u$---either bosonic
(in $V_0$) or fermionic (in $V_1$) away from the insertion point of the
state $m_\beta$ (i.e., $|z_1| > |z_2|$) no matter how small $z_0$ is. 
When we bring the insertion point of the state $u$ closer to the origin $z=0$
of the radial quantization past the insertion point of the state $m_\beta$,
however,
there must be some complex phase (${\rm phase}_{s_0,s_\beta}$) associated with
the statistics of the particles $u \in V_{s_0}$ and $m_\beta \in M_{\beta s_\beta}$.
Although we have not specified the statistics of $m_\beta$, we should have
${\rm phase}_{0,s_\beta}=1$, and $({\rm phase}_{1,s_\beta})^2=1$, because two
fermions combined---think of $Y(u_1,z_1)Y(u_2,z'_1)$ with $u_1,u_2 \in V_1$
passing ${\cal Y}(m_\beta,z_2)$---should behave as a boson. Furthermore,
${\rm phase}_{1,1 \cdot s_\beta} = - {\rm phase}_{1,s_\beta}$ for any $s_\beta
\in S_\beta$, as we may think of the statistical phase when 
$Y(u_1, z_1)$ passes $Y(u_2,z'_1) {\cal Y}(m_\beta,z_2)$. 
It thus follows that $\{ {\rm phase}_{1,s_\beta} \; | \; s_\beta \in S_\beta\}
= \{ \pm 1\}$. 

In (\ref{eq:def-vct-sp-int.op-S}), we have made an artificial choice of
the set-theoretical map $\{ {\rm phase}_{1,s_\beta} \} \cong \{ \pm 1\}$
and $i_\beta: S_\beta \cong \Z/2\Z$, by setting
${\rm phase}_{s_0,s_\beta} = c(s_0, i_\beta(s_\beta)) = (-1)^{+ s_0 i_\beta (s_\beta)}$.
The vector space
${\cal VV}_{\beta,\gamma}^\alpha$ is defined with an artificial(arbitrary)
choice of $i_\beta$. For the two choices, $i_\beta, i'_\beta:
S_\beta \cong \Z/2\Z$ such that $i'_\beta(s_\beta) = i_\beta(s_\beta) + \delta s_0$
for some $\delta s_0 \in \Z/2\Z$,
however, there are isomorphisms; for example, for
any choice of $i_\gamma: S_\gamma \cong \Z/2\Z$, 
\begin{align}
  {\cal VV}_{\beta,\gamma}^\alpha(i_\beta) \ni {\cal YY}=
  & \; \{ {\cal Y}_{s_\beta,s_\gamma} \}
  \longmapsto   \label{eq:temp-iso-vvct-sp-int.op-S-diff-StoZ2} \\
  & \; \{ c(i_\gamma(s_\gamma), \delta s_0)^{-1}
 {\cal Y}_{s_\beta,s_\gamma} =
  c(\delta s_0, i_\gamma(s_\gamma)) 
  {\cal Y}_{s_\beta, s_\gamma} \}
  \in {\cal VV}_{\beta,\gamma}^\alpha(i'_\beta).
   \nonumber 
\end{align}
So an arbitrary choice of $i_\beta: S_\beta \cong Z/2\Z$ is a necessary evil
in defining ${\cal VV}_{\beta,\gamma}^\alpha$, auxiliary in nature, but that
is not terribly toxic.
We will not retain explicit reference to the choice of
$i_\beta: S_\beta \cong \Z/2\Z$ in the notation ${\cal VV}_{\beta,\gamma}^\alpha$
in the following when a choice does not matter. 

The isomorphism (\ref{eq:temp-iso-vvct-sp-int.op-S-diff-StoZ2}) is not
meant as something canonical (authentic, in a more colloquial term)
in the context of SVOA; not only do we have two choices of
$i_\gamma:S_\gamma \cong \Z/2\Z$, but any complex multiple of the isomorphism
(\ref{eq:temp-iso-vvct-sp-int.op-S-diff-StoZ2}) is also an isomorphism
between the two vector spaces ${\cal VV}_{\beta,\gamma}^\alpha(i_\beta)$
and ${\cal VV}_{\beta,\gamma}^\alpha(i'_\beta)$. That does not pose any problem
in the context of SVOA, because we only deal with the set of
whole solutions to the super-Jacobi identity (\ref{eq:def-vct-sp-int.op-S}),
without picking one particular element from
${\cal VV}_{\beta,\gamma}^\alpha(i_\beta)$ or asking which one in
${\cal VV}_{\beta,\gamma}^\alpha(i'_\beta)$ {\it corresponds} to
a given element in ${\cal VV}_{\beta,\gamma}^{\alpha}(i_\beta)$. 
As we move to the context of fermionic 2D CFT (or 2D SCFT) in
section \ref{ssec:fCFT}, however, we will choose a CFT vertex operator,
just like we did in section \ref{sec:CFT} by imposing the condition (t). 
We will see at the end of \ref{statmnt:data-4-fCFT-2} that the
arbitrariness in the choice of $i_\gamma$ for the
isomorphism (\ref{eq:temp-iso-vvct-sp-int.op-S-diff-StoZ2})
cancels in that context, and will have a way to compare descriptions that
use different set-theoretical identifications $i_\beta$ and $i'_\beta$. 
\end{anythng}

\begin{rmk}
\label{rmk:no-systematic-Z2}  
For a given SVOA and a set ${\cal II}$ of its irreducible distinct
representations, it is not guaranteed whether there is a set of
set-theoretical identifications $\{ i_\alpha: S_\alpha \cong Z/2\Z \; |
\; \alpha \in {\cal II}\}$ so that all the set-theoretical maps
$i_\alpha \cdot f[\alpha;\beta,\gamma] \cdot
(i_\beta^{-1} \times i_\gamma^{-1}): Z/2\Z \times \Z/2\Z \rightarrow \Z/2\Z$
agree with the group law (ie., $i_\alpha \cdot f[\alpha;\beta,\gamma] =
i_\beta + i_\gamma$ as maps $S_\beta \times S_\gamma \rightarrow \Z/2\Z$)
for all the triples $(\alpha;\beta,\gamma)$
whose $f[\alpha;\beta,\gamma]$ exist. Indeed, given a fusion rule,
the number of group-law tests to pass, which is the number of triples
$(\alpha;\beta,\gamma)$ with positive fusion constants, tends
to be more than the number of freedoms (choices of $i_\alpha$'s)
we can exploit. 

If one wants to use an SVOA for a Type II CFT (gauging of a chiral fermion
parity rather than that of a diagonal fermion parity for Type 0 CFT), it is
a necessary condition that there exists such a systematic choice of
$\{ i_\beta \; | \beta \in {\cal II}^{\hat{0}}((V,Y))\}$. 
In the case of an SVOA that has a U(1) symmetry, with all the states
in the representations in ${\cal II}^{\hat{0}}$ carrying integral U(1) charges, 
there is a chance, for example, that the even/odd-ness of the charges
can be used in setting the set-theoretical identification of $S_\alpha$
with $\Z/2\Z$. 
On the other hand, in the case of level-$k$ $N=2$ minimal model SVOA
(Virasoro central charge $c=3k/(k+2)$), the U(1) charges of the states of
non-trivial representations are not necessarily integral, so this argument
is not applicable. Indeed, in the case $k$ is even,\footnote{
The author thanks K. Nishikawa for discussion. 
} %
there is no set of
identifications
$\{ i_\alpha: S_\alpha \cong \Z/2\Z \}$
such that $i_\alpha \cdot f[\alpha;\beta,\gamma] = i_\beta + i_\gamma$
for all the triples with non-zero fusion constants. 
\end{rmk}

\begin{anythng}
\label{statmnt:skew-iso-4-NS}  
{\bf the skew-symmetry isomorphisms:} We set the SVOA version of the
skews-symmetry isomorphisms $\Omega_r$ ($r \in \Z$) as follows (cf.
\cite{MR1841356}, \cite[(13)]{bakalov2006generalizedvertexalgebras})\footnote{
Note that $c$ vs $c^{-1}$ matters for a general ${\cal G}$. 
}):
\begin{align}
  {\cal VV}_{\beta,\gamma}^\alpha \ni {\cal YY} = \{ {\cal Y}_{s_\beta, s_\gamma} \}
  \longmapsto
  \{ c(i_\beta(s_\beta), i_\gamma(s_\gamma))^{-1} \;\; 
  \Omega_r {\cal Y}_{s_\beta,s_\gamma} \}
  =: \Omega_r{\cal YY} 
  \in {\cal VV}_{\gamma,\beta}^\alpha.
  \label{eq:def-skew-iso-S}  
\end{align}
We will use the same notation $\Omega_r: {\cal VV}_{\beta,\gamma}^\alpha
\rightarrow {\cal VV}_{\gamma,\beta}^\alpha$ for this. Although it is obvious
that this
assignment maps a quartet ${\cal YY} = \{ {\cal Y}_{s_\beta,s_\gamma} \}$ into
$\oplus_{s_\gamma,s_\beta} {\cal V}_{\gamma s_\gamma,\beta s_\beta}^{\alpha f(s_\beta,s_\gamma)}$,
we still need to prove that this $\Omega_r{\cal YY}$ satisfies the defining
condition (\ref{eq:def-vct-sp-int.op-S}) with $s_0 \neq 0 \in {\cal G}$.
The proof is just to repeat the argument in Lemma \ref{lemma:skew-sym-iso}
with sign factors inserted; the three terms
in (\ref{eq:temp-skew-iso-inttw-prf}) are multiplied by 
\begin{align*}
  & \; c(i_\beta(s_\beta), \; i_\gamma(s_\gamma))^{-1},  \\
  c(s_0, i_\gamma(s_\gamma)) \cdot c(i_\beta(s_0 \cdot s_\beta), \;
  i_\gamma(s_\gamma))^{-1} = & \; c(i_\beta(s_\beta), i_\gamma(s_\gamma))^{-1}, \\
  c(i_\beta(s_\beta), i_\gamma(s_0 \cdot s_\gamma))^{-1}  = & \; 
  c(i_\beta(s_\beta), i_\gamma(s_\gamma))^{-1} \cdot c(s_0, i_\beta(s_\beta)) 
\end{align*}
as desired.

Although the isomorphism $\Omega_r:{\cal VV}_{\beta,\gamma}^\alpha \rightarrow
{\cal VV}_{\gamma,\beta}^\alpha$ above depends on an artificial choice of
the set-theoretical identification $i_\beta: S_\beta \cong \Z/2\Z$
and $i_\gamma: S_\gamma \cong \Z/2\Z$, that is not an issue, because
both ${\cal VV}_{\beta,\gamma}^\alpha$ and ${\cal VV}_{\gamma,\beta}^\alpha$
are defined with a choice of $i_\beta$ and $i_\gamma$, respectively. 
One will note further that $\Omega_r:
{\cal VV}_{\beta,\gamma}^\alpha(i_\beta) \rightarrow
{\cal VV}_{\gamma,\beta}^\alpha(i_\gamma)$ is the same as the composition
of the isomorphism ${\cal VV}_{\beta,\gamma}^\alpha(i_\beta) \rightarrow
{\cal VV}_{\beta,\gamma}^\alpha(i'_\beta)$
in (\ref{eq:temp-iso-vvct-sp-int.op-S-diff-StoZ2}) followed by
$\Omega_r: {\cal VV}_{\beta,\gamma}^\alpha(i'_\beta) \rightarrow
{\cal VV}_{\gamma,\beta}^\alpha(i_\gamma)$.  

The skew-symmetry condition (i-SVOA) on the data $YY$ of an SVOA is also
expressed as invariance under the isomorphism,
$\Omega(YY) = YY$ for $YY \in {\cal VV}_{0,0}^0$. 
\end{anythng}

\begin{anythng}
  \label{statmnt:BnF-S}
{\bf the braiding and fusion isomorphisms :} The braiding isomorphisms
${\cal B}_{-1-r}: {\cal VV}_{(\beta_1 \cdot (\beta_2 \cdot\gamma))}^\alpha
\rightarrow {\cal VV}_{(\beta_2 \cdot (\beta_1 \cdot \gamma))}^\alpha$ for
$\alpha,\beta_1,\beta_2,\gamma \in {\cal II}$, with $r \in \Z$, are
characterized as follows: the underlying VOA $(V_0, Y_{0,0}, \cdots)$
already determines the braiding isomorphisms ${\cal B}_{-1-r}$ of the
intertwining operators of the VOA, but we need to keep track of correlated
choices---due to (\ref{eq:def-vct-sp-int.op-S})---of the eight components 
of $({\cal YY}\otimes {\cal YY}) = \{
{\cal Y}_{\beta_1 s_{\beta 1}, \delta f(s_{\beta 2},s_\gamma)}^{\alpha f(s_{\beta 1}, f(s_{\beta 2}, s_\gamma))} \otimes
{\cal Y}_{\beta_2 s_{\beta 2}, \gamma s_\gamma}^{\delta f(s_{\beta 2},s_\gamma)} |
s_{\beta_1} \in S_{\beta_1}, \; s_{\beta_2} \in S_{\beta_2}, \; s_\gamma \in
S_\gamma \}$ in
${\cal VV}_{(\beta_1\cdot (\beta_2 \cdot \gamma))}^\alpha := \oplus_\delta
({\cal VV}_{\beta_1, \delta}^\alpha \otimes {\cal VV}_{\beta_2,\gamma}^\delta)$.
By introducing a basis $\{ ({\cal YY}_{i1} \otimes {\cal YY}_{i2}) \}$
and $\{ ({\cal YY}_{j1} \otimes {\cal YY}_{j2}) \}$ of the vector spaces 
${\cal VV}_{(\beta_1 \cdot (\beta_2 \cdot \gamma))}^\alpha$ and
${\cal VV}_{(\beta_2 \cdot (\beta_1 \cdot \gamma))}^\alpha$, we characterize
the isomorphism ${\cal B}_{-1-r}$ in the matrix representation
by the relation
\begin{align}
  & \langle m'_{\alpha',f(s_{\beta 1}, f(s_{\beta 2}, s_\gamma))} , \;
  {\cal Y}_{s_{\beta 1}, \; \; f(s_{\beta 2},s_\gamma) ; i1}(m_{\beta_1},z_1) \cdot 
  {\cal Y}_{s_{\beta 2}, \;\; s_\gamma; \; i2}(m_{\beta_2}, z_2) \cdot m_\gamma \rangle
  |_{z_1-z_2\rightarrow e^{\pi i (2r+1)}(z_2-z_1)} \nonumber \\
  & \quad =  \sum_{b}  
          \langle m'_{\alpha',f(s_{\beta 2}, f(s_{\beta 1}, s_\gamma))} , \;
          {\cal Y}_{s_{\beta 2}, \;\; f(s_{\beta 1},s_\gamma); \; j1}(m_{\beta_2},z_2) \cdot 
          {\cal Y}_{s_{\beta 1}, \;\; s_\gamma; j2}(m_{\beta_1}, z_1) \cdot m_\gamma
          \rangle \nonumber \\
  &  \qquad \qquad\qquad \qquad\qquad \qquad
          c(i_{\beta_1}(s_{\beta_1}), \; i_{\beta_2}(s_{\beta_2})) \;\; 
          ({\cal B}_{-1-r})^j_{\;\; i} ,
   \label{eq:chartz-Biso-SVOA}       
\end{align}
where $m_{\beta_1} \in M_{\beta_1, s_{\beta_1}}$, $m_{\beta_2} \in M_{\beta_2, s_{\beta_2}}$,
$m_\gamma \in M_{\gamma,s_\gamma}$. 

Because the characterization above involves an artificial choice of the
identifications $i_{\beta_1}: S_{\beta_1} \cong \Z/2\Z$ and
$i_{\beta_2}: S_{\beta_2} \cong \Z/2\Z$, we need to state in which sense
the isomorphism ${\cal B}_{-1-r}$ has been determined. We make it a rule
to use the isomorphism
\begin{align}
  \oplus_\delta \left( {\cal VV}_{\rho,\delta}^\alpha \otimes
        {\cal VV}_{\sigma,\gamma}^\delta(i_\sigma \right) 
  & \; \ni \{ {\cal Y}_{s_\rho,f(s_\sigma,s_\gamma)} \otimes {\cal Y}_{s_\sigma, s_\gamma} \} \\
        \longmapsto & \; \{ {\cal Y}_{s_\rho, f(s_\sigma,s_\gamma)} \otimes {\cal Y}_{s_\sigma, s_\gamma} \;\; 
      c(i_\gamma(s_\gamma), \; \delta s_0)^{-1}
        \} \in \oplus_\delta \left(
          {\cal VV}_{\rho,\delta}^\alpha \otimes {\cal VV}_{\sigma,\gamma}^\delta(i'_\sigma) \right), \nonumber \\
  \oplus_\delta \left( {\cal VV}_{\rho,\delta}^\alpha(i_\rho) \otimes {\cal VV}_{\sigma,\gamma}^\delta \right) 
  & \; \ni \{ {\cal Y}_{s_\rho,f(s_\sigma,s_\gamma)} \otimes {\cal Y}_{s_\sigma, s_\gamma} \} \\
  \longmapsto \{ {\cal Y}_{s_\rho, f(s_\sigma,s_\gamma)} & \; \otimes {\cal Y}_{s_\sigma, s_\gamma}  \;\; 
 c(i_\sigma(s_\sigma)+i_\gamma(s_\gamma), \; \delta s_0)^{-1}
  \} \in \oplus_\delta \left(
          {\cal VV}_{\rho,\delta}^\alpha(i'_\rho) \otimes {\cal VV}_{\sigma,\gamma}^\delta \right), \nonumber 
\end{align}
when $i'_\sigma= i_\sigma + \delta s_0$ or
$i'_\rho = i_\rho + \delta s_0$ for some $\delta s_0 \in \Z/2\Z$.
Then 
\begin{equation}
    \xymatrix{
   \oplus_\delta
     ({\cal VV}_{\beta_1,\delta}^\alpha \otimes
      {\cal VV}_{\beta_2,\gamma}^\delta(i_{\beta_2})) \ar[r] \ar[d]^{{\cal B}_{-1-r}(i_{\beta_1}, i_{\beta_2})} & 
  \oplus_\delta ({\cal VV}_{\beta_1,\delta}^\alpha \otimes
     {\cal VV}_{\beta_2,\gamma}^\delta(i'_{\beta_2})) \ar[d]^{{\cal B}_{-1-r}(i_{\beta_1},i'_{\beta_2})}  \\
  \oplus_\epsilon ({\cal VV}_{\beta_2,\epsilon}^\alpha(i_{\beta_2}) \otimes {\cal VV}_{\beta_1,\gamma}^\epsilon) \ar[r] 
  &  \oplus_\epsilon ({\cal VV}_{\beta_2,\epsilon}^\alpha(i'_{\beta_2}) \otimes {\cal VV}_{\beta_1,\gamma}^\epsilon)
  }
\end{equation}
and
\begin{equation}
    \xymatrix{
   \oplus_\delta
     ({\cal VV}_{\beta_1,\delta}^\alpha(i_{\beta_1}) \otimes
      {\cal VV}_{\beta_2,\gamma}^\delta) \ar[r] \ar[d]^{{\cal B}_{-1-r}(i_{\beta_1}, i_{\beta_2})} & 
  \oplus_\delta ({\cal VV}_{\beta_1,\delta}^\alpha(i'_{\beta_1}) \otimes
     {\cal VV}_{\beta_2,\gamma}^\delta) \ar[d]^{{\cal B}_{-1-r}(i'_{\beta_1},i_{\beta_2})}  \\
  \oplus_\epsilon ({\cal VV}_{\beta_2,\epsilon}^\alpha \otimes {\cal VV}_{\beta_1,\gamma}^\epsilon(i_{\beta_1})) \ar[r] 
  &  \oplus_\epsilon ({\cal VV}_{\beta_2,\epsilon}^\alpha \otimes {\cal VV}_{\beta_1,\gamma}^\epsilon(i'_{\beta_1}))
  }
\end{equation}
both commute. 

The fusion isomorphism ${\cal F}: {\cal VV}_{(\beta_1 \cdot (\beta_2 \cdot \gamma))}^\alpha \rightarrow {\cal VV}_{((\beta_1 \cdot \beta_2) \cdot \gamma)}^\alpha$ for $\alpha,\beta_1,\beta_2,\gamma \in {\cal II}$ also has a similar characterization.
It is best to use $\oplus_\epsilon ({\cal VV}_{\beta_1,\beta_2}^\epsilon(i_{\beta_1})
\otimes {\cal V}_{\epsilon,\gamma}^\alpha(i_\epsilon))$ with
$i_\epsilon: S_\epsilon\cong \Z/2\Z$ so that $i_\epsilon \cdot
f[\epsilon;\beta_1,\beta_2]=i_{\beta_1}+i_{\beta_2}$ for that purpose
(as we see below). With this choice of $i_\epsilon$, and a choice of a basis
$\{ ({\cal YY}_{k2} \otimes {\cal YY}_{k1}) \}$ of $\oplus_{\epsilon}
\left({\cal VV}_{\beta_1,\beta_2}^\epsilon(i_{\beta_1}) \otimes
{\cal VV}_{\epsilon, \gamma}^\alpha(i_\epsilon)\right)$, the matrix representation
of ${\cal F}$ is characterized by the relation 
\begin{align}
   & \langle m'_{\alpha',f(s_{\beta 1}, f(s_{\beta 2}, s_\gamma))} , \;
   {\cal Y}_{s_{\beta 1}, \;\; f(s_{\beta 2},s_\gamma); \; i1}(m_{\beta_1},z_1) \cdot 
   {\cal Y}_{s_{\beta 2}, \;\; s_\gamma; \; i2}(m_{\beta_2}, z_2) \cdot m_\gamma
   \rangle|_{z_1 \rightarrow z_0+z_2}  \label{eq:chartz-Fiso-SVOA} \\
   & \; \quad = \sum_k 
   \langle m'_{\Upsilon^{\cal I}(\alpha, f(f(s_{\beta 1},s_{\beta 2}),s_\gamma))}, \;
   {\cal Y}_{f(s_{\beta 1}, s_{\beta 2}), \;\; s_\gamma; k1}\left( 
   {\cal Y}_{s_{\beta 1}, \;\; s_{\beta 2}; k2}(m_{\beta_1}, z_0) \cdot m_{\beta_2}, z_2
   \right)     \cdot m_\gamma \rangle \; ({\cal F})^k_{\;\; i} \nonumber
\end{align}
for all the eight components with $s_{\beta_1} \in S_{\beta_1}$, $s_{\beta_2}
\in S_{\beta_2}$ and $s_\gamma \in S_\gamma$. The relation does not look
as nice as the one above if we are to use $i_\epsilon:S_\epsilon \cong \Z/2\Z$
other than the one determined from $i_{\beta_1}$ and $i_{\beta_2}$. 
\end{anythng}

\begin{anythng}
\label{statmnt:Ramond-repr}
{\bf Ramond ($\hat{g}$-twisted) representations:}
As a reminder, $\hat{g}$ is an automorphism of an SVOA $(V, YY, \cdots)$,
reviewed already at the end of \ref{statmnt:def-alg-data-SVOA}.
To refer to a general element of the automorphism group $\hat{\cal G}
\cong \{ \hat{0}, \hat{g}\}$,
we will use $\hat{h}$. The representations that are
$\hat{g}$-twisted are called {\it Ramond} representations in physics
literature; {\it Neveu--Schwarz} (NS) representations in physics
literature are untwisted representations of an SVOA (as reviewed
in \ref{statmnt:NS-repr}). 

We restrict our attention in this article to a class of $\hat{g}$-twisted
modules introduced as follows. It consists of $(S_M, M, YY_M)$ where $S_M$
is a set of two elements that has a non-trivial action of
${\cal G} = S_0 = \Z/2\Z$, $M = \oplus_{s \in S_M} M_s$ is a vector space
made out as the direct sum of the representation spaces of those in
${\cal I}((V_0, Y_{0,0}))$, and
the state-operator correspondence $Y_M = \{ Y_{s_0, Ms} \; | \;
s_0 \in {\cal G}, \; s \in S_M \}$ is a quartet of homomorphisms
$V_{s_0} \rightarrow {\rm Hom}(M_s, \; M_{s_0 \cdot s})[[z,z^{-1}]] z^{-s_0/2}$.
The shift $-s_0/2$ in the power of the formal power series
may well be regarded as $z^{-\hat{g}(s_0)}$, with
$\hat{g}(s_0) \in 2^{-1}\Z/\Z \subset \R/\Z$.
The mode expansion of the field operator is of this
form---in physics literature---for a homogeneous state $u \in V_{s_0,h}$:
\begin{align}
  YY_M(u,z) = \sum_{\nu \in -h + \hat{g}(s_0)+\Z} [u]_\nu z^{-h-\nu},  
\end{align}
so $(h+\nu) \in \hat{g}(s_0) \in \R/\Z$.

The super-Jacobi identity is imposed on $YY_M$ for $\hat{g}$-twisted
representations in a form a little more general than
(\ref{eq:Jacobi-module-S}) (cf \cite{MR1233387}):
\begin{align}
 & \frac{1}{z_0} \delta \left( \frac{z_1-z_2}{z_0} \right)
  YY_M(u_1, z_1) \cdot
  YY_M (u_2, z_2) \cdot m \nonumber \\
 & - \frac{c(s_1,s_2)}{z_0} \delta \left( \frac{z_2-z_1}{-z_0}\right)
  YY_M(u_2,z_2) \cdot YY_M(u_1,z_1) \cdot m  \label{eq:Jacobi-module-S-wR} \\
 & \quad  = \frac{1}{z_2} \delta \left( \frac{z_1-z_0}{z_2}\right) YY_M\left(
  YY(u_1,z_0) \cdot u_2, z_2\right)\cdot \left( \frac{z_1-z_0}{z_2}\right)^{-\hat{h}(s_1)} m, \nonumber 
\end{align}
for $u_1 \in V_{s_1}$, $u_2 \in V_{s_2}$ with $s_1, s_2 \in S_0 = {\cal G}
= \Z/2\Z$, and $\hat{h} \in {\rm Hom}({\cal G}, \R/\Z)$, and $m$ in an 
$\hat{h}$-twisted representation of the SVOA. All the three terms have
$z_0^\Z z_1^{-\hat{h}(s_1)} z_2^{-\hat{h}(s_2)}$.
Each of $M_s$ ($s \in S_M$) is a module of the underlying VOA $(V_0, Y_{0,0})$. 

It is possible, in fact, to think of a broader class of $\hat{g}$-twisted
representations than we do above; it is not necessary as a representation
of an SVOA to have the structure $M = \oplus_{s \in S_M} M_s$ 
with a non-trivial action of ${\cal G}$ on the set $S_M$ of \underline{two}
elements.
A crucial difference between NS modules and Ramond modules
is that the conformal weights of $[u]_\nu \cdot m$ of a homogeneous state
$m \in M$ and $u \in V_1$ differ from that of $m$ by $+1/2 +\Z$ in an
NS module, while the difference is in $+\Z$ in a Ramond
module. So, an untwisted module $M$ of an SVOA should
contain at least two irreducible modules of the underlying VOA $(V_0, Y_{0,0})$,
while one cannot make this argument for a $\hat{g}$-twisted module
(or more generally, for an $\hat{h}$-twisted sector where
$(\hat{\phi}-\hat{h}): {\cal G} \rightarrow \R/\Z$ has a non-trivial
kernel). 
We restrict our attention in this article, however, only to those with
a two-element ${\cal G}$-set $S_M$ (faithful action of ${\cal G}$ on $S_M$).
One needs to find non-trivial generalizations to upgrade the following
arguments to be able to cover the case where the action of ${\cal G}$ on $S_M$ is not faithful.\footnote{
There are simple current extensions where there are fixed points in
${\cal I}((V_0, Y_{0,0}))$ under the action of ${\cal G}$
(e.g., \cite[\S6]{Schellekens:1990xy},
\cite{Fuchs:1996dd}, the VOA $(\mathfrak{su}(2),k)$ with $4|k$);
cf also \cite{MR1935493}, \cite{MR2038578}, for example.
} %
\end{anythng}

\begin{anythng}
\label{statmnt:int.op-S-wR}
  {\bf intertwining operators of states in a $\hat{g}$-twisted module:}
  For a given SVOA $(V,YY,{\bf 1},\omega,\cdots)$, let
  ${\cal II}={\cal II}^{\hat{0}}$
[resp. ${\cal II}^{\hat{g}}$] be a set of distinct simple untwisted
[resp. $\hat{g}$-twisted] modules of the SVOA. Let us formulate the notion
of intertwining operators of the SVOA for a triple $(\alpha;\beta,\gamma)$
where $\beta \in {\cal II}^{\hat{h}_\beta}$, $\gamma \in {\cal II}^{\hat{h}_\gamma}$
and $\alpha \in {\cal II}^{\hat{h}_\beta+\hat{h}_\gamma}$; here
$\hat{h}_\beta, \hat{h}_\gamma \in \hat{\cal G} = \{ \hat{0}, \hat{g} \} \cong
{\rm Hom}({\cal G}, \R/\Z)$. 

The argument at the beginning of \ref{statmnt:int.op-S-NSonly} is applicable
regardless of
whether the representations $\beta$, $\gamma$ are untwisted or twisted.
So there is a set-theoretical map $f[\alpha;\beta,\gamma]: S_\beta \times
S_\gamma \rightarrow S_\alpha$ for irreducible modules $\alpha$, $\beta$,
$\gamma$, compatible with the action of ${\cal G} = S_0 \cong \Z/2\Z$,
when there are non-trivial intertwining operators of the underlying
VOA $(V_0, Y_{0,0})$. 

An intertwining operator ${\cal YY} \in {\cal VV}_{\beta,\gamma}^\alpha$
consists of a quartet ${\cal YY} = \{ {\cal Y}_{s_\beta,s_\gamma} |
s_\beta \in S_\beta, \; s_\gamma \in S_\gamma \}$, each of which
is of the form (\ref{eq:int.op-S-formal-expr}), but the non-integral part
of the power series follows the pattern
\begin{align}
   (h_{\beta s_\beta} + h_{\gamma s_\gamma} - h_{\alpha f(s_\beta, s_\gamma)})
  \equiv_{+\Z} \Delta_{\alpha;\beta,\gamma} + 
     \hat{h}_\beta(i_\gamma(s_\gamma)) + \hat{h}_\gamma( i_\beta(s_\beta))  ;
  \label{eq:int.op-pwr-pattern-wRxR}
\end{align}
with this, we meant to express how the non-integral part
of the power depends on $s_\beta$ and $s_\gamma$, so one may use any
set-theoretical identification 
$i_\beta$, $i_\gamma$, so long as
one chooses an appropriate value of $\Delta_{\alpha;\beta,\gamma} \in \R/\Z$
independent of $s_\beta$ and $s_\gamma$. 
The appropriate value of $\Delta_{\alpha;\beta,\gamma} \in \R/\Z$
may depend on the choice of 
$i_\beta$, $i_\gamma$.
We set the super-Jacobi identity characterizing a quartet of ${\cal YY}$ as
(cf \cite{MR1233387}, \cite{MR1408523})
\begin{align}
  & \; \frac{1}{z_0} \delta \left( \frac{z_1-z_2}{z_0} \right)
  \left( \frac{z_1-z_2}{z_0}\right)^{+ \hat{h}_\beta(s_0)}
  Y_{s_0, \; \alpha f(s_\beta,s_\gamma)}(u ,z_1) \cdot
  {\cal Y}_{s_\beta, \;\; s_\gamma}(m_\beta, z_2) \cdot m_\gamma \nonumber \\
  & \; - \frac{c(s_0, i_\beta(s_\beta))}{z_0}  
  \delta \left( \frac{z_2-z_1}{ e^{- \pi i(2r+1)} z_0}\right)
  \left( \frac{z_2-z_1}{e^{- \pi i(2r+1)} z_0} \right)^{+ \hat{h}_\beta(s_0)}
    \!\!\!\!\!\! {\cal Y}_{s_\beta, \;\; s_0 \cdot s_\gamma}(m_\beta, z_2) \cdot
       Y_{s_0, \gamma s_\gamma}(u,z_1) \cdot m_\gamma
     \nonumber  \\
  & \;\quad
   = \frac{1}{z_2} \delta \left( \frac{z_1-z_0}{z_2}\right)
   {\cal Y}_{s_0 \cdot s_\beta, \;\; s_\gamma}\left( Y_{s_0, \; \beta s_\beta}(u ,z_0)
   \cdot m_\beta, z_2 \right) \cdot
   \left( \frac{z_1-z_0}{z_2} \right)^{- \hat{h}_\gamma(s_0)} m_\gamma,
     \label{eq:def-vct-sp-int.op-S-wR}
\end{align}
where $u \in V_{s_0}$, $m_\beta \in M_{\beta s_\beta}$ and
$m_\gamma \in M_{\gamma s_\gamma}$. The vector space of quartets ${\cal YY}$
satisfying this relation is denoted by ${\cal VV}_{\beta,\gamma}^\alpha(i_\beta,r)$.  We should keep in mind that the second term
of (\ref{eq:def-vct-sp-int.op-S-wR})---and hence the definition of the
vector space ${\cal VV}_{\beta,\gamma}^\alpha$---involves a choice of
set-theoretical $i_\beta: S_\beta \cong \Z/2\Z$. 

%
The power of such terms as
$(z_{12}/z_0)^{+\hat{h}_\beta(s_0)}$,
$(z_{21}/(-z_0))^{+\hat{h}_\beta(s_0)}$ and
$(z_{10}/z_2)^{-\hat{h}_\gamma(s_0)}$ are well-defined only mod $+\Z$
(this is along \cite[(12.118), (12.106, 12.121)]{MR1233387}), instead of
formulating the power $+2\Z$ as
in \cite[\S6, \S9, \S11, half of \S12]{MR1233387}, \cite[\S3]{MR1408523},
and \cite{MR1841356}.  
That makes it necessary
to have the identical argument for $\delta(-)$ and $(-)^{\nu_0+\Z}$ for
a given $\nu_0+\Z \in \R/\Z$, as in (\ref{eq:def-vct-sp-int.op-S-wR}). 
There is still an ambiguity in the expression $\delta(-z) (-z)^{1/2}
= \sum_{\nu \in 1/2+\Z}(-z)^{\nu}$ when $z$ is a dummy variable of
formal power series, and we need to face this issue when comparing
the second term of (\ref{eq:def-vct-sp-int.op-S-wR}) with other terms
if $\hat{h}_\beta \in \hat{\cal G}$ is non-trivial.
Suppose that $T$ is a positive integer so that all the characters
$\hat{h} \in {\rm Hom}({\cal G}, \R/\Z)$ take values in $T^{-1}\Z/\Z$;
$T=2$ in the context of SVOA. Then we need to specify
${\rm Arg}((-1)) \in \R/2\pi T\Z$, not just as a phase mod $+2\pi \Z$.
The $T$ different choices of ${\rm Arg}((-1))$ are labeled by
$r +T\Z \in \Z/T\Z$ in (\ref{eq:def-vct-sp-int.op-S-wR}). 
We accept introducing $T$ separate vector spaces
${\cal VV}_{\beta,\gamma}^\alpha(i_\beta,r)$ labeled by $r \in \Z/T\Z$
in this article ($r \in \Z/T\Z$ is another auxiliary parameter); we will
come back to the $r$-dependence toward the end
of this \ref{statmnt:int.op-S-wR}. 

Here are a couple of consistency checks on (\ref{eq:def-vct-sp-int.op-S-wR}).
First, all the three terms have the power of the dummy variables
\begin{align}
  z_0^{- \hat{h}_\beta(s_0) } \; z_1^{-\hat{h}_\gamma(s_0)/2 } \;
  z_2^{-\Delta_{\alpha;\beta,\gamma}-\hat{h}_\beta(i_\gamma(s_\gamma))
    -\hat{h}_\gamma(i_\beta(s_\beta))}. 
\end{align}
Second, the super-Jacobi identities we have reviewed earlier are reproduced
as special cases of (\ref{eq:def-vct-sp-int.op-S-wR}); 
the relation (\ref{eq:Jacobi-module-S-wR}) of Ramond representations
is the case $\beta = 0$ (so $\hat{h}_\beta =\hat{0} \in \hat{\cal G}$);
the relation (\ref{eq:def-vct-sp-int.op-S}) that defines intertwining
operators when all of $\alpha,\beta,\gamma$ are Neveu--Schwarz
representations is the case $\hat{h}_\beta = \hat{h}_\gamma = \hat{0} \in
\hat{\cal G}$.
Finally, the reasoning on the statistical phase factor for the second term
in \ref{statmnt:int.op-S-NSonly} remains valid, independent of
whether the state $m_\beta$ is from a twisted representation or not;
this is why we use the same sign factor $c(s_0, i_\beta(s_\beta)) =
(-1)^{+s_0 i_\beta(s_\beta)}$.
An implicit assumption in \ref{statmnt:int.op-S-NSonly} is that the phase
factor remains the same
within an irreducible module of the underlying VOA $(V_0,Y_{0,0})$;
then the argument there implies that an irreducible module of the SVOA must
contain at least two irreducible modules of $(V_0, Y_{0,0})$ whose statistical
phase in the commutation with the states in $V_1$ are different.\footnote{
In more general terms, an issue arises when the bimultiplicative form $c$
on ${\cal G}$ is non-trivial when restricted to the stabilizer (isotropy)
subgroup ${\cal G}_{s_*}$ of the action on $S$. This issue will have to be
treated together with the fact that $f: S_\beta \times S_\gamma \rightarrow
S_\alpha$ is a correspondence, not necessarily a set-theoretical map. 
} %
This is why we only deal with $\hat{g}$-twisted irreducible modules that have
the
structure $M = \oplus_{s \in S}M_s$ with $\hat{\cal G}$ acting faithfully on $S$
in this article. 

The role played by the parameter $r \in \Z/T\Z$ is clear when we
translate the super-Jacobi identity (\ref{eq:def-vct-sp-int.op-S-wR})
into the correlation of the quartets ${\cal YY} = \{ {\cal Y}_{s_\beta,s_\gamma}
\}$. One of the consequences of (\ref{eq:def-vct-sp-int.op-S-wR}) is that
(cf \cite[\S7]{MR1233387})  
\begin{align}
  (z_0+z_2)^{\nu_\gamma} & \; \bra{m_\alpha} Y_{s_0, f(s_\beta,s_\gamma)}(u,z_0+z_2)
  \cdot {\cal Y}_{s_\beta, \; \; s_\gamma}(m_\beta, z_2) \cdot m_\gamma \rangle
  \nonumber \\
  & \; = (z_2+z_0)^{\nu_\gamma} \; \bra{m_\alpha} {\cal Y}_{s_0 \cdot s_\beta, \; \;
    s_\gamma}
  \left( Y_{s_0, \; s_\beta}(u, z_0) \cdot m_\beta, z_2 \right) \cdot m_\gamma
  \rangle 
\end{align}
for large enough $\nu_\gamma \in \hat{h}_\gamma(s_0)$; this means that 
the fusion isomorphism of the underlying VOA $(V_0, Y_{0,0})$ describes
the correlation between the components 
\begin{align*}
  {\cal V}_{(s_0 \cdot (\beta s_\beta \cdot \gamma s_\gamma))}^{\alpha s_0 f(s_\beta,s_\gamma)} \ni Y_{s_0, f(s_\beta,s_\gamma)} \otimes
  {\cal Y}_{s_\beta, \;\; s_\gamma}
  \longmapsto {\cal Y}_{s_0 \cdot s_\beta, \; \; s_\gamma} \otimes
  Y_{s_0, \; s_\beta} \in {\cal V}_{((0 s_0 \cdot \beta s_\beta)\cdot \gamma s_\gamma)}^{\alpha s_0 \cdot f(s_\beta,s_\gamma)};  
\end{align*}
the other consequence of (\ref{eq:def-vct-sp-int.op-S-wR}) is that
\cite[\S7]{MR1233387}
\begin{align}
  (z_1-z_2)^{\nu_\beta} & \; \bra{m_\alpha} Y_{s_0, \; f(s_\beta,s_\gamma)}(u,z_1)
  \cdot {\cal Y}_{s_\beta, \;\; s_\gamma}(m_\beta, z_2) \cdot m_\gamma \rangle
  \label{eq:temp-int.op-S-wR-byB-inME}  \\
  = & \; c(s_0, i_\beta(s_\beta)) \;\; 
  (e^{\pi i(2r+1)}(z_2-z_1))^{\nu_\beta} \bra{m_\alpha}
  {\cal Y}_{s_\beta, \;\; s_0 \cdot s_\gamma}(m_\beta,z_2) \cdot
  Y_{s_0, s_\gamma}(u,z_1) \cdot m_\gamma \rangle   \nonumber
\end{align}
for large enough $\nu_\beta \in \hat{h}_\beta(s_0)$; this means that the
braiding isomorphism ${\cal B}_{-1-r}$ of the underlying VOA $(V_0, Y_{0,0})$ 
describes the correlation between the components of
${\cal VV}_{\beta,\gamma}^\alpha(i_\beta,r)$: 
\begin{align*}
  {\cal V}_{(0 s_0 \cdot (\beta s_\beta \cdot \gamma s_\gamma))}^{\alpha s_0 \cdot f(s_\beta,s_\gamma)} \ni & \; 
  Y_{s_0, f(s_\beta,s_\gamma)} \otimes {\cal Y}_{s_\beta,s_\gamma} \longmapsto 
  c(s_0, i_\beta(s_\beta)) \; 
  {\cal Y}_{s_\beta, \;\; s_0 \cdot s_\gamma} \otimes Y_{s_0, s_\gamma}  \in
  {\cal V}_{(\beta s_\beta \cdot (0 s_0 \cdot \gamma s_\gamma))}^{\alpha s_0 f(s_\beta,s_\gamma)} .
\end{align*}
The vector space ${\cal VV}_{\beta,\gamma}^\alpha(i_\beta,r)$ of quartets
${\cal YY}$ is a way to keep track of correlations among
$\{ {\cal V}_{\beta s_\beta, \; \gamma s_\gamma}^{\alpha f(s_\beta,s_\gamma)} \; | \;
s_\beta \in S_\beta, \; s_\gamma \in S_\gamma \}$ by using ${\cal F}$ and
${\cal B}_{-1-r}$; any one of $r \in \Z/T\Z$ serves the purpose.  

There is an isomorphism between the vector spaces
${\cal VV}_{\beta,\gamma}^\alpha(i_\beta,r)$ with different $r \in \Z/T\Z$;
we have seen at the end of \ref{statmnt:int.op-S-NSonly}
that there are between those with different $i_\beta$'s. Such an isomorphism
is found easily by staring at (\ref{eq:def-vct-sp-int.op-S-wR})
or (\ref{eq:temp-int.op-S-wR-byB-inME}): 
\begin{align}
  {\cal VV}_{\beta,\gamma}^\alpha(i_\beta,r) \ni \{ {\cal Y}_{s_\beta,s_\gamma} \}
  \longmapsto \{ {\cal Y}_{s_\beta,s_\gamma}
  e^{-2\pi i \; \Delta r \; \hat{h}_\beta(i_\gamma(s_\gamma))} \} \in
  {\cal VV}_{\beta,\gamma}^\alpha(i_\beta,r+\Delta r).
  \label{eq:temp-iso-vvct-sp-int.op-S-diff-brnch}   
\end{align}
This isomorphism is only meant as an example of isomorphisms between
the vector spaces of linear equations (\ref{eq:def-vct-sp-int.op-S-wR})
in the context of SVOA. The remark at the very end
of \ref{statmnt:int.op-S-NSonly} is also applied here; 
the isomorphism (\ref{eq:temp-iso-vvct-sp-int.op-S-diff-brnch}) is used
in \ref{statmnt:data-4-fCFT-2}, however, to find canonical isomorphisms
between the vector spaces of intertwining operators of fermionic
\underline{CFT}s with different $r \in \Z/T \Z$. 
\end{anythng}

\begin{anythng}   
\label{statmnt:dbl-cov-by-Rsct}
{\bf the $\hat{\cal G}$-bundle twisted around the points of twisted-sector
 insertion:} Think of $n$ points $\{ p_1, p_2, \cdots \}$ in $\Sigma = \P^1$
(or a Riemann surface of higher genus), and assign twists $\hat{h}_i \in
\hat{\cal G}$. Then one can assign a principal $\hat{\cal G}$-bundle
${\cal P} \rightarrow B = (\Sigma \backslash \{ p_{1,\cdots,n}\})$ where a
continuation of a section $\hat{s}$ along a loop in $B$ around $p_i$ in
the clockwise direction once changes $\hat{s}$ to $\hat{s} \cdot \hat{h}_i$. 
An associated bundle denoted by ${\rm Bdl}V$ is
${\cal P}\times_{\hat{\cal G}} V$. The bundle ${\rm Bdl}V$ has a trivialization
in a small neighborhood $U$ of $p \in B$, ${\rm Bdl}V|_U = U \times V$;
when one tracks a section that starts as $U \times  v \in U \times V$
around $p \in B$ along a clockwise loop in $B$ around $p_i$ (by once
going out of $U$), and comes back to $p$, we have a section
$U \times \hat{h}_i \cdot v \in U \times V$ in the trivialization
we have started with (cf \cite[\S5.2]{MR2074176}).  

The homomorphism $Y_M$ of an $\hat{h}$-twisted SVOA representation
$(S_M, M, Y_M)$ formulated in algebra offers a homomorphism from the bundle
$({\rm Bdl}V \otimes M)$ over $B = \C \backslash \{ z=0\}$ to the trivial
bundle $M \times B$ over $B$; this is to read in the relation
$Y_M(u)_{z@ e^{-2\pi i}z(p)} = Y_M( \hat{h} \cdot u)_{z@z(p)}$ for $u \in V$
that the section $(e^{-2\pi i} z(p), u) \in U \times V$ is
$(z(p), \hat{h}\cdot u)$. In the context of the super-Jacobi identity
(\ref{eq:def-vct-sp-int.op-S-wR}) defining the vector space of intertwining
operators, we have $B = \C \backslash \{ p_\beta, p_\gamma\}$,
where $z(p_\beta) = z_2$ and $z(p_\gamma) =0$, with the twist $\hat{h}_\beta$
and $\hat{h}_\gamma \in \hat{\cal G}$, respectively.
Continuation of a section $(p, u) \in U \times V$ along a clockwise
loop around $p_\beta$ converts the section to $(p, \hat{h}_\beta \cdot u)$
in all the three terms of (\ref{eq:def-vct-sp-int.op-S-wR}); the same
change is reproduced in all the three terms by replacing $z_0$ by
$z_0 e^{-2\pi i}$. Continuation along a clockwise loop encircling both $p_\beta$
and $p_\gamma$ converts the section to $(p,
\hat{h}_\beta \hat{h}_\gamma \cdot u)$. The same change is reproduced
in all the three terms by replacing $z_0$ by $z_0 e^{-2\pi i}$ and
$z_1$ by $z_1e^{-2\pi i}$. 
\end{anythng}
\begin{anythng}
{\bf skew-symmetry isomorphisms:} The skewsymmetry isomorphisms
$\Omega_s$ ($s\in \Z$) on intertwining operators of an SVOA involving Ramond
representations can be generalized from the one for NS representations
in \ref{statmnt:skew-iso-4-NS}, as follows.
To start off, focus on the case $2s+1=\mp 1$. Note that the following diagram
commutes because of the hexagon identities:
\begin{equation}
  \xymatrix{
    {\cal V}_{0s_0, \alpha f(s_\beta,s_\gamma)}^{\alpha s_0 f(s_\beta,s_\gamma)} \otimes
    {\cal V}_{\beta s_\beta, \gamma s_\gamma}^{\alpha f(s_\beta,s_\gamma)} \ar[r]^{\Omega_{0/-1}}
    \ar[d]^{\cal F} &
    {\cal V}_{0s_0, \alpha f(s_\beta,s_\gamma)}^{\alpha s_0 f(s_\beta,s_\gamma)} \otimes
    {\cal V}_{\gamma s_\gamma, \beta s_\beta}^{\alpha f(s_\beta,s_\gamma)}
    \ar[d]^{{\cal B}_{-1/0}}
    \\
      {\cal V}_{0 s_0, \beta s_\beta}^{\beta s_0 \cdot s_\beta} \otimes
      {\cal V}_{\beta s_0 \cdot s_\beta, \gamma s_\gamma}^{\alpha s_0 f(s_\beta,s_\gamma)} 
      \ar[r]^{\Omega_{0/-1}} & 
         {\cal V}_{\gamma s_\gamma, \beta s_0 \cdot s_\beta}^{\alpha s_0 f(s_\beta,s_\gamma)}
         \otimes {\cal V}_{0 s_0, \beta s_\beta}^{\beta s_0 \cdot s_\beta}, 
  } 
\end{equation}
\begin{equation}
  \xymatrix{
    {\cal V}_{0s_0, \alpha f(s_\beta,s_\gamma)}^{\alpha s_0 f(s_\beta,s_\gamma)} \otimes
    {\cal V}_{\beta s_\beta, \gamma s_\gamma}^{\alpha f(s_\beta,s_\gamma)} \ar[r]^{\Omega_{0/-1}}
    \ar[d]^{{\cal B}_{0/-1}} &
    {\cal V}_{0s_0, \alpha f(s_\beta,s_\gamma)}^{\alpha s_0 f(s_\beta,s_\gamma)} \otimes
    {\cal V}_{\gamma s_\gamma, \beta s_\beta}^{\alpha f(s_\beta,s_\gamma)} \ar[d]^{\cal F}
    \\
      {\cal V}_{\beta s_\beta, \gamma s_0 \cdot s_\gamma}^{\alpha s_0 f(s_\beta,s_\gamma)}
      \otimes {\cal V}_{0 s_0, \gamma s_\gamma}^{\gamma s_0 \cdot s_\gamma} 
      \ar[r]^{\Omega_{0/-1}} & 
         {\cal V}_{0 s_0, \gamma s_\gamma}^{\gamma s_0 \cdot s_\gamma} \otimes
         {\cal V}_{\gamma s_0 \cdot s_\gamma, \beta s_\beta}^{\alpha s_0 f(s_\beta,s_\gamma)}. 
  }
\end{equation}
This means that $\Omega_0 {\cal YY}$ and $\Omega_{-1}{\cal YY}$ as defined
in (\ref{eq:def-skew-iso-S}) are found in 
\begin{align}
  {\cal VV}_{\beta,\gamma}^\alpha(i_\beta, r)|_{r=-1/0} \ni {\cal YY} \mapsto
  \Omega_{0/-1}({\cal YY}) \in
  {\cal VV}_{\gamma,\beta}^\alpha(i_\gamma, r)|_{r=0/-1} ; 
\label{eq:temp-pattern-domainTarget-skewSymIso-S-wR}
\end{align}          
both the domain and target need to be chosen appropriately for
$\Omega_s {\cal YY}$ with $s \in \{0,-1\}$. This pattern generalizes
to $\Omega_s: {\cal VV}_{\beta,\gamma}^\alpha(i_\beta,-1-s) \rightarrow
{\cal VV}_{\gamma,\beta}^\alpha(i_\gamma,s)$ for $s \in \Z$, as one can use
the pattern (\ref{eq:int.op-pwr-pattern-wRxR}) of the non-integral power
of intertwining operators and the isomorphisms
(\ref{eq:temp-iso-vvct-sp-int.op-S-diff-brnch}).
Two isomorphisms $\Omega_s$ and $\Omega_{s+T}$ share the domain and target,
and are related by $\Omega_{s+T}{\cal YY} = \Omega_s {\cal YY} e^{- 2\pi i T \Delta_{\alpha;\beta,\gamma}}$, where $\Delta_{\alpha;\beta,\gamma}$ was introduced
in (\ref{eq:int.op-pwr-pattern-wRxR}).

It is possible to use the
isomorphisms (\ref{eq:temp-iso-vvct-sp-int.op-S-diff-brnch}) to define
skew-symmetry isomorphisms where the domain and target share the same
value of $r \in \Z/T\Z$. The pair of isomorphisms, 
\begin{align}
 {\cal VV}_{\beta,\gamma}^\alpha(i_\beta,r) \ni & \; {\cal YY} \mapsto \left(
  \Omega_s \left({\cal YY} e^{2\pi i (r+s+1)\hat{h}_\beta(i_\gamma(s_\gamma))} \right)
  \right) e^{-2\pi i (r-s) \hat{h}_\gamma(i_\beta (s_\beta))} \in
        {\cal VV}_{\gamma,\beta}^\alpha(i_\gamma,r)  \nonumber \\
 & \; {\cal YY} \mapsto \left(
  \Omega_{-1-s} \left({\cal YY} e^{2\pi i (r-s)\hat{h}_\beta(i_\gamma(s_\gamma))} \right)
  \right) e^{-2\pi i (r+s+1) \hat{h}_\gamma(i_\beta (s_\beta))} 
\end{align}
are the inverse of the other. 
\end{anythng}

\begin{anythng}
{\bf the fusion and braiding isomorphisms:} To define a fusion isomorphism
{\it naturally}, as in (\ref{eq:chartz-Fiso-SVOA}),
between the vector spaces of intertwining operators of an SVOA
involving a SVOA-twisted (Ramond) representation,
\begin{align}
  {\cal F}: \oplus_{\delta} {\cal VV}_{\beta_1,\delta}^\alpha(i_{\beta_1},r_1)
  \otimes {\cal VV}_{\beta_2,\gamma}^\delta(i_{\beta_2},r_2) \rightarrow
  \oplus_{\epsilon} {\cal VV}_{\beta_1,\beta_2}^{\epsilon}(i_{\beta_1}, r_3)
  \otimes {\cal VV}_{\epsilon,\gamma}^\alpha(i_{\beta_1}+i_{\beta_2}, r_4),  
\end{align}
the auxiliary parameters $r$ of the domain and target need to be chosen
appropriately, just like we have seen in the skew-symmetry isomorphisms.
We restrict our attention to the choices $r_{1,2,3,4}$ within $\{0, -1\}$,
so that the hexagon identity may be used. The two conditions for the hexagon
identity to be applicable is that (i) $r_1 = r_3 \in \{ 0,-1\}$ if $\beta_1$
is a twisted representation, and (ii) $r_2=r_4 \in \{ 0,-1\}$ if
$\beta_2$ is a twisted representation but $\beta_1$ is not; when
both $\beta_1$ and $\beta_2$ are twisted representations, we would
need both $r_1=r_3 \in \{ 0,-1\}$ and $r_1=r_2= r_4 \in \{ 0,-1\}$ to be
able to apply
the hexagon identity formally; in the case of SVOA, however, it is enough to
set
$r_1=r_2=r_3 \in \{0,-1\}$ because $\hat{h}_\epsilon = \hat{0} \in \Z/2\Z$. 

To define the braiding isomorphisms {\it naturally}, as
in (\ref{eq:chartz-Biso-SVOA}), on intertwining operators
involving Ramond representations, 
\begin{align}
  {\cal B}_{-1-r}: \oplus_\delta {\cal VV}_{\beta_1,\delta}^\alpha(i_{\beta_1}, r_1)
  \otimes
  {\cal VV}_{\beta_2,\gamma}^\delta(i_{\beta_2},r_2) \rightarrow
  \oplus_\epsilon {\cal VV}_{\beta_2,\epsilon}^\alpha(i_{\beta_2}, r_3) \otimes
  {\cal VV}_{\beta_1,\gamma}^\epsilon(i_{\beta_1},r_4)
\end{align}
we also need to set the auxiliary parameters $r_{1,2,3,4}$ appropriately.
Let us restrict our attention to the cases $r,r_{1,2,3,4} \in \{ 0,-1\}$.
For the hexagon/box identity to be applicable formally, we need
(i) $r=r_3$ if $\beta_2$ is a twisted representation, (ii)
$r = -1-r_1$ if $\beta_1$ is a twisted representation, and also
(iii) $r_1=r_4$ and $r_2=r_3$ of both $\beta_1$ and $\beta_2$ are both
twisted representations. 

The box identity ${\cal B}_{-1-r} = {\cal F}^{-1} \cdot \Omega_r \cdot {\cal F}$
is therefore expected not to work on intertwining operators as supermultiplets
on general ground; the composition ${\cal F}^{-1} \cdot \Omega_r \cdot
{\cal F}$ does not appear to be well-defined naturally. Indeed, 
think of starting off from ${\cal F}:
{\cal VV}_{\beta_1,\delta}^\alpha(r_1)\otimes {\cal VV}_{\beta_2,\gamma}^\delta(r_2)$
to ${\cal VV}_{\beta_1,\beta_2}^\epsilon(r_3)\otimes
{\cal VV}_{\epsilon,\gamma}^\alpha(r_4)$, with both $\beta_1$ and $\beta_2$
a twisted representation. For the next $\Omega_r$ to be applicable,
we want to use $r_3 = -1-r$, which requires that $r_1=r_2=r_4=-1-r$ in
the fusion isomorphism, at least for a formal applicability of the hexagon
identity. The skewsymmetry isomorphism $\Omega_r$ maps
${\cal VV}_{\beta_1,\beta_2}^\epsilon(-1-r)$ into
${\cal VV}_{\beta_1,\beta_2}^\epsilon(r)$. The fusion isomorphism ${\cal F}^{-1}$
for the supermultiplets described above is not ready to map
${\cal VV}_{\beta_1,\beta_2}^\epsilon(r) \otimes
{\cal VV}_{\epsilon,\gamma}^\alpha(-1-r)$ back to something like
${\cal VV}_{\beta_2,\delta}^\alpha(r'_2) \otimes
{\cal VV}_{\beta_1,\gamma}^\delta(r'_1)$. In the case of SVOA, with ${\cal G}
\cong \Z/2\Z$, however, $\hat{h}_{\beta_1} = \hat{h}_{\beta_2} = \hat{g}$ implies
that $\hat{h}_\epsilon = 2\hat{g} = \hat{0}$, so we do not have to specify
$r_4$. So, we can define ${\cal F}^{-1} \cdot \Omega_r \cdot {\cal F}$
naturally from the choice $r_1=r_2=-1-r$ to $r'_2 = r'_1 = r$. Even then,
the natural definition of ${\cal B}_{-1-r}$ requires $r_2=r'_2= r$ and
$r_1=r'_1 = -1-r$, so the two isomorphisms ${\cal F}^{-1} \cdot \Omega_r
\cdot {\cal F}$ and ${\cal B}_{-1-r}$ do not share the same domain and
target to be compared against each other. 
\end{anythng}

\begin{anythng}
\label{statmnt:xing-S}
{\bf the crossing symmetry of SVOA and crossing symmetry isomorphisms:}
Let us start off by going back to the data $(V, YY, \cdots)$ of an
{\bf SVOA itself}; the relation $\lambda_{s_0} \cdot \Theta Y_{0,s_0} \cdot
\lambda_{s_0}^{-1} = Y_{0, -s_0}$ that facilitates the equivalence between
the representations of the underlying VOA $(V_0, Y_{0,0})$ is the defining
condition of the isomorphisms $\lambda_{s_0}: V_{s_0}^\vee \rightarrow V_{-s_0}$.
In the context of SVOA, however, this relation/condition is not symmetric
or systematic wrpt the two subscripts of $Y_{s_{0,1},s_{0,2}}$. It is therefore
natural to wonder how the relation should be expanded to the one for
$\Theta Y_{s_{0,1}, s_{0,2}}$. Note that this question is relevant to
SVOA without unitarity, because the isomorphisms
$\{ \lambda_{s_0} \;| \;s_0 \in {\cal G}\}$ are well-defined without
a Hermitian inner product. 

A hint is available by staring at 
the condition (p-SVOA) in (\ref{eq:cond-o-4VOA-onUps-S}). 
The relation
(\ref{eq:cond-o-4VOA-onUps-S}) can be decomposed,
as we discuss in the next paragraph, into (A) non-trivial
relations that only involve $\{ Y_{s_{0,1},\; s_{0,2}}\}$, $\lambda_{s_0}$ and the
crossing symmetry isomorphisms, and (B) additional constraints that are
relevant to unitarity. So we can extract the (A) part as a relation
that is also relevant to a non-unitary SVOA
(cf \ref{statmnt:unitarity-refined-VOA}, \ref{statmnt:cntrG-mod}). 

Let us first rewrite the LHS of (\ref{eq:cond-o-4VOA-onUps-S}).
To simplify notation, we use $s$ for $s_{0,1} \in {\cal G}=S_0$ and
$t$ for $s_{0,2}$, respectively, in this paragraph. 
\begin{align*}
 &
  \langle Y_{-s, \; s+t}(\Upsilon_{-s,s}(a_s), z) \cdot u_{s+t} | v_t \rangle_{t} \\
  &  =  \langle Y_{-s, \; s+t}(\Upsilon_{-s,s}(a_s), z) \cdot
  \Upsilon_{s+t,-(s+t)}\Upsilon_{-(s+t),(s+t)} (u_{s+t}) | v_t \rangle_{t} \; \;
  (\epsilon_{-s-t})^{cc} \\
  &  = \langle \Upsilon_{t,-t} \cdot Y_{s,-(s+t)}(a_s,z) \cdot
  \Upsilon_{-(s+t),(s+t)} (u_{s+t}) | v_t \rangle_{t} \;\; \epsilon_{s+t} \; \;
  c(s,-s-t)^{-1} \\ 
  & = \langle \Upsilon_{t,-t}^{-1}(v_t) | Y_{s,-(s+t)}(a_s,z) \cdot \Upsilon_{-(s+t),(s+t)}(u) \rangle_{-t} \; \epsilon_{s+t} \; c(s,-s-t)^{-1} \\ 
  & = \langle \Upsilon_{-t}^{-1}\lambda_{-t}^{-1}(v_t) |
  Y_{s,-(s+t)}(a_s,z) \cdot \Upsilon_{-(s+t),(s+t)}(u) \rangle_{-t} \;\;
  \epsilon_{s+t} \;\; c(s,-s-t)^{-1} \\ 
  & = \langle \lambda_{-t}^{-1}(v_t), \;\;
  Y_{s,-(s+t)}(a_s,z) \cdot \Upsilon_{-(s+t),(s+t)}(u_{s+t}) \rangle_{(-t)^\vee,(-t)}
  \;\; \epsilon_{s+t} \;\; c(s,-s-t)^{-1} . 
\end{align*}
Note that we used unitarity of $\lambda$ at the first and third equality
($(\epsilon_{-s-t})^{cc} = \epsilon_{-s-t}^{-1} = \epsilon_{s+t}$ then),
and the condition (q-SVOA) and also $c(s,-s-t) \in S^1$ at the second
equality. The final expression does not use a Hermitian inner product, and
the anti-unitary maps $\Upsilon$ are involved on only one state
$u_{s+t} \in V_{s+t}$. Let us also rewrite the RHS
of (\ref{eq:cond-o-4VOA-onUps-S}):
\begin{align*}
&  \langle u_{s+t} | Y_{s,t} \left( e^{zL_1} (e^{- \pi i}/z^2)^{L_0} \cdot a_s, 1/z
  \right) \cdot v_t \rangle_{s+t} \\
  & = \langle \Upsilon_{(s+t)}(u_{s+t}), \; Y_{s,t}\left(
  e^{zL_1} (e^{-\pi i}/z^2)^{L_0} \cdot a_s, 1/z
  \right) \cdot v_t \rangle_{(s+t)^\vee, \; (s+t)} \\
 & = \langle \lambda_{-(s+t)}^T\Upsilon_{(s+t)}(u_{s+t}), \; \lambda_{-(s+t)}^{-1}\cdot Y_{s,t} \left(
  e^{zL_1} (e^{-\pi i}/z^2)^{L_0} \cdot a_s, 1/z
  \right) \cdot v_t \rangle_{(-s-t), \; (-s-t)^\vee} \\
  & = \epsilon_{s+t} \; \langle \Upsilon_{(-s-t),(s+t)}(u_{s+t}), \;
  (\lambda_{-s-t}^{-1}Y_{s,t}\lambda_{-t})
  \left( e^{zL_1} (e^{-\pi i}/z^2)^{L_0} \cdot a_s, 1/z  \right) \cdot
  \lambda_{-t}^{-1}(v_t) \rangle_{(-s-t), \; (-s-t)^\vee} \\
  & = \langle \Theta_{-1}\left(\lambda_{-s-t}^{-1}Y_{s,t}\lambda_{-t} \right)
  (a_s, z) \cdot \Upsilon_{(-s-t),(s+t)}(u_{s+t}), \;
  \lambda_{-t}^{-1}(v_t) \rangle_{-t, (-t)^\vee} \; \epsilon_{s+t}.
\end{align*}
It is therefore necessary for a unitary structure to exist, that
an SVOA $(V,YY,\cdots)$ along with a set of isomorphisms
$\{\lambda_{s_0} : V_{s_0}^\vee \rightarrow
V_{-s_0}\; | \; s_0 \in {\cal G}\}$ satisfy a relation 
(see (\ref{eq:xing-SVOA-Y-vs-Y-vs-lambda-m1}) below)
that does not involve the CPT isomorphisms $\Upsilon_{-s_0,s_0}$ or Hermitian
inner products $\langle -|-\rangle_{s_0}$. 
The condition (p-SVOA) can be decomposed into (A) this relation involving
$YY$ and $\{\lambda_{s_0}\}$, and (B) the requirement that
$\{ \lambda_{s_0}\}$ are unitary wrpt the Hermitian inner products on $V_{s_0}$.

Inspired by the reanalysis of the condition (p-SVOA) above, one may reformulate
(not-necessarily unitary) SVOA as follows: its defining algebraic data
consists of $(V, YY, {\bf 1}, \omega, \{ \lambda_{s_0} : V_{s_0}^\vee \rightarrow V_{-s_0} \; | \; s_0 \in {\cal G}\})$; along with all the conditions
(a-SVOA)--($\ell$-SVOA), impose 
\begin{align}
  Y_{s_{0,1},s_{0,2}} = \lambda_{-s_{0,1}-s_{0,2}} \cdot
  (\Theta_{0}Y_{s_{0,1},-s_{0,1}-s_{0,2}}) \cdot \lambda_{-s_{0,2}}^{-1} \;
  c(s_{0,1}, -s_{0,1}-s_{0,2})^{-1}.  
 \label{eq:xing-SVOA-Y-vs-Y-vs-lambda-m1}
\end{align}
Equivalently, we may also write this as
\begin{align}
  Y_{s_{0,1},-s_{0,1}-s_{0,2}} & \; =
  \lambda_{-s_{0,2}}^T \cdot (\Theta_{-1} Y_{s_{0,1},s_{0,2}})
  \cdot (\lambda_{-s_{0,1}-s_{0,2}}^{-1})^T  \;\; 
  c(s_{0,1}, -s_{0,1}-s_{0,2}) ,
  \label{eq:xing-SVOA-Y-vs-Y-vs-lambda-m1-temp2}  \\ 
    & \; =   \Theta_{-1} (\lambda_{-s_{0,1}-s_{0,2}}^{-1} \cdot Y_{s_{0,1},s_{0,2}}
    \cdot \lambda_{-s_{0,2}}) \; \;
    c(s_{0,1}, -s_{0,1}-s_{0,2}).  
    \nonumber 
\end{align}
For an SVOA in this formulation to be unitary, additional data
is the Hermitian inner products on $V_{s_0}$, and the condition
(p-SVOA) can be replaced by the unitarity of $\lambda_{s_0}$. 

This condition (\ref{eq:xing-SVOA-Y-vs-Y-vs-lambda-m1}) does not yield
constraints on $YY$ when there are enough freedom of $\{ \lambda_{s_0} \; | \;
s_0 \in {\cal G}\}$. It is quite likely so for ${\cal G}=\Z/2\Z$,
even when an SVOA in question is not necessarily fermionic
(${\cal V}_{1,1}^0 \geq$ in particular); we should read this relation
with $s_{0,1} = 1 \in {\cal G}=\Z/2\Z$ as an extra constraint on
$\lambda_1$ for a given $Y_{1,1}$ (and for a given $Y_{0,0}$, $Y_{0,1}$ and
$Y_{1,0}=\Omega Y_{0,1}$).

There are two more lessons to be extracted from this study, relevant when
we move on from SVOA to its representations and intertwining operators.
One is that a nice form of the crossing-symmetry transformation
on $(\#{\cal G})^2$ components of $YY$ is of the form
\begin{align}
  YY = \left\{ Y_{s_{0,1},s_{0,2}} \right\} \mapsto
  \lambda (\Theta_{0}YY) \lambda^{-1} := \left\{ \lambda_{s_{0,2}} \cdot
  (\Theta_{0} Y_{s_{0,1},s_{0,2}}) \cdot \lambda_{s_{0,1}+s_{0,2}}^{-1} \;
  c(s_{0,1}, s_{0,2})^{-1} 
  \right\}.
  \label{eq:xing-sym-iso-S}
\end{align}
To be fair, however, arguments so far cannot determine whether we should use
$c(s_{0,1}, s_{0,2})^{-1}$ as above, or $c(s_{0,1}, s_{0,2})$ instead;  
if we had used $c(s_{0,1}, s_{0,2})$ in (\ref{eq:Hconj-VOA-wOPE-S})
instead of $c(s_{0,1}, s_{0,2})^{-1}$ there, we would have endded up
with $c(s_{0,1}, s_{0,2})$ in (\ref{eq:xing-SVOA-Y-vs-Y-vs-lambda-m1})
(this difference is practically unimportant for $c(s_1,s_2) =
(-1)^{s_1s_2}$ for $s_1,s_2 \in \Z/2\Z$, of course). One can see that the
choice above continues to fit in the  discussion in the rest
of \ref{statmnt:xing-S}, whereas the other one does not; that was how
we determined the expression in (\ref{eq:Hconj-VOA-wOPE-S}). 

The other lesson is an extra constraint on $[h_{s_0}] \in \R/\Z$ and
$\epsilon_{s_0} \in S^1$ for $s_0 \in {\cal G}$ to be able to impose
the condition (\ref{eq:xing-SVOA-Y-vs-Y-vs-lambda-m1}). To see that there
is a constraint, let us rewrite (\ref{eq:xing-SVOA-Y-vs-Y-vs-lambda-m1-temp2})
by using $\lambda^T_{s_0} = \epsilon_{-s_0} \lambda_{-s_0}$, and substituting
$s_{0,2} \rightarrow -(s_{0,1}+s_{0,2})$, $-(s_{0,1}+s_{0,2}) \rightarrow s_{0,2}$. 
Then we have two relations between $Y_{s_{0,1},s_{0,2}}$ and
$\Theta Y_{s_{0,1}, -s_{0,1}-s_{0,2}}$, one from
(\ref{eq:xing-SVOA-Y-vs-Y-vs-lambda-m1}) and the other from
(\ref{eq:xing-SVOA-Y-vs-Y-vs-lambda-m1-temp2}). For the two relations
to be satisfied simultaneously, we need
\begin{align*}
  c(s_{0,1}, s_{0,1}+s_{0,2}) \epsilon_{s_{0,1}+s_{0,2}} 
  = \epsilon_{s_{0,2}} \; c(s_{0,1},s_{0,2}) \; e^{-2\pi i h_{s_{0,1}}} 
\end{align*}
Given the bimultiplicativity of $c$ on ${\cal G} \times {\cal G}$,
we have the relation
\begin{align}
  \frac{\epsilon_{s_{0,1}+s_{0,2}}}{\epsilon_{s_{0,2}}} = c(s_{0,1}, \; s_{0,1})^{-1}
  \; e^{-2\pi i h_{s_{0,1}}}.
  \label{eq:reltn-eps-c-h}
\end{align}
A couple of consequences follow from this relation. First, the map
$\epsilon: {\cal G} \rightarrow S^1$ (or $\C^\times$ if we do not care
about unitarity), is a homomorphism, because the RHS is. Second,
$\epsilon_{s_0} \in \{ \pm 1\}$; to see this, note that $c(-s_{0,1},-s_{0,1}) =
c(s_{0,1},s_{0,1})$ and $[h_{s_0}] = [h_{-s_0}]$ in $\R/\Z$, so $\epsilon_{-s_0} =
\epsilon_{s_0}$; this is combined with $\epsilon_{-s_0} \epsilon_{s_0} = 1$. 
Thirdly, $[h_{s_0}] \in \{ 0, 1/2\}_{+\Z} \subset \R/\Z$, because
$\epsilon_{s_0} \in \{ \pm 1\}$ and $c(s_0, s_0) \in \{ \pm 1\}$
(skew-symmetry of $c$) for all $s_0 \in {\cal G}$. To summarize,
$[h_{s_0}] \in \R/\Z$ is severely constrained (at most half-integral),
$c$ is not, but $\epsilon_{s_0}$ is completely determined by $[h_{s_0}]$
and $c(s_0, s_0)$. In the case of SVOA (${\cal G} = \Z/2\Z$ and $c$ given
by (\ref{eq:fermion-exch-phase})), for example, there is no choice but
$\epsilon_1 = +1$. 

Given a simple {\bf representation} $(S_\alpha, M_\alpha, YY_\alpha)$ of
an SVOA $(V, YY, \cdots)$, a contragradient representation of the SVOA can
be constructed as follows, by using the dual vector space
$\oplus_{s_\alpha \in S_\alpha} (M_{\alpha s_\alpha})^\vee$ as the representation space.
By choosing one set-theoretical identification $j_\alpha: S_\alpha \cong
{\cal G}$, set
\begin{align}
   \Theta_s YY_\alpha^{(j_\alpha)} := \left\{ \left.
   c(s_0, j_\alpha(s_\alpha))^{-1}
   (\Theta_s Y_{0s_0, \alpha s_\alpha}) \in
      {\cal V}_{0 s_0, \; (s_0 \cdot s_\alpha)^\vee}^{(\alpha s_\alpha)^\vee} \right|
   s_0 \in {\cal G}, \; s_\alpha \in S_\alpha    \right\}. 
\end{align}
To see that this gives a solution to (\ref{eq:Jacobi-module-S},
\ref{eq:Jacobi-module-S-wR}), let us first focus on the case $2s+1=\mp 1$
and verify that 
\begin{align}
  {\cal F}: & \;
  {\cal V}^{(\alpha s_\alpha)^\vee}_{(0s_{0,1} \cdot (0s_{0,2} \cdot (\alpha s_{0,1}s_{0,2}\cdot s_\alpha)^\vee))}
  \ni 
  (\Theta[\mp]YY_\alpha^{(j_\alpha)})_{s_{0,1}, \; (s_{0,1} \cdot s_\alpha)^\vee}
  \otimes
  (\Theta[\mp]YY_\alpha^{(j_\alpha)})_{s_{0,2}, \; (s_{0,1}s_{0,2} \cdot s_\alpha)^\vee}
   \nonumber \\
  & \mapsto  Y_{0s_{0,1},0s_{0,2}} \otimes
  (\Theta[\mp]YY_\alpha^{(j_\alpha)})_{s_{0,1}s_{0,2}, (s_{0,1}s_{0,2} \cdot s_\alpha)^\vee} 
    \in {\cal V}^{(\alpha s_\alpha)^\vee}_{((0s_{0,1}\cdot 0s_{0,2})\cdot (\alpha s_{0,1}s_{0,2} \cdot s_\alpha)^\vee)} .
\end{align}
Indeed, the fact that $YY_\alpha$ is an SVOA representation
is combined with \ref{statmnt:ThetaTheta-vs-fusion} to see that 
\begin{align}
  {\cal F}: \Theta[\mp](Y_{s_{0,1},\alpha s_\alpha}) \otimes
  \Theta[\mp](Y_{s_{0,2}, \; \alpha s_{0,1}\cdot s_\alpha}) \mapsto
  \Theta[\mp](Y_{s_{0,2}\cdot s_{0,1}, \; \alpha s_\alpha})
  \otimes \Omega(Y_{s_{0,2}, \; s_{0,1}}); 
\end{align}
one may multiply
\[
  c(s_{0,1}, j_\alpha(s_\alpha))^{-1} \cdot c(s_{0,2},j_\alpha(s_{0,1}\cdot s_\alpha))^{-1} 
  = c(s_{0,1}+s_{0,2}, j_\alpha(s_\alpha))^{-1} \cdot
  c(s_{0,2}, s_{0,1})^{-1} 
\]
on both sides; 
finally the condition (i-SVOA) (generalized skew-symmetry) of the SVOA data
is used to verify the claim. To verify that 
\begin{align*}
  {\cal B}: & \; 
  {\cal V}_{(0s_{0,1} \cdot (0s_{0,2} \cdot ( \alpha
    s_{0,1}s_{0,2}\cdot s_\alpha)^\vee))}^{(\alpha s_\alpha)^\vee}
  \ni 
    (\Theta[\mp]YY_\alpha^{(j_\alpha)})_{s_{0,1}, (s_{0,1} \cdot s_\alpha)^\vee} \otimes
  (\Theta[\mp]YY_\alpha^{(j_\alpha)})_{s_{0,2}, (s_{0,1}s_{0,2} \cdot s_\alpha)^\vee}  \\
 & \mapsto c(s_{0,1}, s_{0,2})  
  (\Theta[\mp]YY_\alpha^{(j_\alpha)})_{s_{0,2}, (s_{0,2} \cdot s_\alpha)^\vee} \otimes
  (\Theta[\mp]Y_\alpha^{(j_\alpha)})_{s_{0,1}, (s_{0,1}s_{0,2} \cdot s_\alpha)^\vee}, 
\end{align*}
we only need to use ${\cal B} = {\cal F}^{-1} \Omega {\cal F}$,
the result above, and the generalized skew-symmetry of the SVOA.

The construction above involves a choice of 
$j_\alpha: S_\alpha \cong {\cal G}$ and $s \in \Z/T\Z$;
the resulting representations with
different $j_\alpha$'s and $2s+1$ are equivalent to each other, in the sense
that there is a unitary map $\lambda$ from
$M_\alpha^\vee := \oplus_{s \in S_\alpha} (M_{\alpha s_\alpha})^\vee$ to itself so that
$\lambda \cdot \Theta_s YY_\alpha^{(j'_\alpha)} \cdot \lambda^{-1} =
\Theta_{0}YY_\alpha^{(j_\alpha)}$ for each choice of $j'_\alpha$ and $s$.
We call this equivalence class the {\it representation of the SVOA
contragradient to} $(S_\alpha, M_\alpha, YY_\alpha)$. When one contragradient
representation $(S_{\Upsilon^{\cal II}(\alpha)}, M_{\Upsilon^{\cal II}(\alpha)},
YY_{\Upsilon^{\cal II}(\gamma)})$ is chosen from its equivalence class,
let $\lambda_{\epsilon,j_\alpha}: M_{\alpha}^\vee \rightarrow
M_{\Upsilon^{\cal II}(\alpha)}$ so that 
\begin{align}
  \lambda_{2s+1,j_\alpha} \cdot (\Theta_sYY_\alpha^{(j_\alpha)}) \cdot
  \lambda_{2s+1,j_\alpha}^{-1} = YY_{\Upsilon^{\cal I}(\alpha)}. 
\end{align}
Once an isomorphism $\lambda_{2s_*+1,j_{\alpha *}}: M_\alpha^\vee \rightarrow
M_{\Upsilon^{\cal II}(\alpha)}$ is found for one choice of $s=s_*$ and
$j_\alpha = j_{\alpha *}$, then $\lambda_{2s+1,j_\alpha}$ with other
$s$ or $j_\alpha$ can be chosen as 
\begin{align}
  \lambda_{\epsilon,j_{\alpha *} + \delta s_0} & \; \propto \lambda_{\epsilon,j_\alpha}
  c(j_\alpha(s_\alpha), \; \delta s_0 )^{-1}, \\
  \lambda_{\epsilon+2\Delta s,j_{\alpha *}} & \; \propto \lambda_{\epsilon,j_\alpha}
  e^{2\pi i (\Delta s)
    \left( h_{\Upsilon(s_\alpha)} - \hat{h}_{\Upsilon(\alpha)}(j_\alpha(s_\alpha))\right) }, \label{eq:temp-pattern-on-s4xingSymTrnsf}
\end{align}
where $\propto$ allows $s_\alpha$-independent constant.

Let ${\cal YY}$ be an {\bf intertwining operator} in
${\cal VV}_{\beta,\gamma}^\alpha(i_\beta, r)$; we hope to introduce
generalized crossing-symmetry isomorphisms on intertwining operators of
an SVOA $(V, YY, {\bf 1}, \cdots)$. Let us first focus on the case
$2r+1 = \epsilon \in \{ \pm 1\}$.

To construct a crossing symmetry isomorphism, let us begin with
preparation. Choose any $j_\alpha: S_\alpha \cong \Z/2\Z$,
$j_\gamma: S_\gamma \cong \Z/2\Z$.
Our proposal is to set ${\cal YY}' :=
\{ {\cal Y}'_{s_\beta, \Upsilon^{\cal I}(s_\alpha)} \}$ as
\begin{align}
  {\cal Y}'_{s_\beta, \Upsilon^{\cal I}(s_\alpha)} =
  \lambda_{\epsilon,j_\gamma} \cdot
  \left( \Theta[-\epsilon]{\cal Y}_{s_\beta, s_\gamma} \right)
  \cdot \lambda_{\epsilon,j_\alpha}^{-1} \;\; 
  c(i_\beta(s_\beta), j_\gamma(s_\gamma))^{-1}  ;  
  \label{eq:gen-xing-iso-S-wR}
\end{align}
where $s_\alpha = f(s_\beta,s_\gamma)$.
We claim that ${\cal YY}' \in {\cal VV}_{\beta,\Upsilon^{\cal II}(\alpha)}^{\Upsilon^{\cal II}(\gamma)}(i'_\beta, r')$ with $2r' +1 = -\epsilon$ and
$i'_\beta:S_\beta \cong \Z/2\Z$ given by the map
$(j_\alpha\cdot f) - j_\gamma : S_\beta \times S_\gamma \rightarrow \Z/2\Z$ 
independently from $s_\gamma \in S_\gamma$. 
 
To see that ${\cal YY}' \in
{\cal VV}_{\beta,,\Upsilon^{\cal II}(\alpha)}^{\Upsilon^{\cal II}(\gamma)}(i'_\beta, r')$, 
we use the following commuting diagram:
\begin{equation}
 \xymatrix{
   {\cal V}^{00}_{(\Upsilon^{\cal I}(\alpha s_0 \cdot f(s_\beta, s_\gamma)) \cdot (\beta s_\beta \cdot (0 s_0 \cdot \gamma s_\gamma)))} 
   \ar[r]^{(\Omega \Omega){\cal F}\Omega {\cal F}} &
  {\cal V}^{00}_{(\gamma s_\gamma \cdot (0 s_0 \cdot (\beta s_\beta \cdot
    \Upsilon^{\cal I}(\alpha s_0 \cdot f(s_\beta , s_\gamma)))))}
               \ar[d]^{{\cal B}(-\epsilon_2)} \\ 
       {\cal V}^{00}_{(\Upsilon^{\cal I}(\alpha s_0 \cdot f(s_\beta, s_\gamma)) \cdot (0 s_0 \cdot (\beta s_\beta\cdot \gamma s_\gamma)))} \ar[d]^{\cal F}
       \ar[r]^{\Omega\Omega {\cal F}\Omega {\cal F}} \ar[u]^{{\cal B}(\epsilon_2)} &
         {\cal V}^{00}_{(\gamma s_\gamma \cdot (\beta s_\beta \cdot (0 s_0 \cdot
           \Upsilon^{\cal I}(\alpha s_0 \cdot f(s_\beta , s_\gamma)))))}
         \ar[d]^{{\cal F}\cdot {\cal B}(\epsilon_1)=\Omega(-\epsilon_1)\cdot {\cal F}} \\
  {\cal V}^{00}_{(\Upsilon^{\cal I}(\alpha s_0 \cdot f(s_\beta, s_\gamma)) \cdot ((0 s_0 \cdot \beta s_\beta)\cdot \gamma s_\gamma))} \ar[r]^{(\Omega\Omega){\cal F}}
  &   {\cal V}^{00}_{(\gamma s_\gamma \cdot ((0 s_0 \cdot \beta s_\beta) \cdot \Upsilon^{\cal I}(\alpha s_0 \cdot f(s_\beta, s_\gamma))))}
  }
\end{equation}
Here, the isomorphisms in the first two rows are that
of (\ref{eq:Theta-Theta-a}) using $\Omega(\epsilon_1)$ with
$\epsilon_1 \in \{ \pm \}$,  and the one in the bottom row that of
(\ref{eq:xing-iso-by-FnOmega-b}) with $\Omega_s = \Omega(\epsilon_1)$. 
Although the diagram commutes for arbitrary pair of $\epsilon_2 \in \{\pm 1\}$
and $\epsilon_1 \in \{\pm \}$,  we will use the diagram with
$\epsilon_1=\epsilon_2$ (so the composition simplifies
${\cal F}\cdot {\cal B}(\epsilon_1)\cdot
{\cal B}(-\epsilon_2)={\cal F}$ in the right column), and set
$\epsilon_2=\epsilon$ so that we can deal with the components of 
${\cal YY} \in {\cal V}_{\beta,\gamma}^\alpha(i_\beta,r)|_{2r+1=\epsilon}$
on the left column.

On the left column, we should use
$c(s_0,  i_\beta(s_\beta)) \;  {\cal Y}_{s_\beta, s_0 \cdot s_\gamma} \otimes
Y_{s_0, s_\gamma}$ in the top row,
$Y_{s_0, f(s_\beta,s_\gamma)} \otimes {\cal Y}_{s_\beta,s_\gamma}$ in the middle row
and
$Y_{s_0,s_\beta} \otimes {\cal Y}_{s_0 \cdot s_\beta, s_\gamma}$ in the bottom row,
respectively. They are mapped into the right column, respectively, as
$(\Theta[-\epsilon]{\cal Y}_{s_\beta, s_0 \cdot s_\gamma}) \otimes
(\Theta[-\epsilon]Y_{s_0, s_\gamma}) \; c(s_0, i_\beta(s_\beta))$,
$(\Theta[-\epsilon]{\cal Y}_{s_\beta,s_\gamma}) \otimes
(\Theta[-\epsilon]Y_{s_0,f(s_\beta,s_\gamma)})$ and
$(\Theta[-\epsilon]{\cal Y}_{s_0 \cdot s_\beta, s_\gamma}) \otimes Y_{s_0,s_\beta}$.
This means, by adjoining $\lambda_{\epsilon,j_\gamma}^{-1}$, $\lambda_{\epsilon,j_\alpha}$, $c(i_\beta(s_\beta), j_\gamma(s_\gamma))^{-1}$, that 
\begin{align}
  Y_{s_0, \Upsilon^{\cal I}(s_0 \cdot s_\gamma)} \otimes
  {\cal Y}'_{s_\beta,\Upsilon^{\cal I}(s_0 \cdot s_\alpha)}
  \;\; c(s_0, j_\gamma(s_\gamma)) 
\end{align}
in the top row is mapped by ${\cal B}(-\epsilon)$ to
\begin{align}
  {\cal Y}'_{s_\beta,\Upsilon^{\cal I}(s_\alpha)} \otimes
  Y_{s_0, \Upsilon^{\cal I}(s_\alpha)}
  \;\; c(s_0 , j_\alpha(s_\alpha))
\end{align}
in the middle row, and also mapped by ${\cal F}$ to 
\begin{align}
  Y_{s_0, s_\beta} \otimes {\cal Y}'_{s_0 \cdot s_\beta,\Upsilon^{\cal I}(s_\alpha)}
  \;\; c(s_0, j_\gamma(s_\gamma)) 
\end{align}
in the bottom row. So, ${\cal YY}'$ is indeed in ${\cal VV}_{\beta,\Upsilon^{\cal II}(\alpha)}^{\Upsilon^{\cal II}(\gamma)}(i'_\beta, r')$ with
$2r'+1=-\epsilon$ when $i'_\beta = 
(j_\alpha \cdot f[\alpha;\beta,\gamma]) - j_\gamma :
S_\beta \cong \Z/2\Z$.  \qed

Different choices of $j_\gamma$ and $j_\alpha$ change the vector space
that ${\cal YY}'$ ends up with. As we have already discussed isomorphisms
among ${\cal VV}_{\beta,\gamma}^\alpha(i_\beta,r)$ with different $i_\beta$,
we may use those isomorphisms to adjust the
crossing symmetry isomorphism (\ref{eq:gen-xing-iso-S-wR}) so that
$i_\beta$ before and after the crossing symmetry isomorphism
remain the same.

We have introduced the crossing symmetry isomorphisms on the
intertwining oprators in (\ref{eq:gen-xing-iso-S-wR}), which is denoted
by $\lambda \cdot (\Theta_s {\cal YY}) \cdot \lambda^{-1} =
\{ {\cal Y}'_{s_\beta, \; \Upsilon^{\cal I}(s_\alpha)} \}$, only for $s=0,-1$.
This is virtually enough for SVOA, because the conformal weights
in a given simple representation of an SVOA differ only by $T^{-1}\Z/\Z$
with $T=2$. For a better understanding of what is going on here
(with a view toward generalization to the cases with a more general
finite abelian group ${\cal G}$ in \ref{statmnt:genG-w2cond}), however,
let us explore what happens for $s \in \Z/T\Z$. To start off,
we see that $\lambda \Theta_{0/-1} \lambda^{-1}: {\cal VV}(r=0/-1) \rightarrow
{\cal VV}(r=-1/0)$; this pattern is in parallel\footnote{
Probably it was better to define $\Theta_0$ [resp. $\Theta_s$]
in this article as $\Theta_{-1}$ [resp. $\Theta_{-1-s}$] to have a complete
parallel.  
} %
with (\ref{eq:temp-pattern-domainTarget-skewSymIso-S-wR}). Moreover,
we may use the $s$-dependence in (\ref{eq:temp-pattern-on-s4xingSymTrnsf})
along with (\ref{eq:int.op-pwr-pattern-wRxR}) to see that
\begin{align*}
 & \left[ \lambda_{2s+1,j_\gamma} \cdot (\Theta_s {\cal YY}) \cdot
    \lambda_{2s+1,j_\alpha}^{-1}\right]_{s_\beta,\Upsilon^{\cal I}(s_\alpha)} \\
 & \qquad \propto
  \left[ \lambda_{1,j_\gamma} \cdot (\Theta_0 {\cal YY}) \cdot
    \lambda_{1,j_\alpha}^{-1}\right]_{s_\beta,\Upsilon^{\cal I}(s_\alpha)} \;
  e^{2\pi i s \left( h_{s_\gamma} + h_{s_\beta} - h_{s_\alpha} + \hat{h}_\gamma(j_\gamma(s_\gamma)) - \hat{h}_\alpha(j_\alpha(s_\alpha)) \right)} \\
 & \qquad  = \left[ \lambda_{1,j_\gamma} \cdot (\Theta_0 {\cal YY}) \cdot
    \lambda_{1,j_\alpha}^{-1}\right]_{s_\beta,\Upsilon^{\cal I}(s_\alpha)} \;
 e^{2\pi i s \left( \Delta_{\alpha;\beta,\gamma} + \hat{h}_{\beta}(j_\gamma(s_\gamma))
   + \hat{h}_\gamma(i_\beta(s_\beta)+j_\gamma(s_\gamma))
   - \hat{h}_\alpha(j_\alpha(s_\alpha)) \right)} \\
 & \qquad  \propto \left[ \lambda_{1,j_\gamma} \cdot (\Theta_0 {\cal YY}) \cdot
    \lambda_{1,j_\alpha}^{-1}\right]_{s_\beta,\Upsilon^{\cal I}(s_\alpha)} \;
 e^{2\pi i s \hat{h}_{\beta}(j_\gamma(s_\gamma))} \;  
 e^{- 2\pi i s \hat{h}_\beta(j_\alpha(s_\alpha))} . 
\end{align*}
As we may see that $e^{2\pi i s \hat{h}_\beta(j_\gamma(s_\gamma))}$ is an isomorphism
${\cal VV}_{\beta,\gamma}^{\alpha}(r=s)\rightarrow
{\cal VV}_{\beta,\gamma}^\alpha(r=0)$, and
$e^{2\pi i s\hat{h}_\beta(-j_\alpha(s_\alpha))}$ an isomorphism 
${\cal VV}_{\beta,\Upsilon^{\cal II}(\alpha)}^{\Upsilon^{\cal II}(\gamma)}(r=-1)
\rightarrow {\cal VV}_{\beta,\Upsilon^{\cal II}(\alpha)}^{\Upsilon^{\cal II}(\gamma)}
(r=-1-s)$, we see that the crossing symmetry isomorphism
${\cal YY} \mapsto \lambda_{2s+1,j_\gamma} \cdot (\Theta_s{\cal YY}) \cdot
\lambda_{2s+1,j_\alpha}^{-1}$ is from ${\cal VV}_{\beta,\gamma}^\alpha(i_\beta,s)$
to ${\cal VV}_{\beta,\Upsilon^{\cal II}(\alpha)}^{\Upsilon^{\cal II}(\gamma)}
(j_\alpha -j_\gamma,-1-s)$.

One might call them (with or without isomorphisms)
generalized crossing symmetry isomorphisms. 

\end{anythng}

\begin{anythng}
\label{statmnt:genG-w2cond}
{\bf a few layers of generalization of SVOA:}
SVOA with an extra assumption in Rmk. \ref{rmk:SVOA-fermionic}
is a special property (not generally expected, but one may impose) on
a pair of VOA $(V_0, Y_{0,0}, {\bf 1},\cdots)$ and a set of its simple, unitary\footnote{
Many papers on simple current extension were written primarily for the case the VOA $(V_0, Y_{0,0}, \cdots)$ is rational,
so implicitly the set ${\cal I}((V_0, Y_{0,0}))$ of \underline{all} the simple modules is under consideration. In the case $(V_0, Y_{0,0})$ is not rational,
we have to specify a set ${\cal I}((V_0, Y_{0,0}))$ of simple $(V_0, Y_{0,0})$-modules (e.g., as in footnote \ref{fn:ex-I(VY)-torus}), 
for the definition of simple currents of $(V_0, Y_{0,0})$ in the main text to make sense.
} %
and distinct representations ${\cal I}((V_0, Y_{0,0}))$.
A special property more general than SVOA is {\it simple current extension}, 
and there is a sequence of intermediate ones in between.
cf \cite{Schellekens:1989am},  
\cite{MR1233387},  
\cite{MR1408523}. 
A large fraction of this \ref{statmnt:genG-w2cond} (possibly except the last 1 page or so) is material well-established in the literature. 

Let us begin with simple current extension.
An irreducible representation $s \in {\cal I}((V_0, Y_{0,0}))$ of
$(V_0, Y_{0,0},\cdots)$ is a {\it simple current}
\cite{Schellekens:1989am}, if its fusion $([e_s] \cdot):
{\cal I}((V_0,Y_{0,0})) \rightarrow \Z[{\cal I}((V_0,Y_{0,0}))]$ is interpreted
as a permutation on the set ${\cal I}((V_0,Y_{0,0}))$; in particular, 
for any $s' \in {\cal I}((V_0,Y_{0,0}))$, there is just one
$s''\in {\cal I}((V_0, Y_{0,0}))$ with $N_{s, s'}^{s''}=1$ while
$N_{s,s'}^{s^{""}}=0$ for all other $s^{""} \in {\cal I}((V_0, Y_{0,0}))$. 
Simple currents within 
${\cal I}((V_0,Y_{0,0}))$ form an abelian group with respect to the fusion.
One chooses a finite group ${\cal G}$ as a subgroup of these simple currents,
and introduces a ${\cal G}$-graded vector space
$V = \oplus_{s_0 \in {\cal G}} V_{s_0}$.
One may choose a basis $Y_{s_{0,1},\; s_{0,2}}$ arbitrarily in the 1-dimensional
vector space ${\cal V}_{s_{0,1}, \; \; s_{0,2}}^{s_{0,1}\cdot s_{0,2}}$ of intertwining
operators of the VOA $(V_0, Y_{0,0})$. Then a set of complex numbers
$\{ h(s_{0,1},s_{0,2}, s_{0,3}) \; | \; s_{0,1}, s_{0,2}, s_{0,3} \in {\cal G} \}$
are determined by the fusion isomorphisms of the following form:  
\begin{align}
 {\cal F}: {\cal V}_{(s_{0,1} \cdot (s_{0,2} \cdot s_{0,3}))}^{s_{0,1}\cdot s_{0,2}\cdot s_{0,3}}
 & \; \ni Y_{s_{0,1}, \;\; s_{0,2}\cdot s_{0,3}} \otimes Y_{s_{0,2}, \;\; s_{0,3}} \longmapsto \\
 & Y_{s_{0,1}\cdot s_{0,2} , \; s_{0,3}} \otimes Y_{s_{0,1}, \; s_{0,2}} \;\; 
 h(s_{0,1},s_{0,2},s_{0,3})
 \in {\cal V}_{((s_{0,1}\cdot s_{0,2})\cdot s_{0,3})}^{s_{0,1}\cdot s_{0,2}\cdot s_{0,3}}. 
  \nonumber 
\end{align}
Another set of complex numbers
$\{ c_{\mp}(s_{0,1},s_{0,2},s_{0,3}) \; | \;
s_{0,1}, s_{0,2}, s_{0,3} \in {\cal G}\}$ are extracted from the braiding
isomorphisms by using the same set of basis elements: 
\begin{align}
 {\cal B}(\pm): {\cal V}_{(s_{0,1} \cdot (s_{0,2} \cdot s_{0,3}))}^{s_{0,1}\cdot s_{0,2}\cdot s_{0,3}}
 & \; \ni Y_{s_{0,1}, \; \; s_{0,2}\cdot s_{0,3}} \otimes Y_{s_{0,2}, \;\; s_{0,3}}
 \longmapsto \\
 & Y_{s_{0,2}, \;\; s_{0,1}\cdot s_{0,3}} \otimes Y_{s_{0,1}, \;\; s_{0,3}} \;
 c_{\mp}(s_{0,1},s_{0,2}, s_{0,3})
 \in {\cal V}_{(s_{0,2} \cdot (s_{0,1}\cdot s_{0,3}))}^{s_{0,1}\cdot s_{0,2}\cdot s_{0,3}}.
 \nonumber 
\end{align}
The pentagon identity indicates that $h$ is a 3-cocycle
(eg., \cite[\S12]{MR1233387}, \cite[p. 696]{MR1408523}).
The complex numbers
$c_{\mp}$ are constrained by the hexagon identity for a given
$h$. Given the ${\cal G}$-grading, there is a group of automorphisms
$\hat{\cal G} = {\rm Hom}({\cal G}, \C^\times) \cong
{\rm Hom}({\cal G}, \R/\Z)$ on the structure $(V, \{ Y_{s_{0,1}, \;\; s_{0,2}}\})$
no matter how we choose the basis element $Y_{s_{0,1}, \; s_{0,2}}$ of
${\cal V}_{s_{0,1}, \;\; s_{0,2}}^{s_{0,1}\cdot s_{0,2}}$. We will use the notation
$\hat{h}|_{V_{s_0}} = e^{2\pi i \hat{h}(s_0)}$, with $\hat{h}(s_0) \in \R/\Z$. 

There are {\bf two independent conditions} one may impose on
$(V_0, Y_{0,0},\cdots)$, 
${\cal I}((V_0,Y_{0,0}))$ and ${\cal G}$ to define notions/structures that are
less general than simple current extension and more general than SVOA.
One is to require that the cohomology class $[h] \in H^3({\cal G}, \C^\times)$
is trivial. Suppose that the 3-cocycle $h$ is the coboundary of a 2-cochain
$\{ \alpha(s_{0,1},s_{0,2}) \}$ of ${\cal G}$;  then one can rescale
$Y_{s_{0,1}, \; \; s_{0,2}}$ by $\alpha(s_{0,1},s_{0,2})$ so that $h=1$ for all
$s_{0,1}$, $s_{0,2}$, $s_{0,3}$ in ${\cal G}$. Now, we assume that the
basis elements $Y_{s_{0,1},s_{0,2}}$ are chosen\footnote{
In the super-Jacobi identity (\ref{eq:super-Jacobi-SVOA}) for SVOA, we have
imposed this condition (including the rescaling of the basis elements)
by setting the coefficients of the first term and the RHS the same ($h=1$).
} %
so that $h=1$. The hexagon 
identity then implies that $c: {\cal G} \times {\cal G}\rightarrow \C^\times$
is bimultiplicative:
\begin{align}
  c(s_{0,1} + s_{0,2}, s_{0,3}) = c(s_{0,1},s_{0,3}) c(s_{0,2}, s_{0,3}), \quad
  c(s_{0,1}, s_{0,2}+s_{0,3}) = c(s_{0,1}, s_{0,2}) c(s_{0,1}, s_{0,3}).
  \label{eq:cond-bimultipl}  
\end{align}
The value $c(s_{0,1}, s_{0,2})$ is a complex phase, in particular,
as we only think of a finite group ${\cal G}$. 
The box identity ${\cal B}(\pm) = {\cal F}^{-1} \cdot
\Omega(\mp) \cdot {\cal F}$, along with a choice for $h=1$, implies that the complex numbers $c_\mp(s_{0,1},s_{0,2},s_{0,3})$ are independent of $s_{0,3} \in {\cal G}$, and are equal to the
coefficients in $\Omega(\mp):Y_{s_{0,1}, \;\; s_{0,2}} \mapsto
Y_{s_{0,2}, \;\; s_{0,1}} \; c_{\mp}(s_{0,1},s_{0,2})$. Because
$\Omega(\mp) \Omega(\mp)$ maps $Y_{s_{0,1},s_{0,2}}(z)$ to
$Y_{s_{0,1}, \;\; s_{0,2}}(e^{\pm 2\pi i} z)$ \cite{Schellekens:1989am},
\begin{align}
  c_\mp(s_{0,1}, s_{0,2}) c_{\mp}(s_{0,2},s_{0,1}) =
  e^{\pm 2\pi i (h_{s_{0,1}\cdot s_{0,2}} - h_{s_{0,1}} - h_{s_{0,2}})}. 
\end{align}

The other condition one may impose is that $Y_{s_{0,1}, \; \; s_{0,2}} \in
{\cal V}_{s_{0,1}, \;\; s_{0,2}}^{s_{0,1}\cdot s_{0,2}}$ have only integral powers of the
dummy variable $z$ for all $s_{0,1}, s_{0,2} \in {\cal G}$. It follows then
that the assignment $\hat{\phi}: {\cal G} \ni s_0 \mapsto [h_{s_0}]
\in \R/\Z$ is a group homomorphism. In the case of SVOA, we set 
$\hat{\phi}: {\cal G} \ni s_0 \mapsto [h_{s_0}] = s_0/2 +\Z \in \R/\Z$
at the very beginning.

Suppose that both conditions are imposed, and that the basis
$Y_{s_{0,1}, \;\; s_{0,2}}$
is chosen so that $h=1$; this still leaves freedom to rescale the basis
elements by a 2-cocycle (not by a general 2-cochain anymore
eg., \cite[\S3.4]{bakalov2006generalizedvertexalgebras}).  
Now, $c_-=c_+$ because $\Omega(\mp) \cdot \Omega(\pm ) = {\rm id}$;
the complex numbers $c(s_{0,1}, s_{0,2})$ depend only on permutations, not
on braidings.\footnote{
This is why we do not have to care about branch cuts and
braidings in making well-defined the second term of the super-Jacobi
identity (\ref{eq:super-Jacobi-SVOA}) for SVOA.
} %
The bimultiplicative form $c= c_-=c_+:{\cal G} \times {\cal G} \rightarrow
S^1$ is also skewsymmetric: 
\begin{align}
  c(s_{0,1},s_{0,2}) c(s_{0,2},s_{0,1})=1 . 
  \label{eq:cond-skewsym-C}
\end{align}
The residual freedom of rescaling the basis elements $Y_{s_{0,1},s_{0,2}}$---by
any one in $H^2({\cal G},\C^\times)$---changes the bimultiplicative and
skewsymmetric\footnote{
There is no guarantee (for a general finite abelian group ${\cal G}$)
that we can find an appropriate rescaling by $H^2({\cal G},\C^\times)$
to make it symmetric ($c(s_{0,1},s_{0,2}) = c(s_{0,2},s_{0,1})$).
} %
form $c$. 
Rescaling on the basis elements $Y_{s_{0,1},s_{0,2}}$ by 
the coboundary of any 1-cocycle $\psi$, changing $Y_{s_{0,1}, \; s_{0,2}}$ by
$\psi(s_{0,1})\psi(s_{0,2})/\psi(s_{0,1}\cdot s_{0,2})$, does not affect the
super-Jacobi identity.
When ${\cal G}$ is a cyclic group $\Z/T\Z$, $H^2({\cal G},\C^\times)$ is
trivial, and any bimultiplicative skewsymmetric form $c$ goes through a
homomorphism ${\cal G} \rightarrow \Z/2\Z$; $c=1$ if $N$ is odd, in particular.

Let us continue for a little more for the case both of the two independent
conditions are imposed. This category offers a class of close relatives of
fermionic SVOA.

A {\bf representation} of the structure $(V, YY)$, where
$YY = \{ Y_{s_{0,1}, s_{0,2}}\}$, has a data set
$(S_M, M, YY_M)$; $S_M$ is a set with an action of ${\cal G}$, and
a vector space $M$ has a structure $\oplus_{s \in S_M} M_{M s}$, with
each of $M_s$ being the direct sum of modules of ${\cal I}((V_0, Y_{0,0}))$. 
$YY_M = \{ Y_{s_0, s} \; | \; s_0 \in {\cal G}, \; s \in S_M \}$, where
$Y_{s_0, s}: V_{s_0} \rightarrow {\rm Hom}(M_s, \; M_{s_0 \cdot s})[[z,z^{-1}]]
z^{-\hat{h}(s_0)}$ for some $\hat{h} \in \hat{\cal G}$.
We will {\bf only deal with the case ${\cal G}$ acts faithfully on} $S_M$
in this article for multiple technical reasons, although 
the ${\cal G}$ action on $S_M$ does not have to be faithful in general.  
In the super-Jacobi identity,  
the ratio between the first term and the RHS governs the fusion process
${\cal F}: {\cal V}_{0s_{0,1}, M s_{0,2}\cdot s}^{M s_{0,1} \cdot s_{0,2}\cdot s}\otimes
{\cal V}_{0s_{0,2},Ms}^{Ms_{0,2}\cdot s} \rightarrow
{\cal V}_{0s_{0,1},0s_{0,2}}^{0s_{0,1}s_{0,2}} \otimes
{\cal V}_{0s_{0,1}s_{0,2}, Ms}^{M s_{0,1}s_{0,2}s}$, where all the vector spaces
involved are of 1-dimension. So, with a choice of a basis element
$Y_{0s_0, s} \in {\cal V}_{0s_0, Ms}^{M s_0 \cdot s}$, those isomorphisms can be described by complex numbers $h^M: {\cal G} \times {\cal G} \times S_M
\rightarrow \C^\times$. Similarly, the ratio between the first and second
terms is governed by the braiding isomorphisms ${\cal B}(\pm)$,
which are described by the complex numbers
$c_\mp^M: {\cal G} \times {\cal G} \times S_M \rightarrow \C^\times$
as we have chosen a basis of the 1-dimensional vector spaces. 
We impose the condition that $[h]=0 \in H^3({\cal G},\C^\times)$
and assume that $YY$ is chosen so that $h=1$; 
the pentagon identity then indicates that there is a way to rescale
$Y_{0s_0, s}$ so that $h^M=1$ for all of ${\cal G} \times {\cal G} \times S_M$,
if ${\cal G}$ acts on $S_M$ faithfully. 
This rescaling has been done for SVOA already in (\ref{eq:Jacobi-module-S-wR})
(the first factor of the RHS is $h^M/z_2= 1/z_2$).
Suppose that we have made this choice of basis elements. Then the maps
$c^M_{\mp}: {\cal G} \times {\cal G} \times S_M \rightarrow
\C^\times$ are independent of the last argument $s \in S_M$, and agree with
that of the structure $(V, \{ Y_{s_{0,1}, \;\; s_{0,2}}\})$, as we can use
the relation ${\cal B}(\pm) =
{\cal F}^{-1} \cdot \Omega(\mp) \cdot {\cal F}$.
With the other condition ($YY$ takes value in ${\rm End}(V)[[z,z^{-1}]]$
with integral power of $z$) imposed, the distinction between $c^M_-$ and $c^M_+$
is also gone.  In the case of SVOA, the first factor $(-1)^{s_{0,1}s_{0,2}}/z_0$
of the second term of (\ref{eq:Jacobi-module-S-wR}) is the same as that of
(\ref{eq:super-Jacobi-SVOA}).

Let us move on to {\bf intertwining operators} of the structure
$(V, \{ Y_{s_{0,1}, \; s_{0,2}}\})$. We only think of simple representations
$\alpha$, $\beta$, $\gamma$ where ${\cal G}$ acts faithfully on $S_\alpha$,
$S_\beta$, $S_\gamma$ here to avoid technical problems. So we have
a set-theoretical map $f[\alpha;\beta,\gamma]: S_\beta \times S_\gamma
\rightarrow S_\alpha$ compatible with the action of ${\cal G}$, which
governs in the way ${\cal V}_{\beta s_\beta, \gamma s_\gamma}^{\alpha f(s_\beta,s_\gamma)}$
is non-empty. In the super-Jacobi identity defining ${\cal YY} = \{
{\cal Y}_{s_\beta, s_\gamma} \}$, we do not have a complex number
$h^{\alpha;\beta,\gamma}(s_0, s_\beta, s_\gamma)$ for
$(s_0,s_\beta,s_\gamma) \in {\cal G} \times S_\beta \times S_\gamma$, but a
vector-space isomorphism
${\cal V}_{\beta s_\beta,\gamma s_\gamma}^{\alpha f(s_\beta,s_\gamma)} \ni {\cal Y} \mapsto
h^{\alpha;\beta,\gamma}({\cal Y}) \in {\cal V}_{\beta s_0 \cdot s_\beta,\gamma s_\gamma}^{\alpha s_0 \cdot f(s_\beta,s_\gamma)}$ given by ${\cal F}: {\cal V}_{0 s_0, \alpha f(s_\beta,s_\gamma)}^{\alpha s_0 f(s_\beta,s_\gamma)} \otimes {\cal V}_{\beta s_\beta, \gamma s_\gamma}^{\alpha f(s_\beta,s_\gamma)} \ni Y_{s_0, f(s_\beta,s_\gamma)} \otimes {\cal Y} \mapsto
Y_{s_0, s_\beta} \otimes h^{\alpha;\beta,\gamma}({\cal Y}) \in {\cal V}_{0 s_0, \beta s_\beta}^{\beta s_0 \cdot s_\beta} \otimes {\cal V}_{\beta s_0 \cdot s_\beta, \; \gamma s_\gamma}^{\alpha s_0 f(s_\beta, s_\gamma)}$.
When $[h]=0 \in H^3({\cal G},\C^\times)$ and $\{ Y_{s_{0,1}, \; s_{0,2}} \}$ and
$\{ Y_{s_0, s} \; | \; s_0 \in {\cal G}, \; s \in S_M\}$ are chosen so that
$h=h^M=1$ (as we have assumed), the pentagon identity implies that
the network of the vector-space isomorphisms $h^{\alpha;\beta,\gamma}(-,-,s_\gamma)$
is consistent in the composition,
\begin{align}
h^{\alpha;\beta,\gamma}(s_{0,1}, s_{0,2} s_\beta, s_\gamma) \cdot h^{\alpha;\beta,\gamma}(s_{0,2}, s_\beta, s_\gamma)  
=
h^{\alpha;\beta,\gamma}(s_{0,1} s_{0,2}, s_\beta, s_\gamma)
\end{align}
for each $s_\gamma \in S_\gamma$. 
Moreover, $c_{\mp}^{\alpha;\beta,\gamma}(s_0, s_\beta, s_\gamma)$ for
$(s_0, s_\beta, s_\gamma) \in
{\cal G} \times S_\beta \times S_\gamma$ is not a complex number, but
a vector-space isomorphism
${\cal V}_{\beta s_\beta, \; \gamma s_\gamma}^{\alpha f(s_\beta,s_\gamma)} \rightarrow
{\cal V}_{\beta s_\beta, \; \gamma s_0 \cdot s_\gamma}^{\alpha s_0 \cdot f(s_\beta,s_\gamma)}$. 
That this diagram 
\begin{equation}
  \xymatrix{
  {\cal V}_{(0s_{0,1} \cdot (0s_{0,2} \cdot (\beta s_\beta \cdot \gamma s_\gamma)))}
  \ar[r]^{\cal F} \ar[dr]^{{\cal B}(\pm)} &
     {\cal V}_{((0s_{0,1}\cdot 0s_{0,2})\cdot (\beta s_\beta \cdot \gamma s_\gamma))}
     \ar[r]^{{\cal B}(\pm)} &
        {\cal V}_{(\beta s_\beta \cdot ((0s_{0,1}\cdot 0s_{0,2})\cdot \gamma s_\gamma))}
  \ar[d]^{{\cal F}^{-1}}  \\
   &  {\cal V}_{(0s_{0,1} \cdot (\beta s_\beta \cdot (0s_{0,2} \cdot \gamma s_\gamma)))}
     \ar[r]^{{\cal B}(\pm)} & 
  {\cal V}_{(\beta s_\beta \cdot (0s_{0,1} \cdot (0s_{0,2}\cdot \gamma s_\gamma)))}
  }
\end{equation}
commutes because of the hexagon identity implies, under the same
set of assumptions, that 
\begin{align}
  c^{\alpha;\beta,\gamma}_{\mp} (s_{0,1}\cdot s_{0,2}, s_\beta, s_\gamma) =
  c^{\alpha;\beta,\gamma}_{\mp} (s_{0,1}, s_\beta, s_{0,2} \cdot s_\gamma) \cdot
  c^{\alpha;\beta,\gamma}_{\mp} (s_{0,2},s_\beta, s_\gamma). 
\end{align}
The network of vector-space isomorphisms
$c^{\alpha;\beta,\gamma}_{\mp}(-,s_\beta,-)$ for each
$s_\beta \in S_\beta$ is consistent in the composition.

An idea for intertwining operators ${\cal YY}$ of the structure
$(V, \{ Y_{s_{0,1}, \;\; s_{0,2}}\})$ is to identify correlation among
the $(\#{\cal G})^2$ components of ${\cal YY} =
\{ {\cal Y}_{s_\beta,s_\gamma} \in {\cal V}_{\beta s_\beta, \gamma s_\gamma}^{\alpha f(s_\beta, s_\gamma)} \;
| \; s_\beta \in S_\beta, \; s_\gamma \in S_\gamma\}$. The network of
isomorphisms above are therefore useful for this purpose, but the full
network of $c^{\alpha;\beta,\gamma}_-$ (or alternatively $c^{\alpha;\beta,\gamma}_+$)
and $h^{\alpha;\beta,\gamma}$ has a mismatch, however. This commuting diagram 
\begin{equation}
  \xymatrix{
    {\cal V}_{(0 s_{0,1} \cdot (0s_{0,2} \cdot (\beta s_\beta \cdot \gamma s_\gamma)))}
    \ar[d]^{{\cal F}} \ar[r]^{{\cal B}(\pm)} &
       {\cal V}_{(0 s_{0,2} \cdot (0s_{0,1} \cdot (\beta s_\beta \cdot \gamma s_\gamma)))}
       \ar[r]^{{\cal B}(\pm)} &
          {\cal V}_{(0 s_{0,2} \cdot (\beta s_\beta \cdot ( 0s_{0,1}\cdot \gamma s_\gamma)))}
          \ar[dl]^{{\cal F}}  \\
   {\cal V}_{(0 s_{0,1} \cdot ((0s_{0,2} \cdot \beta s_\beta) \cdot \gamma s_\gamma))}
    \ar[r]^{{\cal B}(\pm)} &
       {\cal V}_{((0s_{0,2} \cdot \beta s_\beta) \cdot (0 s_{0,1} \cdot \gamma s_\gamma))}
  }
\end{equation}
implies that the two routes of vector-space isomorphisms
${\cal V}_{\beta s_\beta, \gamma s_\gamma}^{\alpha f(s_\beta, s_\gamma)}
\rightarrow {\cal V}_{\beta s_{0,2} \cdot s_\beta, \gamma s_{0,1}\cdot s_\gamma}^{\alpha s_{0,1}s_{0,2}f(s_\beta,s_\gamma)}$ using the network of isomorphisms $c^{\alpha;\beta,\gamma}$
and $h^{\alpha;\beta,\gamma}$ may disagree by an overall scalar factor: 
\begin{align}
  c(s_{0,1}, s_{0,2}) & \; 
  \left( h^{\alpha;\beta,\gamma}(s_{0,2}, s_\beta, s_{0,1} \cdot s_\gamma) \cdot
  c^{\alpha;\beta,\gamma}_{\mp}(s_{0,1}, s_\beta, s_\gamma) \right)
   \label{eq:smpl-crnt-ch-mismatch}  \\
   & \quad
   = c^{\alpha;\beta,\gamma}_{\mp} (s_{0,1}, s_{0,2} \cdot s_\beta, s_\gamma) \cdot
  h^{\alpha;\beta,\gamma} (s_{0,2}, s_\beta, s_\gamma).  \nonumber
\end{align}
So, the correlation among the $(\#{\cal G})^2$ components of ${\cal YY} \in
{\cal V}_{\beta,\gamma}^\alpha(r)|_{2r+1= \mp 1}$ is set by choosing $s_{\beta *}
\in S_\beta$ as a reference point, using the network of isomorphisms
$c^{\alpha;\beta,\gamma}_{\mp}(-, s_{\beta *},-)$ among
${\cal V}_{\beta s_{\beta *}, \gamma s_\gamma}^{\alpha f(s_{\beta *}, s_\gamma)}$
for ${}^\forall s_\gamma \in S_\gamma$, and finally by using
the network of isomorphisms $h^{\alpha;\beta,\gamma}(-,-,s_\gamma)$ among
${\cal V}_{\beta s_\beta, \gamma s_\gamma}^{\alpha f(s_\beta,s_\gamma)}$ with
${}^\forall s_\beta \in S_\beta$ for each one of $s_\gamma \in S_\gamma$.
If one is to track the network of
isomorphisms in a different route, by changing $s_\beta$ from $s_{\beta *}$
to $s_{0,2} \cdot s_{\beta *}$ by $h^{\alpha;\beta,\gamma} (-,-,s_\gamma)$ first
and then changing $s_\gamma$ to $s_{0,1} \cdot s_\gamma$ by
$c_{\mp}^{\alpha;\beta,\gamma}$ later, then the component
${\cal Y}_{s_\beta,s_\gamma}$ of
${\cal YY} \in {\cal VV}_{\beta,\gamma}^\alpha(r)|_{2r+1=\mp 1}$ is mapped
to $c(s_{0,1}, s_{0,2})$ times the component
${\cal Y}_{s_\beta, s_{0,1} \cdot s_\gamma}$ of the same ${\cal YY}$. 
We have seen this already in (\ref{eq:temp-int.op-S-wR-byB-inME}),
and as the coefficient $(-1)^{+s_0 i_\beta (s_\beta)}$ in
(\ref{eq:def-vct-sp-int.op-S}). The choice of $s_{\beta *} \in S_\beta$
has introduced a set-theoretical identification
$i_\beta: S_\beta  \cong {\cal G}$ via $s_\beta = s_0 \cdot s_{\beta *}$, 
and the mismatch factor $c_\mp(s_{0,1}, s_{0,2}) \in \C^\times$ can be written as
$c_{\mp}(s_{0,1}, i_\beta(s_\beta))$. 
Artificial choice of $i_\beta: S_\beta \rightarrow {\cal G}$ and $r$
in the definition of intertwining operators ${\cal VV}_{\beta, \gamma}^\alpha$
therefore do not go away in this category with a general group ${\cal G}$.

It is likely that the discussions on skew-symmetry, crossing-symmetry,
fusion and braiding isomorphisms for SVOA can be translated into this
category (with the two conditions imposed, and faithful action of ${\cal G}$
on $S_M$). Most of the discussions prior
to \ref{statmnt:genG-w2cond} can also be read for this class of structures
(close relatives of SVOA), with a general finite abelian group ${\cal G}$
and a bimultiplicative and skewsymmetric form $c$. 

The condition (\ref{eq:xing-SVOA-Y-vs-Y-vs-lambda-m1}) on
$\{ Y_{s_{0,1}, \; s_{0,2}} \}$ and $\{ \lambda_{s_0}\}$ can be implemented
systematically in this class of structures. Representations of the
underlying VOA $(V_0, Y_{0,0})$ determines $Y_{0,s_{0,2}}$ with ${}^\forall
 s_{0,2} \in {\cal G}$, and the generalized skew-symmetry (i-SVOA) sets
 $Y_{s_{0,1},0} = (\Omega Y_{0,s_{0,1}})$. A part of the crossing symmetry
 condition (\ref{eq:xing-SVOA-Y-vs-Y-vs-lambda-m1}),
 $Y_{s_{0}, -s_0} \propto \lambda_0 \cdot (\Theta_{0} Y_{s_{0}, 0}) \cdot
 \lambda_{s_0}^{-1}$, does not imply very much, because the vector space
 ${\cal V}_{s_0, -s_0}^0$ is of 1-dimension in any simple current extension,
 and we know that it is generated by $(\Theta_{0} \Omega Y_{0,s_0})$; no matter
 how we choose $Y_{s_0, -s_0}$ (in a way the super-Jacobi identity for
 ${\cal G}$ and $c$ is satisfied), $\lambda_{s_0}$ is determined accordingly
 (the arbitrariness of scalar multiplication on $\lambda_{s}$ that remains
   in $\lambda_s (\Theta Y_{0,s}) \lambda_s^{-1} = Y_{0,-s}$ is fixed). 
 Once we know that $Y_{s,-s}$ is chosen this way, then we can use this as
 a part of input of \ref{statmnt:Theta-by-FnOmega}, along with generalized
 skew-symmetry (i-``SVOA'') and
 generalized ``super''-Jacobi identity ($\ell$/k-``SVOA''), to see that
 the rest of the
 relation (\ref{eq:xing-SVOA-Y-vs-Y-vs-lambda-m1}) follows automatically.
 The condition (\ref{eq:xing-SVOA-Y-vs-Y-vs-lambda-m1}) determines
 $\{\lambda \}$ completely for a given $YY$, but does not constrain $YY$
 more than other conditions do.

The condition (\ref{eq:xing-SVOA-Y-vs-Y-vs-lambda-m1})---if we impose
it---demands that $[h_{s_0}]$ for $s_0 \in {\cal G}$ is either integral or
half-odd (in $2^{-1}\Z/\Z$) and $\epsilon_{s_0}$ not arbitrary, as we have
seen already at the end of the SVOA section of \ref{statmnt:xing-S}.
The presentation here therefore fails to explain how to introduce
crossing-symmetry isomorphisms and invariance for the whole general
simple current extensions subject to the two conditions
(vanishing 3rd cohomology and $\hat{\phi}: {\cal G} \rightarrow \R/\Z$
as a homomorphism). This also means that the combination of the
(p-SVOA) and (q-SVOA) conditions (and its equivalent translations)
offers a way to introduce unitarity structure only for the subclass
of simple current extensions with this special property of
$[h_{s_0}]$ and $\epsilon_{s_0}$. 
\end{anythng}   

\subsection{Fermionic 2D CFT and its Type 0 (Gauged) Version}
\label{ssec:fCFT}

In section \ref{ssec:fCFT}, descriptions may sometimes be written for
fermionic 2D CFTs and SVOAs, but notations/expressions are valid for
a broader class of simple current extensions with a general finite abelian
group ${\cal G}$ characterized by the vanishing 3rd cohomology,
bi-multiplicative and skewsymmetric form $c$ (equivalently,
$\hat{\phi}: {\cal G} \rightarrow \R/\Z$ being a homomorphism),
fixed point free action on $S_\gamma$'s, and (\ref{eq:reltn-eps-c-h}). 
\begin{anythng}
\label{statmnt:data-4-fCFT}
{\bf defining data of a fermionic modular invariant 2D CFT:}
To construct a {\it 2D fermionic modular invariant CFT}, we used
a set of algebraic data of the following form:
\begin{align}
  & {\cal A} = (V, Y, {\bf 1}, \omega, \cdots), \quad
  \widetilde{\cal A} =
  (\widetilde{V}, \widetilde{Y}, \tilde{\bf 1}, \tilde{\omega}, \cdots ), \quad
  \amalg_{\hat{h} \in \hat{\cal G}} {\cal II}^{\hat{h}}
  \quad \amalg_{\tilde{\hat{h}} \in \hat{\widetilde{\cal G}}}
  \widetilde{\cal II}^{\tilde{\hat{h}}}, \\
  &
  \{ \sigma_\alpha : S_\alpha \rightarrow \widetilde{S}_{\pi(\alpha)} \; |
  \alpha \in {\cal II}^{\hat{0}} \amalg {\cal II}^{\hat{g}} \}, \quad
  \pi: {\cal II}^{\sigma_0^*(\hat{\tilde{h}})} \rightarrow
  \widetilde{\cal II}^{\hat{\tilde{h}}}, 
\quad
  \{ YY_{\rm tot} \}, \nonumber \\
  & \left\{ (S_\alpha, M_\alpha, YY_\alpha, \langle - | - \rangle_\alpha, \lambda_\alpha) \; | \; \alpha \in {\cal II}^{\hat{h}} \right\}, \quad
  \left\{ (\widetilde{S}_{\tilde{\alpha}}, \widetilde{M}_{\tilde{\alpha}}, \widetilde{YY}_{\tilde{\alpha}}, \langle - | - \rangle_{\tilde{\alpha}}, \tilde{\lambda}_{\tilde{\alpha}}) \; | \; \alpha \in \widetilde{\cal II}^{\tilde{\hat{h}}} \right\}. 
  \nonumber 
\end{align}
Here, $(V,Y, {\bf 1},\cdots)$ and $(\widetilde{V}, \widetilde{Y},
\tilde{\bf 1},\cdots)$ are left-moving and right-moving fermionic SVOA of
central charge $c$ and $\tilde{c}$, respectively.
${\cal II}^{\hat{0}}$ and $\widetilde{\cal II}^{\tilde{\hat{0}}}$ are a set
  of distinct equivalence classes of unitary and untwisted (NS)
  representations of the two SVOAs, while ${\cal II}^{\hat{g}}$
[resp. $\widetilde{\cal II}^{\tilde{\hat{g}}}$]
a set of distinct classes of unitary $\hat{g}$-
[resp. $\tilde{\hat{g}}$-]twisted
(Ramond) representations of the corresponding SVOA (or twisted representations
more generally for $\hat{h} \in \hat{\cal G}$).
For each equivalence class, choose one representative along with its
unitary structure $(\langle -|-\rangle_\alpha, \lambda_\alpha)$.
We adopt a description along the line of the observation in
Rmk. \ref{rmk:Moore-Seiberg}, so we do not include a multiplicity matrix
(as in \ref{statmnt:data-4-CFT} and eq. (\ref{eq:str-Htot-bos-byVOA})),
but set-theoretical one-to-one maps $\pi$, where $\pi(0) = \tilde{0}
\in \widetilde{\cal II}^{\tilde{\hat{0}}}$ and $\pi(\Upsilon^{\cal II}(\alpha)) =
\widetilde{\Upsilon}^{\widetilde{\cal II}}(\pi(\alpha))$, as in the argument around
eq. (\ref{eq:str-Htot-bos-byChiAlg}).
Each of $\{ \sigma_\alpha \}$ in the data is a set-theoretical
map between $\#{\cal G}$-element sets; one of them $\sigma_0:
S_0 \rightarrow \widetilde{S}_{\tilde{0}}$ should be the group
isomorphism ${\cal G} \cong \widetilde{\cal G}$ where
the pull-back of the bi-multiplicative skewsymmetric form $c$
(\ref{eq:fermion-exch-phase}) of $\widetilde{\cal G}$ by $\sigma_0$ agrees
with that of ${\cal G}$;  
the map $\sigma_\alpha$ needs to be compatible with the group action,
$\sigma_\alpha( s_0 \cdot s_\alpha) = \sigma_0(s_0) \cdot \sigma_\alpha(s_\alpha)$;
the set of maps $\{\sigma_\alpha\}$ should be compatible with the
fusion algebra on both sides (cf Rmk. \ref{rmk:no-systematic-Z2})
\begin{align}
  \sigma_\alpha \left( f[\alpha;\beta,\gamma](s_\beta,s_\gamma) \right) =
  \tilde{f}[\pi(\alpha);\pi(\beta),\pi(\gamma)]
  (\sigma_\beta(s_\beta),\sigma_\gamma(s_\gamma))
\end{align}
for all the triples $\alpha, \beta,\gamma$ in $\amalg_{\hat{h}}
{\cal II}^{\hat{h}}$ with non-empty ${\cal VV}_{\beta,\gamma}^\alpha$
and $\widetilde{\cal VV}_{\pi(\beta),\pi(\gamma)}^{\pi(\alpha)}$. 
$\{ YY_{\rm tot}\}$ is a choice of intertwining operators,
which we will elaborate more later in \ref{statmnt:data-4-fCFT-3} and
\ref{statmnt:data-4-fCFT-4}. 

With one set of defining algebraic data, we may construct two different
{\it Hilbert space}s:\footnote{
To be rigorous, we should distinguish the completion of ${\cal H}_{\rm tot}$
in (\ref{eq:Htot-SCFT-wF}, \ref{eq:Htot-SCFT-wR}) wrpt the positive definite
Hermitian inner product from the direct sum; the lower-boundedness property of
the direct sum is lost after the completion. 
} %
one is 
\begin{align}
  {\cal H}_{\rm tot}^{\rm pre} \cong \oplus_{\alpha \in {\cal II}^0} \left(
  M_\alpha \otimes \widetilde{M}_{\pi(\alpha)} \right).
  \label{eq:Htot-SCFT-wF}
\end{align}
States in $M_{\alpha s_\alpha} \otimes \widetilde{M}_{\pi(\alpha) \sigma_\alpha(s_\alpha)}$
are regarded bosonic, and those in $M_{\alpha 1 \cdot s_\alpha} \otimes
\widetilde{M}_{\pi(\alpha) \sigma_\alpha(s_\alpha)}$ regarded fermionic. The other is 
\begin{align}
  {\cal H}_{\rm tot}^{\rm post} \cong \oplus_{\hat{h} \in \hat{\cal G}}
  \left( \oplus_{\alpha \in {\cal II}^{\hat{h}}} \oplus_{s_\alpha \in S_\alpha} \left(
  M_{\alpha s_\alpha} \otimes \widetilde{M}_{\pi(\alpha) \sigma_\alpha(s_\alpha)} \right) \right).
  \label{eq:Htot-SCFT-wR}
\end{align}

In physics terminology, ${\cal H}_{\rm tot}^{\rm pre}$ is the one before
gauging the diagonal fermion parity, and ${\cal H}_{\rm tot}^{\rm post}$ the one
after gauging it. One will also note that one may change the set of
maps $\{ \sigma_\alpha\}$ by replacing all of $\sigma_{\alpha}$ for $\alpha$ in
the Ramond sector (${}^\forall \alpha \in {\cal II}^{\hat{g}}$) by the opposite
one, and yet the new set of maps is compatible with the fusion algebra.
The resulting
${\cal H}_{\rm tot}^{\rm post}$ is different from the one before. The relation
between the two choices of $\{ \sigma_\alpha \}$ (and of
${\cal H}_{\rm tot}^{\rm post}$) is known as the distinction between Type 0A and
Type 0B CFTs. The generalization of this Type 0A-vs-0B variation
is parametrized by ${\rm vr} \in {\rm Hom}(\hat{\cal G}, \widetilde{\cal G})$
for a general finite abelian group ${\cal G}$; $\sigma_\alpha$ for
$\alpha \in {\cal II}^{\hat{h}}$ is shifted by ${\rm vr}(\hat{h})$.

A Hermitian inner product on ${\cal H}_{\rm tot}^{\rm pre/post}$ and a
CPT isomorphism $\Upsilon_{\rm tot}$ on ${\cal H}_{\rm tot}^{\rm pre/post}$
are supposed to be given by those on $M_\alpha$ and
$\widetilde{M}_{\pi(\alpha)}$; naively, that would be just like in the case
of 2D CFTs. In fact, there are a couple of subtleties involved, so
we need to postpone this discussion and come back to this issue
at \ref{statmnt:data-4-fCFT-2}, and also at
\ref{statmnt:out-going-states+can-iso}. 

The condition (s) of a 2D modular invariant CFT is modified as follows:
the four character functionals on $V_0 \otimes \widetilde{V}_0$,
\begin{align}
  \sum_{\alpha \in {\cal II}^0, s_\alpha\in S_\alpha} \sum_{s_0 \in {\cal G}} & \; 
  \chi_{\alpha s_0 \cdot s_\alpha} (v, \tau)
  \tilde{\chi}_{\pi(\alpha) \sigma_\alpha(s_\alpha)}(\tilde{v},\tau),  \nonumber \\
  \sum_{\alpha \in {\cal II}^0, s_\alpha\in S_\alpha} \sum_{s_0 \in {\cal G}} & \; 
  (-1)^{s_0} \; \chi_{\alpha s_0 \cdot s_\alpha}(v, \tau)
  \tilde{\chi}_{\pi(\alpha) \sigma_\alpha(s_\alpha)}(\tilde{v},\tau), \nonumber \\
  \sum_{\alpha \in {\cal II}^{\hat{g}} s_\alpha \in S_\alpha} \sum_{s_0 \in {\cal G}} & \; 
  \chi_{\alpha s_0 \cdot s_\alpha} (v, \tau)
  \tilde{\chi}_{\pi(\alpha) \sigma_\alpha(s_\alpha)}(\tilde{v},\tau), \nonumber \\
  \sum_{\alpha \in {\cal II}^{\hat{g}} s_\alpha \in S_\alpha} \sum_{s_0 \in {\cal G}} & \; 
  (-1)^{s_0} \; \chi_{\alpha s_0 \cdot s_\alpha}(v, \tau)
  \tilde{\chi}_{\pi(\alpha) \sigma_\alpha(s_\alpha)}(\tilde{v},\tau),  
\end{align}
for $v \in V_0$, $\tilde{v} \in \widetilde{V}_0$ and $\tau \in \mathfrak{H}$, 
transform under the $({\rm perm} \otimes (\chi_\eta)^{-(c-\tilde{c})})$
representation of ${\rm SL}_2\Z$; $\chi_\eta$ has been explained
below (\ref{eq:cond-mod-inv}) in \ref{statmnt:data-4-CFT},
and $({\rm perm})$ is the permutation represenation of
${\rm SL}_2\Z/\Gamma(2) \cong {\rm SL}_2(\Z/2\Z)$ on the set $(\Z/2\Z)^2$.
It is an option to demand only the invariance of the first one
under $\Gamma(2)$ and the $S$-transformation, if we are
interested only in ${\cal H}_{\rm tot}^{\rm pre}$, not in
${\cal H}_{\rm tot}^{\rm post}$ (then we do not need to include
the set ${\rm II}^{\hat{g}}$, the map $\pi: {\cal II}^{\hat{g}}\rightarrow
\widetilde{\cal II}^{\tilde{\hat{g}}}$, and $\{\sigma_\alpha \; | \; \alpha
\in {\cal II}^{\hat{g}} \}$).
For a general finite abelian group ${\cal G}$, we have $(\#\hat{\cal G})^2$
character functionals of $V_0 \otimes \widetilde{V}_0$; the sum over
$\alpha \in {\cal II}^{\hat{0}}$ or ${\cal II}^{\hat{g}}$ is replaced by
$\alpha \in {\cal II}^{\hat{h}}$ for $\hat{h} \in \hat{\cal G}$,
and the insertion of $(\pm 1)^{s_0}$ by $e^{2\pi i \hat{k}(s_0)}$ for
$\hat{k} \in \hat{\cal G}$. The permutation representation of
${\rm SL}_2\Z$ acts on $\hat{\cal G} \times \hat{\cal G}$, and
factors through ${\rm SL}_2(\Z/T\Z)$ when all the elements of $\hat{\cal G}$
are $T$-torsion. 
\end{anythng}

\begin{anythng}
A fermionic 2D CFT is said to be $N \geq (1,1)$ 2D {\it superconformal
 field theory} (SCFT) when a special element $[g] \in V_1$
and $[\tilde{g}] \in \widetilde{V}_1$ are identified, so that
$(V, YY, {\bf 1}, \omega, g)$ is a left-mover $N\geq 1$ SVOA and
$(\widetilde{V}, \widetilde{YY}, \tilde{\bf 1}, \tilde{\omega}, \tilde{g})$
also a right-mover $N\geq 1$ SVOA. 

Physics literature only demands the existence of an element of the property
required for $g$ and $\tilde{g}$, but does not identify and fix them as
a part of defining data of an $N\geq (1,1)$ SCFT, contrary to the math-style
definition above. Those two different usage
of terminology (and the ways to grasp objects) have their own advantages. 
\end{anythng}

\begin{anythng}
{\bf the diag($\hat{\cal G} \times \widetilde{\hat{\cal G}}$)-bundles:}
Let us take a moment to slow down, as we have not been explicit enough
about conventions associated with the right-moving SVOA (or superchiral algebra
$\widetilde{\cal A}$). The group $\widetilde{\cal G}$ governs
the group-grading of $\widetilde{V}$; the conformal weights
$[\tilde{h}_{\tilde{s}_0}] = \hat{\tilde{\phi}}(\tilde{s}_0) \in \R/\Z$ must be
the same as that $[h_{s_0}] = \hat{\phi}(s_0)$ when $\sigma_0(s_0) =
\tilde{s}_0$. The automorphism group of the right-mover SVOA has 
$\widetilde{\hat{\cal G}} = {\rm Hom}(\widetilde{\cal G},\C^\times) \cong
{\rm Hom}(\widetilde{\cal G}, \R/\Z)$, and its element $\tilde{\hat{h}}$
multiplies
$e^{2\pi i \tilde{\hat{h}}(\tilde{s}_0)}$ on $\widetilde{V}_{\tilde{s}_0}$.
In an $\tilde{\hat{h}}$-twisted representation $(\widetilde{S}_M,
\widetilde{M},\widetilde{Y}_M)$, $\widetilde{Y}_M$ maps
$\widetilde{V}_{\tilde{s}_0}$
to $\oplus_{\tilde{s} \in \widetilde{S}_M} {\rm Hom}(\widetilde{M}_{\tilde{s}} ,
\widetilde{M}_{\tilde{s}_0 \cdot \tilde{s}})[[\bar{z},\bar{z}^{-1}]]\bar{z}^{-\tilde{\hat{h}}(\tilde{s}_0)}$. All of those conventions should have been mentioned
explicitly
to give an unambiguous meaning in the condition that $\sigma_0: S_0 \rightarrow \widetilde{S}_0$ should be the group isomorphism
${\cal G} \cong \widetilde{\cal G}$, and in the condition
that the set-theoretical map $\pi$ is from
${\cal II}^{\hat{h}}$ to $\widetilde{\cal II}^{\sigma_0(\hat{h})}$ 
for $\hat{h} \in \hat{\cal G}$ (we meant by $\sigma_0(\hat{h})$
what should be written as $(\sigma_0^*)^{-1}(\hat{h})$). 

In the theory with ${\cal H}_{\rm tot}^{\rm pre}$, there are no twisted states.
So, the bundle ${\cal P}\otimes \widetilde{\cal P}$ on a Riemann surface
with the structure group ${\cal G}\times \widetilde{\cal G}$ is trivial. 
We have a well-defined notion of inserting a state in the vector space
${\cal H}_{\rm tot}^{\rm pre}$ (rather than a section of a vector bundle).
Now, for any $\hat{h} \in \hat{\cal G}$ and any $\tilde{\hat{h}} \in
\widetilde{\hat{\cal G}}$, think of inserting
one state in the $(\hat{h}, \tilde{\hat{h}})$-twisted
representation at a point $p$ of a Riemann surface. We then
have ${\rm Bdl}(V\otimes \widetilde{V})$ on the Riemann surface from which
the point $p$ is removed; we can track a section associated with a state
$u\otimes \tilde{u} \in V \otimes \widetilde{V}$ along a clockwise loop
around $p$ to come back to the section that is $\hat{h}\tilde{\hat{h}}^{-1}$
times the section we started with; this is because\footnote{
Whether the twist is $\hat{h}\tilde{\hat{h}}^{-1}$ or
$\hat{h}\tilde{\hat{h}}$ does not make a practical difference
so long as one is interested only in ${\cal G} \cong \Z/2\Z$ as in SVOA,
although they are different for a more general ${\cal G}$.
} %
$\widetilde{Y}(\tilde{u})_{z@(e^{-2\pi i}(z_0-z(p)))}
= \widetilde{Y}(\tilde{u})_{\bar{z}@(e^{2\pi i}(\bar{z}_0-\bar{z}(p)))}
= \widetilde{Y}(\tilde{u})_{\bar{z}@(\bar{z}_0-\bar{z}(p))}
e^{-2\pi i\tilde{\hat{h}}(\tilde{s}_0)}$ when
$\tilde{u} \in \widetilde{V}_{\tilde{s}_0}$. 
So, the fact that the vector space
${\cal H}_{\rm tot}^{\rm post}$ involves only a pair of twists of the form
$(\hat{h}, \tilde{\hat{h}}) = (\hat{h}, \sigma_0(\hat{h}))$ implies
that we only have bundles whose structure group is the diagonal group of
$(\hat{\cal G} \times \widetilde{\hat{\cal G}})$, and 
the subbundle ${\rm Bdl}(\oplus_{s_0 \in {\cal G}} V_{s_0} \otimes
\widetilde{V}_{\sigma_0(s_0)})$ of ${\rm Bdl}(V\otimes \widetilde{V})$ 
is a trivial bundle; other subbundles
${\rm Bdl}(\oplus_{s_0 \in {\cal G}} (V_{s'_0 \cdot s_0} \otimes
\widetilde{V}_{\sigma_0(s_0)}))$ are non-trivial ($s'_0 \neq 0 \in {\cal G}$).
The Hilbert space ${\cal H}_{\rm tot}^{\rm post}$ is
designed to retain only the part where the diagonal
$(\hat{\cal G} \times \widetilde{\hat{\cal G}})$-bundle is trivial. 

As we have stated already in \ref{statmnt:data-4-fCFT}, the bi-multiplicative
skew-symmetric form $c$ on $\widetilde{\cal G}$ in the super-Jacobi identity of the right-mover SVOA $(\widetilde{V}, \widetilde{YY},\cdots)$ is required to agree with that of the left-mover SVOA, when pulled back by $\sigma_0$;
the super-Jacobi identity for the right mover SVOA differs from that of
the left-mover only by the use of dummy variables $\bar{z}$ instead of $z$. 
This means that the generalized skew-symmetry isomorphism (and the invariance
condition (i-SVOA)) for the right-mover should also use the same
$c(\tilde{s}_{0,1},\tilde{s}_{0,2})^{-1}$  or
$c(\tilde{\imath}_{\pi(\beta)}(\tilde{s}_{\pi(\beta)}),
\tilde{\imath}_{\pi(\gamma)}(\tilde{s}_{\pi(\gamma)}))^{-1}$
as in (\ref{eq:def-skew-iso-S}). The discussion on the crossing
symmetry in \ref{statmnt:xing-S} also works for the right-mover without
a modification, when we use 
$c(\tilde{\imath}_{\pi(\beta)}(\tilde{s}_{\pi(\beta)}), \tilde{\imath}_{\pi(\gamma)}(\tilde{s}_{\pi(\gamma)}))^{-1}$ in (\ref{eq:xing-sym-iso-S},
\ref{eq:gen-xing-iso-S-wR}) (cf also
footnote \ref{fn:right-mover-opp-sign-4-z/i}).
\end{anythng}

\begin{anythng}
\label{statmnt:data-4-fCFT-2}
{\bf preparation for \ref{statmnt:data-4-fCFT-3}:} 
There are a few more details that need to be spelled out
in dealing with the algebra of both the left-mover and right-mover SVOAs
simultaneously. We begin with representation, and move on to intertwining
operators. 

Let $(S_\gamma, M_\gamma, Y_\gamma)$ and
$(\widetilde{S}_{\pi(\gamma)}, \widetilde{M}_{\pi(\gamma)},
\widetilde{Y}_{\pi(\gamma)})$ be a pair of representations
of the left-mover and right-mover SVOAs. Choose one set-theoretical
identification $i_\gamma: S_\gamma \cong {\cal G}$; we make it a rule
in this article to choose a set-theoretical identification
$\tilde{\imath}_{\tilde{\gamma}}: \widetilde{S}_{\tilde{\gamma}} \cong {\cal G}$
so that 
$\tilde{\imath}_{\pi(\gamma)} \cdot \sigma_\gamma = i_\gamma$. 
Then a representation $(Y_\gamma \otimes \widetilde{Y}_{\pi(\gamma)})^{(i_\gamma)}$
of $V\otimes_\C \widetilde{V}$ is built as follows on a vector space
denoted by $(M_\gamma \otimes \widetilde{M}_{\pi(\gamma)})^{(i_\gamma)}$ that is
isomorphic to $M_\gamma \otimes_\C \widetilde{M}_{\pi(\gamma)}$;
for an element $(m \otimes \tilde{m})^{(i_\gamma)}$
 in $(M_\gamma \otimes \widetilde{M}_{\pi(\gamma)})^{(i_\gamma)}$ labeled by
$m \in M_{\gamma s_\gamma}$ and $\tilde{m} \in
  \widetilde{M}_{\pi(\gamma) \tilde{s}_\gamma}$, set 
\begin{align}
  (Y_\gamma\otimes \widetilde{Y}_{\pi(\gamma)})^{(i_\gamma)}
  (v \otimes \tilde{v}, z) & \; \cdot
  (m \otimes \tilde{m})^{(i_\gamma)} := \\
  & \left[ \left( Y_\gamma(v)_{z@z}\cdot m\right)
  \otimes
  \left(\widetilde{Y}_{\pi(\gamma)}( \tilde{v})  \cdot \tilde{m} \right)
  \right]^{(i_\gamma)} \; \; c(\tilde{s}_0, \; i_\gamma(s_\gamma))^{-1} 
   \nonumber 
\end{align}
for $v \otimes \tilde{v} \in V_{s_0} \otimes \widetilde{V}_{\tilde{s}_0}$,
where $s_0 \in {\cal G}$ and $\tilde{s}_0 \in \widetilde{\cal G}$; 
the skew-symmetric bimultiplicative forms of ${\cal G}$ and
$\widetilde{\cal G}$ are identified under $\sigma_0$, we just use
one common notation $c$, and $\sigma_0$ has been omitted from the
arguments of $c$.
For the two different set-theoretical identifications $i_\gamma$ and
$i'_\gamma = i_\gamma + \delta s_0$
($\tilde{\imath}_\gamma$ and $\tilde{\imath}'_\gamma$ come along with them),  
the two representations are equivalent: one may choose 
\begin{align}
  P_{i'_\gamma,i_\gamma}: (M_\gamma \otimes \widetilde{M}_{\pi(\gamma)})^{(i_\gamma)}
  & \;  \ni (m\otimes \tilde{m})^{(i_\gamma)} \mapsto
  \label{eq:temp-equiv-VVtild-repr-iso} \\
 &  (m \otimes \tilde{m})^{(i'_\gamma)} \;
  c(\tilde{\imath}_{\pi(\gamma)}(\tilde{s}_{\pi(\gamma)}), \; \delta s_0)^{-1} 
  \in  (M_\gamma \otimes \widetilde{M}_{\pi(\gamma)})^{(i_\gamma)},  \nonumber 
\end{align}
for example, to have $P_{i'_\gamma,i_\gamma} \cdot (Y_\gamma \otimes
\widetilde{Y}_{\pi(\gamma)})^{(i_\gamma)} =
(Y_\gamma \otimes \widetilde{Y}_{\pi(\gamma)})^{(i'_\gamma)} \cdot
P_{i'_\gamma,i_\gamma}$ indeed. There is no canonical isomorphism between
the two representation spaces $(M_\gamma \otimes \widetilde{M}_{\pi(\gamma)})^{(i_\gamma)}$ and $(M_\gamma \otimes \widetilde{M}_{\pi(\gamma)})^{(i'_\gamma)}$,
as we can see from the fact that we could have used
$c(\delta s_0, \; \tilde{\imath}'_{\pi(\gamma)}(\tilde{s}_{\pi(\gamma)}))^{-1}$
%
%
instead of $c(\delta s_0, \; \tilde{\imath}_{\pi(\gamma)}(\tilde{s}_{\pi(\gamma)}))^{-1}$
%
in $P_{i'_\gamma,i_\gamma}$ above. We will come back to this issue
in \ref{statmnt:out-going-states+can-iso}. 

This (equivalence class of) representation(s) satisfies a super-Jacobi
identity similar to (\ref{eq:Jacobi-module-S})
or (\ref{eq:Jacobi-module-S-wR}); the commutation phase $c(s_{0,1}, s_{0,2})$
is replaced by
$c(s_{0,1}-\tilde{s}_{0,1}, s_{0,2}-\tilde{s}_{0,2})$, 
so the states in 
$\oplus_{s_0} (V_{s_0}\otimes \widetilde{V}_{\sigma_0(s_0)}) \subset V\otimes
\widetilde{V}$ are bosons, while---in the case of SVOA/fermionic CFT---those
in $\oplus_{s_0} V_{1 \cdot s_0} \otimes \widetilde{V}_{\sigma_0(s_0)}$ are fermions
indeed. The factor with a branch cut $((z_1-z_0)/z_2)^{-\hat{h}(s_0)}$
in the RHS of (\ref{eq:Jacobi-module-S-wR}) shows up along with
$((\bar{z}_1-\bar{z}_0)/\bar{z}_2)^{-\tilde{\hat{h}}(\tilde{s}_0)}$; because
the states in ${\cal H}_{\rm tot}^{\rm post}$ are in the combination of
$\sigma_0^*(\tilde{\hat{h}}) = \hat{h}$ and $\tilde{s}_0 = \sigma_0(s_0)$,
there is no branch cut (or multiple cover), despite the possible
fractional powers of $(z_1-z_0)$ and $(\bar{z}_1-\bar{z}_0)$. 

We set the Hermitian inner product on the representation space
$(M_\gamma \otimes \widetilde{M}_{\pi(\gamma)})^{(i_\gamma)}$ by
\begin{align}
  \langle (m'_\gamma \otimes \tilde{m}'_{\pi(\gamma)})^{(i_\gamma)} |
  (m_\gamma \otimes \tilde{m}_{\pi(\gamma)})^{(i_\gamma)} \rangle :=
  \bra{m'_\gamma } \; m_\gamma \rangle_{M_\gamma} \;\;
  \bra{\tilde{m}'_\gamma } \; \tilde{m}_\gamma \rangle_{\widetilde{M}_{\pi(\gamma)}} .
\end{align}
This is positive definite, because those of the left-mover and right-mover
SVOAs are. The invariance of the Hermitian inner
product (\ref{eq:cond-o-4VOA-onUps-S})
under the action of the left-mover and right-mover SVOA has the
following consequence: for $v \in V_{s_0}$ and
$\tilde{v} \in \widetilde{V}_{\tilde{s}_0}$, 
\begin{align}
&
  \bra{(m'_\gamma \otimes \tilde{m}'_{\pi(\gamma)})^{(i_\gamma)} } \;
  (Y_\gamma \widetilde{Y}_{\pi(\gamma)})^{(i_\gamma)}\left(
  (e^{zL_1}(\frac{e^{-\pi i}}{z^2})^{L_0} \cdot v)\otimes
  (e^{\bar{z}\tilde{L}_1}(\frac{e^{+\pi i}}{\bar{z}^2})^{\tilde{L}_0} \cdot \tilde{v}), \frac{1}{z} \right) \cdot (m_\gamma \otimes \tilde{m}_{\pi(\gamma)})^{(i_\gamma)} \rangle \nonumber \\
  &  = \bra{Y_\gamma(\Upsilon(v),z) \cdot m'_\gamma} m_\gamma \rangle
  \bra{\widetilde{Y}_{\pi(\gamma)}(\widetilde{\Upsilon}(\tilde{v}),\bar{z})
    \cdot \tilde{m}'_{\pi(\gamma)} } \tilde{m}_{\pi(\gamma)} \rangle \;\; 
  c(\tilde{s}_0, i_\gamma(s_\gamma))^{-1}  \nonumber \\
 &  = \bra{Y_\gamma(\Upsilon(v),z) \cdot m'_\gamma} m_\gamma \rangle
  \bra{\widetilde{Y}_{\pi(\gamma)}(\widetilde{\Upsilon}(\tilde{v}),\bar{z})
    \cdot \tilde{m}'_{\pi(\gamma)} } \tilde{m}_{\pi(\gamma)} \rangle \;\; 
  \left( c(- \tilde{s}_0, i_\gamma(s'_\gamma))^{-1} \cdot c( s_0, \tilde{s}_0 ) \right)^{cc}  \nonumber \\
  & = \bra{ (Y_{\gamma}\widetilde{Y}_{\pi(\gamma)})^{(i_\gamma)} \left( \Upsilon(v)\otimes \widetilde{\Upsilon}(\tilde{v}),z \right) \cdot (m'_\gamma \otimes \tilde{m}'_{\pi(\gamma)})^{(i_\gamma)} } (m_\gamma \otimes \tilde{m}_{\pi(\gamma)})^{(i_\gamma)} \rangle \;\; c( s_0, \tilde{s}_0 )^{cc},
  \label{eq:cond-o-4CFT-onUps-S}
\end{align}
where $m_\gamma \in M_{\gamma s_\gamma}$, $m'_\gamma \in M_{\gamma s'_\gamma}$,
$\tilde{m}_{\pi(\gamma)} \in \widetilde{M}_{\pi(\gamma) \tilde{s}_{\pi(\gamma)}}$,
$\tilde{m}'_{\pi(\gamma)} \in \widetilde{M}_{\pi(\gamma) \tilde{s}'_{\pi(\gamma)}}$,
so
$s'_\gamma = s_0 \cdot s_\gamma$, $\tilde{s}'_{\pi(\gamma)} = \tilde{s}_0 \cdot
\tilde{s}_{\pi(\gamma)}$. We used the fact that the commutation phase $c$ is
in $S^1$ at the equality in the middle.  

For an algebraic data
${\cal YY}\in {\cal VV}_{\beta,\gamma}^\alpha(i_\beta,r)$ and
$\widetilde{\cal YY} \in
\widetilde{\cal VV}_{\pi(\beta),\pi(\gamma)}^{\pi(\alpha)}
(\tilde{\imath}_{\pi(\beta)},-1-r)$ of the left- and right-mover SVOAs, we define 
an intertwining operator of $V \otimes_\C \widetilde{V}$
\begin{align}
  \left( {\cal YY}\otimes \widetilde{\cal YY} \right)^{(i_\beta,i_\gamma)}: & \; 
  \left( M_\beta \otimes \widetilde{M}_{\pi(\beta)}\right)^{(i_\beta)} \rightarrow
   \nonumber \\
 & \qquad  {\rm Hom}\left( \left( M_\gamma \otimes \widetilde{M}_{\pi(\gamma)}\right)^{(i_\gamma)} , \; \left( M_\alpha \otimes \widetilde{M}_{\pi(\alpha)}\right)^{(i_\beta+i_\gamma)}\right)  \\
 & \qquad \qquad \qquad \qquad  [[z,z^{-1},\bar{z},\bar{z}^{-1}]]
       z^{-h_{s_\beta}-h_{s_\gamma}+h_{f(s_\beta,s_\gamma)}}
       \bar{z}^{-\tilde{h}_{\tilde{s}_\beta}-\tilde{h}_{\tilde{s}_\gamma}+\tilde{h}_{f(\tilde{s}_\beta,\tilde{s}_\gamma)}}  \nonumber
\end{align}
as follows; here, the set-theoretical identification $(i_\beta+i_\gamma):
S_\alpha \rightarrow {\cal G}$ is determined from $i_\beta$ and $i_\gamma$
so that $(i_\beta + i_\gamma) \cdot f
= i_\beta + i_\gamma: S_\beta \times S_\gamma \rightarrow {\cal G}$, and 
\begin{align}
  &\left( {\cal YY}\otimes \widetilde{\cal YY} \right)^{(i_\beta,i_\gamma)} \left(
  (m_\beta \otimes \tilde{m}_{\pi(\beta)})^{(i_\beta)}, z \right) \cdot (m_\gamma \otimes \tilde{m}_{\pi(\gamma)})^{(i_\gamma)}
  \label{eq:def-int.op-action-fCFT-detail} \\
 & := \left(\left( {\cal YY}(m_\beta)_{z@z} \cdot m_\gamma \right) \otimes
  \left( \widetilde{\cal YY}( \tilde{m}_{\pi(\beta)})_{\bar{z}@\bar{z}} \cdot
  \tilde{m}_{\pi(\gamma)} \right) \right)^{(i_\beta + i_\gamma)}
  \!\!\!\!\!\!  c(\tilde{\imath}_{\pi(\beta)}(\tilde{s}_{\pi(\beta)}), \;
   i_\gamma(s_\gamma))^{-1} .   \nonumber 
\end{align}
This intertwining operator satisfies the super-Jacobi identity similar to
(\ref{eq:def-vct-sp-int.op-S}) and (\ref{eq:def-vct-sp-int.op-S-wR}), but
with the commutation phase $c(s_0, i_\beta(s_\beta))$ replaced by
$c(s_0-\tilde{s}_0, \;
i_\beta(s_\beta) - \tilde{\imath}_{\pi(\beta)}(\tilde{s}_{\pi(\beta)}))$, which
captures the sign upon fermion exchange in the theory with
${\cal H}_{\rm tot}^{\rm pre}$, and is trivial in the theory with
${\cal H}_{\rm tot}^{\rm post}$. Note also that the branch cut and the $r$-dependence
in the second term of the super-Jacobi identity is also gone, because
$\sigma_0(\hat{h})(\sigma_\beta(s_\beta)) = \hat{h}(s_\beta)$ on
${\cal H}_{\rm tot}^{\rm post}$. 

Although there is no canonical isomorphism between vector spaces
${\cal VV}_{\beta,\gamma}^\alpha(i_\beta,r)$ with different auxiliary
parameters $r \in \Z/T\Z$ (cf \ref{statmnt:int.op-S-NSonly}),
we \underline{can} build canonical ones between the vector spaces
${\cal VV}_{\beta,\gamma}^\alpha(i_\beta,r)\otimes
\widetilde{\cal VV}_{\pi(\beta),\pi(\gamma)}^{\pi(\alpha)}
(\tilde{\imath}_{\pi(\beta)},-1-r)$ with different $r \in \Z/T\Z$.
To see this, recall that fairly nice choices of isomorphisms
${\cal VV}_{\beta,\gamma}^\alpha(r) \rightarrow {\cal VV}_{\beta,\gamma}^\alpha(r+\Delta r)$ in (\ref{eq:temp-iso-vvct-sp-int.op-S-diff-brnch}) are labeled by
choices $i_\gamma: S_\gamma \cong {\cal G}$. For different $i_\gamma$, the corresponding isomorphisms are different. This difference is canceled, however,
when we make it a rule to use the isomorphism
$\widetilde{\cal VV}_{\pi(\beta),\pi(\gamma)}^{\pi(\alpha)}(-1-r) \rightarrow
\widetilde{\cal VV}_{\pi(\beta),\pi(\gamma)}^{\pi(\alpha)}(-1-r-\Delta r)$
corresponding to the label $\tilde{\imath}_\gamma$ such that
$\tilde{\imath}_\gamma \cdot \sigma_\gamma = i_\gamma$;
that is because the factor $e^{-2 \pi i \hat{h}_\beta(i_\gamma(s_\gamma)) \Delta r} \cdot
e^{-2\pi i\tilde{\hat{h}}_{\tilde{\beta}}(\tilde{\imath}_{\pi(\gamma)}(\tilde{s}_\gamma))(-\Delta r)}$
changes under $i'_\gamma(s_\gamma) = i_\gamma(s_\gamma) + (\delta s_0)$
and $\tilde{\imath}'_{\pi(\gamma)}(\tilde{s}_{\pi(\gamma)}) =
\tilde{\imath}_{\pi(\gamma)}(\tilde{s}_{\pi(\gamma)}) + (\delta s_0)$
by $e^{-2\pi i (\hat{h}_\beta - \tilde{\hat{h}}_{\tilde{\beta}})(\delta s_0)}$, even when
$(M_\beta \otimes \widetilde{M}_{\pi(\beta)})$ is in the Ramond--Ramond sector
of ${\cal H}_{\rm tot}^{\rm post}$. 
We also call it the canonical isomorphism.

Given a pair of $({\cal YY}\otimes \widetilde{\cal YY})^{(r)} \in
{\cal VV}_{\beta,\gamma}^\alpha(r) \otimes \widetilde{\cal VV}_{\pi(\beta),\pi(\gamma)}^{\pi(\alpha)}(-1-r)$ and $({\cal YY}\otimes \widetilde{\cal YY})^{(r')}$ in
${\cal VV}_{\beta,\gamma}^\alpha(r') \otimes \widetilde{\cal VV}_{\pi(\beta),\pi(\gamma)}^{\pi(\alpha)}(-1-r')$ that are identified under the canonical isomorphism
above, one may wonder if they give rise to the same action of
a state $(m_\beta \otimes \tilde{m}_{\pi(\beta)})^{(i_\beta)} \in
(M_\beta \otimes \widetilde{M}_{\pi(\beta)})^{(i_\beta)}$. They do, in fact. 
If this state is in the untwisted (NS--NS) sector, there is no $r$-dependence
in the intertwining operator from the beginning.  If this state is in the
$\hat{h}$--$\sigma_0(\hat{h})$-twisted (Ramond--Ramond sector), not all the
$(\#{\cal G})^2 \times (\#{\cal G})^2$ components
$({\cal Y}_{s_\beta,s_\gamma} \otimes \widetilde{\cal Y}_{\tilde{s}_{\pi(\beta)},\tilde{s}_{\pi(\gamma)}})^{(r)} \in {\cal V}_{\beta s_\beta, \gamma s_\gamma}^{\alpha f(s_\beta,s_\gamma)}\otimes \widetilde{\cal V}_{\pi(\beta)\tilde{s}_{\pi(\beta)},\pi(\gamma)\tilde{s}_{\pi(\gamma)}}^{\pi(\alpha)\tilde{f}(\tilde{s}_{\pi(\beta)},\tilde{s}_{\pi(\gamma)})}$ of
$({\cal YY}\otimes \widetilde{\cal YY})^{(r)}$ are identical to
those of $({\cal YY}\otimes \widetilde{\cal YY})^{(r')}$; because
of the structure of ${\cal H}_{\rm tot}^{\rm post}$, however, only
$(\#{\cal G})^2$ out of $(\#{\cal G})^{2 \cdot 2}$ components are used, where
$({\cal Y}\otimes \widetilde{\cal Y})^{(r)}$
agrees with $({\cal Y}\otimes \widetilde{\cal Y})^{(r')}$. 

The situation is subtler in the relation between the intertwining operators
labeled by different set-theoretical identifications $i_\beta: S_\beta \cong
{\cal G}$. Recall that fairly nice choices of an isomorphism
${\cal VV}_{\beta,\gamma}^\alpha(i_\beta) \rightarrow
{\cal VV}_{\beta,\gamma}^\alpha(i'_\beta)$
in (\ref{eq:temp-iso-vvct-sp-int.op-S-diff-StoZ2}) are labeled by 
choices of an identification $i_\gamma: S_\gamma \cong {\cal G}$;
for different $i_\gamma$, the corresponding isomorphisms are different. 
Let us make it a rule to use the isomorphism
$\widetilde{\cal VV}_{\pi(\beta),\pi(\gamma)}^{\pi(\alpha)}(\tilde{\imath}_\beta)
\rightarrow \widetilde{\cal VV}_{\pi(\beta),\pi(\gamma)}^{\pi(\alpha)}
(\tilde{\imath}'_\beta)$ in (\ref{eq:temp-iso-vvct-sp-int.op-S-diff-StoZ2})
corresponding to the label
$\tilde{\imath}_\gamma: \widetilde{S}_{\pi(\gamma)}\rightarrow \widetilde{\cal G}$
such that $\tilde{\imath}_\gamma \cdot \sigma_\gamma = i_\gamma$.
Then the isomorphisms ${\cal VV}_{\beta,\gamma}^\alpha(i_\beta)\otimes \widetilde{\cal VV}_{\pi(\beta),\pi(\gamma)}^{\pi(\alpha)}(\tilde{\imath}_{\pi(\beta)})$
to that with $i'_\beta = i_\beta + \delta s_0$ do not depend on
the choice of $i_\gamma$ (and $\tilde{\imath}_{\pi(\gamma)}$ accordingly), 
when ${\cal G}$ is 2-torsion (every element of ${\cal G}$ except the identity
is of order 2), as in the case of SVOA; the isomorphisms with $i_\gamma$ and
$i'_\gamma = i_\gamma + \Delta s_0$ differ by the phase factor
$c(\Delta s_0, \delta s_0)^{-2}$, which is 1 when $c$ is $\{ \pm 1\}$-valued on ${\cal G} \times {\cal G}$.
In such cases (including SVOA), we call it the canonical isomorphism. 
The action of ${\cal YY} \otimes \widetilde{\cal YY} \in
{\cal VV}_{\beta,\gamma}^{\alpha}(i'_\beta) \otimes \widetilde{\cal VV}_{\pi(\beta),\pi(\gamma)}^{\pi(\alpha)}(\tilde{\imath}_\beta')$ on states in
$(M_\beta \otimes \widetilde{M}_{\pi(\beta)})^{(i_\beta)}$ and
$(M_\gamma \otimes \widetilde{M}_{\pi(\gamma)})^{(i_\gamma)}$ is determined by
translating ${\cal YY} \otimes \widetilde{\cal YY}$ into an element
in ${\cal VV}_{\beta,\gamma}^\alpha(i_\beta) \otimes
\widetilde{\cal VV}_{\pi(\beta),\pi(\gamma)}^{\pi(\alpha)}
(\tilde{\imath}_{\pi(\beta)})$ by this canonial isomorphism before applying
the prescription in (\ref{eq:def-int.op-action-fCFT-detail}).

As we do not have a canonical isomorphism between the representation spaces $(M_\beta \otimes \widetilde{M}_{\pi(\beta)})^{(i_\beta)}$ with different identifications $i_\beta$ at this moment (even when $c$ is $\{\pm 1\}$-valued),
we have a trouble in talking about ``the'' normalized correlation function of $x_1\otimes \cdots \otimes x_n \in ({\cal H}_{\rm tot}^{\rm pre/post})^{\otimes n}$ as we did in section \ref{ssec:CFT-obs};
when $x_i$ and $x_j$ are both in the same simple representation $(\beta,\pi(\beta)) \in {\cal II}^{\hat{h}}({\cal A}) \times \widetilde{\cal II}^{\sigma_0(\hat{h})}(\widetilde{\cal A})$, 
we cannot think of having $x_i$ from $(M_\beta \otimes \widetilde{M}_{\pi(\beta)})^{(i_\beta)}$ and $x_j$ from $(M_\beta \otimes \widetilde{M}_{\pi(\beta)})^{(i_\beta  + s_{\neq 0})}$ because we cannot compare $x_i$ and $x_j$ in the absence of a canonical
isomorphism between the two representation spaces.
To formulate normalized correlation functions of a 2D fermionic CFT,
it is not enough at this moment to give ${\cal H}_{\rm tot}^{\rm pre/post}$ as an isomorphism class of representations of ${\cal A} \times \widetilde{\cal A}$ as in (\ref{eq:Htot-SCFT-wF}, \ref{eq:Htot-SCFT-wR}),
or not even enough to fix a representation $(M_\alpha,Y_\alpha)$ of ${\cal A}$ and a representation $(\widetilde{M}_{\pi(\alpha)}, \widetilde{Y}_{\pi(\alpha)})$ of $\widetilde{\cal A}$.
We need to fix a representation of ${\cal A} \times \widetilde{\cal A}$ by choosing one identification $i_{\alpha *}: S_\alpha \cong {\cal G}$ for each of $\alpha \in  {\cal II}^{\hat{h}}({\cal A})$, and set
\begin{align}
  {\cal H}_{\rm tot}^{\rm pre} := \oplus_{\alpha \in {\cal II}^{\hat{0}}({\cal A})}
  (M_\alpha \otimes \widetilde{M}_{\pi(\alpha)})^{(i_{\alpha *})}
  \label{eq:Htot-def-w-embed-specified-NSNS}
\end{align}
instead of (\ref{eq:Htot-SCFT-wF}); we should do similarly for
${\cal H}_{\rm tot}^{\rm post}$.
\end{anythng}

\begin{anythng}
\label{statmnt:data-4-fCFT-3}
As a part of the defining data of a fermionic 2D CFT, we need to
specify one element 
\begin{align}
  YY_{\rm tot}^{(\alpha;\beta,\gamma)} \in {\cal VV}_{\beta,\gamma}^\alpha \otimes
  {\cal VV}_{\pi(\beta),\pi(\gamma)}^{\pi(\alpha)}. 
\end{align}
Identification under the canonical isomorphisms (at the end
of \ref{statmnt:data-4-fCFT-2}) is already implicit. 
A choice of this {\it fermionic CFT vertex operator}
$YY_{\rm tot}^{(\alpha;\beta,\gamma)}$ is parametrized by
\begin{align}
  Y_{\rm tot}^{(\alpha;\beta,\gamma)} = {\cal YY}_i \otimes
  \widetilde{\cal YY}_{\tilde{\imath}} \;\; d^{i,\tilde{\imath}} \in
            {\cal VV}_{\beta,\gamma}^\alpha \otimes
            \widetilde{\cal VV}_{\pi(\beta),\pi(\gamma)}^{\pi(\alpha)},
\end{align}
using a basis $\{ {\cal YY}^a \}$ of the vector space ${\cal VV}_{\beta,\gamma}^\alpha$ and $\{ \widetilde{\cal YY}^{\tilde{a}} \}$ of $\widetilde{\cal VV}_{\pi(\beta)\pi(\gamma)}^{\pi(\alpha)}$, and complex-valued {\it left-right coupling matrix}
$d$. 

Let us now state the conditions (t-fCFT) on the choice of
$YY_{\rm tot}^{(\alpha;\beta,\gamma)}$. First, similarly to the
condition (t-1), $YY_{\rm tot}^{(\gamma;0,\gamma)}$ should
be the pair of representations $YY_\gamma \otimes \widetilde{YY}_{\pi(\gamma)}$
of the left-mover and right-mover SVOAs (t-1-f).

The skew-symmetry condition (t-2sk-f) is that 
\begin{align}
  \Omega_r \widetilde{\Omega}_{-1-r} \left(  YY_{\rm tot}^{(\alpha;\beta,\gamma)}
  \right) = YY_{\rm tot}^{(\alpha;\gamma,\beta)},
 \label{eq:cond-Ytot-skew-f}
\end{align}
where we use $\Omega_r$ on ${\cal VV}_{\beta,\gamma}^\alpha$ and
$\widetilde{\Omega}_{-1-r}$ on
$\widetilde{\cal VV}_{\pi(\beta),\pi(\gamma)}^{\pi(\alpha)}$ as
in (\ref{eq:def-skew-iso-S}).
The consequence is that the analytic continuation of
the $(M_\alpha \otimes \widetilde{M}_{\pi(\alpha)})^{(i_\beta+i_\gamma)}$-valued
formal power series 
\begin{align*}
  \left( Y_{\rm tot}^{(\alpha;\beta,\gamma)} \right)^{(i_\beta,i_\gamma)}
  \left( (m_\beta \otimes \tilde{m}_{\pi(\beta)})^{(i_\beta)} , \; z \right)
    \cdot (m_\gamma \otimes \tilde{m}_{\pi(\gamma)})^{(i_\gamma)}   
\end{align*}
from real positive $z$ to real negative $z=e^{\pi i(2r+1)}\underline{z}$
while increacing its phase by $(2r+1) \pi$ is equal to 
\begin{align}
 e^{zL_{-1}}e^{\bar{z}\tilde{L}_{-1}}
 \cdot \left( Y_{\rm tot}^{(\alpha;\gamma,\beta)}\right)^{(i_\gamma,i_\beta)} \left(
 (m_\gamma \otimes \tilde{m}_{\pi(\gamma)})^{(i_\gamma)} , \; \underline{z} \right)
 \cdot (m_\beta \otimes \tilde{m}_{\pi(\beta)})^{(i_\beta)} \rangle
 \label{eq:cond-skew-fCFT-inCrrFcn} \\
 c(i_\beta(s_\beta)-\tilde{\imath}_{\pi(\beta)}(\tilde{s}_{\pi(\beta)}), \;
  i_\gamma(s_\gamma) - \tilde{\imath}_{\pi(\gamma)}(\tilde{s}_{\pi(\gamma)}) )
 \nonumber
\end{align}
for $m_\beta \in M_{\beta s_\beta}$, $\tilde{m}_{\pi(\beta)} \in
\widetilde{M}_{\pi(\beta) \tilde{s}_{\pi(\beta)}}$,
$m_\gamma \in M_{\gamma s_\gamma}$, $\tilde{m}_{\pi(\gamma)} \in
\widetilde{M}_{\pi(\gamma) \tilde{s}_{\pi(\gamma)}}$. 
The sign at the end only reflects whether the states
$(m_\beta \otimes \tilde{m}_{\pi(\beta)})$ and $(m_\gamma \otimes
\tilde{m}_{\pi(\gamma)})$ are fermionic or bosonic in the theory with
${\cal H}_{\rm tot}^{\rm pre}$, and is trivial in the theory with
${\cal H}_{\rm tot}^{\rm post}$. 

The associativity condition (t-3-f) is that the isomorphism 
\begin{align*}
 & ({\cal F} \otimes \widetilde{\cal F}): \oplus_{\delta}
  \left({\cal VV}_{\beta_1,\delta}^{\alpha}(i_{\beta_1}) \otimes
  \widetilde{\cal VV}_{\pi(\beta_1),\pi(\delta)}^{\pi(\alpha)}(\tilde{\imath}_{\beta_1})
  \right) \otimes \left( {\cal VV}_{\beta_2,\gamma}^{\delta}(i_{\beta_2}) \otimes
  \widetilde{\cal VV}_{\pi(\beta_2),\pi(\gamma)}^{\pi(\delta)}(\tilde{\imath}_{\beta_2})
  \right) \\
 & \rightarrow
  \oplus_{\epsilon}
  \left({\cal VV}_{\beta_1,\beta_2}^{\epsilon}(i_{\beta_1}) \otimes
  \widetilde{\cal VV}_{\pi(\beta_1),\pi(\beta_2)}^{\pi(\epsilon)}
  (\tilde{\imath}_{\beta_1})
  \right) \otimes \left( {\cal VV}_{\epsilon,\gamma}^{\alpha}(i_{\beta_1} + i_{\beta_2}) \otimes
  \widetilde{\cal VV}_{\pi(\epsilon),\pi(\gamma)}^{\pi(\alpha)}(
  \tilde{\imath}_{\beta_1} + \tilde{\imath}_{\beta_2})  \right)
\end{align*}
preserves $YY_{\rm tot}$:
\begin{align}
  {\cal F} \otimes \widetilde{\cal F}: 
  \left( YY_{\rm tot} \otimes YY_{\rm tot}\right) \mapsto
  \left( YY_{\rm tot} \otimes YY_{\rm tot} \right)
\end{align}
for all $\beta_1$, $\beta_2$, $\gamma$. Recall that, on one hand, we have
formulated the fusion isomorphism for SVOA in \ref{statmnt:BnF-S} only for
a special choice $i_\epsilon= i_{\beta_1}+i_{\beta_2}$; on the other hand,
we have formulated the action of intertwining operators of fermionic
CFTs only in a way that the outgoing states are in $(M_\alpha \otimes
\widetilde{M}_{\pi(\alpha)})^{(i_\alpha)}$ for a special choice
$i_\alpha = i_\beta + i_\gamma$. Those two limitations do not do any harm,
however, as the consequence of the condition can be written in
this more intuitive way; 
\begin{align}
  \sum_\delta & \; 
  Y_{\rm tot}^{(\alpha;\beta_1,\delta)}\left( (m_{\beta_1}\otimes \tilde{m}_{\pi(\beta_1)})^{(i_{\beta_1})}, z_1\right) \cdot
  Y_{\rm tot}^{(\delta;\beta_2,\gamma)} \left( (m_{\beta_2} \otimes \tilde{m}_{\pi(\beta_2)})^{(i_{\beta_2})},z_2\right)  \\
 &  = \sum_\epsilon Y_{\rm tot}^{(\alpha;\epsilon,\gamma)}\left(
   Y_{\rm tot}^{(\epsilon;\beta_1,\beta_2)}( (m_{\beta_1}\otimes \tilde{m}_{\pi(\beta_1)})^{(i_{\beta_1})}, z_1-z_2) \cdot (m_{\beta_2}\otimes \tilde{m}_{\pi(\beta_2)})^{(i_{\beta_2})} , z_2 \right) \nonumber 
\end{align}
the outgoing states from $Y_{\rm tot}^{(\epsilon;\beta_1,\beta_2)}$ is
in the representation space
$(M_\epsilon \otimes \widetilde{M}_{\pi(\epsilon)})^{(i_{\beta_1}+i_{\beta_2})}$,
so we can use $Y_{\rm tot}^{(\alpha;\epsilon,\gamma)} \in {\cal VV}_{\epsilon,\gamma}^\alpha(i_{\beta_1}+i_{\beta_2})\otimes
\widetilde{\cal VV}_{\pi(\epsilon),\pi(\gamma)}^{\pi(\alpha)}$ without
invoking a canonical isomorphism. 

The condition (t-f) on the fermionic CFT vertex operators $\{YY_{\rm tot}\}$
consists of (t-1-f), (t-2sk-f) and (t-3-f) described above, and
one more: the crossing symmetry condition (t-2cr-f). Let us pause
for a moment, however, for a little more preparation before stating
the condition (t-2cr-f). 
\end{anythng}

\begin{rmk}
  \label{rmk:Type0-bosonic-dict}
{\bf the perspective of section \ref{sec:CFT} for the theory with
  ${\cal H}_{\rm tot}^{\rm post}$:} The Hilbert space ${\cal H}_{\rm tot}^{\rm pre}$
contains states whose $(L_0-\tilde{L}_0)$ eigenvalues are not integers,
but the eigenvalues are integers for all the states in
${\cal H}_{\rm tot}^{\rm post}$. By focusing on the action of the bosonic part
$(V_0, Y_{0,0})$ and $(\widetilde{V}_0,\widetilde{Y}_{0,0})$ of the superchiral
algebras of the fermionic CFT, we may think of $\{YY_{\rm tot}\}$ on
${\cal H}_{\rm tot}^{\rm post}$ for the states in ${\cal H}_{\rm tot}^{\rm post}$
as that of CFT vertex operators reviewed in section \ref{sec:CFT}. This
perspective is useful in making sense of what is being done
in \ref{statmnt:data-4-fCFT-2} and \ref{statmnt:data-4-fCFT-3}.  

Here, in Rmk. \ref{rmk:Type0-bosonic-dict}, we only deal with states
and the representation spaces in ${\cal H}_{\rm tot}^{\rm post}$ (rather than
${\cal H}_{\rm tot}^{\rm pre}$), and omit $\sigma_\gamma$, $\pi$, and sometimes
$\widetilde{~}$ that indicates an object associated with right-mover algebra
(if that is not essential). 

First, the role played by a fixed choice of set-theoretical identification
$i_{\gamma *}$ (for $\gamma \in {\cal II}^{\hat{h}}$) is clear in the relation
\begin{align}
  Y_{\rm tot}^{(\gamma s_0 \cdot s_\gamma; s_0, \gamma s_\gamma)} =
  Y_{s_0, s_\gamma} \otimes \widetilde{Y}_{s_0, s_\gamma} \;
  c(s_0, i_{\gamma *}(s_\gamma))^{-1} 
\end{align}
acting on $M_{\gamma s_\gamma} \otimes \widetilde{M}_{\gamma s_\gamma} \subset
(M_\gamma \otimes \widetilde{M}_{\gamma})^{(i_{\gamma *})}$, 
where $Y_{\rm tot}^{(\gamma s_0 \cdot s_\gamma; s_0, \gamma s_\gamma)}$ is the
CFT vertex operator in the sense of section \ref{sec:CFT}, and
$Y_{s_0, s_\gamma}$ and $\widetilde{Y}_{s_0, s_\gamma}$ on the RHS
the $(\#{\cal G})^2$ components of $YY \in {\cal VV}_{0,\gamma}^{\gamma}$
and $\widetilde{YY} \in \widetilde{\cal VV}_{0,\gamma}^\gamma$.
To put it colloquially, the supermultiplets $YY$ and $\widetilde{YY}$
are determined from $Y_{\rm tot}$ by comparing the components with
$s_\gamma = s_{\gamma *} \in S_\gamma$ so that
$i_{\gamma *}(s_{\gamma *}) = 0 \in {\cal G}$; all the other components
of the supermultiplets $YY$ and $\widetilde{YY}$ are determined by
the fusion and braiding. 

For a more general type $(\alpha;\beta,\gamma)$, the relation between the
CFT vertex operator on ${\cal H}_{\rm tot}^{\rm post}$ and the components
of the supermultiplet of intertwining operators is 
\begin{align}
  Y_{\rm tot}^{(\alpha s_\alpha; \beta s_\beta, \gamma s_\gamma)} =
  Y_{s_\beta, s_\gamma} \otimes \widetilde{Y}_{s_\beta,s_\gamma} \;
  c(i_{\beta }(s_\beta), i_\gamma(s_\gamma))^{-1}, 
\end{align}
where $\alpha, \beta, \gamma$ are simple representations of the left-mover
SVOA. Here, the CFT vertex operator $Y_{\rm tot}$ on the LHS is meant to act
on $M_{\gamma s_\gamma} \otimes \widetilde{M}_{\gamma s_\gamma} \subset
(M_\gamma \otimes \widetilde{M}_{\gamma})^{(i_\gamma)}$. 
The skew-symmetry invariance (t-2sk) condition
on the CFT vertex operator on the LHS and the condition (t-2sk-f) on
the supermultiplets $YY_{\rm tot}^{(\alpha;\beta,\gamma)} = {\cal YY} \otimes
\widetilde{\cal YY}$ on the RHS are equivalent. 
\end{rmk}

\begin{anythng}
{\bf normalized correlation functions of 2D fermionic CFTs:}
We have reviewed two classes of observables of 2D CFTs: normalized
correlation functions and unnormalized ones. The exposition there works 
almost as it is for fermionic 2D CFTs as well. There are three changes
we have to make, however, to adapt it for fermionic 2D CFTs.

One of them stems from the fact that there is no canonical isomorphism between
the representation spaces $(M_\beta \otimes \widetilde{M}_{\pi(\beta)})^{(i_\beta)}$
with different identifications $i_\beta$.
The multi-peripheral computation algorithm
(\ref{eq:algrthm-multi-prhrl}) should be more specific
about the CFT vertex operators for computations:
$(Y_{\rm tot}^{(\delta_{k-1};\beta_{k},\delta_{k})})(x_{k})_{@p_k}^{(i_{\beta_k *},i_{\delta_k})}$
instead of $Y_{\rm tot}^{(\delta_{k-1};\beta_{k},\delta_{k})}(x_{k})_{@p_k}$
in (\ref{eq:algrthm-multi-prhrl}), with
$i_{\delta_{k-1}} = i_{\beta_k *} + i_{\delta_k}$
determined recursively, starting from $i_{\delta_n}$ as the obvious $i_{0 *}:
{\cal G} = {\cal G}$. At the end of the product of $n$ CFT supervertex
operators, 
we end up with $(M_0 \otimes \widetilde{M}_{\tilde{0}})^{(i_{\delta_0})}$,
where $i_{\delta_0} = i_{\beta_1 *}+\cdots + i_{\beta_n *}$. This set-theoretical
identification $i_{\delta_0}$ is not necessarily the same as the obvious one
$i_{0 *}$. This is not a problem, however, because there is a canonical
isomorphism
between $(M_0 \otimes \widetilde{M}_{\tilde{0}})^{(i_{\delta_0})}$ and
$(M_0 \otimes \widetilde{M}_{\tilde{0}})^{(i_{0 *})}$, characterized by
$({\bf 1} \otimes \tilde{\bf 1})^{(i_{\delta_0})} \leftrightarrow
{\bf 1}\tilde{\bf 1}$, although there is no such a thing for a general
representation of ${\cal A} \times \widetilde{\cal A}$. So, the normalized
correlation functions are defined in the multi-peripheral algorithm
by inserting this canonical isomorphism $(M_0\otimes \widetilde{M}_{\tilde{0}})^{(i_{\delta_0})} \rightarrow (M_0\otimes \widetilde{M}_{\tilde{0}})^{(i_{0 *})}$
if necessary at the end of the product of $n$ CFT supervertex operators. 

The second change is that the normalized correlation functions involving
${\cal B}(YY_{\rm tot} \otimes YY_{\rm tot})$ is no longer equal to the analytic
continuation of those involving $YY_{\rm tot}\otimes YY_{\rm tot}$ in the theory
with ${\cal H}_{\rm tot}^{\rm pre}$. They differ by the phase factor
(the fermionic sign factor for SVOA) in (\ref{eq:cond-skew-fCFT-inCrrFcn}); 
this factor is determined by permutation rather than braiding.

One more change is about the choice of the local coordinate $z_i$ around the
point $p_i$ at which a CFT supervertex operator is inserted in the theory with
${\cal H}_{\rm tot}^{\rm pre}$.
It is not enough to describe the relation between $z_i$ and the global
coordinate $z$ through a function $z = f_i(z_i)$ (as we did in section
\ref{ssec:CFT-obs}); in the conversion
$R_{f_i@z_i(p_i)}^{-1}\tilde{R}_{f_i@z_i(p_i)}^{-1}$, we need to have a well-defined
$(f_0)^{-h_{\beta s_\beta}}(f_c^{cc})^{-\tilde{h}_{\pi(\beta) \tilde{s}_{\pi(\beta)}}}
= |f_0|^{-2\tilde{h}_{\pi(\beta) \tilde{s}_{\pi(\beta)}}} f_0^{-(h_{\beta s_\beta} - \tilde{h}_{\pi(\beta) \tilde{s}_{\pi(\beta)}})}$, whereas $h_{\beta s_\beta} - \tilde{h}_{\pi(\beta) \tilde{s}_{\pi(\beta)}}
\in T^{-1}\Z$ in general in the theory with ${\cal H}_{\rm tot}^{\rm pre}$.
So, the choice of a local coordinate needs to be specified by
$(f_i(z_i), {}^T\sqrt{\partial f_i/\partial z_i})$, including one out of the
$T$ $T$-th roots chosen. 
\end{anythng}

\begin{anythng}
\label{statmnt:data-4-fCFT-4}
  {\bf the crossing symmetry condition on fermionic CFTs:}
We set the crossing symmetry condition (t-2cr-f) in this article as 
\begin{align}
  \lambda_{1,i_{\gamma*}} \tilde{\lambda}_{-1,\tilde{\imath}_{\pi(\gamma)*}} \cdot
  \Theta_{0}\widetilde{\Theta}_{-1}\left( YY_{\rm tot}^{(\alpha;\beta,\gamma)} \right)
  \cdot (\lambda_{1,i_{\beta *}+i_{\gamma *}}
  \tilde{\lambda}_{-1,\tilde{\imath}_{\pi(\beta)*}+\tilde{\imath}_{\pi(\gamma)*} }
  )^{-1} =  YY_{\rm tot}^{(\Upsilon^{\cal II}(\gamma);\beta,\Upsilon^{\cal II}(\alpha))}. 
 \label{eq:cond-Ytot-xing-f}   
\end{align}
This is imposed on the algebraic data using the language of the left-moving
and right-moving fermionic SVOAs. Just like in section \ref{ssec:CFT-data},
we can compare both sides
(and demand that they should be equal) only after specifying
$\lambda_\alpha$'s and $\tilde{\lambda}_{\tilde{\alpha}}$'s.
We make it a rule, inspired by (\ref{eq:temp-pattern-on-s4xingSymTrnsf}),
to set 
\begin{align}
\lambda_{1,\delta s + i_{\gamma *}} = \lambda_{1,i_{\gamma *}}
c(i_{\gamma *}(s), \delta s)^{-1}, \qquad 
\tilde{\lambda}_{-1,\delta \tilde{s} + \tilde{\imath}_{\pi(\gamma)*}}
= \tilde{\lambda}_{-1,\tilde{\imath}_{\pi(\gamma)*}} c(\tilde{\imath}_{\pi(\gamma)*}(\tilde{s}), \delta \tilde{s})^{-1}.
 \label{eq:temp-choice-on-s4xingSymTrnsf}
\end{align}

We use
the generalized crossing-symmetry isomorphism $\Theta_{0} = \Theta[-]$ of
left-moving intertwining operators in (\ref{eq:gen-xing-iso-S-wR}) here.  
Recall that the choice of $\Theta_s$ with $s=0$ is traced back to the
choice of branch $(-1/z^2)^{L_0} \Rightarrow (e^{-\pi i}/z^2)^{L_0}$
in (\ref{eq:cond-o-4VOA-onUps-S}) for the condition (p-SVOA); the (p-SVOA)
condition for the right-mover fermionic SVOA would use
$(-1/\bar{z}^2)^{\tilde{L}_0} \Rightarrow (e^{+\pi i}/\bar{z}^2)^{\tilde{L}_0}$
instead, which means that the discussion in \ref{statmnt:xing-S} for a
right-mover SVOA ends up with the generalized crossing symmetry isomorphism
using $\widetilde{\Theta}_{s=-1} = \widetilde{\Theta}[+]$. 
The isomorphism on the LHS of (\ref{eq:cond-Ytot-xing-f}) maps
$YY_{\rm tot}^{(\alpha;\beta,\gamma)}$ in
${\cal VV}_{\beta,\gamma}^\alpha(r=0)\otimes
\widetilde{\cal VV}_{\pi(\beta),\pi(\gamma)}^{\pi(\alpha)}(\tilde{r}=-1)$
to ${\cal VV}_{\beta,\Upsilon^{\cal II}(\alpha)}^{\Upsilon^{\cal II}(\gamma)}(r=-1)\otimes
\widetilde{\cal VV}_{\pi(\beta),\widetilde{\Upsilon}^{\cal II}(\pi(\alpha))}^{\widetilde{\Upsilon}^{\cal II}(\pi(\gamma))}(\tilde{r}=0)$ (see \ref{statmnt:xing-S}),
so the canonical isomorphism needs to
be used before comparing the LHS with the RHS. 

It is debatable whether this is a good choice of the condition.
Certainly the expressions on both sides of (\ref{eq:cond-Ytot-xing-f})
are in the same vector space, so the condition (\ref{eq:cond-Ytot-xing-f}, 
\ref{eq:temp-choice-on-s4xingSymTrnsf}) is not nonsense, at least.  

A large fraction of the condition (t-2cr-f) is redundant, in fact.
We only have to impose (\ref{eq:cond-Ytot-xing-f}) for the types
of the form $(\gamma; \gamma,0)$; the condition for this type determines
the overall normalization of $\lambda_{1,i_{\gamma *}} =
\{ \lambda_{1,i_{\gamma*};s_\gamma} \; | \; s_\gamma \in S_\gamma\}$ for
$\gamma \in {\cal II}^{\hat{h}}$, just like\footnote{
The relation between $\lambda_{\gamma s_\gamma}$ in the perspective of
section \ref{sec:CFT} and $\lambda_{1,i_{\gamma *};s_\gamma}$ here are related
by $\lambda_{\gamma s_\gamma}\tilde{\lambda}_{\gamma s_\gamma} =
\lambda_{1;i_{\gamma *};s_\gamma} \tilde{\lambda}_{-1,i_{\gamma *};s_\gamma}
\; c(i_{\gamma *}(s_\gamma), i_{\Upsilon^{\cal II}(\gamma)*}(\Upsilon^{\cal I}(s_\gamma))
)^{-1}$. This relation follows both from (t-2cr-f) for the types
$(\gamma;\gamma,0)$ and $(\gamma;0,\gamma)$. 
} %
the condition (t-2cr) determines
the normalization of $\lambda_{\gamma s_\gamma}$ for $\gamma s_\gamma \in
{\cal I}$. All the rest of the relation 
(\ref{eq:cond-Ytot-xing-f}) follows from the combination of
the relation with the types $(\gamma;\gamma,0)$, (t-2sk-f) and (t-3-f),
similarly to what we have seen in section \ref{ssec:CFT-data}. 
\end{anythng}

\begin{anythng}
\label{statmnt:out-going-states+can-iso}
{\bf matrix elements:} There is an extra issue in fermionic 2D unitary
CFTs when one wants to formulate matrix elements rather than normalized
correlation functions ending with the vacuum states ${\bf 1}\tilde{\bf 1}$.
A product of $n$ CFT vertex operators may end up with an irreducible
representation $\alpha \in {\cal II}^{\hat{h}}$, but in the representation
space $(M_\alpha \otimes \widetilde{M}_{\pi(\alpha)})^{(i_{\delta_0})}$ with
$i_{\delta_0}:= i_{\beta_1 *} + \cdots + i_{\beta_n *}$ not necessarily equal to
$i_{\alpha *}$ that is used to set ${\cal H}_{\rm tot}^{{\rm pre/post}}$
in (\ref{eq:Htot-def-w-embed-specified-NSNS}). We have neither formulated
a canonical isomorphism between those equivalent representation spaces
with different set-theoretical identifications, nor the anti-unitary
CPT isomorphism $\Upsilon_{\rm tot}: {\cal H}_{\rm tot}^{\rm pre/post} \rightarrow
{\cal H}_{\rm tot}^{\rm pre/post}$. 

We can exploit the fact that a 2D non-holomorphic CFT is not just
a representation theory of $V\otimes_\C \widetilde{V}$ (of the left-mover
and right-mover (super-) chiral algebras; ${\cal H}_{\rm tot}^{\rm pre/post}$
is not only a representation space, but all the elements (not just those
in $V \otimes_\C \widetilde{V}$) have their representation through the
CFT vertex operator $Y_{\rm tot}$. We start off by finding a generalization
of the invariance property (\ref{eq:cond-o-4CFT-onUps-S}), from
$(v\otimes \tilde{v}) \in V \otimes_\C \widetilde{V}$ there to
$(m_\beta \otimes \tilde{m}_{\pi(\beta)})^{(i_{\beta *})} \in
(M_\beta \otimes \widetilde{M}_{\pi(\beta)})^{(i_{\beta *})}$. 

Let us begin with studying the following expression, with
$\beta = \Upsilon^{\cal II}(\delta)$, $i_{\beta*}: S_\beta \cong {\cal G}$
used in setting ${\cal H}_{\rm tot}^{\rm pre/post}$
in (\ref{eq:Htot-def-w-embed-specified-NSNS}) but $i_\delta: S_\delta
\cong {\cal G}$ not necessarily the one $i_{\Upsilon^{\cal II}(\beta)*}
:S_\delta \cong {\cal G}$ used in setting ${\cal H}_{\rm tot}^{\rm pre/post}$
 in (\ref{eq:Htot-def-w-embed-specified-NSNS}):  
\begin{align}
  &  \bra{{\bf 1}\tilde{\bf 1}} (Y_{\rm tot}^{(0;\beta,\delta)})^{(i_{\beta*},i_\delta)}
  \left(
  (e^{zL_1}(\frac{e^{- \pi i}}{z^2})^{L_0} m_\beta\otimes
  e^{\bar{z}\tilde{L}_1} (\frac{e^{+\pi i}}{\bar{z}^2} )^{\tilde{L}_0}
  \tilde{m}_{\pi(\beta)})^{(i_{\beta*})} ,\frac{1}{z} \right)
  \cdot (m_\delta \otimes \tilde{m}_{\pi(\delta)})^{(i_\delta)} \rangle  \nonumber \\
&  = \bra{\bf 1} \lambda_0 (\Theta_{0}Y_{s_\beta, -s_0}) \lambda_\beta^{-1}
  \left( e^{zL_1} (\frac{e^{-\pi i}}{z^2})^{L_0} m_\beta, \frac{1}{z} \right)
  \cdot m_\delta \rangle \label{eq:temp-fCFT-2ptME-endWvac}   \\
& \quad \bra{\tilde{\bf 1}} \tilde{\lambda}_{\tilde{0}} (\widetilde{\Theta}_{-1}
  \widetilde{Y}_{\tilde{s}_{\pi(\beta)}, -\tilde{s}_0})
  \tilde{\lambda}_{\pi(\beta)}^{-1} \left( e^{\bar{z}\tilde{L}_1}
  (\frac{e^{+\pi i}}{\bar{z}^2})^{\tilde{L}_0} \tilde{m}_{\pi(\beta)}, \frac{1}{\bar{z}} \right) \cdot \tilde{m}_{\pi(\delta)} \rangle \; \;
  c(\tilde{\imath}_{\pi(\beta)*} (\tilde{s}_{\pi(\beta)}), i_\delta (s_\delta))^{-1},
  \nonumber 
\end{align}
where $s_0 := f[0;\beta,\delta](s_\beta,s_\delta)$ and $\tilde{s}_0 := \tilde{f}
(\tilde{s}_{\pi(\beta)},\tilde{s}_{\pi(\delta)})$, 
the crossing-symmetry invariance of the algebraic data 
$Y_{\rm tot}$ is used, along with $c(i_{\beta *}(s_\beta),-s_0)^{-1} = 1$
for $s_0 = 0 \in {\cal G}$. The crossing symmetry isomorphism
$\Theta_0$ maps ${\cal VV}_{\beta,0}^0(i_{\beta *},r=0)$ to
${\cal VV}_{\beta,\delta}^0(i_{\beta *},r=-1)$, so we had to include
the consequence of the canonical isomorphism of $\Delta r = +1$ back
to ${\cal VV}_{\beta,\delta}^0(i_{\beta *},r=0)$; the phase
$e^{-2\pi i\hat{h}_\beta(i_{\Upsilon^{\cal II}(\beta)*}(\Upsilon^{\cal I}(s_\beta)))}$
times $e^{+2\pi i \tilde{\hat{h}}_{\pi(\beta)}(\tilde{\imath}_{\pi \Upsilon(\beta)*}(\widetilde{\Upsilon}^{\cal I}(s_\beta)))}$ is always 1 on ${\cal H}_{\rm tot}^{\rm pre}$ or
${\cal H}_{\rm tot}^{\rm post}$, so we did not include this phase in the
expression above. 
The two factors on the RHS of (\ref{eq:temp-fCFT-2ptME-endWvac})
are rewritten as follows by using the properties of SVOA. 
\begin{align*}
& \bra{\bf 1} \lambda_0 (\Theta_{0} Y_{s_\beta, s_0}) \lambda_\beta^{-1}
  \left( e^{zL_1} (\frac{e^{-\pi i}}{z^2})^{L_0} m_\beta, \frac{1}{z} \right)
  \cdot m_\delta \rangle \\
&  = \langle Y_{s_\beta,0}(m_\beta,z) \cdot {\bf 1}, \; \lambda_\beta^{-1}(m_\delta) \rangle =
  \langle (\lambda_\beta^{-1})^T \cdot Y_{s_\beta,0}(m_\beta,z) \cdot {\bf 1}, \; m_\delta \rangle \\
&  = \epsilon_\delta^{-1} \bra{ \Upsilon_\delta^{-1} \lambda_\delta^{-1} \cdot
    Y_{s_\beta,0}(m_\beta,z) \cdot {\bf 1} } m_\delta \rangle
  =  \bra{ Y_{\Upsilon^{\cal I}(s_\beta),0}(\Upsilon_{s_\delta, s_\beta}(m_\beta),z) \cdot
    {\bf 1} }   m_\delta \rangle. 
\end{align*}
In evaluating the factor 
$c(\tilde{\imath}_{\pi(\beta)*}(\tilde{s}_{\pi(\beta)}), i_\gamma(s_\gamma))^{-1}$,
use the fact that those expressions are non-zero only when
$i_0 \cdot f(s_\beta,s_\gamma) = 0$,
$\tilde{\imath}_0 \cdot \tilde{f}(\tilde{s}_{\beta}, \tilde{s}_\gamma)=0$.
So, we can use a dictionary
\[
i_\delta(s_\delta) = -i_{\beta *}(s_\beta)
+ \Delta s_{i_0;i_{\beta *}, i_\delta}, \qquad
\Delta s_{i_\alpha; i_{\beta},i_\gamma} := i_\beta + i_\gamma - i_\alpha \cdot f[\alpha;\beta,\gamma] \in {\cal G}
\]
to see that 
\begin{align*}
  c(\tilde{\imath}_{\pi(\beta)*}(\tilde{s}_{\pi(\beta)}), \; i_\delta (s_\delta))^{-1}
  = c(\tilde{\imath}_{\pi(\beta)*}(\tilde{s}_{\pi(\beta)}), \;
  \Delta s_{i_0;i_{\beta *}, i_\delta})^{cc}
  \cdot c(i_{\beta *}(s_\beta), \; 
  \tilde{\imath}_{\pi(\beta)*}(\tilde{s}_{\pi(\beta)})   )^{cc}. 
\end{align*}

The first lesson from the evaluation above is that we can set the
CPT isomorphism $\Upsilon_{\rm tot}$ from $(M_\beta \otimes
\widetilde{M}_{\pi(\beta)})^{(i_{\beta *})} \subset {\cal H}_{\rm tot}^{\rm pre/post}$
to $(M_{\Upsilon^{\cal II}(\beta)} \otimes
\widetilde{M}_{\widetilde{\Upsilon}^{\cal II}\pi(\beta)})^{(i_{\Upsilon^{\cal II}(\beta) *})}
\subset {\cal H}_{\rm tot}^{\rm pre/post}$ by 
\begin{align}
  \Upsilon_{\rm tot}: & \; 
  (m_\beta \otimes \tilde{m}_{\pi(\beta)})^{(i_{\beta *})} \longmapsto \\
&  \qquad \quad \left(
  \Upsilon_{\Upsilon^{\cal II}(\beta),\beta}(m_\beta) \otimes
  \widetilde{\Upsilon}_{\widetilde{\Upsilon}^{\cal II}\pi(\beta),\pi(\beta)}
  (\tilde{m}_{\pi(\beta)}) \right)^{(i_{\Upsilon^{\cal II}(\beta)*})} 
  c(\tilde{\imath}_{\pi(\beta)*}(\tilde{s}_{\pi(\beta)}), \;
  \Delta s_{i_0;i_{\beta *},i_{\Upsilon^{\cal II}(\beta)*}}) .  \nonumber 
\end{align}
With this definition, and with $i_\delta = i_{\Upsilon^{\cal II}(\beta)*}$ in the
evaluation above, we have the relation
\begin{align}
  \bra{{\bf 1}\tilde{\bf 1}} & \; (Y_{\rm tot}^{(0;\beta,\Upsilon^{\cal II}(\beta))})^{(i_{\beta*},i_{\Upsilon^{\cal II}(\beta)*})} \left( (e^{zL_1} (\frac{e^{-\pi i}}{z^2})^{L_0} m_\beta \otimes
  e^{\bar{z}\tilde{L}_1} (\frac{e^{+\pi i}}{\bar{z}^2})^{\tilde{L}_0} \tilde{m}_{\pi(\beta)})^{(i_{\beta *})}, \frac{1}{z} \right) \cdot  \nonumber \\
  & \qquad \qquad \qquad \qquad \qquad \qquad 
  (m_{\Upsilon^{\cal II}(\beta)}\otimes
  \tilde{m}_{\widetilde{\Upsilon}^{\cal II}(\beta)})^{(i_{\Upsilon^{\cal II}(\beta)*})} \rangle
  \nonumber  \\
  & = \langle (Y_{\rm tot}^{(\Upsilon^{\cal II}(\beta);\Upsilon^{\cal II}(\beta),0)})^{(i_{\Upsilon^{\cal II}(\beta)*},i_0)} \left( \Upsilon_{\rm tot}((m_\beta \otimes \tilde{m}_{\pi(\beta)})^{(i_\beta)}),z \right) \cdot {\bf 1}\tilde{\bf 1} |
  \label{eq:cond-o-4CFT-onUps-S-intOp-temp} \\
  & \qquad \qquad \qquad \qquad \qquad \qquad
  (m_{\Upsilon^{\cal II}(\beta)}\otimes
  \tilde{m}_{\widetilde{\Upsilon}^{\cal II}\pi(\beta)})^{(i_{\Upsilon^{\cal II}(\beta)*})} \rangle
  \cdot c(i_\beta(s_\beta), 
  \tilde{\imath}_{\pi(\beta)}(\tilde{s}_{\pi(\beta)}   )^{cc}.
  \nonumber 
\end{align}
This is an analogue of (p) and (p-SVOA) in (\ref{eq:cond-o-4VOA-onUps})
and (\ref{eq:cond-o-4VOA-onUps-S}), and a generalization from 
(\ref{eq:cond-o-4CFT-onUps-S}).

The second lesson can be extracted by thinking of the case $i_\delta \neq
i_{\Upsilon^{\cal II}(\beta)}$; when
$(m_\delta \otimes \tilde{m}_{\pi(\delta)})^{(i_\delta)}$
in (\ref{eq:temp-fCFT-2ptME-endWvac}) comes out of a product of $n$
CFT vertex operators, we have to accept that $i_\delta$ is not necessarily
equal to $i_{\Upsilon^{\cal II}(\beta)*}$. We set the following isomorphism
from $(M_{\gamma}\otimes \widetilde{M}_{\pi(\gamma)})^{(i_{\gamma 1})}$ to 
$(M_{\gamma}\otimes \widetilde{M}_{\pi(\gamma)})^{(i_{\gamma 2})}$ with
$i_{\gamma 2} =:i_{\gamma 1} + \delta s_0$ as canonical:
\begin{align}
  P_{i_{\gamma 2}, i_{\gamma 1}}^{\rm can.}:
  (m_\gamma \otimes \tilde{m}_{\pi(\gamma)})^{(i_{\gamma 1})} \mapsto
  (m_\gamma \otimes \tilde{m}_{\pi(\gamma)})^{(i_{\gamma 2})} \;
  c(\tilde{\imath}_{\pi(\gamma)*}(\tilde{s}_{\pi(\gamma)})
  - \Delta s_{i_0;i_{\Upsilon^{\cal II}(\gamma)*}, i_{\gamma *}}, \; \delta s_0)^{-1};   
\end{align}
this choice is in the class of isomorphisms discussed below 
(\ref{eq:temp-equiv-VVtild-repr-iso}); the relevant set-thoretical
identification $\widetilde{S}_{\pi(\gamma)} \cong \widetilde{\cal G}$
is not so much as $\tilde{\imath}_{\pi(\gamma)1}$, 
$\tilde{\imath}_{\pi(\gamma)2}$, or even independent of the choice
of $\tilde{\imath}_{\pi(\gamma)*}$, but is determined by
the choice of $\tilde{i}_{\widetilde{\Upsilon}^{\cal II}(\gamma)*}$
through $\Delta s_{i_0;i_{\Upsilon^{\cal II}(\gamma)*}, i_{\gamma *}} \in {\cal G}$. 
The benefit of this choice as the canonical isomorphism is that
(for $\Upsilon^{\cal II}(\delta) = \beta$ and $i_\delta = i_{\beta_2}+i_{\delta_2}$)
\begin{align}
  \bra{{\bf 1}\tilde{\bf 1}} & \;
  (Y_{\rm tot}^{(0;\beta,\delta)})^{(i_{\beta*},i_{\delta})}
  \left( (e^{zL_1} (\frac{e^{-\pi i}}{z^2})^{L_0} m_\beta \otimes
  e^{\bar{z}\tilde{L}_1} (\frac{e^{+\pi i}}{\bar{z}^2})^{\tilde{L}_0} \tilde{m}_{\pi(\beta)})^{(i_{\beta *})}, \frac{1}{z} \right) \cdot  \nonumber \\
  & \qquad \qquad \qquad 
  (Y_{\rm tot}^{(\delta;\beta_2,\delta_2)})^{(i_{\beta_2},i_{\delta_2})}
   (m_{\beta_2}\otimes \tilde{m}_{\pi(\beta_2)})^{(i_{\beta_2 *})} ,z_2) \cdots \rangle
  \nonumber  \\
  & = \langle P_{i_\delta, i_{\Upsilon^{\cal II}(\beta) *}}^{\rm can.} \cdot
  (Y_{\rm tot}^{(\Upsilon^{\cal II}(\beta);\Upsilon^{\cal II}(\beta),0)})^{(i_{\Upsilon^{\cal II}(\beta)*},i_0)} \left( \Upsilon_{\rm tot}((m_\beta \otimes \tilde{m}_{\pi(\beta)})^{(i_\beta)}),z \right) \cdot {\bf 1}\tilde{\bf 1} | \\
  & \qquad \qquad \qquad 
  (Y_{\rm tot}^{(\delta;\beta_2,\delta_2)})^{(i_{\beta_2},i_{\delta_2})}
   (m_{\beta_2}\otimes \tilde{m}_{\pi(\beta_2)})^{(i_{\beta_2 *})} ,z_2) \cdots \rangle
  \cdot c(i_\beta(s_\beta), 
  \tilde{\imath}_{\pi(\beta)}(\tilde{s}_{\pi(\beta)} )  )^{cc}.
  \nonumber  
\end{align}

One can take one step further, to define matrix elements by using the
canonical isomorphisms as follows:
\begin{align}
  \bra{(m_\alpha \otimes \tilde{m}_{\pi(\alpha)})^{(i_{\alpha *})} }
  Y_{\rm tot}((m_{\beta_1}\otimes \tilde{m}_{\pi(\beta_1)})^{(i_{\beta_1*})},z_1)
  \cdots Y_{\rm tot}((m_{\beta_n} \otimes \tilde{m}_{\pi(\beta_n)})^{(i_{\beta_n *})} \cdot {\bf 1}\tilde{\bf 1} \rangle
\end{align}
is set (in the multi-peripheral algorithm) as
\begin{align}
  \bra{ P_{i_{\delta_0}, i_{\alpha *}}^{\rm can.} (m_\alpha \otimes \tilde{m}_{\pi(\alpha)})^{(i_{\alpha *})} }
  (Y_{\rm tot}^{(\alpha;\beta_1,\delta_1)})^{(i_{\beta_1 *},i_{\delta_1*})} \cdots
  (Y_{\rm tot}^{(\delta_{n-1};\beta_n,0)})^{(i_{\beta_n *},i_{0*})} \cdot {\bf 1}\tilde{\bf 1} \rangle
\end{align}
with $i_{\delta_0}: S_\alpha \cong {\cal G}$ determined from
$i_{\beta_1*} + \cdots + i_{\beta_n *}$. 

Finally, let us derive the invariance property of the Hermitian
inner product of a 2D non-holomorphic fermionic CFT.
According to the calculation below, we have 
\begin{align}
&  \langle P_{i_{\beta*}+i_{\gamma*},i_{\alpha *}}^{\rm can.} \cdot (m_\alpha \otimes \tilde{m}_{\pi(\alpha)})^{(i_{\alpha *})} | \label{eq:cond-o-4CFT-onUps-S-full-RHS}   \\
&  \qquad (Y_{\rm tot}^{(\alpha;\beta,\gamma)})^{(i_{\beta*},i_{\gamma*})} \left(
  (e^{zL_1} ( \frac{e^{-\pi i}}{z^2})^{L_0} m_\beta \otimes
  e^{\bar{z}\tilde{L}_1} ( \frac{e^{+\pi i}}{\bar{z}^2})^{\tilde{L}_0} \tilde{m}_{\pi(\beta)})^{(i_{\beta *})}, \frac{1}{z} \right) 
  \cdot (m_\gamma \otimes \tilde{m}_{\pi(\gamma)})^{(i_{\gamma *})} \rangle \nonumber \\
 & =   \langle P^{\rm can.}_{i_{\gamma*},i_{\Upsilon^{\cal II}(\beta)*}+i_{\alpha *}}
  Y_{\rm tot}^{(\gamma;\beta,\Upsilon^{\cal II}(\alpha))}\left(
    \Upsilon_{\rm tot}[(m_\beta \otimes \tilde{m}_{\pi(\beta)})^{(i_{\beta *})}], z
    \right) \cdot
    (m_\alpha \otimes \tilde{m}_{\pi(\alpha)})^{(i_{\alpha *})} \nonumber \\
  & \qquad \qquad 
     (m_\gamma \otimes \tilde{m}_{\pi(\gamma)})^{(i_{\gamma *})}  \rangle \;\; 
    c(i_{\beta *}(s_\beta), \; \tilde{\imath}_{\pi(\beta)*}(\tilde{s}_{\pi(\beta)}))^{cc} \nonumber \\
    & \qquad \qquad \quad 
    c(i_{\alpha *}(s_\alpha) + \tilde{\imath}_{\pi(\alpha)*}(\tilde{s}_{\pi(\alpha)}), \; \Delta s_{i_0;i_{\beta *}, i_{\Upsilon^{\cal II}(\beta)*}} )
    \nonumber \\
    & \qquad \qquad \quad c(i_{\gamma *}(s_\gamma)+\tilde{\imath}_{\pi(\gamma)*}(\tilde{s}_{\pi(\gamma)}) - 2\Delta s_{i_0;i_{\gamma*},i_{\Upsilon^{\cal II}(\gamma)*}}, \nonumber \\
    & \qquad \qquad\qquad \qquad \qquad \qquad \qquad
    \; \Delta s_{i_{\alpha*};i_{\beta*},i_{\gamma *}} + \Delta s_{i_0;i_{\alpha*},i_{\Upsilon^{\cal II}(\alpha)*}} - \Delta s_{i_0;i_{\gamma *},i_{\Upsilon^{\cal II}(\gamma)*}})
    \nonumber \\
    & \qquad \qquad \quad c(\Delta s_{i_{\alpha*};i_{\beta*},i_{\gamma *}}, \;
    \Delta s_{i_{\alpha*};i_{\beta*},i_{\gamma *}})^{-1} \;
  c(\Delta s_{i_0;i_{\beta *},i_{\Upsilon^{\cal II}(\beta)*}} + 2\Delta s_{i_0;i_{\alpha*},i_{\Upsilon^{\cal II}(\alpha)*}}, \Delta s_{i_0;i_{\gamma *},i_{\Upsilon^{\cal II}(\gamma)*}}) \nonumber \\
  & \qquad \qquad \quad
  c(\Delta s_{i_{\alpha *};i_{\beta *},i_{\gamma *}}, \; \Delta s_{i_0;i_{\alpha *},i_{\Upsilon^{\cal II}(\alpha)*}} + \Delta s_{i_0;i_{\beta *},i_{\Upsilon^{\cal II}(\beta)*}} -
    \Delta s_{i_0;i_{\gamma *},i_{\Upsilon^{\cal II}(\gamma)*}}).  \nonumber
\end{align}
The relation (\ref{eq:cond-o-4CFT-onUps-S-intOp-temp}) is reproduced by
setting $\alpha = 0$, $s_\alpha = \tilde{s}_{\pi(\alpha)} = 0$, because the
factors on the 3rd, 4th and 6th lines of the RHS are 1, and those on the
5th line are multiplied to be 1. The relation (\ref{eq:cond-o-4CFT-onUps-S})
for
modules is available by setting $\beta = 0$, when all the factors in the
3rd--6th lines of the RHS are 1. The phase factors in the last three lines
are 1 also in a theory with a systematic choice of set-theoretical
identification compatible with the group law/fusion.

The formula (\ref{eq:cond-o-4CFT-onUps-S-full-RHS}) is unexpectedly ugly;
the present author feels uncomfortable especially about the fact that factors
on the 3rd and 4th lines of the RHS depend on
$i_{\alpha*}(s_\alpha) + \tilde{\imath}_{\pi(\alpha)*}(\tilde{s}_{\pi(\alpha)})$
than $i_{\alpha*}(s_\alpha) - \tilde{\imath}_{\pi(\alpha)*}(\tilde{s}_{\pi(\alpha)})$, 
or on $i_{\gamma*}(s_\gamma) - \tilde{\imath}_{\pi(\gamma)*}(\tilde{s}_{\pi(\gamma)})$
in the 4th line, which would vanish on ${\cal H}_{\rm tot}^{\rm post}$. 
It is likely that there are mistakes in the calculation below, or
there is a better formulation.  We record the calculation leading to
(\ref{eq:cond-o-4CFT-onUps-S-full-RHS}) below for expert scrutiny. 

\proof The LHS of (\ref{eq:cond-o-4CFT-onUps-S-full-RHS}) is 
\begin{align*}
  \bra{ m_\alpha } {\cal Y}^{(\alpha;\beta,\gamma)}_{i; s_\beta,s_\gamma}
  \left( e^{zL_1}(\frac{e^{-\pi i}}{z^2})^{L_0} m_\beta , \frac{1}{z} \right)
  \cdot m_\gamma \rangle
  \bra{ \tilde{m}_{\pi(\alpha)} }
  \widetilde{\cal Y}^{(\alpha;\beta,\gamma)}_{\tilde{\imath}; \tilde{s}_{\pi(\beta)},\tilde{s}_{\pi(\gamma)}}
  \left( e^{\bar{z}\tilde{L}_1}(\frac{e^{+\pi i}}{\bar{z}^2})^{\tilde{L}_0} \tilde{m}_{\pi(\beta)} , \frac{1}{\bar{z}} \right)
  \cdot \tilde{m}_{\pi(\gamma)} \rangle \\
  d^{i\tilde{\imath}} \; c(\tilde{\imath}_{\pi(\beta) *}(\tilde{s}_{\pi(\beta)}), \;
  i_{\gamma *}(s_\gamma))^{-1}  \;
  c( (\tilde{\imath}_0 \tilde{f}_{[0;\Upsilon^{\cal II}(\pi(\alpha)),\pi(\alpha)]} - \tilde{\imath}_{\Upsilon^{\cal II}(\pi(\alpha))*})(\tilde{s}_{\pi(\alpha)}), 
 \;  \Delta s_{i_{\alpha *};i_{\beta *}, i_{\gamma *}})^{-1,cc} .  
\end{align*}
%
The invariance of $YY_{\rm tot}$ under the crossing symmetry isomorphism
allows us to replace
\begin{align*}
 d^{i,\tilde{\imath}} {\cal Y}_{i;s_\beta,s_\gamma} & \; \otimes 
  \widetilde{\cal Y}_{\tilde{\imath};\tilde{s}_{\pi(\beta)},\tilde{s}_{\pi(\gamma)}}
  \rightarrow \\
  d^{j,\tilde{\jmath}} & \; \left(
  \lambda_{1,i_{\Upsilon^{\cal II}(\alpha)*};\Upsilon^{\cal I}(s_\alpha)}
  (\Theta_0{\cal Y}_{j;s_\beta,\Upsilon^{\cal I}(s_\alpha)})
  \lambda^{-1}_{1,i_{\Upsilon^{\cal II}(\gamma)*};\Upsilon^{\cal I}(s_\gamma)}\right) \\
&   \otimes 
  \left(\tilde{\lambda}_{-1,\tilde{\imath}_{\widetilde{\Upsilon}(\pi(\alpha))*}; \widetilde{\Upsilon}^{\cal I}(\tilde{s}_{\pi(\alpha)})}
  (\widetilde{\Theta}_{-1} \widetilde{\cal Y}_{\tilde{\jmath};\tilde{s}_{\pi(\beta)}, \widetilde{\Upsilon}^{\cal I}(\tilde{s}_{\pi(\alpha)})})
  \tilde{\lambda}^{-1}_{-1, \tilde{\imath}_{\widetilde{\Upsilon}^{\cal II}(\gamma)*};
    \widetilde{\Upsilon}^{\cal I}(\pi(\gamma))} \right) \; \; \\
  &  c \left( i_{\beta *}(s_\beta), \;\;
  i_{\Upsilon^{\cal II}(\alpha)*}(\Upsilon^{\cal I}(s_\alpha))   \right)^{-1} \;\;
  c\left(\tilde{\imath}_{\pi(\beta) *}(\tilde{s}_{\pi(\beta)}), \;\;
  \tilde{\imath}_{\widetilde{\Upsilon}^{\cal II}(\pi(\alpha))*}
  (\widetilde{\Upsilon}^{\cal I}(\tilde{s}_{\pi(\alpha)})) \right)^{-1} \\
  & 
  c \left( i_{\Upsilon^{\cal II}(\gamma) *}(\Upsilon^{\cal I}(s_\gamma)), \Delta s_{i_{\Upsilon^{\cal II}(\gamma)*}; i_{\beta *}, i_{\Upsilon^{\cal II}(\alpha)*}} \right) ^{-1} \\
  & c \left( \tilde{\imath}_{\widetilde{\Upsilon}^{\cal II}\pi(\gamma)*}(\widetilde{\Upsilon}^{\cal I}(\tilde{s}_{\pi(\gamma)})),
  \Delta s_{i_{\Upsilon^{\cal II}(\gamma)*}; i_{\beta *}, i_{\Upsilon^{\cal II}(\alpha)*}} \right) ^{-1};
\end{align*}
the last two factors are due to the prescription
(\ref{eq:temp-choice-on-s4xingSymTrnsf}).
The factors attributed to the left-moving SVOA are therefore
\begin{align*}
  \bra{m_\alpha} & \; \lambda_{\Upsilon^{\cal I}(s_\alpha)} \cdot
  (\Theta_0{\cal Y}_{j;s_\beta,\Upsilon^{\cal I}(s_\alpha)}) \cdot
  \lambda_{\Upsilon^{\cal I}(s_\gamma)}^{-1} \left(
  e^{zL_1} (\frac{e^{-\pi i}}{z^2})^{L_0} m_\beta, \frac{1}{z} \right)
  \cdot m_\gamma \rangle_{M_{s_\alpha}}  \; \\
  & \qquad \qquad \qquad
  c(i_{\beta *}(s_\beta), \; i_{\Upsilon^{\cal II}(\alpha)*}(\Upsilon^{\cal I}(s_\alpha))
  )^{-1}  c \left( i_{\Upsilon^{\cal II}(\gamma) *}(\Upsilon^{\cal I}(s_\gamma)), \Delta s\right)^{-1}\\
  & \;  = \epsilon_{s_\alpha} \langle
  \Upsilon_{\Upsilon^{\cal I}(s_\alpha),s_\alpha}(m_\alpha), \;
  (\Theta_0 {\cal Y}_{j;s_\beta,\Upsilon^{\cal I}(s_\alpha)}) \left(
  e^{zL_1} (\frac{e^{-\pi i}}{z^2})^{L_0} m_\beta, \frac{1}{z} \right) \cdot
  \lambda_{\Upsilon^{\cal I}(s_\gamma)}^{-1}(m_\gamma) \rangle \\
 & \qquad \qquad \qquad 
  \; c(i_{\beta *}(s_\beta), \;
  i_{\Upsilon^{\cal II}(\alpha)*}(\Upsilon^{\cal I}(s_\alpha)))^{-1}
  c \left( i_{\Upsilon^{\cal II}(\gamma) *}(\Upsilon^{\cal I}(s_\gamma)), \Delta s\right)^{-1}
  \\
  & \;  = \epsilon_{s_\alpha} \langle {\cal Y}_{j;s_\beta,\Upsilon^{\cal I}(s_\alpha)}
  (m_\beta,z) \cdot \Upsilon_{\Upsilon^{\cal I}(s_\alpha),s_\alpha}(m_\alpha), \;
    \lambda_{\Upsilon^{\cal I}(s_\gamma)}^{-1}(m_\gamma) \rangle \\
    &  \qquad \qquad \qquad
    c(i_{\beta *}(s_\beta), \;
    i_{\Upsilon^{\cal II}(\alpha)*}(\Upsilon^{\cal I}(s_\alpha))  )^{-1} c \left( i_{\Upsilon^{\cal II}(\gamma) *}(\Upsilon^{\cal I}(s_\gamma)), \Delta s\right)^{-1} \\
  & \; = \epsilon_{s_\alpha} \langle \Upsilon_{s_\gamma,\Upsilon^{\cal I}(s_\gamma)} \cdot
  {\cal Y}_{j;s_\beta,\Upsilon^{\cal I}(s_\alpha)}
  (m_\beta,z) \cdot \Upsilon_{\Upsilon^{\cal I}(s_\alpha),s_\alpha}(m_\alpha) |
  m_\gamma \rangle_{M_{s_\gamma}}  \; \\
  & \qquad \qquad \qquad c(i_{\beta *}(s_\beta), \;
  i_{\Upsilon^{\cal II}(\alpha)*}(\Upsilon^{\cal I}(s_\alpha))   )^{-1}
  c \left( i_{\Upsilon^{\cal II}(\gamma) *}(\Upsilon^{\cal I}(s_\gamma)), \Delta s\right)^{-1}
  \\
 & \; = \langle {\cal Y}_{j;s_\beta,\Upsilon^{\cal I}(s_\alpha)}
  (\Upsilon_{\Upsilon^{\cal I}(s_\beta),s_\beta}(m_\beta),z) \cdot m_\alpha | m_\gamma \rangle_{M_{s_\gamma}}
  c \left( i_{\Upsilon^{\cal II}(\gamma) *}(\Upsilon^{\cal I}(s_\gamma)), \Delta s \right)^{-1} 
  . 
\end{align*}

The Hermitian inner products of the left-mover and right-mover SVOAs
need to be translated to that of ${\cal H}_{\rm tot}^{\rm pre/post}$ as follows: 
\begin{align*}
&  \left[
    \left({\cal Y}_{j;\Upsilon(s_\beta),s_\alpha}(\Upsilon(m_\beta),z) \cdot m_\alpha
    \right)   \otimes
    \left( \widetilde{\cal Y}_{\tilde{\jmath};
      \tilde{s}_{\Upsilon^{\cal I}\pi(\beta)},\tilde{s}_{\pi(\alpha)}}
    (\widetilde{\Upsilon}(\tilde{m}_{\pi(\beta)}),\bar{z}) \cdot
    \tilde{m}_{\pi(\alpha)} \right)    \right]^{(i_{\gamma *})} \\
  & =
   c(\tilde{\imath}_{\pi(\gamma)*}\tilde{f}(\widetilde{\Upsilon}^{\cal I}(\tilde{s}_{\pi(\beta)}),\tilde{s}_{\pi(\alpha)}) - \Delta s_{i_0;i_{\Upsilon^{\cal II}(\gamma)*},i_{\gamma *}}, \; \Delta s_{i_{\gamma *};i_{\Upsilon^{\cal II}(\beta) *},i_{\alpha *}})^{-1} \;\; 
   P_{i_{\gamma *}, i_{\Upsilon^{\cal II}(\beta)*}+i_{\alpha *}}^{\rm can.} \cdot \\
  & \qquad 
   \left[
    \left({\cal Y}_{j;\Upsilon(s_\beta),s_\alpha}(\Upsilon(m_\beta),z) \cdot m_\alpha
    \right)   \otimes
    \left( \widetilde{\cal Y}_{\tilde{\jmath};
      \tilde{s}_{\Upsilon^{\cal I}\pi(\beta)},\tilde{s}_{\pi(\alpha)}}
    (\widetilde{\Upsilon}(\tilde{m}_{\pi(\beta)}),\bar{z}) \cdot
    \tilde{m}_{\pi(\alpha)} \right)    \right]^{(i_{\Upsilon^{\cal II}(\beta)*} + i_{\alpha *})},
\end{align*}
and 
\begin{align*}
 &   \left[
    \left({\cal Y}_{j;\Upsilon(s_\beta),s_\alpha}(\Upsilon(m_\beta),z) \cdot m_\alpha
    \right)   \otimes
    \left( \widetilde{\cal Y}_{\tilde{\jmath};
      \tilde{s}_{\Upsilon^{\cal I}\pi(\beta)},\tilde{s}_{\pi(\alpha)}}
    (\widetilde{\Upsilon}(\tilde{m}_{\pi(\beta)}),\bar{z}) \cdot
    \tilde{m}_{\pi(\alpha)} \right)    \right]^{(i_{\Upsilon^{\cal II}(\beta)*} + i_{\alpha *})}
    \\
 &\quad  = ({\cal Y}_j\otimes \widetilde{\cal Y}_{\tilde{\jmath}})^{(i_{\Upsilon^{\cal II}(\beta)*}, i_{\alpha *})} \left( \Upsilon_{\rm tot}(
  (m_\beta \otimes \tilde{m}_{\pi(\beta)})^{(i_{\beta *})}  ) , z \right) \cdot
  (m_\alpha \otimes \tilde{m}_{\pi(\alpha)})^{(i_{\alpha *})} \\
  & \qquad \qquad \qquad 
    c(\tilde{\imath}_{\widetilde{\Upsilon}^{\cal II}(\pi(\beta)) *}
  (\Upsilon^{\cal I}(\tilde{s}_{\pi(\beta)}) ), \; i_{\alpha *}(s_\alpha) ) \;
  c(\tilde{\imath}_{\pi(\beta)*}(\tilde{s}_{\pi(\beta)}),
  \Delta s_{i_0;i_{\beta *},i_{\Upsilon^{\cal II}(\beta) *}})^{-1} .
\end{align*}
So, the LHS of (\ref{eq:cond-o-4CFT-onUps-S-full-RHS}) is equal to 
\begin{align*}
&  \langle P^{\rm can.}_{i_{\gamma*},i_{\Upsilon^{\cal II}(\beta)*}+i_{\alpha *}}
  Y_{\rm tot}^{(\gamma;\beta,\Upsilon^{\cal II}(\alpha))}\left(
    \Upsilon_{\rm tot}[(m_\beta \otimes \tilde{m}_{\pi(\beta)})^{(i_{\beta *})}], z
    \right) \cdot
    (m_\alpha \otimes \tilde{m}_{\pi(\alpha)})^{(i_{\alpha *})} \\
    & \qquad \qquad \qquad \qquad    | 
    (m_\gamma \otimes \tilde{m}_{\pi(\gamma)})^{(i_{\gamma *})} \rangle  \\
& c(\tilde{\imath}_{\pi(\beta) *}(\tilde{s}_{\pi(\beta)}), \;
  i_{\gamma *}(s_\gamma))^{-1}  \;
  c( (\tilde{\imath}_0 \tilde{f}_{[0;\Upsilon^{\cal II}(\pi(\alpha)),\pi(\alpha)]} - \tilde{\imath}_{\Upsilon^{\cal II}(\pi(\alpha))*})(\tilde{s}_{\pi(\alpha)}), 
  \;  \Delta s_{i_{\alpha *};i_{\beta *}, i_{\gamma *}})^{-1,cc},  \\
  &   c(\tilde{\imath}_{\pi(\gamma)*}\tilde{f}(\widetilde{\Upsilon}^{\cal I}(\tilde{s}_{\pi(\beta)}),\tilde{s}_{\pi(\alpha)}) - \Delta s_{i_0;i_{\Upsilon^{\cal II}(\gamma)*},i_{\gamma *}}, \; \Delta s_{i_{\gamma *};i_{\Upsilon^{\cal II}(\beta) *},i_{\alpha *}})^{-1,cc} \;\; \\
 &     c(\tilde{\imath}_{\widetilde{\Upsilon}^{\cal II}(\pi(\beta)) *}
  (\Upsilon^{\cal I}(\tilde{s}_{\pi(\beta)}) ), \; i_{\alpha *}(s_\alpha) )^{cc} \;
  c(\tilde{\imath}_{\pi(\beta)*}(\tilde{s}_{\pi(\beta)}),
  \Delta s_{i_0;i_{\beta *},i_{\Upsilon^{\cal II}(\beta) *}})^{-1,cc} \\
  &
  c \left( i_{\Upsilon^{\cal II}(\gamma) *}(\Upsilon^{\cal I}(s_\gamma))
  + \tilde{\imath}_{\widetilde{\Upsilon}^{\cal II}\pi(\gamma)*}(\widetilde{\Upsilon}^{\cal I}(\tilde{s}_{\pi(\gamma)}))
  , \; \Delta s_{i_{\Upsilon^{\cal II}(\gamma)*}; i_{\beta *}, i_{\Upsilon^{\cal II}(\alpha)*}} \right) ^{-1}. 
\end{align*}

Among the six phase factors in the expression right above, pick up
the 1st and 4th one, first. The factors that depend
both on $\tilde{s}_{\pi(\beta)}$ and $s_\alpha$ cancel: 
\begin{align*}
& c(\tilde{\imath}_{\pi(\beta) *}(\tilde{s}_{\pi(\beta)}), \;
 i_{\gamma *}(s_\gamma))^{-1} = c( \tilde{\imath}_{\pi(\beta)*}(\tilde{s}_{\pi(\beta)}), \; i_{\alpha *}(s_\alpha) - i_{\beta *}(s_\beta) + \Delta s_{i_{\alpha *};i_{\beta *},i_{\gamma *}} )^{-1}, \\
&  c(\tilde{\imath}_{\widetilde{\Upsilon}^{\cal II}(\pi(\beta)) *}
  (\Upsilon^{\cal I}(\tilde{s}_{\pi(\beta)}) ), \; i_{\alpha *}(s_\alpha) )^{-1}
  = c(-\tilde{\imath}_{\pi(\beta)*}(\tilde{s}_{\pi(\beta)}) + \Delta s_{i_0;i_{\Upsilon^{\cal II}(\beta)*},i_{\beta *}},   \; i_{\alpha *}(s_\alpha))^{-1}, 
\end{align*}
while the factor
$c(i_{\beta *}(s_\beta), \; \tilde{\imath}_{\pi(\beta)*}(\tilde{s}_{\pi(\beta)}))^{cc}$
remains to be the phase factor on the 2nd line of the RHS
of (\ref{eq:cond-o-4CFT-onUps-S-full-RHS}). 

The rest of the proof is to show that the remaining phase factors 
\begin{align*}
&   c( \tilde{\imath}_{\pi(\beta)*}(\tilde{s}_{\pi(\beta)}), \; \Delta s_{i_{\alpha *};i_{\beta *},i_{\gamma *}} )^{-1} \;\; 
  c( (\tilde{\imath}_0 \tilde{f}_{[0;\Upsilon^{\cal II}(\pi(\alpha)),\pi(\alpha)]} - \tilde{\imath}_{\Upsilon^{\cal II}(\pi(\alpha))*})(\tilde{s}_{\pi(\alpha)}), 
    \;  \Delta s_{i_{\alpha *};i_{\beta *}, i_{\gamma *}}) \\
&    c(\tilde{\imath}_{\pi(\gamma)*}\tilde{f}(\widetilde{\Upsilon}^{\cal I}(\tilde{s}_{\pi(\beta)}),\tilde{s}_{\pi(\alpha)}) - \Delta s_{i_0;i_{\Upsilon^{\cal II}(\gamma)*},i_{\gamma *}}, \; \Delta s_{i_{\gamma *};i_{\Upsilon^{\cal II}(\beta) *},i_{\alpha *}}) \;\; \\
&   c(\Delta s_{i_0;i_{\Upsilon^{\cal II}(\beta)*},i_{\beta *}},   \; i_{\alpha *}(s_\alpha))^{-1}\;\;   c(\tilde{\imath}_{\pi(\beta)*}(\tilde{s}_{\pi(\beta)}),
    \Delta s_{i_0;i_{\beta *},i_{\Upsilon^{\cal II}(\beta) *}}) \\
&   c \left( i_{\Upsilon^{\cal II}(\gamma) *}(\Upsilon^{\cal I}(s_\gamma))
  + \tilde{\imath}_{\widetilde{\Upsilon}^{\cal II}\pi(\gamma)*}(\widetilde{\Upsilon}^{\cal I}(\tilde{s}_{\pi(\gamma)}))
  , \; \Delta s_{i_{\Upsilon^{\cal II}(\gamma)*}; i_{\beta *}, i_{\Upsilon^{\cal II}(\alpha)*}} \right) ^{-1}    
\end{align*}
agree with those in the 3rd--6th lines of the RHS
of (\ref{eq:cond-o-4CFT-onUps-S-full-RHS}).
The last (6th) phase factor is the one in the 4th line of the RHS of
(\ref{eq:cond-o-4CFT-onUps-S-full-RHS}), as we can use
\begin{align*}
 i_{\Upsilon^{\cal II}(\gamma)*}(\Upsilon^{\cal I}(s_\gamma)) = - i_{\gamma*}(s_\gamma) + \Delta s_{i_0;i_{\gamma *},i_{\Upsilon^{\cal II}(\gamma)*}}, 
\end{align*}
and
\begin{align}
  \Delta s_{i_{\gamma *}; i_{\Upsilon^{\cal II}(\beta)*},i_{\alpha *}} & \; = \Delta s_{i_0;i_{\Upsilon^{\cal II}(\beta)*}, i_{\beta *}} - \Delta s_{i_{\alpha *};i_{\beta *},i_{\gamma *}},
  \label{eq:temp-DeltaS-crossing-reltn} \\
  \Delta s_{i_{\Upsilon^{\cal II}(\gamma)*}; i_{\beta *}, i_{\Upsilon^{\cal II}(\alpha)*}} & \; =
  \Delta s_{i_{\alpha*};i_{\beta*},i_{\gamma*}} + \Delta s_{i_0;i_{\alpha*}, i_{\Upsilon^{\cal II}(\alpha)*}} - \Delta s_{i_0; i_{\gamma *}, i_{\Upsilon^{\cal II}(\gamma)*}}.  \nonumber
\end{align}
To rewrite the 2nd and 3rd phase factors, use the following relations: 
\begin{align*}
&  \left( \tilde{i}_0\tilde{f}_{[0;\widetilde{\Upsilon}^{\cal II}(\pi(\alpha)),\pi(\alpha)]}-\tilde{\imath}_{\widetilde{\Upsilon}^{\cal II}(\alpha)*}\right)(\tilde{s}_{\pi(\alpha)})
  = \tilde{\imath}_{\pi(\alpha)*}(\tilde{s}_{\pi(\alpha)})
  - \Delta s_{i_0;i_{\Upsilon^{\cal II}(\alpha)*},i_{\alpha*}},
\end{align*}
\begin{align*}
  &  \tilde{\imath}_{\pi(\gamma)*}\tilde{f}(\widetilde{\Upsilon}^{\cal I}(\tilde{s}_{\pi(\beta)}), \tilde{s}_{\pi(\alpha)}) - \Delta s_{i_0;i_{\Upsilon^{\cal II}(\gamma)*},i_{\gamma *}}
  \\
&  = -\Delta s_{i_{\gamma *};i_{\Upsilon^{\cal II}(\beta)*},i_{\alpha *}}
  + \tilde{\imath}_{\widetilde{\Upsilon}^{\cal II}(\pi(\beta))*}(\widetilde{\Upsilon}^{\cal I}(\tilde{s}_{\pi(\beta)})) + \tilde{\imath}_{\pi(\alpha)*}(\tilde{s}_{\pi(\alpha)})
  - \Delta s_{i_0;i_{\Upsilon^{\cal II}(\gamma)*},i_{\gamma *}} \\
  & = \tilde{\imath}_{\pi(\alpha)*}(\tilde{s}_{\pi(\alpha)})
  - \tilde{\imath}_{\pi(\beta)*}(\tilde{s}_{\pi(\beta)})
  -\Delta s_{i_{\gamma *};i_{\Upsilon^{\cal II}(\beta)*},i_{\alpha *}}
  - \Delta s_{i_0;i_{\Upsilon^{\cal II}(\gamma)*},i_{\gamma *}}
  +\Delta s_{i_0;i_{\Upsilon^{\cal II}(\beta)*},i_{\beta *}}, 
\end{align*}
So, the $\tilde{s}_{\pi(\beta)}$-dependent ones among the remaining
phase factors are 
\begin{align*}
  c(\tilde{\imath}_{\pi(\beta)*}(\tilde{s}_{\pi(\beta)}),
  -\Delta s_{i_{\alpha *}; i_{\beta *},i_{\gamma *}}
  -\Delta s_{i_{\gamma *}; i_{\Upsilon^{\cal II}(\beta)*},i_{\alpha *}}
  +\Delta s_{i_0;i_{\beta *},i_{\Upsilon^{\cal II}(\beta)*}}   ) = 1.
\end{align*}
The $s_\alpha$- or $\tilde{s}_{\pi(\alpha)}$-dependent ones are
\begin{align*}
  c(\tilde{\imath}_{\pi(\alpha)*}(\tilde{s}_{\pi(\alpha)*}),
  \Delta s_{i_{\alpha *}; i_{\beta *},i_{\gamma *}}
  + \Delta s_{i_{\gamma *};i_{\Upsilon^{\cal II}(\beta)*},i_{\alpha *}}
  )
  c(i_{\alpha *}(s_\alpha), \Delta s_{i_0;i_{\Upsilon^{\cal II}(\beta)*},i_{\beta *}} ) \\
  = c(i_{\alpha *}(s_\alpha) + \tilde{\imath}_{\pi(\alpha)*}(\tilde{s}_{\pi(\alpha)}),
  \Delta s_{i_0;i_{\Upsilon^{\cal II}(\beta)*},i_{\beta *}} ),
\end{align*}
which is the 3rd line of the RHS of (\ref{eq:cond-o-4CFT-onUps-S-full-RHS}).

The remaining phase factors independent of $s_{\alpha}$, $\tilde{s}_{\pi(\alpha)}$,
$\tilde{s}_{\pi(\beta)}$ or $s_\gamma$ are 
\begin{align*}
  & c(\Delta s_{i_0;i_{\Upsilon^{\cal II}(\alpha)*},i_{\alpha *}},
  \Delta s_{i_{\alpha *};i_{\beta *},i_{\gamma *}} )^{-1} 
  c(\Delta s_{i_{\alpha *};i_{\beta *}, i_{\gamma*}} - \Delta s_{i_0;i_{\gamma *},i_{\Upsilon^{\cal II}(\gamma)*}}, \;
  \Delta s_{i_{\gamma *}; i_{\Upsilon^{\cal II}(\beta)*}, i_{\alpha *}})
\end{align*}
which can be reorganized into the factors in the 5th--6th lines of the
RHS of (\ref{eq:cond-o-4CFT-onUps-S-full-RHS}) by
using (\ref{eq:temp-DeltaS-crossing-reltn}). \qed 
\end{anythng}

\begin{rmk}
The presentation in section \ref{ssec:fCFT} can be read as that of
a general pair of simple current extensions associated with a finite
abelian group ${\cal G}$ with the four properties: (i) vanishing 3rd
cohomology on ${\cal G}$, (ii) homomorphic assignment of the conformal
weights on $s_0 \in {\cal G}$ (equivalently, skew-symmetry of the bi-multiplicative form $c$ on ${\cal G} \times {\cal G}$), (iii) fixed point free action
of ${\cal G}$ on the ${\cal G}$-set $S_\gamma$ for a simple representation,
and (iv) the relation (\ref{eq:reltn-eps-c-h}). Extra work is necessary
to go beyond those limitations; to include non-trivial $h$ to deal with
the case with non-trivial 3rd cohomology under the fusion (i), to distinguish
$c_-$ and $c_+$ (ii), and to deal with the representation theory
of the stabilizer (isotropy) subgroup of ${\cal G}$ in its action
on $S_\gamma$ (iii), obviously.

It is not obvious to the present author how to go beyond the limit of (iv). 
That might be through re-examining the places where we have made an artificial
(but not strictly necessary) choice for definiteness, and seeking
different routes. Candidates of such places are (p-SVOA, q-SVOA),
(\ref{eq:xing-SVOA-Y-vs-Y-vs-lambda-m1}), 
(\ref{eq:temp-choice-on-s4xingSymTrnsf}), and treatment of the right-moving
sectors relatively to that of the left-moving sector. 
Not enough attention has been paid to discrete torsion in this presentation,
either. 
\end{rmk}

\subsection*{Acknowledgments}   

The present author thanks Scott Carnahan, Kentaro Hori, Kanade Nishikawa, 
Masaki Okada for useful comments and discussions. 
This work is supported in part by WPI Initiative and 
 a Grant-in-Aid for Scientific Research (proj. number 26K07082), MEXT, Japan.


%
%


\bibliographystyle{alpha}
\bibliography{rationalCFT.bib}
\end{document}